\documentclass[traditabstract]{aa}

\usepackage{graphicx}
\usepackage{txfonts}

\begin{document}

\title{The AGN fraction of submm-selected galaxies and contributions to the submm/mm-wave extragalactic background light}

\author{Stephen Serjeant \inst{1}
\and  Mattia Negrello \inst{1}
\and  Chris Pearson \inst{1,2,17}
\and  Angela Mortier \inst{3}
\and  Jason Austermann \inst{4}
\and Itziar Aretxaga \inst{5}
\and David Clements \inst{3}
\and Scott Chapman \inst{18}
\and Simon Dye \inst{6}
\and James Dunlop \inst{7}
\and Loretta Dunne  \inst{8}
\and Duncan Farrah  \inst{9}
\and David Hughes \inst{5}
\and Hyung-Mok Lee \inst{12}
\and Hideo Matsuhara \inst{13}
\and Edo Ibar \inst{10}
\and Myungshin Im \inst{12} 
\and Woong-Seob Jeong \inst{16}
\and Sungeun Kim  \inst{11}
\and Shinki Oyabu \inst{13}
\and Toshinobu Takagi  \inst{13}
\and Takehiko Wada \inst{13}
\and Grant Wilson  \inst{14}
\and Mattia Vaccari \inst{15}
\and Min Yun \inst{14}
}
 \institute{Department of Physics \& Astronomy, The Open University, Milton Keynes MK7 6BJ 
 \and  Rutherford Appleton Laboratory, Chilton, Didcot, Oxfordshire OX11 0QX 
 \and  Astrophysics Group, Dept. of Physics, Imperial College London, Prince Consort Road, London, SW7 2BW 
 \and  Center for Astrophysics and Space Astronomy, University of Colorado, Boulder, CO80309, USA 
 \and  Instituto Nacional de Astrof\'{i}sica, Optica y Electr\'{o}nica (INAOE), Tonantzintla, Puebla, M\'{e}xico 
 \and  Cardiff University, School of Physics \& Astronomy, Queens Buildings, The Parade, Cardiff, CF24 3AA, UK 
 \and Scottish Universities Physics Alliance (SUPA), 
      Institute for Astronomy, The University of Edinburgh, Royal Observatory, Blackford Hill, Edinburgh EH9 3HJ, UK 
 \and School of Physics \& Astronomy, University of Nottingham, University Park, Nottingham, NG7 2RD 
 \and Astronomy Centre, University of Sussex, Falmer, Brighton, BN1 9QH 
 \and UK Astronomy Technology Centre, Royal Observatory, Edinburgh, EH9 3HJ 
 \and Astronomy \& Space Science Department, Sejong University, 143-747, Seoul, South Korea 
 \and Department of Physics and Astronomy, Seoul National University, Shillim-Dong, Kwanak-Gu, Seoul 151-742, South Korea 
 \and  Institute of Space and Astronautical Science, Japan Aerospace Exploration Agency, Sagamihara, Kanagawa, 229 8510 
 \and Department of Astronomy, University of Massachusetts, Amherst, MA01003, USA 
 \and Department of Astronomy, University of Padova, vic. Osservatorio 2, 35122, Padova - I, Italy 
 \and Space Science Divison, KASI, 61-1, Whaam-dong, Yuseong-gu, Deajeon, 305-348, South Korea 
 \and Department of Physics, University of Lethbridge, 4401 University Drive, Lethbridge, Alberta T1J 1B1, Canada 
 \and Institute of Astronomy, University of Cambridge, Madingley Road, Cambridge, CB3 0HA, 
  }

\titlerunning{AGN fraction of submm-selected galaxies \& EBL} 
\authorrunning{S. Serjeant et al.}

\abstract {We present a comparison of the SCUBA Half Degree Extragalactic Survey 
(SHADES) at $450\,\mu$m, $850\,\mu$m and $1100\,\mu$m 
with deep guaranteed time $15\,\mu$m AKARI FU-HYU survey data and Spitzer guaranteed time data at $3.6-24\,\mu$m in the Lockman
Hole East. 
The AKARI data was analysed using 
bespoke 
software 
based in part on the drizzling and minimum-variance matched filtering developed for SHADES,
and was cross-calibrated against Infrared Space Observatory (ISO) fluxes. Our stacking
analyses find AKARI $15\,\mu$m 
galaxies with $\stackrel{>}{_\sim}200\,\mu$Jy contribute $>10\%$ of the $450\,\mu$m
background, but only $<4\%$ of the $1100\,\mu$m
background, suggesting that different populations contribute at
mm-wavelengths. We confirm our earlier result that the ultra-deep $450\,\mu$m SCUBA-2 
Cosmology Survey will be dominated by populations already detected by AKARI and Spitzer mid-infrared surveys.
The superb mid-infrared wavelength coverage afforded by combining Spitzer and AKARI photometry
is an excellent diagnostic of AGN contributions, and 
we find that $(23-52)\%$ of submm-selected galaxies 
have AGN bolometric fractions $f_{\rm AGN}>0.3$.
}


\maketitle

\section{Introduction}
The pioneering submm-wave observations of the Hubble Deep Field North (Hughes et al. 1998, Barger et al. 1998) and galaxy cluster 
gravitational lenses
(Smail et al. 1997) revealed the presence of completely unanticipated populations of submm-luminous galaxies with
fluxes of several mJy at $850\,\mu$m. 
This discovery was one of 
the first in a series that led to the development and ultimately wide acceptance 
of ``downsizing'' phenomenological models of galaxy evolution, that massive galaxies 
formed the bulk of their stars at high redshifts in large starburst events, with the opposite behaviour found in 
lower-mass systems. This is contrary to naive expectations from 
hierarchical models, but nevertheless it was found these observations could be made 
consistent with semi-analytic hierarchical models by adjusting the physical parameters used to characterised feedback
processes (e.g. Granato et al. 2006) and/or with adjustments to the initial mass function (e.g. Baugh et al. 2005, 
Lacey et al. 2008). 

Indeed the existence of strong relationships
between supermassive black hole masses in nearby galaxies and the properties of the host bulges or spheroids
(e.g. the Magorrian relation, Magorrian et al. 1998; see also e.g. Ferrarese \& Merritt 2000) 
all but require a strong link between high-redshift 
black hole accretion and the generation of the stellar mass of the host. This is reinforced by the observations that 
many or perhaps all quasars show evidence
of star formation (e.g. Lehnert et al. 1992, 
Hughes et al. 1997, 
Aretxaga et al. 1998, Brotherton et al. 1999, 
Lutz et al. 2008, 
Mullaney et al. 2009, Shi et al. 2009, 
Veilleux et al. 2009, 
Serjeant \& Hatziminaoglou 2009). While submm and mm-wave direct detections are only made in
a minority of high-redshift quasars (e.g. Carilli et al. 2001, Omont et al. 2001, Isaak et al. 2002, Bertoldi \& Cox 2002, Omont et al. 2003, 
Priddey et al. 2003a,b, Willlot et al. 2003, Bertoldi et al. 2003), stacking the far-infrared to mm-wave non-detections 
yielded significant detections and demonstrated clearly that all quasars are on average ultraluminous starbursts (Serjeant \& Hatziminaoglou 2009). 
If submm-selected galaxies are the violent 
starbursts that mark the sites of the progenitors of present-day giant ellipticals, then 
the presence and nature of AGN in submm-selected galaxies may track the feedback processes regulating the stellar
mass assembly. 

Are the bolometric luminosities of submm galaxies dominated by the 
release of gravitational binding energy around a central supermassive black hole, or are they 
dominated by the release of nuclear binding energy in starbursts? Similar questions were asked of local ultraluminous
infrared galaxies in the 1990s (see e.g. Genzel \& Cezarsky 2000); 
now the issues of feedback and the Magorrian relation add an additional perspective and motivation to
the higher-redshift equivalent questions for submm-selected galaxies. 
Deep $2-10\,$keV imaging of submm-selected galaxies has found that around three-quarters show evidence of an active
nucleus (Alexander et al. 2005), the majority of which are heavily obscured. Nevertheless, the X-ray to far-infrared 
luminosity ratios suggest that star formation rather than active nuclei typically dominate the bolometric output,
assuming the underlying spectral energy distributions of the active nuclei resemble those of unobscured quasars. 
In order to eliminate the possibility that Compton-thick active nuclei in
submm-selected galaxies have different underlying spectral energy distributions than the comparator populations, 
one should sample the spectra around the expected peak of the dust torus output, in the mid-infrared. 

The advent of the Spitzer space telescope led to the discovery of new dust-shrouded galaxy populations 
with starbursts and/or active nuclei
(e.g. Mart\'{i}nez-Sansigre et al. 2005, Dey et al. 2008). 
The mid-IR where the dust torus luminous output peaks is the ideal place to 
test for AGN; we expect star-forming galaxies to have strong PAH and silicate features, while 
AGN dust tori have much more featureless spectra, with the possible sole exception of a $10\,\mu$m rest-frame 
silicate absorption trough. 
We found in Negrello et al. 2009 that with sufficiently comprehensive broad-band photometry in the mid-IR, starburst galaxies
can have photometric redshift determinations from the redshifting of PAH features. 
Furthermore, if the bolometric
AGN fraction is more than $30\%$, the mid-IR SEDs are sufficiently close to featureless power-laws in broad-band
photometry that photometric redshifts are impossible with mid-IR data alone. However in these cases, the power-law
SEDs are a useful signature of bolometrically-significant AGN regardless of redshift. 
In this case, the longer wavelength range obtained by combining AKARI and Spitzer offers significant advantages over
the insights available (albeit useful ones) from broad-band Spitzer photometry alone (e.g. Yun et al. 2008). 
A recent mid-infrared spectroscopic study of $13$ submm-selected galaxies (Pope et al. 2008) found only $2/13$ 
with such power-law mid-infrared spectral energy distributions; similarly, Men\'{e}ndez-Delmestre et al. (2009) found
$4/23$ submm-selected galaxies with continuum-dominated mid-infrared spectra. Clearly there is a need for larger samples
of submm-selected galaxies with constrained AGN bolometric fractions. 

Submm galaxies are also significant contributors to the submm-wave extragalactic background light (EBL). The EBL can
be expressed as an integral over the comoving volume-averaged luminosity density (with an additional $(1+z)$ factor, 
e.g. Peacock 1999), so
at any redshift $z$ the galaxies that dominate the far-IR to mm-wave EBL are also the ones that dominate the star formation 
rate at that $z$. But resolved submm galaxies only contribute a few tens of percent at most (Hughes et al. 1998) to the EBL
at those wavelengths. 
To constrain the populations that contribute the rest one needs to
use stacking analyses. Several populations of galaxies have been shown to contribute significant minorities of the
submm EBL, such as EROs and HEROs, BzK galaxies, (e.g. Coppin et al. 2004, Takagi et al. 2007, Greve et al. 2009), and 
Lyman-break galaxies (e.g. Peacock et al. 2000). 
In Serjeant et al. 2008 we found that the
$24\,\mu$m-selected galaxies contribute the bulk of the $450\,\mu$m background, 
but that at $850\,\mu$m there are unknown new populations that contribute the bulk of that background. 

In this paper we will constrain the AGN bolometric fractions of a sample of submm-selected galaxies in 
the Lockman Hole East, taken from the SCUBA Half Degree Extragalactic Survey (SHADES), using mid-infrared
photometry from the AKARI and Spitzer space telescopes. We also constrain the contributions the mid-infrared
populations make to the submm and mm-wave extragalactic background light using stacking analyses. 
SHADES papers I and II (Mortier et al. 2005 and Coppin
et al. 2006) present the SHADES survey design, data analysis and source counts. 
Paper III (Ivison et al. 2007) gave radio and mid-infrared counterparts of SHADES galaxies. 
SHADES paper IV (Aretxaga et al. 2007) presents 
photometric redshift constraints of submm galaxies derived from 
far-infrared and radio data alone. SHADES paper V (Takagi
et al. 2007) constrains the submm properties of near-infrared-selected galaxies. 
Paper VI (Coppin et al. 2007) gave $350\,\mu$m photometry for SHADES galaxies, while papers VII and VIII
(Dye et al. 2008 and Clements et al. 2008 respectively) analysed the multiwavelength spectral energy distributions
based on data available at the time. 
SHADES paper IX (Serjeant et al. 2008) made submm
stacking analyses of ISO and Spitzer-selected galaxies, and in this paper we will extend
this analysis to include the AKARI catalogues described in the section \ref{sec:data_acquisition}. 
Austermann et al. (2009) extended the SHADES survey to mm-wavelengths and substantially increased the 
areal coverage. The comparison of the $850\,\mu$m and $1100\,\mu$m data, including a critical comparison
of the $1100\,\mu$m data from different instruments and a systematic search for populations of submm-wave
drop-outs, will be given in Negrello et al. (2009, in preparation). In this paper we will use the 
AzTEC $1100\,\mu$m data, rather than the $1100\,\mu$m data from Coppin et al. 2008 which has not been 
corrected for Eddington bias (Eddington 1913). 
We will also use the multi-wavelength data 
from SHADES papers III, VI, VII and VIII (Ivison et
al. 2007, Coppin et
al. 2007, Dye et al. 2008 and Clements et al. 2008 respectively) 
who presented 
spectral energy distribution (SED) fits of
SHADES galaxies. 

\section{Data acquisition}\label{sec:data_acquisition}

\subsection{Submm and mm-wave observations}
The SCUBA Half Degree Extragalactic Survey (SHADES) was the largest project on the James Clerk Maxwell Telescope
from 2003-2005. During the instrument lifetime of SCUBA the survey mapped a total of approximately a quarter of a square
degree at $850\,\mu$m to a typical $3.5\,\sigma$ depth of 8\,mJy, in the Lockman Hole East and the Subaru-XMM Deep
Field (Mortier et al. 2005). The originally-planned areal coverage and more was
completed by the SHADES team with the AzTEC instrument at $1100\,\mu$m (Austermann et al. 2009) to typical $3.5\sigma$ depths 
in the Lockman Hole East of $3.1-4.6$\,mJy. Both SHADES fields benefit from enormous multi-wavelength wide-field
campaigns, and the Lockman Hole East in particular has
some of the deepest Spitzer Space Telescope mid-infrared mapping of any contiguous field 
over hundreds of square arcminutes. More details on the SHADES project, including the survey goals, methodology and
submm data analysis, can be found in SHADES papers I and II (Mortier et al. 2005, Coppin et al. 2006). Details on the
SHADES mm-wave data analysis are given in Austermann et al. (2009). We use the Austermann et al. (2009) mm-wave measurements
in preference to Coppin et al. (2008), since the former but not the latter have been corrected for Eddington bias.

\subsection{Mid-infrared observations}
The Lockman Hole East was targetted by the Spitzer space telescope in guaranteed time. The field was mapped
with the IRAC and MIPS instruments (Fazio et al. 2004, Rieke et al. 2004) and sources 
extracted to depths of $4.47\mu$Jy at $3.6\,\mu$m 
($3\sigma$),
$4.54\mu$Jy at $4.5\,\mu$m ($3\sigma$), $20.9\mu$Jy at $5.8\,\mu$m ($3\sigma$), $12.5\mu$Jy at $8\,\mu$m 
($3.2\sigma$) and $38\mu$Jy at $24\,\mu$m ($4\sigma$). For more details of this catalogue, see Serjeant et al. 2008 
and references therein. 
The Lockman Hole East was also mapped at $15\,\mu$m
with the CAM instrument on the Infrared Space Observatory (ISO). More details of the ISO survey can be found 
in Elbaz et al. 1999 and Rodighiero et al. 2004. 

The Lockman Hole East was also the subject of a $10'\times 30'$ survey in the L15 $15\,\mu$m filter by 
Infrared Camera (IRC) of 
AKARI, as part of the {\it FU-HYU} mission programme (Pearson et al. 2009), to a typical $1\sigma$ noise
level of $20\,\mu$Jy. This programme targetted well-studied
Spitzer fields to capitalize on the comprehensive multi-wavelength data sets available in these fields. Some 
{\it FU-HYU} observations in the GOODS-N field have been presented in Negrello et al. (2009). 

A pipeline analysis of the IRC is available in the Image Reduction and Analysis Facility (IRAF). Despite being an excellent
general-purpose analysis, we found that the processing was not optimal for our purposes of minimum-variance point source
extraction. The standard pipeline locates and
flags bad pixels, then interpolates the image to correct for field distortions. However this interpolation stage creates
Moir\'{e} effects, correlates the noise between pixels and in some cases spreads bad individual pixel readouts over several
pixels in the interpolated image. 
Furthermore, some of the AKARI observations had been mistakenly taken without the requested dithering 
intended to remove bad pixels, so our only option was to use the spacecraft jitter, which necessarily requires a bespoke analysis.

We had previously developed mapping and minimum-variance point source extraction tools in the Interactive Data Language
(IDL) for the SHADES survey, so we opted to use a similar approach here. 
We stopped the IRC pipeline after the flat fielding, imported the data into IDL, then 
determined an approximate pointing solution using catalogues of ISO and Spitzer sources in the field. 
The relative jitter was found by measuring centroids of
bright sources in each data frame, and hence we could determine a World
Coordinate System solution for each frame. Bad pixels were identified and masked using 
the same pixel histogram fitting procedure employed in the SCUBA Hubble 
Deep Field North and SHADES observations (Serjeant et al. 2003a,
Mortier et al. 2005). Several frames were affected by Earthshine, so we 
performed a polynomial fit to each row, then each column, then
identified and corrected a small number of pixels which reported
consistently low counts regardless of target. This latter effect may
indicate the need to improve the flat field determinations, but the offsets induced by these corrections 
are much smaller than our photometric errors. Following the best practice in SCUBA observations
we determined the noise in each detector pixel 
using Gaussian fits to readout histograms. We adopted the field distortion corrections
as coded in the IRAF IRC pipeline, but from a comparison of our initial AKARI catalogues against
Spitzer observations we found evidence for further field distortions. We fit to these
using a second-order polynomial and incorporated the correction into our total field distortion solution. 

To mosaic the individual frames onto a master image we used the 
``zerofootprint'' drizzling technique developed 
originally for SCUBA (Serjeant et al. 2003a): detector pixels are mapped onto much finer sky 
pixels, accounting for field distortion and jitter, with the flux from a detector pixel deposited
in a single sky pixel. 
Where a sky pixel has multiple observations, fluxes are combined using noise-weighted coadds. 
The final coadded map represents a detector pixel's view of the sky, in the sense that at every point 
it reports the noise-weighted mean flux of all detector pixels centred exactly on that point. Notwithstanding
cross-talk between detector pixels (which we neglect), the final coadded map also has statistically
independent pixels, i.e. without covariances. 

We performed a further deglitching stage on the coadded map. The minimum-variance estimator for
the point source flux $F$ at any point in the map can be expressed a noise-weighted point spread function convolution,
as used in submm surveys (Serjeant et al. 2003a, Mortier et al. 2005):
\begin{equation}\label{eqn:bff}
F = \frac{(IW)\otimes P}{W \otimes P^2}
\end{equation}
where $I$ is the coadded image, $W$ is the weight image (the reciprocal of the noise squared), and $P$ is the point
spread function. Propagating errors on equation \ref{eqn:bff} yields an expression for $\Delta F$, the error on $F$:
\begin{equation}\label{eqn:dbff}
\left (\Delta F\right )^2 = \frac{1}{W \otimes P^2}~.
\end{equation}
For the point spread function we assumed a Gaussian with a full width half maximum of $5.47''$.  

\begin{figure}[!ht]
\vspace*{-5cm}
\begin{center}
   \resizebox{1.3\hsize}{!}{
     \includegraphics*[0,0][510,542]{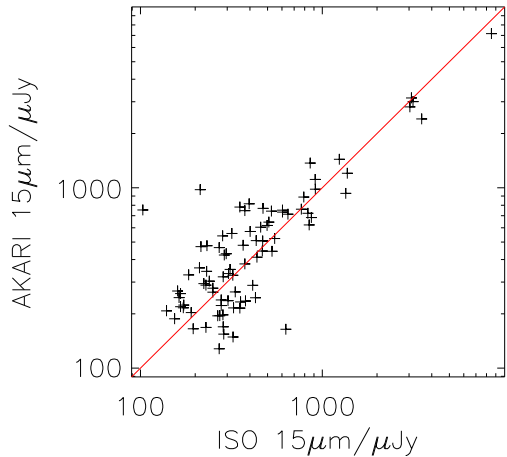}
   }
\end{center}
\caption{
Comparison of $15\,\mu$m fluxes obtained in the Lockman Hole East with the Infrared
Space Observatory (Elbaz et al. 1999, Rodighiero et al. 2004) with those obtained with our AKARI survey (Pearson
et al. 2009). Note
that the mean flux calibration of AKARI has been determined by the cross-matching with the ISO survey
(implicitly using the ISO flux calibration) and
is not determined independently in this study. 
}\label{fig:iso_akari_flux_comparison}
\end{figure}

\begin{figure}[!ht]
\begin{center}
   \resizebox{1.0\hsize}{!}{
     \includegraphics*[0,42][510,542]{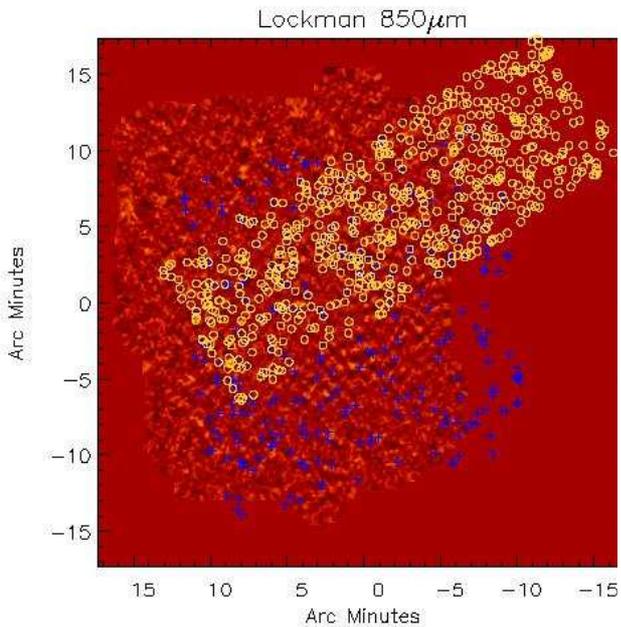}
   }
\end{center}
\caption{
The SHADES SCUBA $850\,\mu$m signal-to-noise image of the Lockman Hole East. The 
AKARI $15\,\mu$m sources are overplotted in yellow (rectangular coverage) and the ISO
$15\,\mu$m sources are overplotted in blue (approximately square coverage). 
}\label{fig:850_field}
\end{figure}

\begin{figure}[!ht]
\begin{center}
   \resizebox{1.0\hsize}{!}{
     \includegraphics*[0,43][510,542]{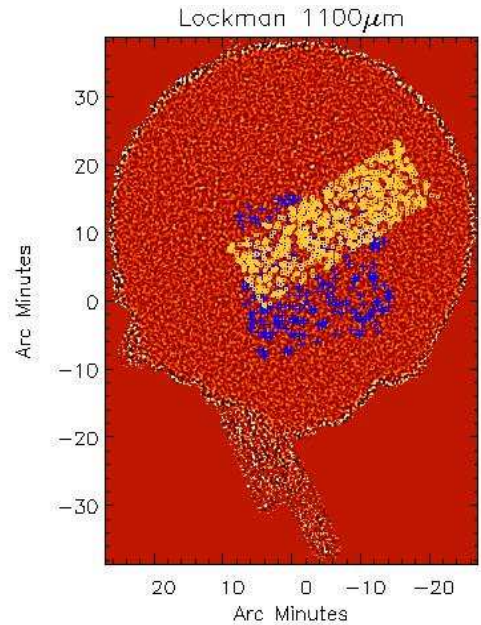}
   }
\end{center}
\caption{
The SHADES AzTEC $1100\,\mu$m image of the Lockman Hole East. The 
AKARI $15\,\mu$m sources are overplotted in yellow (rectangular coverage) and the ISO
$15\,\mu$m sources are overplotted in blue (approximately square coverage). 
}\label{fig:1100_field}
\end{figure}

Our point source detection map is the $F/\Delta F$ image, in which we found objects using a connected pixels algorithm. 
Objects were identified as discrete regions with map values greater than a given threshold. We used a series of thresholds of
$3-10\sigma$ in steps of $1\sigma$, because objects which are blended at a low threshold may be distinct objects at a
brighter threshold (e.g. Mortier et al. 2005), yielding a catalogue of 622 distinct objects. 
The IRC pipeline flux calibration is given only for photometry taken with
the IRC pipeline outputs, which is necessarily systematically different to our pipeline outputs,
so we cross-matched our catalogue with the previous ISO $15\,\mu$m sources 
from Rodighiero et al. (2004) and 
based our calibration on the cross-matchings so the average flux ratio is unity, 
shown in Figs. \ref{fig:iso_akari_flux_comparison}. The final AKARI $15\,\mu$m sources
are shown in Figs. \ref{fig:850_field} and \ref{fig:1100_field}.  

\section{Methods}\label{sec:methods}
For the stacking analyses of submm galaxies we adopted the same methodologies
of Serjeant et al. 2008. 
In brief, we first subtract the catalogued objects from the $850\,\mu$m and $1100\,\mu$m maps (there are
no secure $450\,\mu$m point source detections). 
We used Gaussian point spread functions for the
submm data and the point spread function in Austermann et al. (2009) for the AzTEC data. We refer to these maps as the residual maps. 
The comparison of the histograms of mm-wave fluxes in the map and at the positions of $15\,\mu$m-selected galaxies yielded a possibly
strong signal ($\sim 8\sigma$) but which appeared to be dominated by a non-Gaussian shape to the latter histogram (Fig.
\ref{fig:histograms}). This could be
suggestive of weak but coherent features in parts of the map, but which would necessarily have to have absolute values 
less than $1\sigma$ per beam, making 
investigation difficult. Alternatively, this could be a consequence of having a relatively small number of AzTEC beams in the $15\,\mu$m 
survey coverage. Without passing judgement on these issues, 
we have opted to take a very conservative approach and adopted an additional beamswitching 
sky subtraction, by subtracting from the AzTEC residual map the same map offset by $100$ pixels East, or $300''$. 
We also derived a noise map for this beamswitched residual map by measuring the standard deviation in a $45''$ radius circular
region at every point in the map. 

There is some disagreement
in the literature over the best practice in stacking methodology
Some (such as Peacock et al. 2000 and Serjeant et al. 2004, 2008)
have opted to subtract known point sources, while others (such as Wang et al. 2007, Scott et al. 2008, Marsden et al. 2009) 
opt to use the unsubtracted map. In escense, it depends on what one wants to measure. 
If one requires the total intensity from all galaxies, one should not subtract any flux (or the
subtracted population should be added afterwards). If however one requires the mean flux from 
a particular population, as distinct from all galaxies as a whole, then one needs to clean the map to
avoid any biases in the mean. 

In Serjeant et al. (2008) we found that the 
submm-selected galaxies had very different mid-IR/submm flux ratios to the galaxies
selected at $24\,\mu$m. This could be due to bimodality, i.e. two populations, or we could
be sampling two ends of a continuum; it is difficult to distinguish these possibilities from stacks alone. 
The possibility that a few $15\,\mu$m-selected galaxies are (relative to the rest)
pathological would suggest the use of point-source-subtracted maps, which is the approach we have adopted. 

We take unweighted mean averages of the submm fluxes in the residual maps 
(or for the AzTEC observations, mm-wave fluxes) at the positions of mid-infrared-selected
galaxies. By the Central Limit Theorem the unweighted mean flux is Gaussian distributed with a dispersion
of $\sigma/\sqrt{N}$ where $\sigma$ is the standard deviation of the sample of fluxes and
$N$ is the number of fluxes being averaged. We also use Kolmogorov-Smirnoff tests to compare the 
histogram of fluxes at the positions of mid-infrared-selected galaxies with the histogram of the
submm/mm-wave fluxes in the submm/mm-wave map as a whole, excluding regions with no mid-infrared data. This latter
test inherently incorporates a control comparison. We include the $15\,\mu$m survey from ISO (Elbaz et al. 1999, 
Rodighiero et al. 2004); 
although this survey is not as deep as our AKARI mapping, it increases the areal coverage of the stacking
analysis (Figs. \ref{fig:850_field} and \ref{fig:1100_field}) and therefore also the number of submm 
and mm-wave beams contributing, which improve the stacked signal (Serjeant et al. 2008). 
Unweighted means give cosmetically poor results for stacked postage stamps due to edge effects, so we
used noise-weighted means for the purposes of making stacked postage stamps only. 

For the characterisation of the spectral energy
distributions of submm-selected galaxies, we assume a minimum photometric error of ten percent. 
We use an ensemble of starburst model outputs from Takagi et al. (2003, 2004) and an ensemble of dust torus model outputs from
Efstathiou \& Rowan-Robinson (1995), 
following the methodology of Negrello et al. 2009. We make linear superimpositions of these model 
spectral energy distributions neglecting the small effects of energy transfer between the dust torus
and the starburst dust components. The key physical parameters in these models are: the torus opening
angle $\theta$ (which we set to $45^\circ$), the ratio of outer and inner torus radii (which we set to $20$), 
and viewing angle $\theta_{\rm view}$; starburst compactness parameter $\Theta$, age $T$
and dust composition (Milky Way, SMC or LMC). A star formation timescale of $100\,$Myr was adopted for the Takagi et al. 
models. A wider parameter space of AGN models is not justified by the contraints available from our data. 
We leave the redshift as a further free parameter even where spectroscopic redshifts have been published, to 
avoid dependence on the often-uncertain discussion of identifications. The total number of fitted parameters
(including normalisation, but not counting dust composition) is three, and two more with the addition of AGN, plus
a further fitting parameter for redshift.

\section{Results}
\subsection{Stacking analysis results}
\begin{figure*}[!ht]
\begin{center}
   \resizebox{0.7\hsize}{!}{
     \includegraphics*{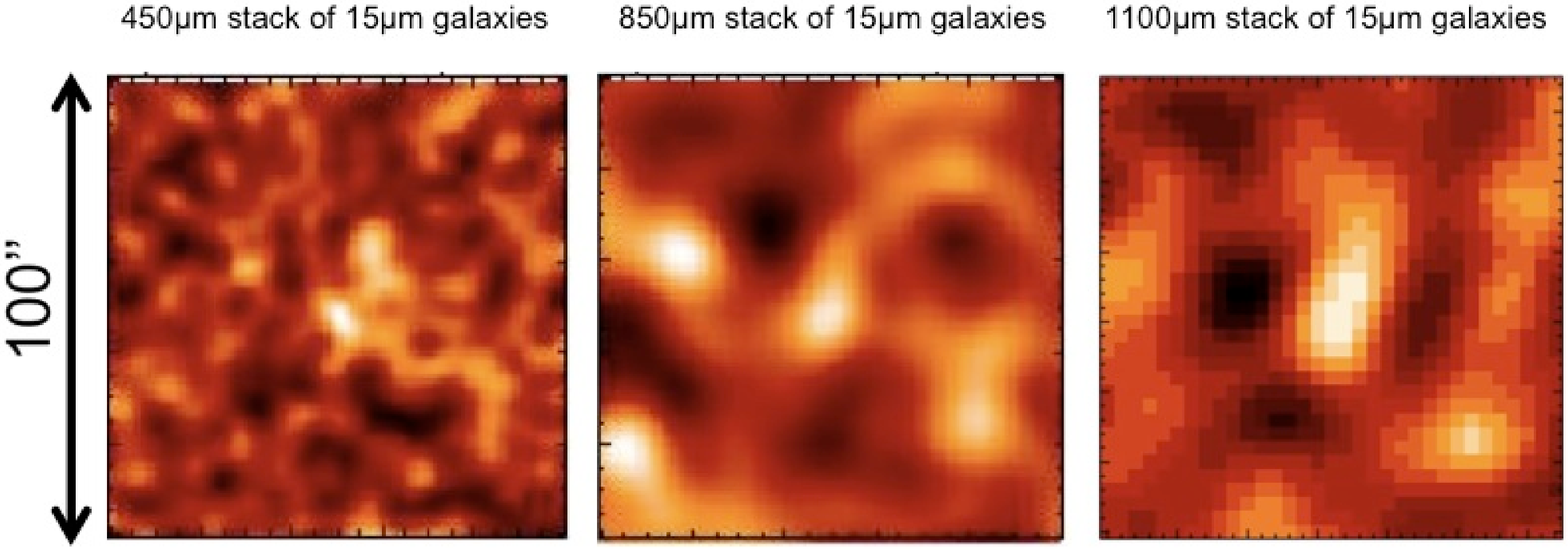}
   }
\end{center}
\caption{
Stacked mm-wave and submm signal-to-noise postage stamps of $15\,\mu$m sources from 
AKARI and ISO, at (from left to right) $450\,\mu$m, $850\,\mu$m and
$1100\,\mu$m. The images are scaled from $-2.81\sigma$ to $3.73\sigma$ ($450\,\mu$m), 
$-2.73\sigma$ to $2.83\sigma$ ($850\,\mu$m) and $-3.01\sigma$ to $2.99\sigma$ ($1100\,\mu$m).
}\label{fig:postage_stacks}
\end{figure*}

Fig. \ref{fig:postage_stacks} shows noise-weighted stacked $450\,\mu$m, $850\,\mu$m and $1100\,\mu$m postage 
stamps of the AKARI and ISO $15\,\mu$m-selected galaxies. This is a significant improvement over the detections in 
Serjeant et al. (2008), and for the first time extends the stacked signal of mid-infrared-selected galaxies 
to mm-wavelengths. The unweighted mean submm-wave and mm-wave fluxes of the $15\,\mu$m-selected galaxies are 
$\langle S_{{450\,\mu}\rm m}\rangle = (3.7\pm 1.1)\,$mJy, 
$\langle S_{{850\,\mu}\rm m}\rangle = (0.206\pm 0.084)\,$mJy and 
$\langle S_{{1100\,\mu}\rm m}\rangle = (0.148\pm 0.052)\,$mJy. 
The $450:1100\,\mu$m flux ratio is consistent with a grey-body index of $\beta=1.5$ on the Rayleigh-Jeans tail, 
though the $850\,\mu$m 
stacked flux is marginally discrepant. In the source count model of Pearson (2005) the predicted median redshift 
of our sample is $z\sim 1$, with of the order $10\%$ at $z>2$. 

\begin{figure*}[!ht]
\begin{center}
\hspace*{-2.8cm}\resizebox{0.5\hsize}{!}{\includegraphics*{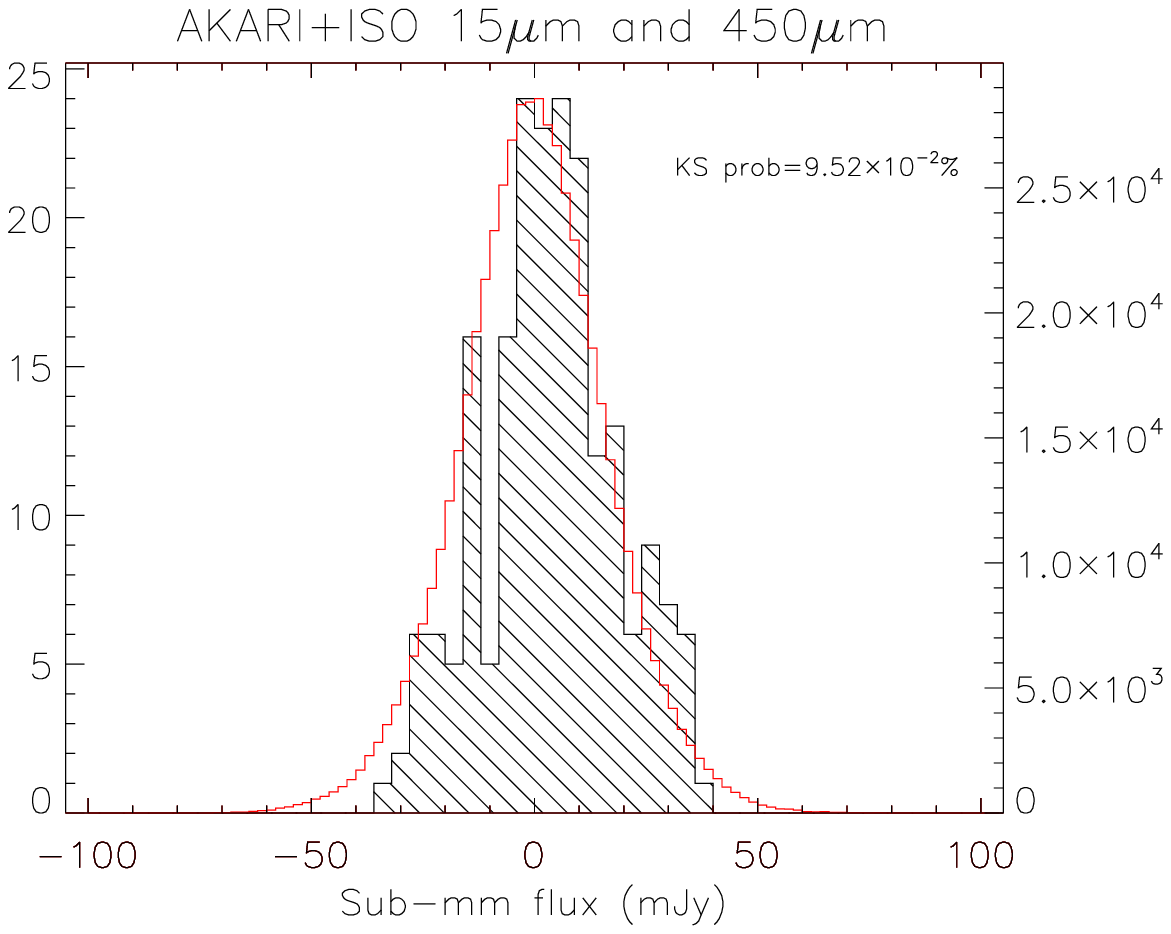}}\nolinebreak
\hspace*{-2.8cm}\resizebox{0.5\hsize}{!}{\includegraphics*{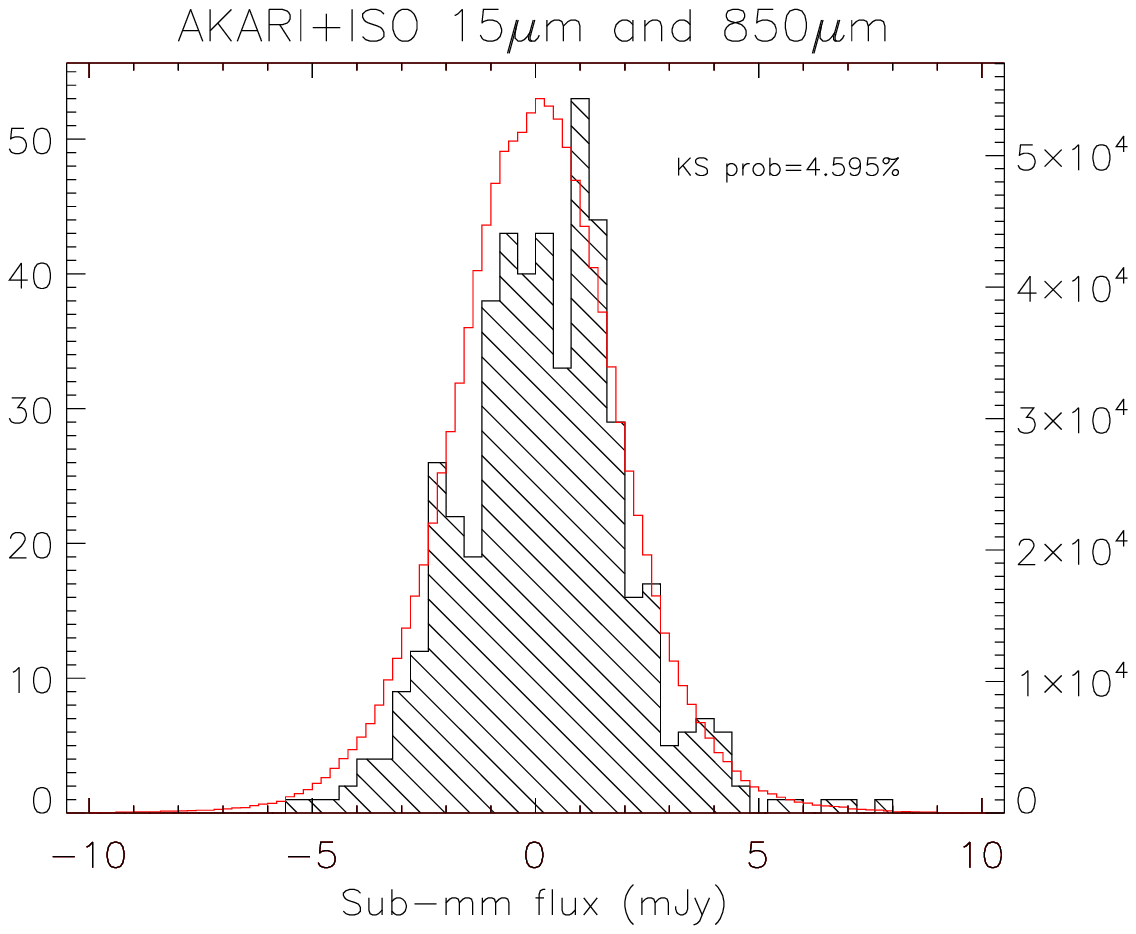}}
\hspace*{-2.8cm}\resizebox{0.5\hsize}{!}{\includegraphics*{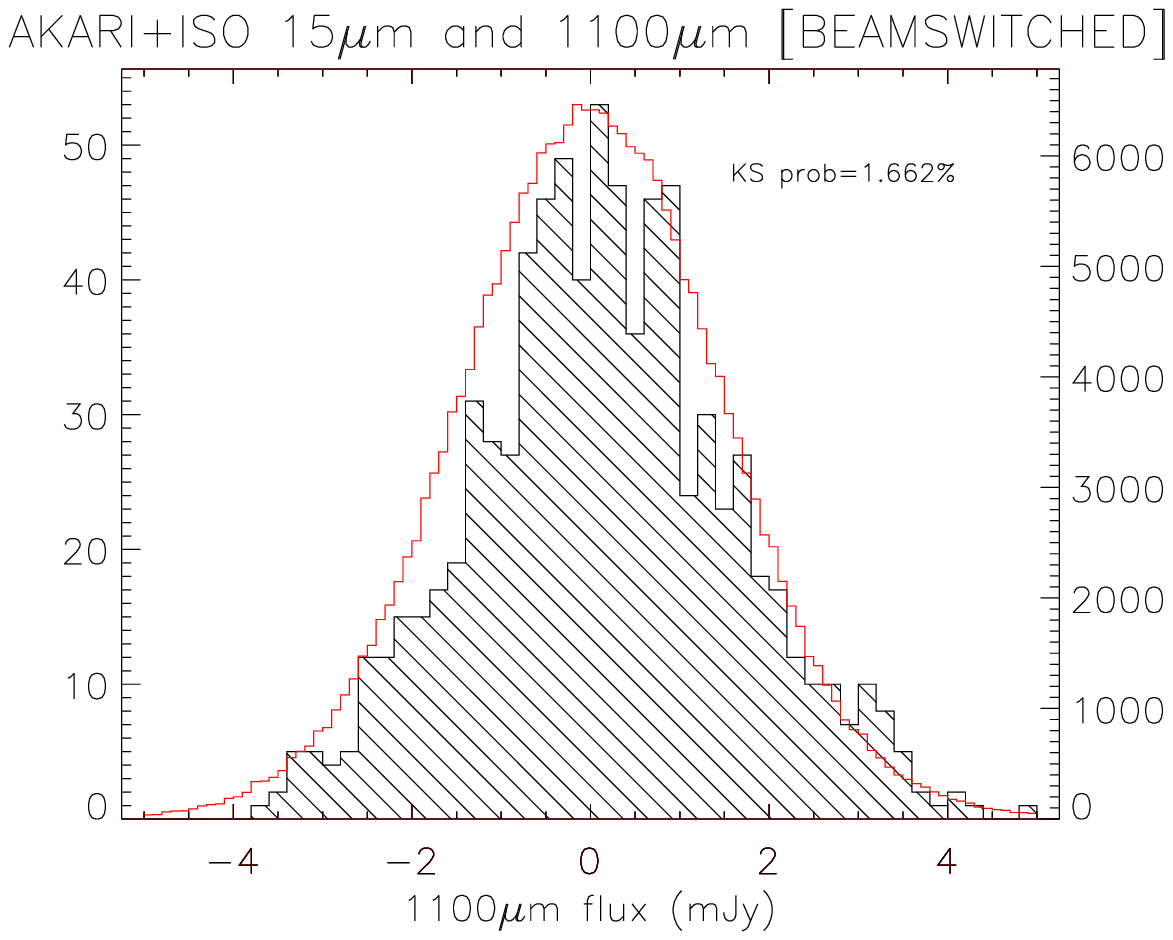}}\nolinebreak
\hspace*{-2.8cm}\resizebox{0.5\hsize}{!}{\includegraphics*{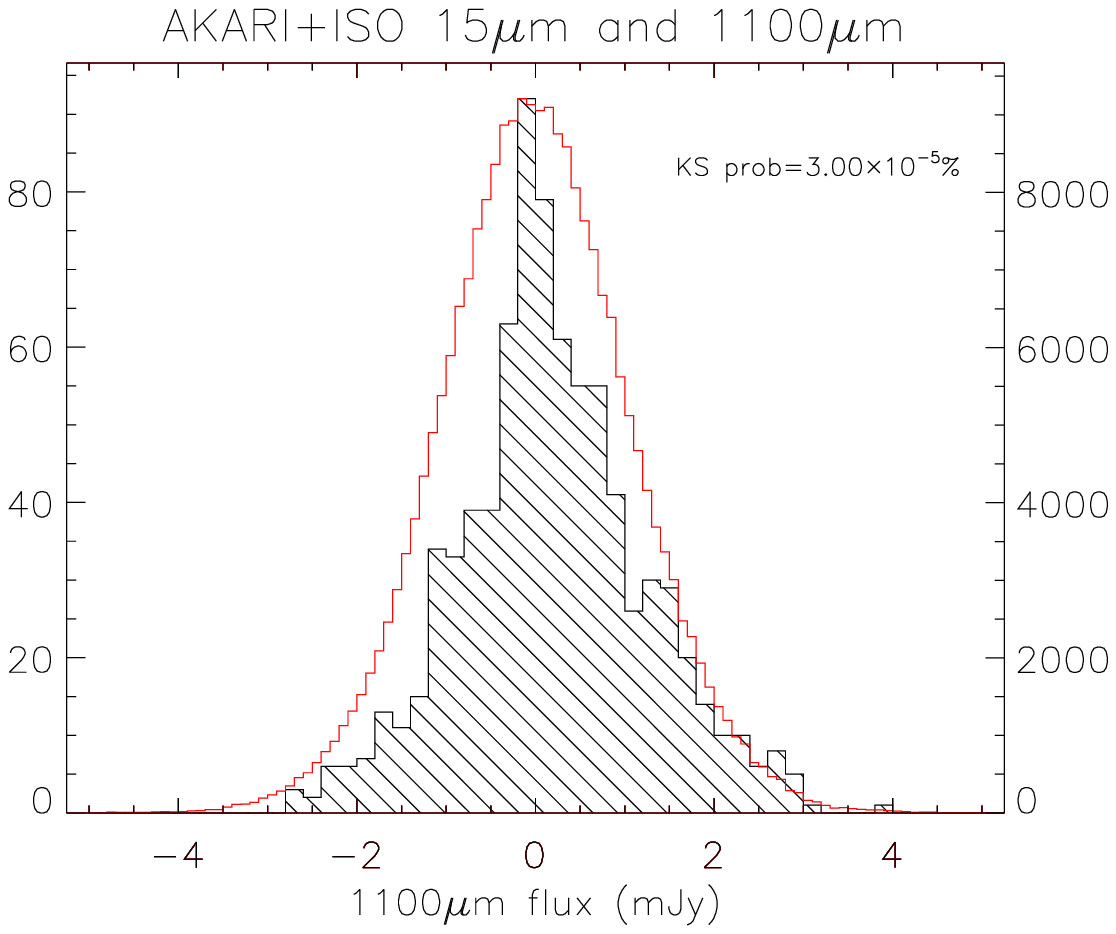}}\nolinebreak
\end{center}
\caption{\label{fig:histograms}
Histograms of the submm and mm-wave fluxes in the regions with $15\,\mu$m data (unhatched
data, right-hand ordinates) compared to the submm and mm-wave flux measurements at the positions of $15\,\mu$m point sources 
(hatched data, left-hand ordinates). The significance values of Kolmogorov-Smirnov tests are quoted in the figures. As noted in
the text, the $1100\,\mu$m map has had an additional sky subtraction from beamswitching. 
}
\end{figure*}

Fig. \ref{fig:histograms} shows the histograms of submm and mm-wave fluxes at the positions of the $15\,\mu$m-selected galaxies, 
compared to the histograms of the maps as a whole were $15\,\mu$m data is available. The
probabilities that the $15\,\mu$m-selected galaxies are drawn randomly from the submm maps are $\sim10^{-3}$ at
$450\,\mu$m, $0.046$ at $850\,\mu$m and $0.016$ at $1100\,\mu$m, i.e. a significant stacking detection. 
These probabilities in all three cases are due to slight asymmetries or positive shifts in the submm/mm-wave flux distributions.

\subsection{SED fitting results}
We found centroids in the $15\,\mu$m AKARI point source detection map at the positions of SHADES Lockman 
galaxies reported in Dye et al. (2008). Where no centroid solution could be obtained, we took the value
of the point source flux $F$ and its associated error $\Delta F$ at the SHADES counterpart position (equations
\ref{eqn:bff} and \ref{eqn:dbff}). Tables \ref{tab:photometry_opt_ir} and \ref{tab:photometry_fir} give a 
compilation of available photometry of the SHADES galaxies from Dye et al. (2008), Ivison et al. (2007) and
Coppin et al. (2007), and incorporating our new $15\,\mu$m photometry. 

\begin{figure*}[!ht]
\begin{center}
\hspace*{-1.8cm}\resizebox{0.37\hsize}{!}{\includegraphics*{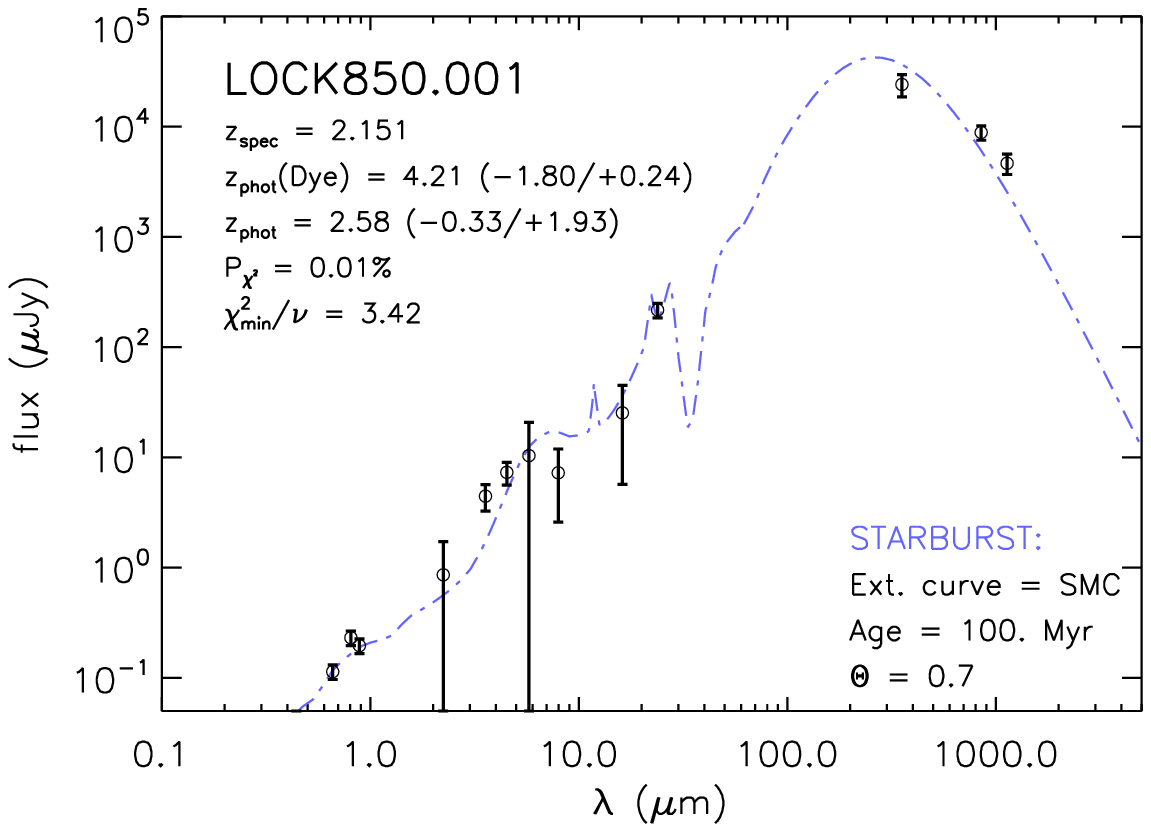}}\nolinebreak
\hspace*{-2.65cm}\resizebox{0.37\hsize}{!}{\includegraphics*{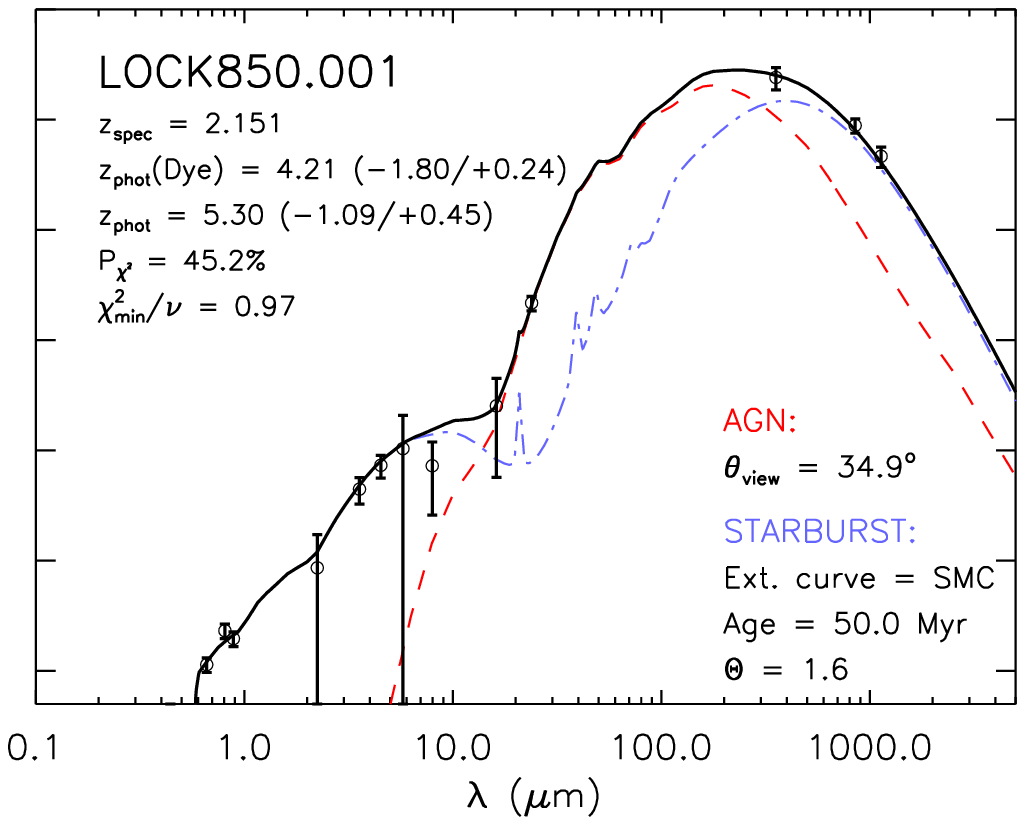}}\nolinebreak
\hspace*{-1.8cm}\resizebox{0.37\hsize}{!}{\includegraphics*{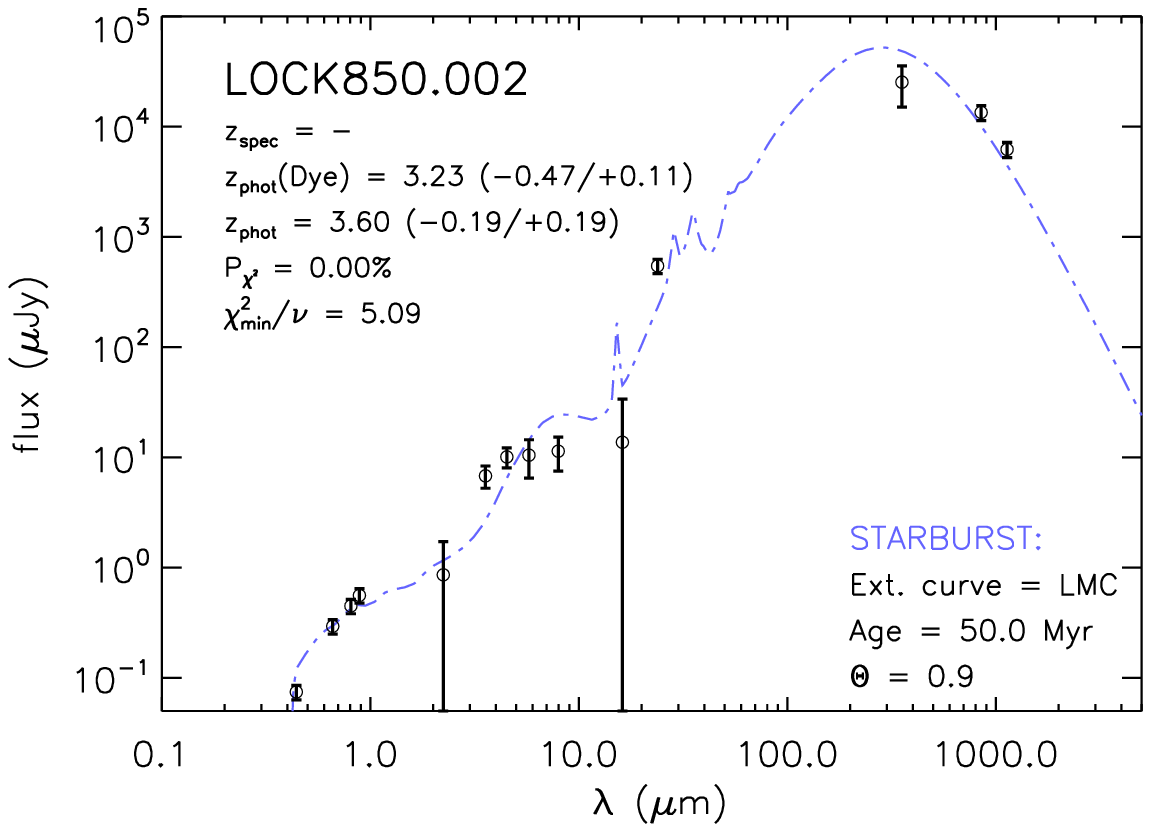}}\nolinebreak
\hspace*{-2.65cm}\resizebox{0.37\hsize}{!}{\includegraphics*{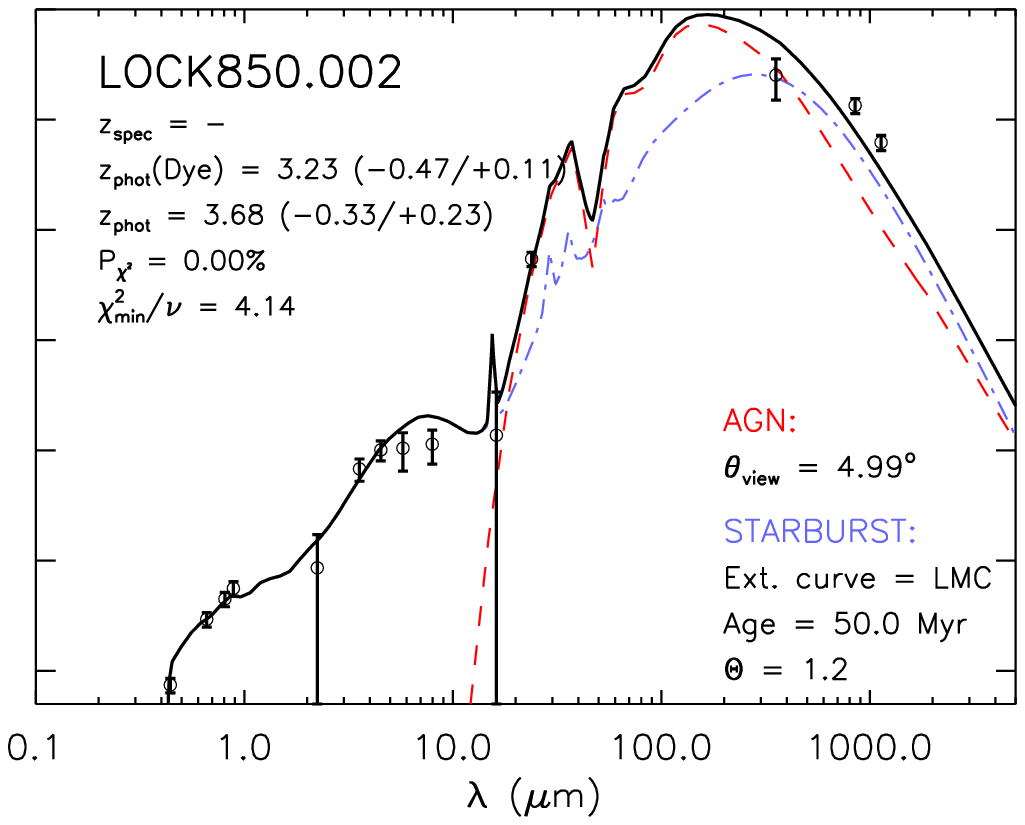}}\nolinebreak
\end{center}\vspace*{-1.3cm}
\begin{center}
\hspace*{-1.8cm}\resizebox{0.37\hsize}{!}{\includegraphics*{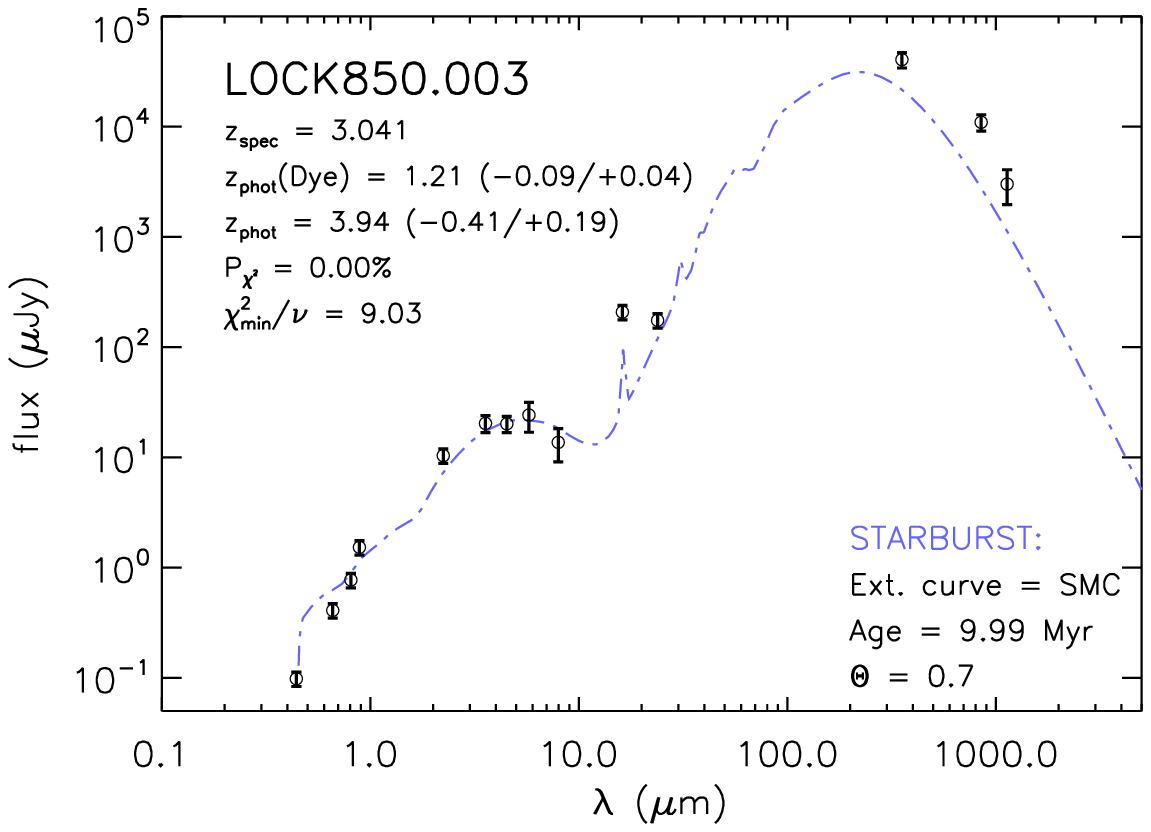}}\nolinebreak
\hspace*{-2.65cm}\resizebox{0.37\hsize}{!}{\includegraphics*{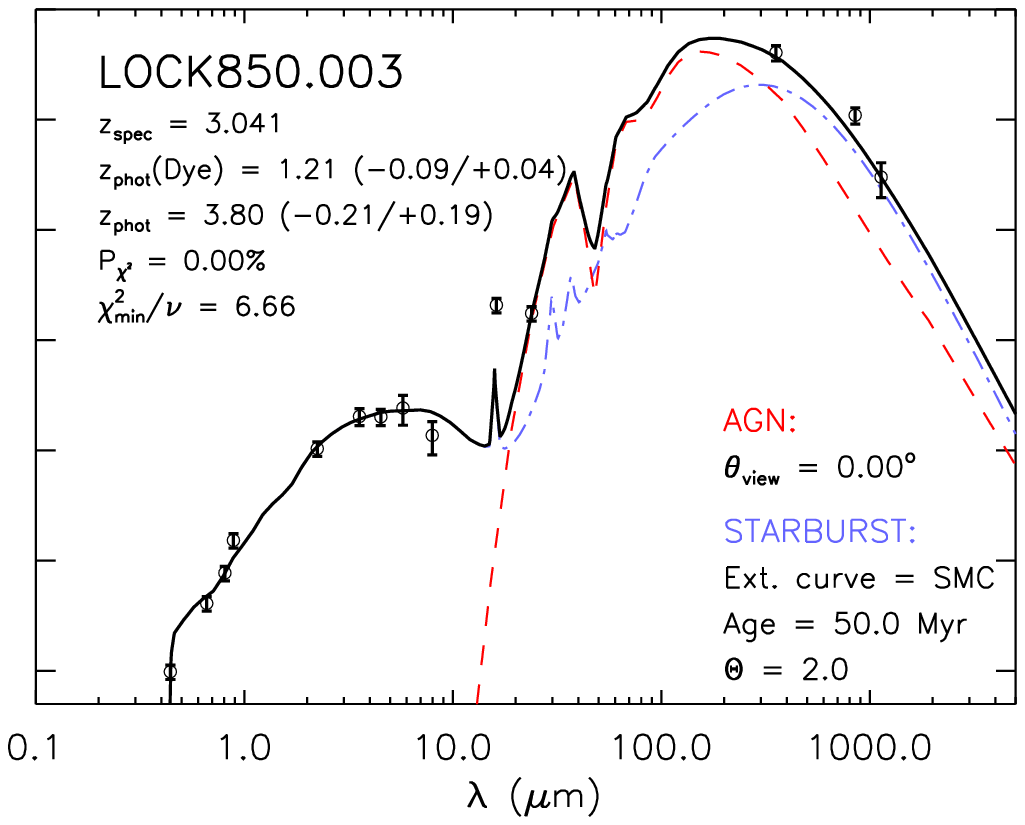}}\nolinebreak
\hspace*{-1.8cm}\resizebox{0.37\hsize}{!}{\includegraphics*{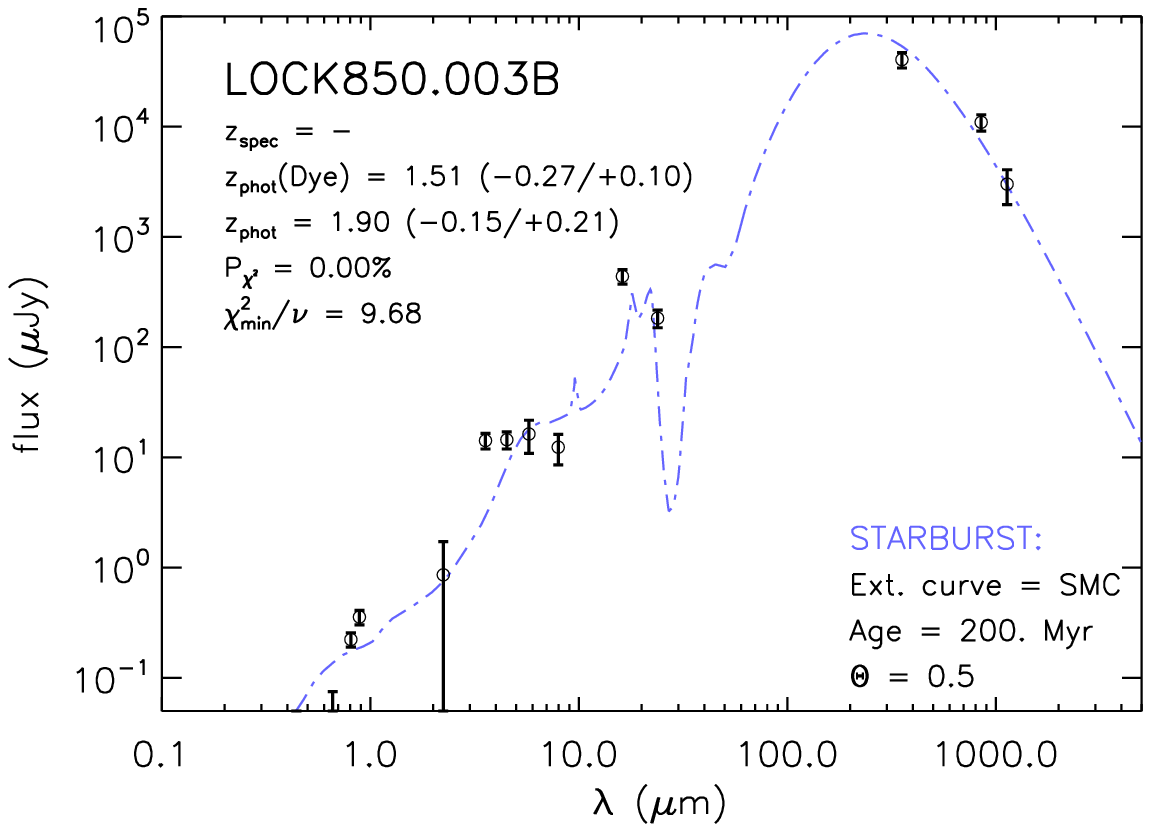}}\nolinebreak
\hspace*{-2.65cm}\resizebox{0.37\hsize}{!}{\includegraphics*{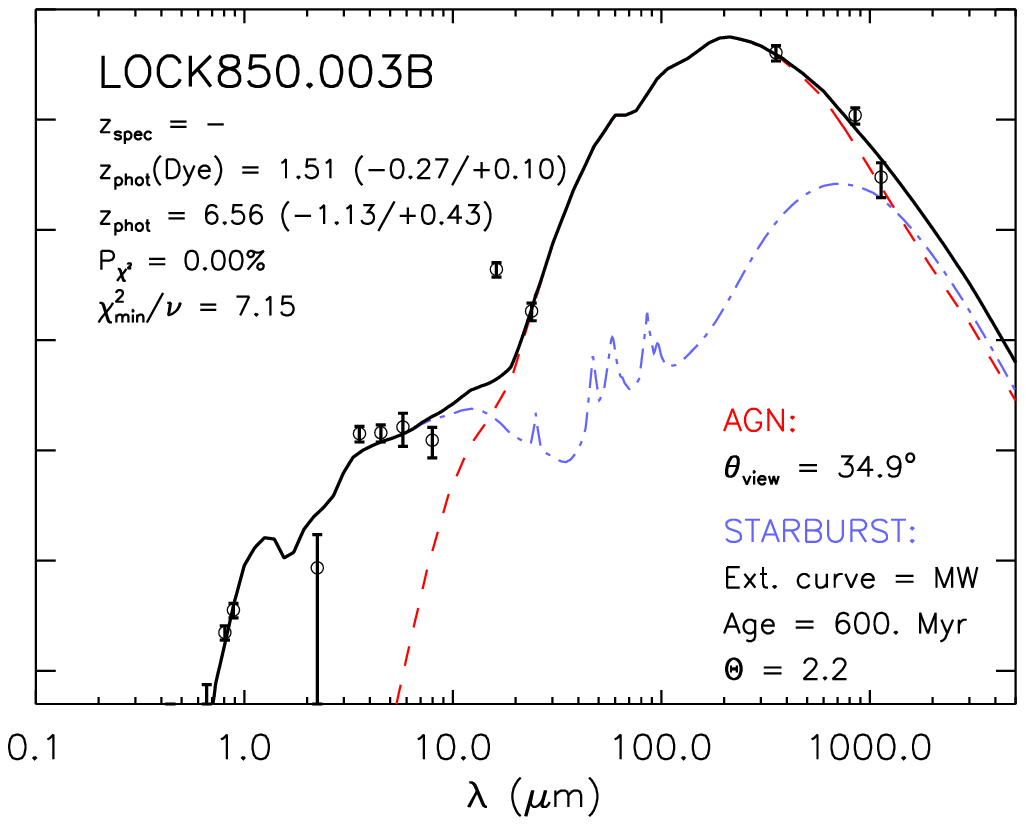}}\nolinebreak
\end{center}\vspace*{-1.3cm}
\begin{center}
\hspace*{-1.8cm}\resizebox{0.37\hsize}{!}{\includegraphics*{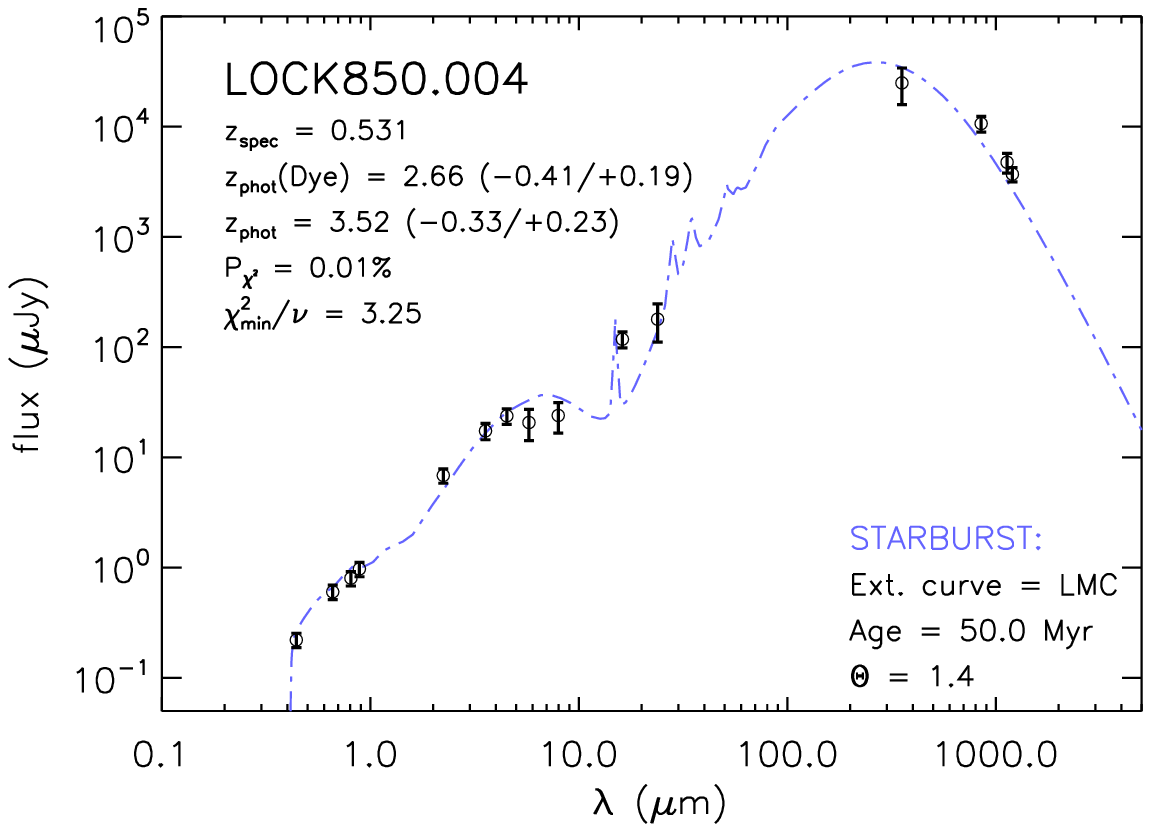}}\nolinebreak
\hspace*{-2.65cm}\resizebox{0.37\hsize}{!}{\includegraphics*{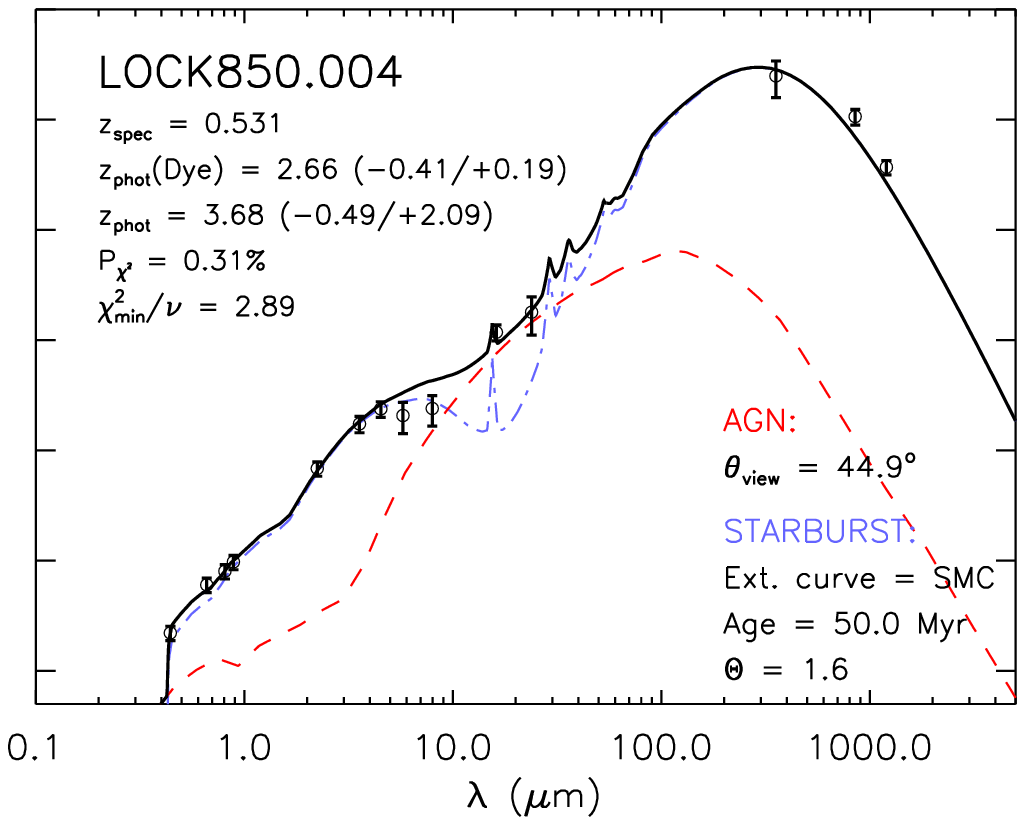}}\nolinebreak
\hspace*{-1.8cm}\resizebox{0.37\hsize}{!}{\includegraphics*{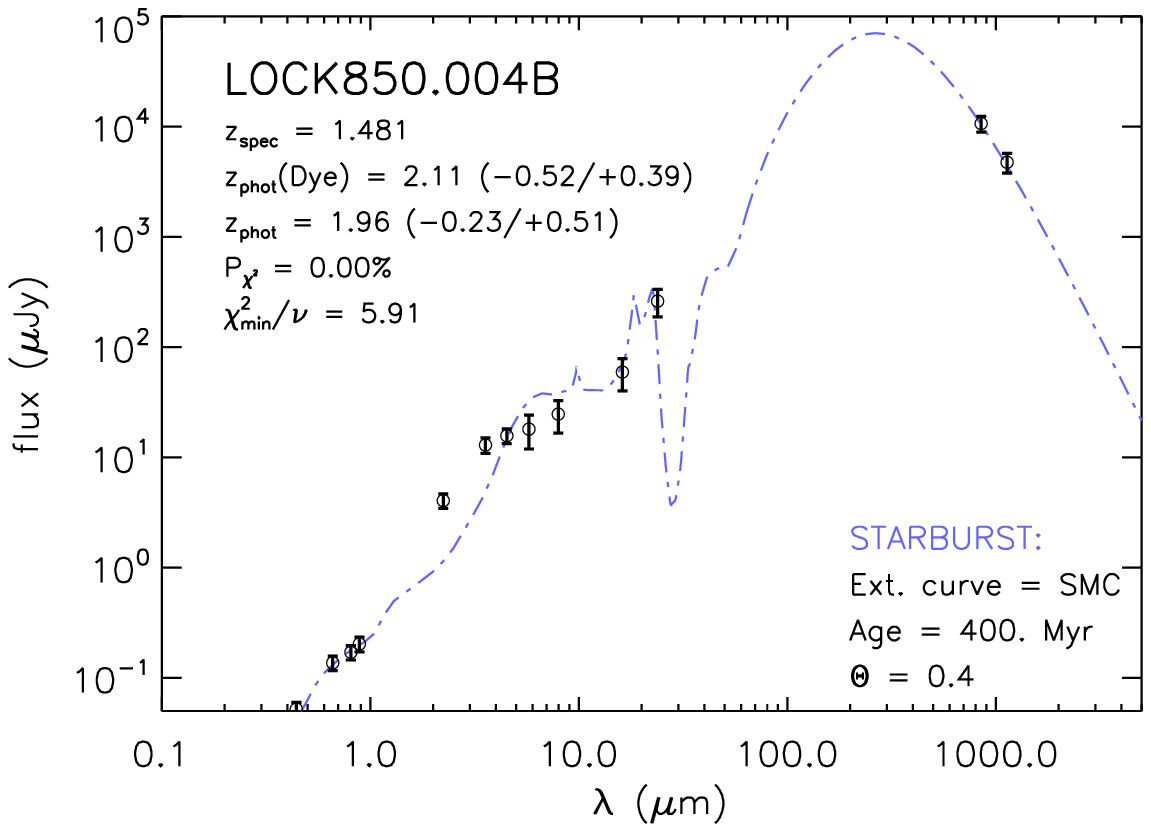}}\nolinebreak
\hspace*{-2.65cm}\resizebox{0.37\hsize}{!}{\includegraphics*{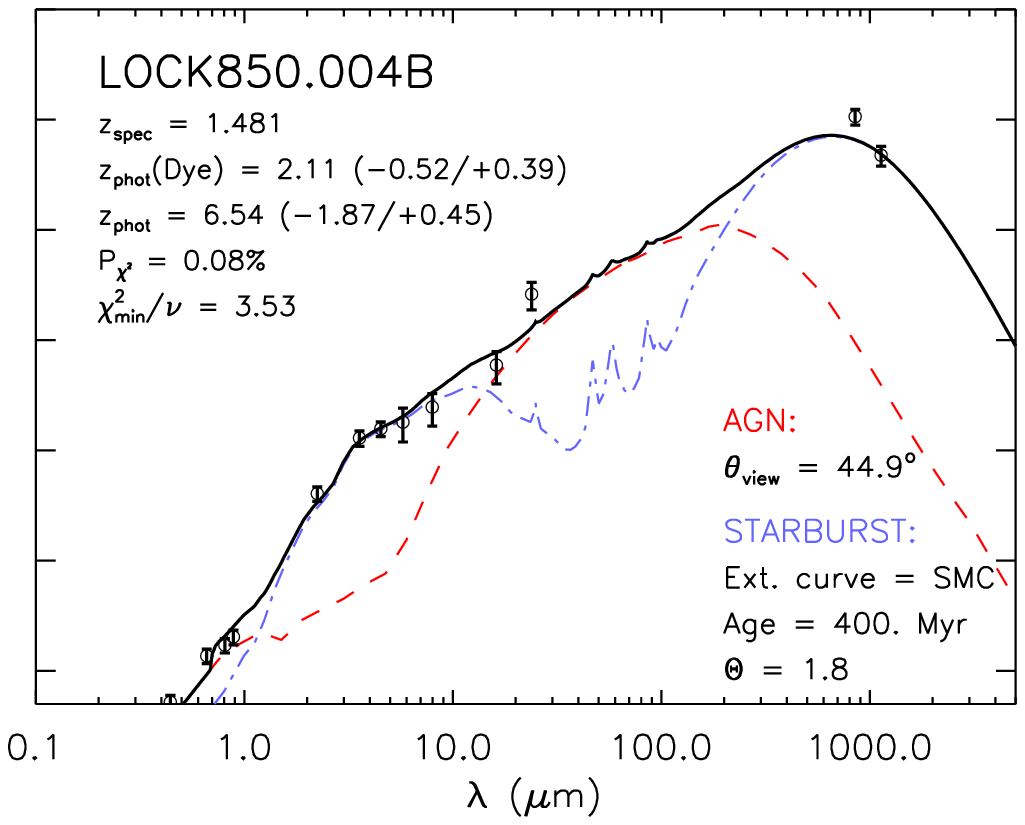}}\nolinebreak
\end{center}\vspace*{-1.3cm}
\begin{center}
\hspace*{-1.8cm}\resizebox{0.37\hsize}{!}{\includegraphics*{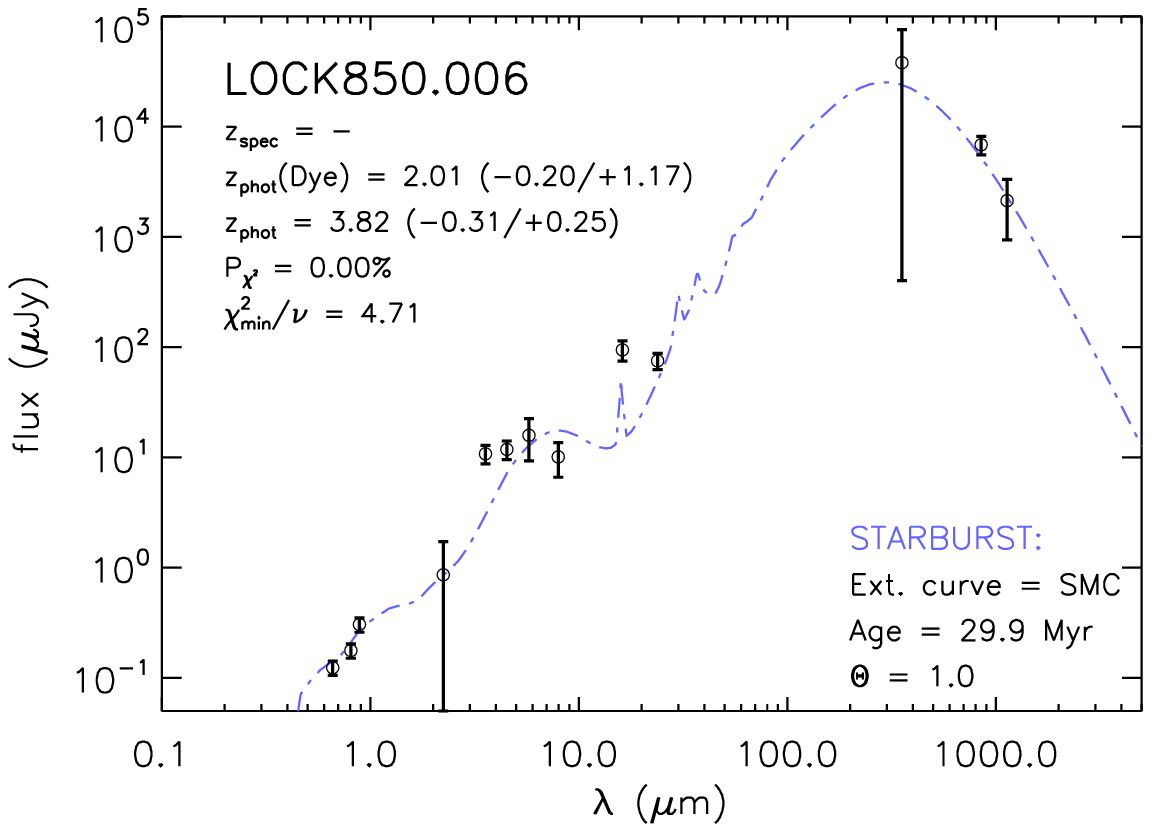}}\nolinebreak
\hspace*{-2.65cm}\resizebox{0.37\hsize}{!}{\includegraphics*{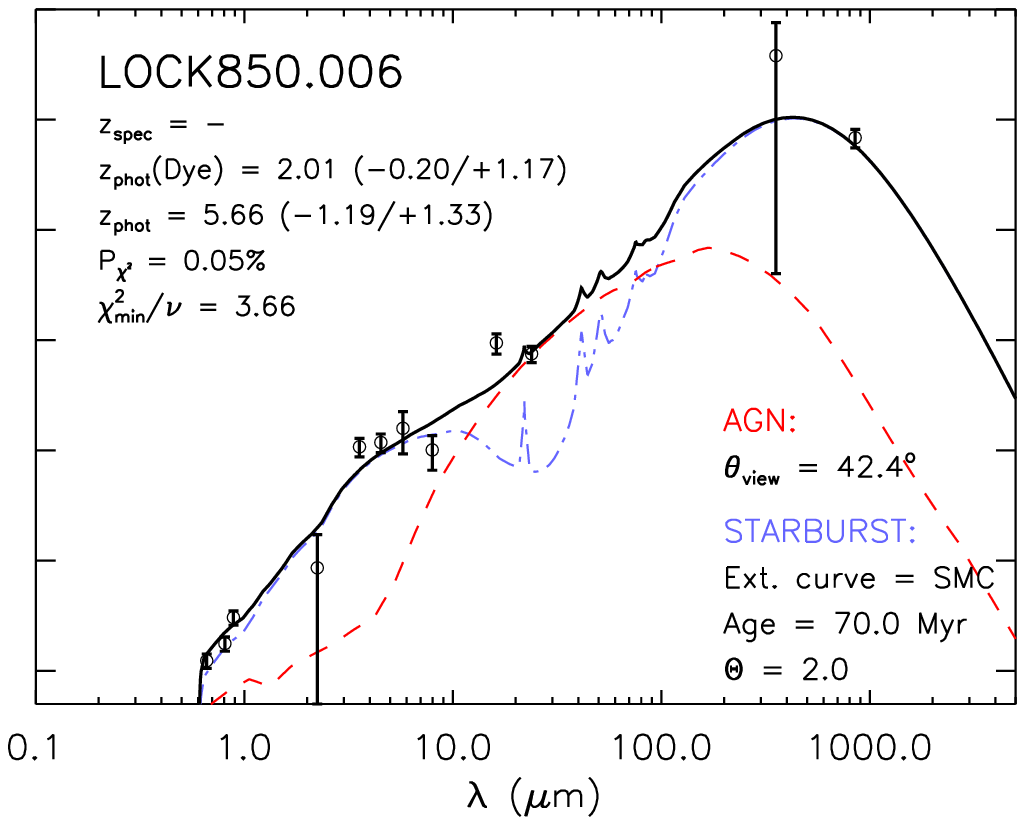}}\nolinebreak
\hspace*{-1.8cm}\resizebox{0.37\hsize}{!}{\includegraphics*{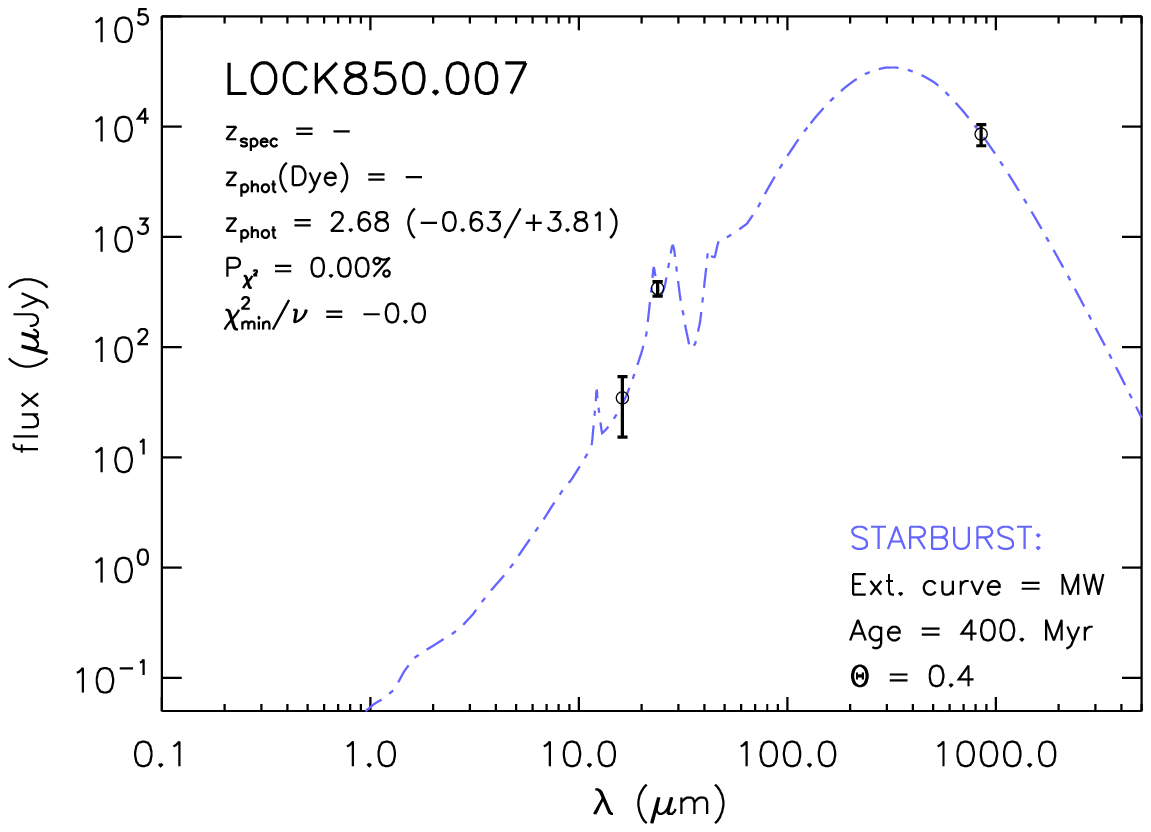}}\nolinebreak
\hspace*{-2.65cm}\resizebox{0.37\hsize}{!}{\includegraphics*{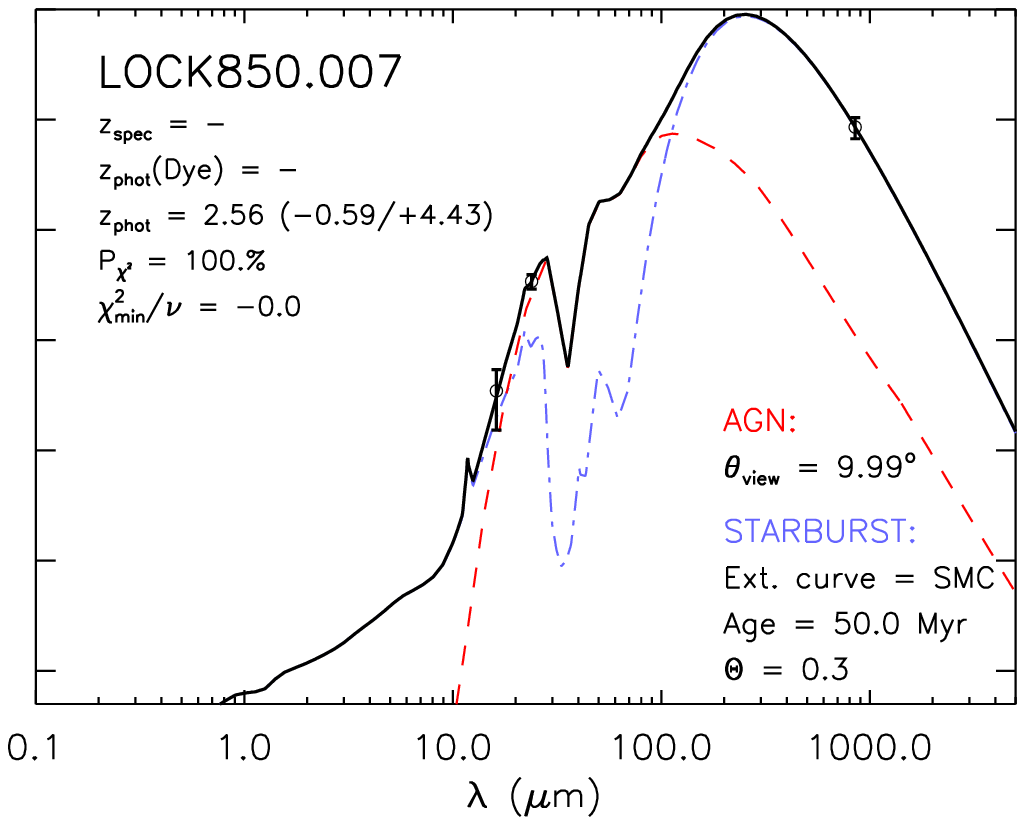}}\nolinebreak
\end{center}\vspace*{-1.3cm}
\begin{center}
\hspace*{-1.8cm}\resizebox{0.37\hsize}{!}{\includegraphics*{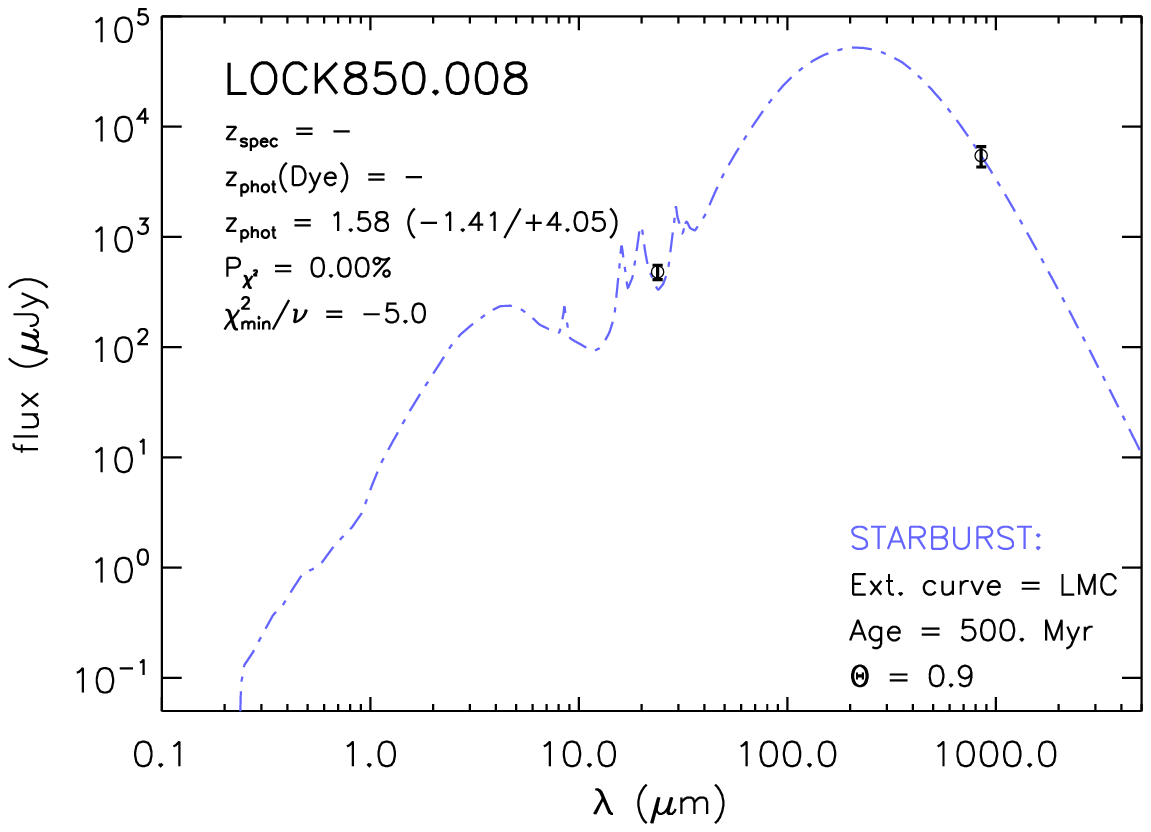}}\nolinebreak
\hspace*{-2.65cm}\resizebox{0.37\hsize}{!}{\includegraphics*{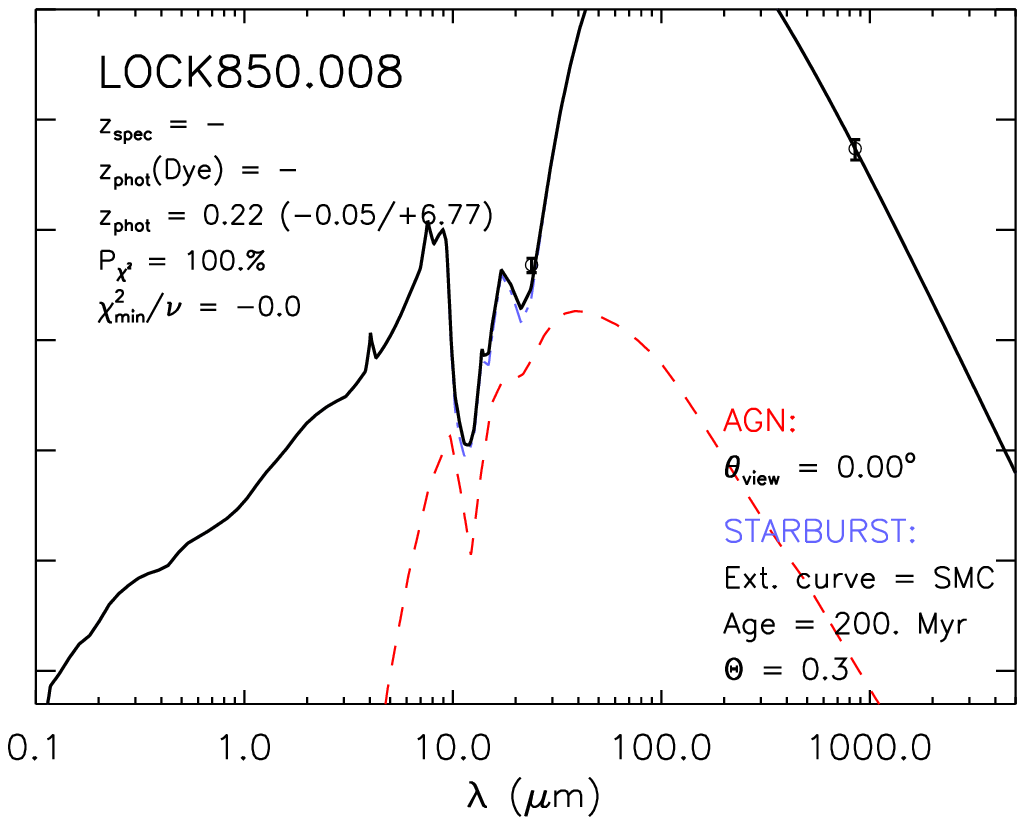}}\nolinebreak
\hspace*{-1.8cm}\resizebox{0.37\hsize}{!}{\includegraphics*{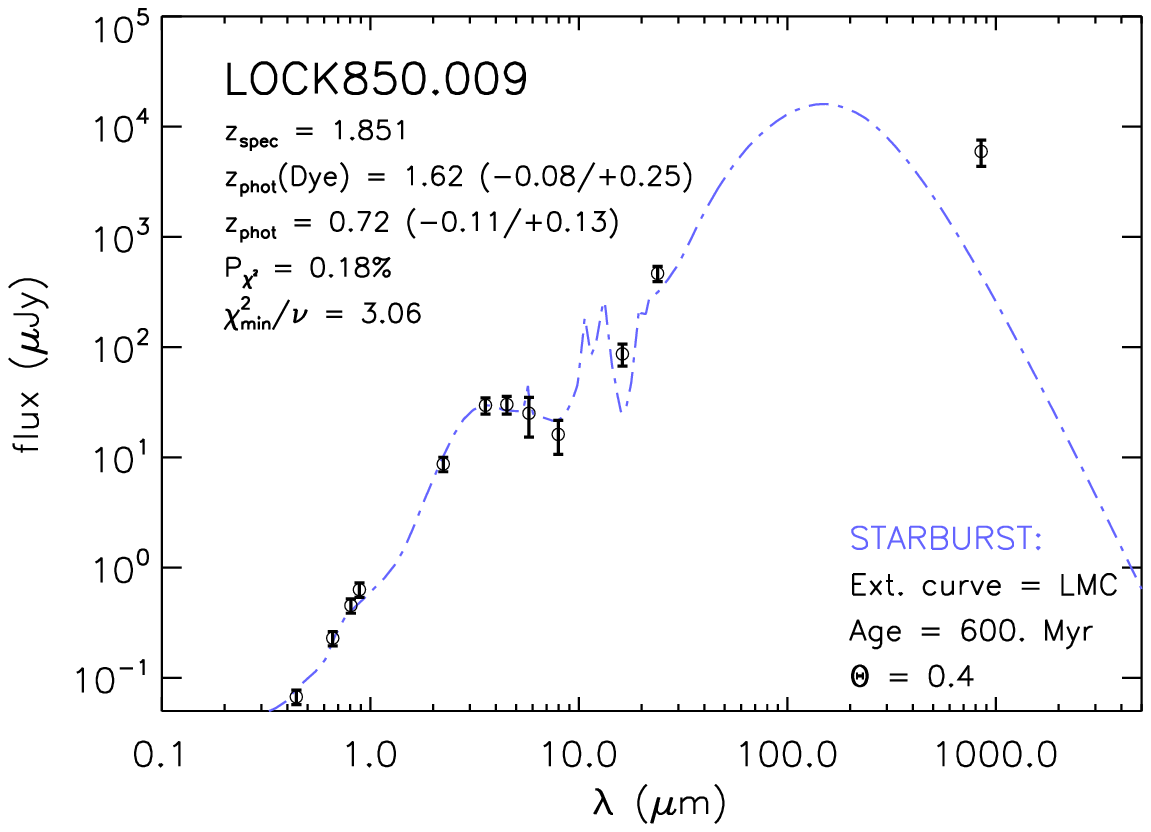}}\nolinebreak
\hspace*{-2.65cm}\resizebox{0.37\hsize}{!}{\includegraphics*{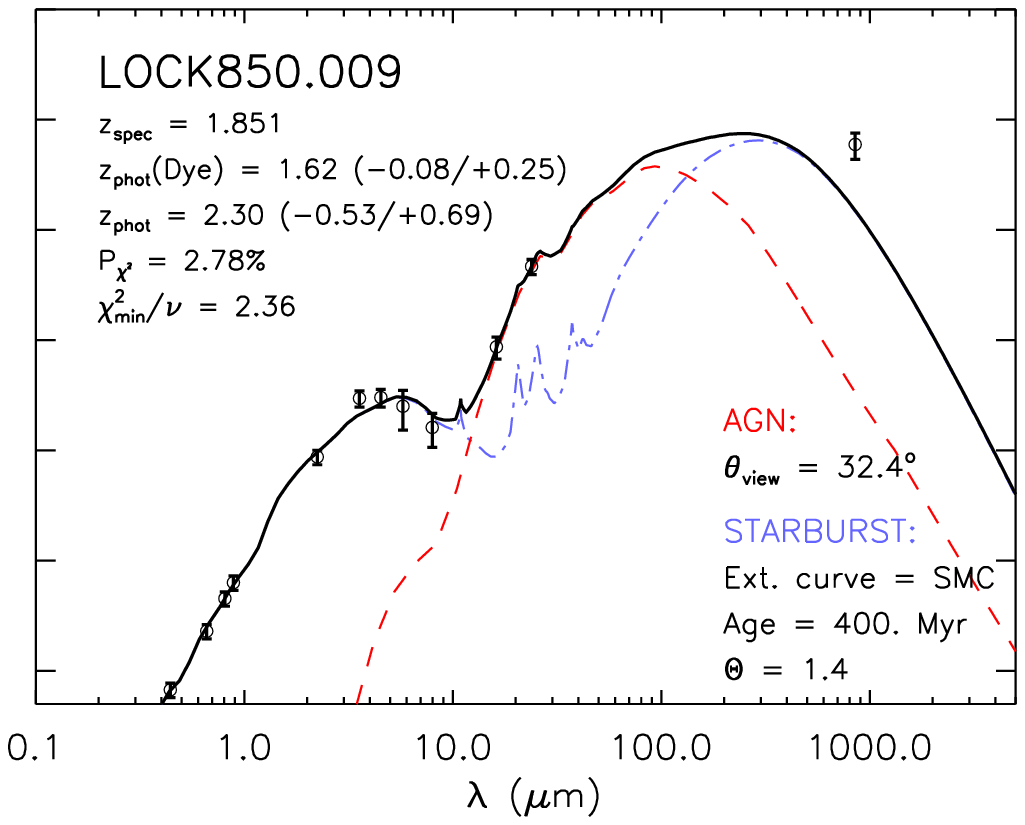}}\nolinebreak
\end{center}\vspace*{-1.3cm}
\begin{center}
\hspace*{-1.8cm}\resizebox{0.37\hsize}{!}{\includegraphics*{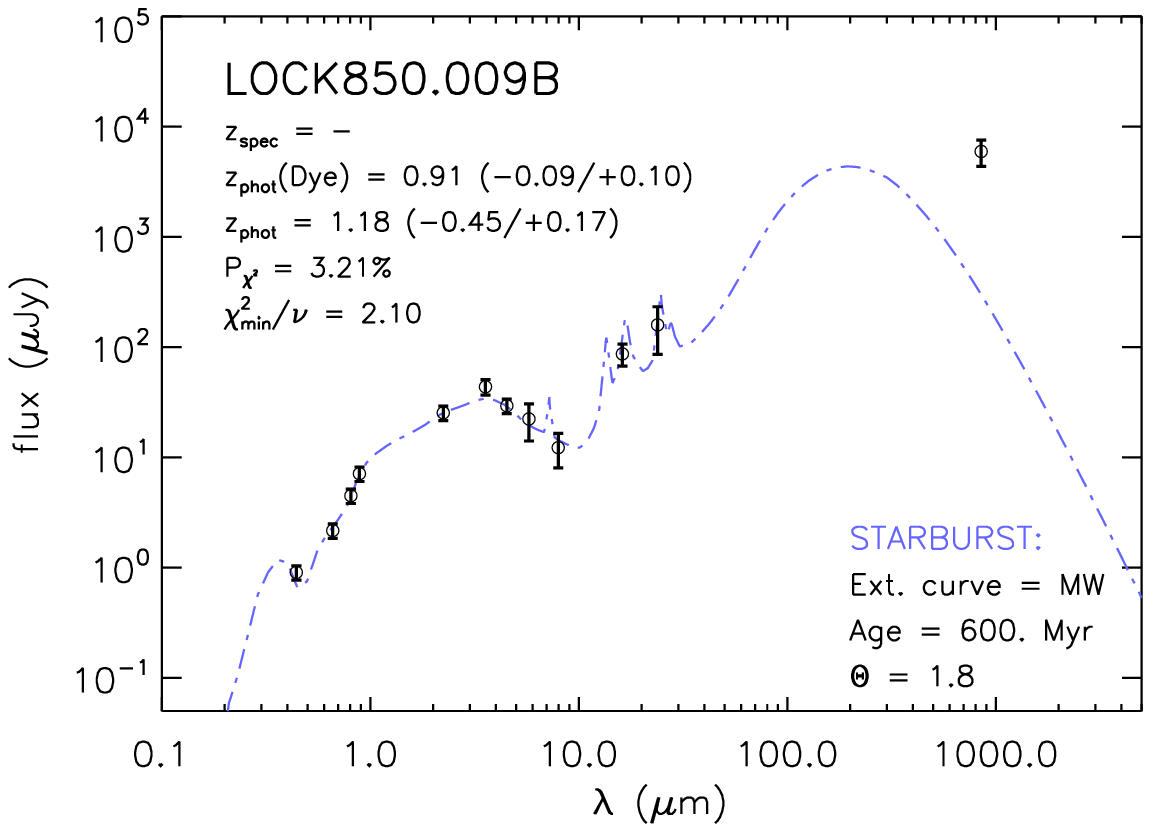}}\nolinebreak
\hspace*{-2.65cm}\resizebox{0.37\hsize}{!}{\includegraphics*{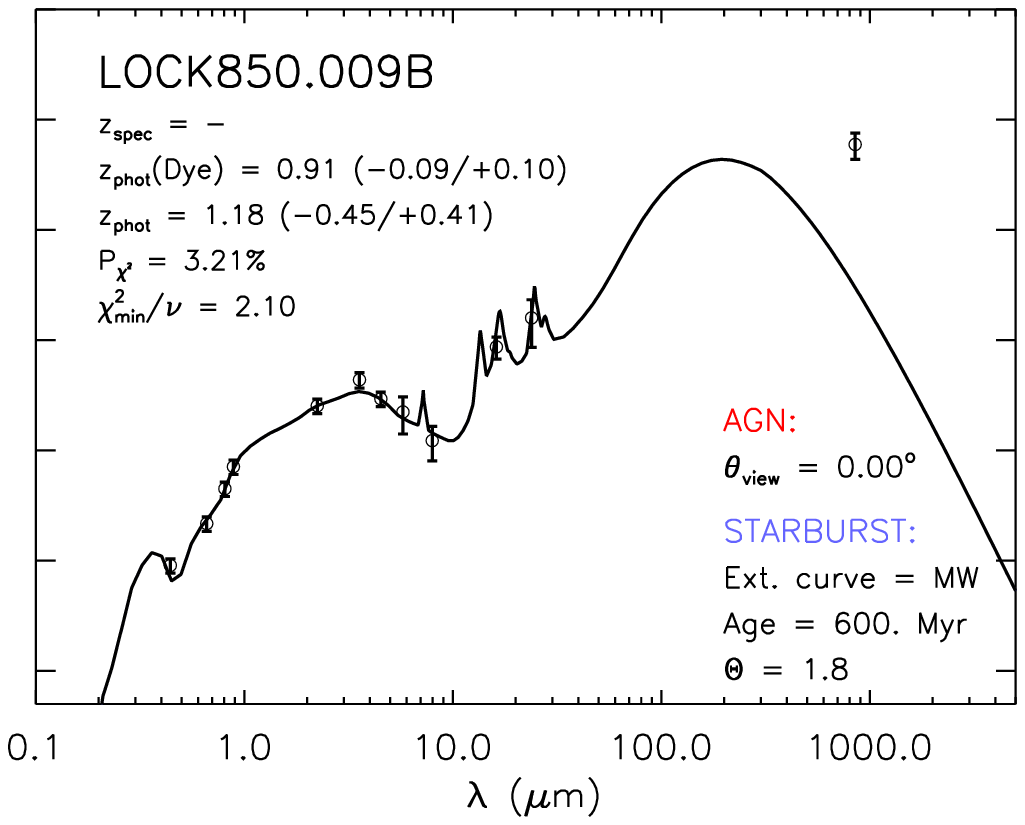}}\nolinebreak
\hspace*{-1.8cm}\resizebox{0.37\hsize}{!}{\includegraphics*{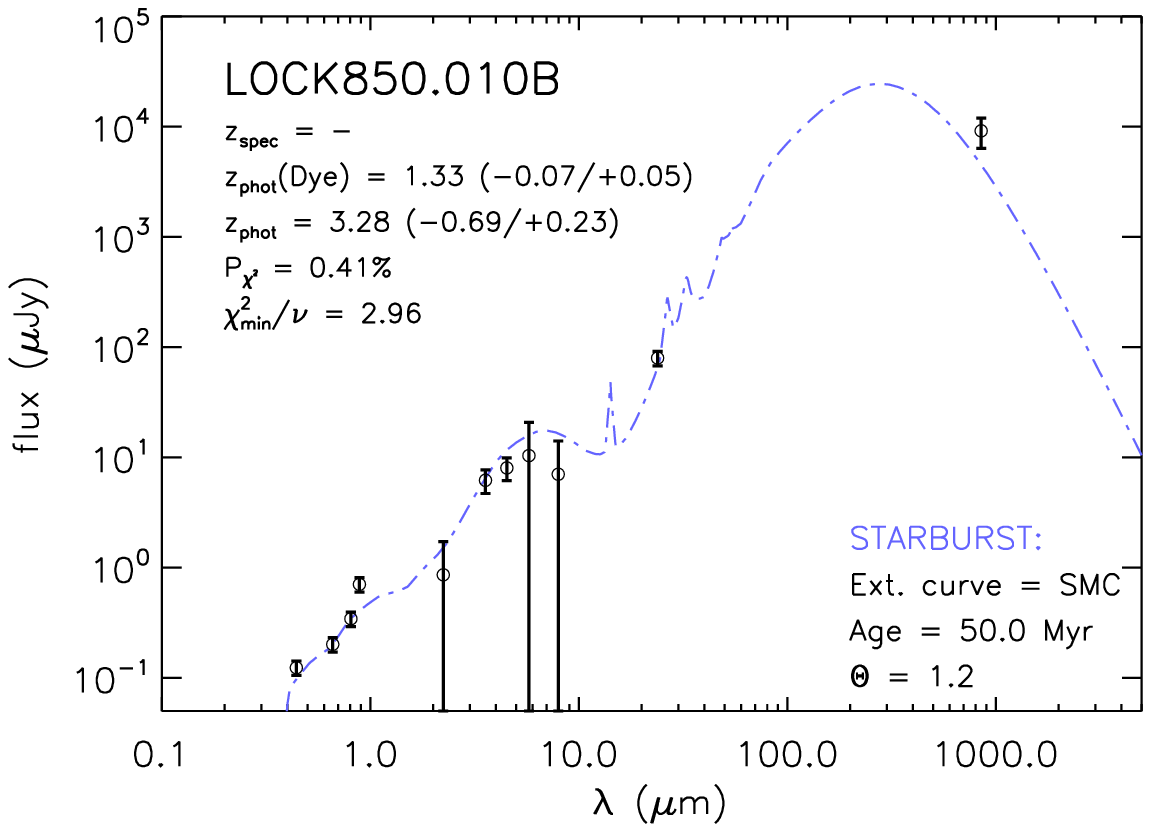}}\nolinebreak
\hspace*{-2.65cm}\resizebox{0.37\hsize}{!}{\includegraphics*{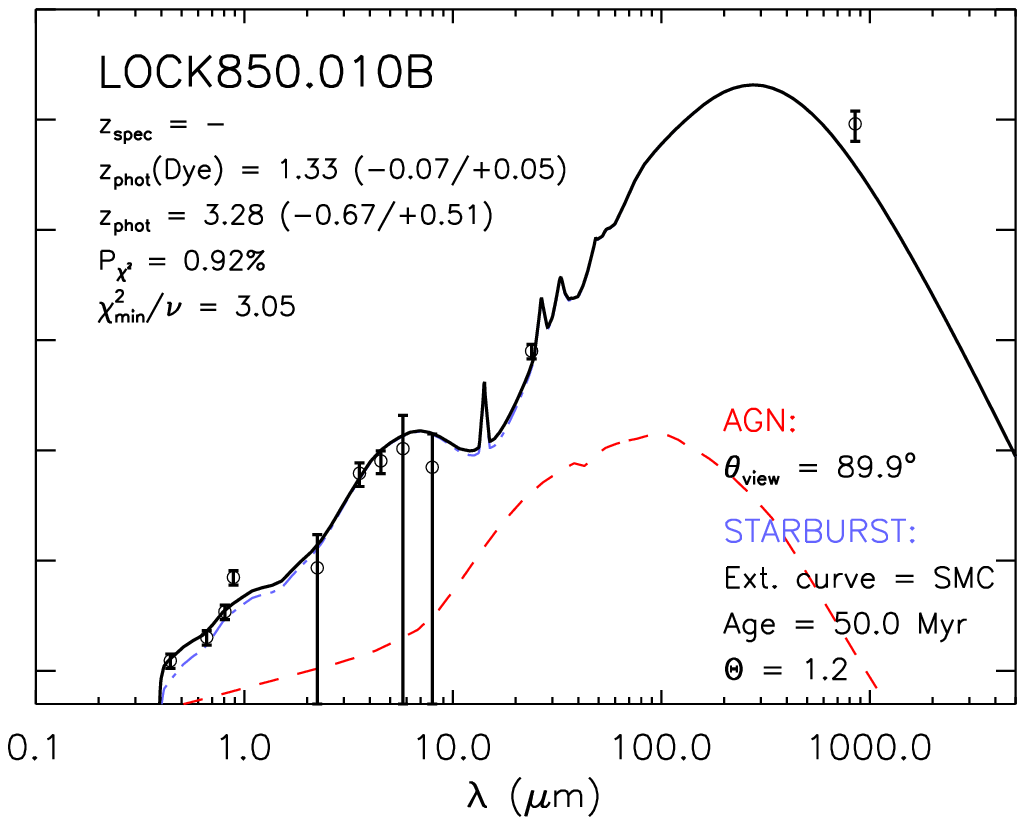}}\nolinebreak
\end{center}\vspace*{-1.3cm}
\begin{center}
\hspace*{-1.8cm}\resizebox{0.37\hsize}{!}{\includegraphics*{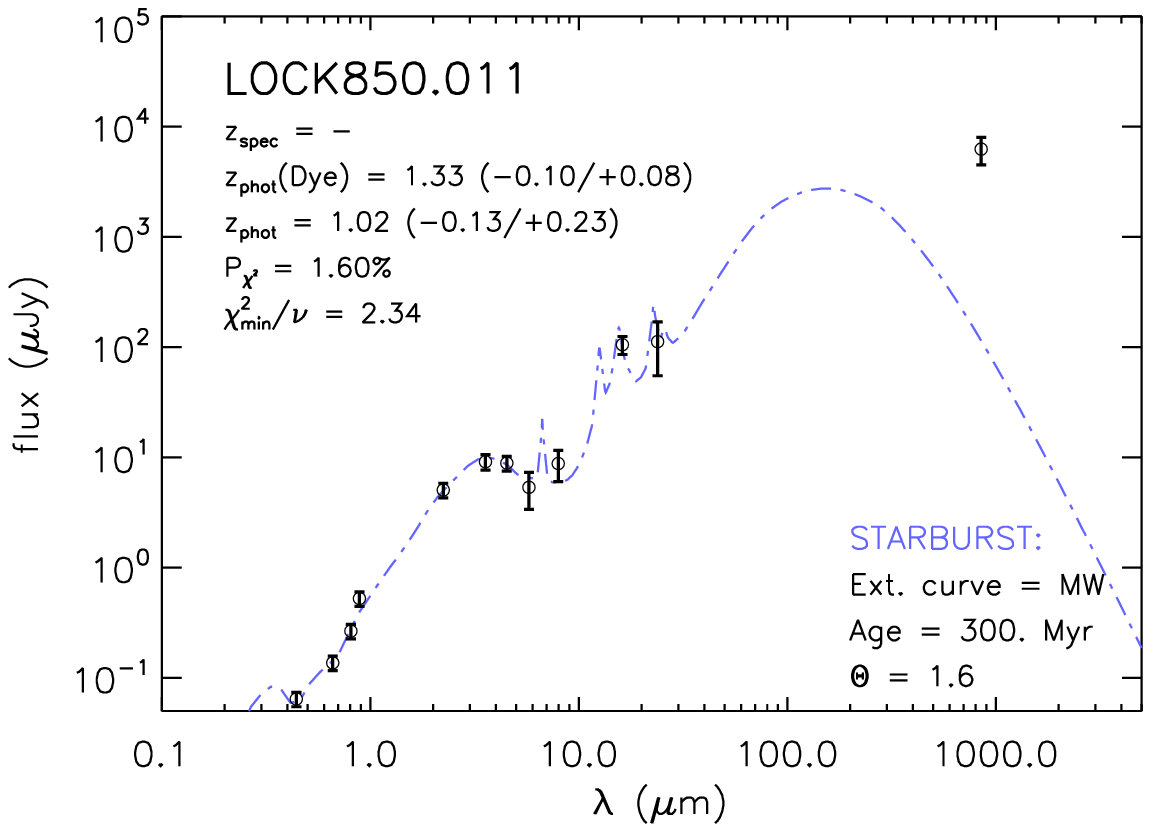}}\nolinebreak
\hspace*{-2.65cm}\resizebox{0.37\hsize}{!}{\includegraphics*{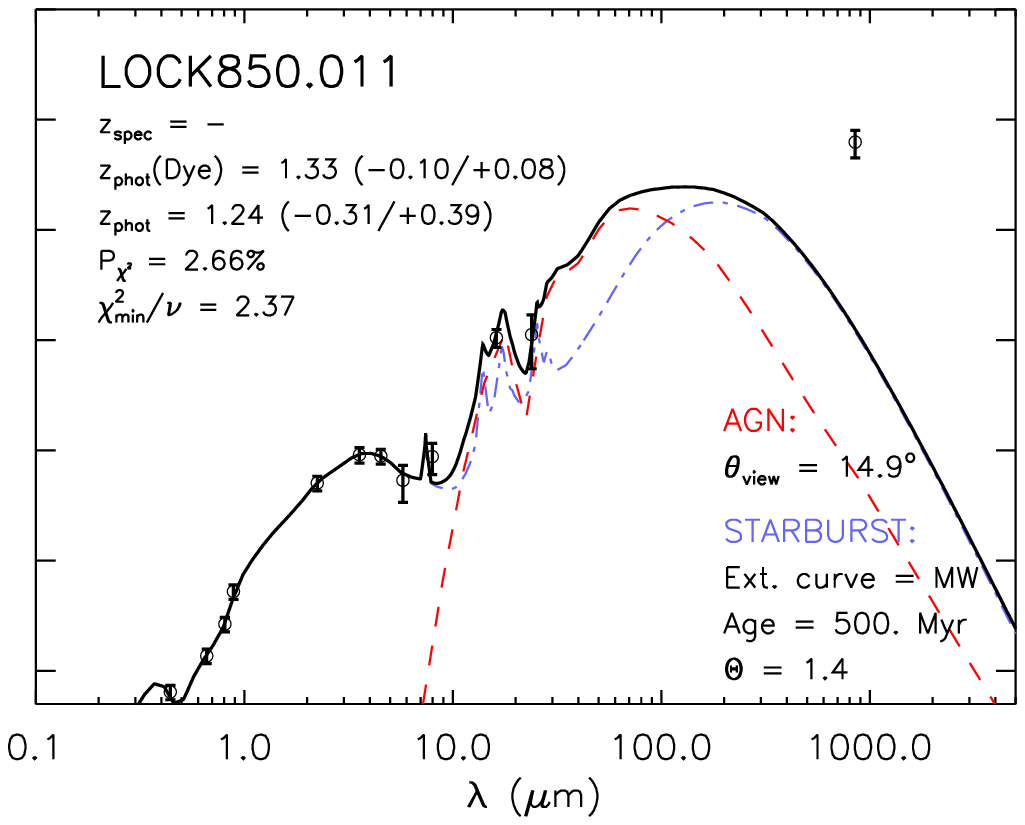}}\nolinebreak
\hspace*{-1.8cm}\resizebox{0.37\hsize}{!}{\includegraphics*{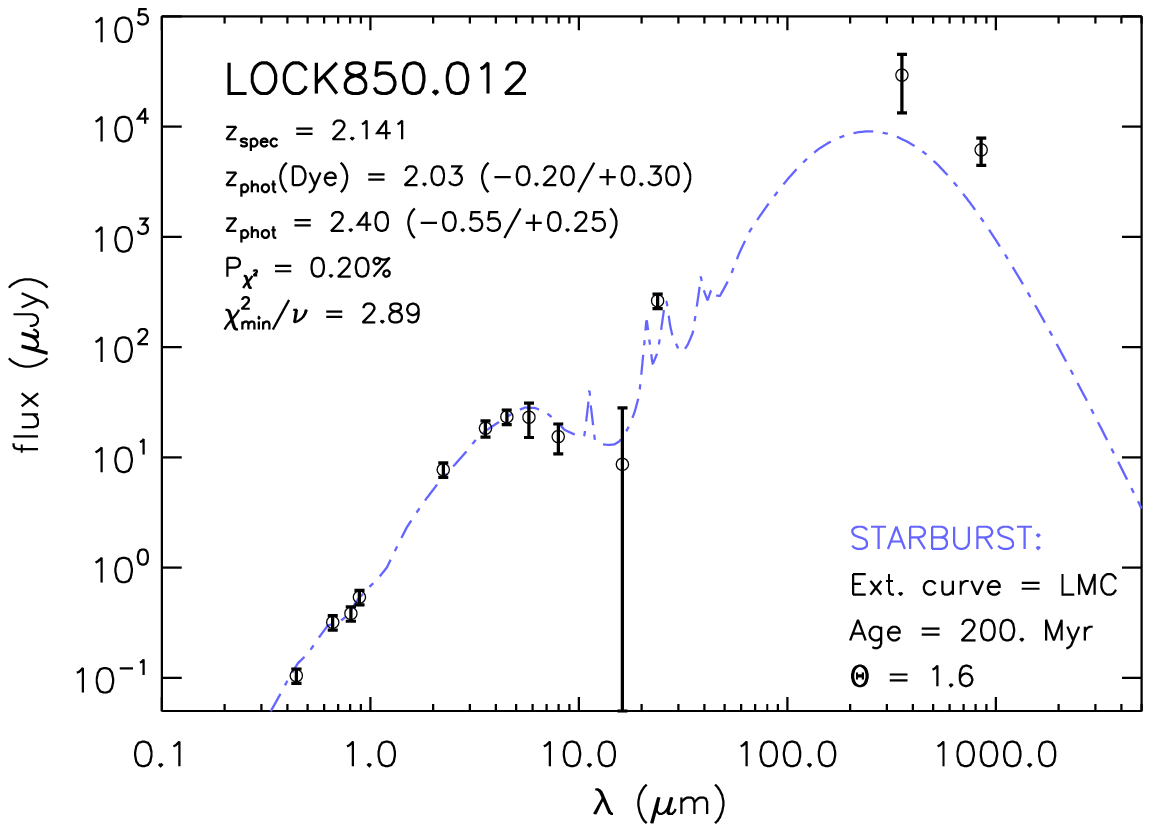}}\nolinebreak
\hspace*{-2.65cm}\resizebox{0.37\hsize}{!}{\includegraphics*{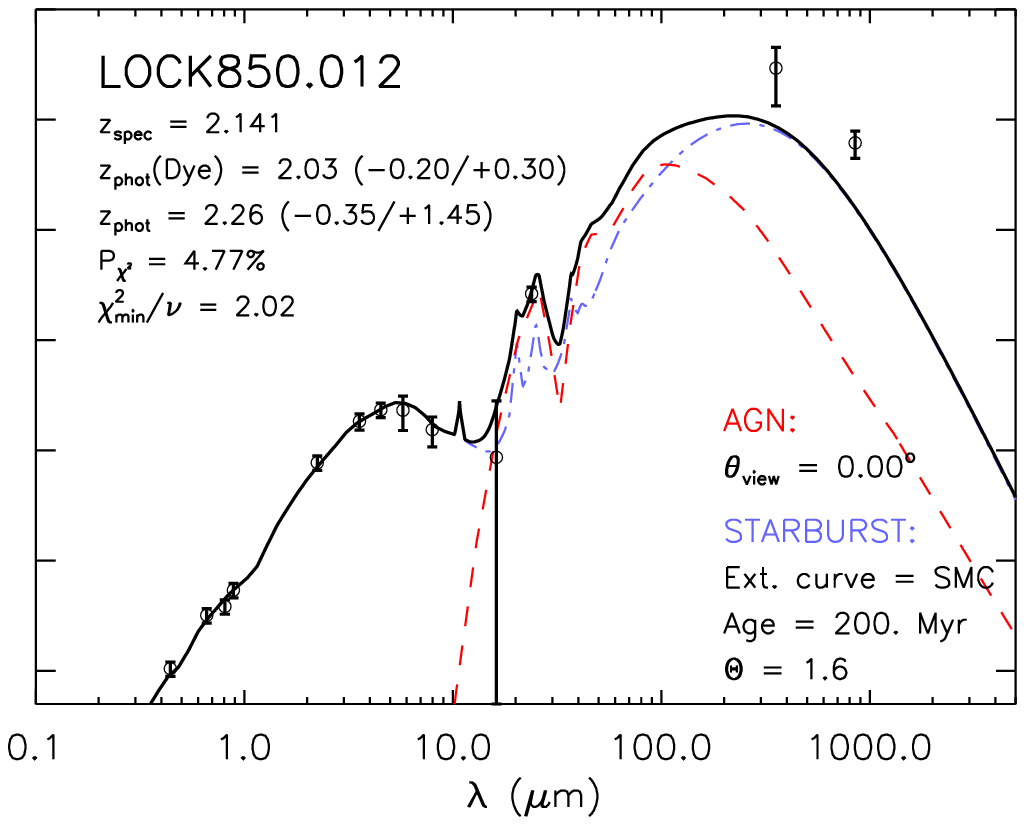}}\nolinebreak
\end{center}\vspace*{-1.3cm}
\vspace*{1.6cm}\caption{SED fits to submm-selected SHADES galaxies in the Lockman Hole East, using models from Takagi et al. (2003, 2004) and Efstathiou \& Rowan-Robinson (1995).}\label{fig:seds1}\end{figure*}

\begin{figure*}[!ht]
\begin{center}
\hspace*{-1.8cm}\resizebox{0.37\hsize}{!}{\includegraphics*{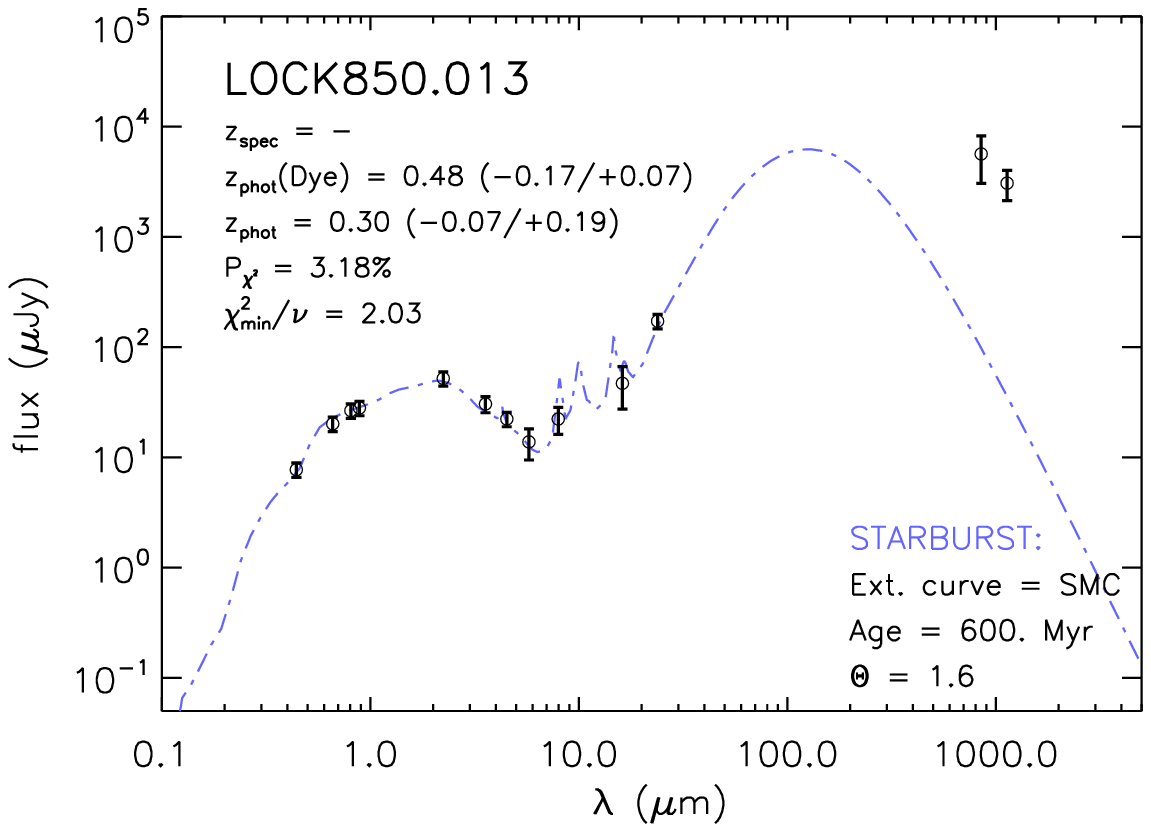}}\nolinebreak
\hspace*{-2.65cm}\resizebox{0.37\hsize}{!}{\includegraphics*{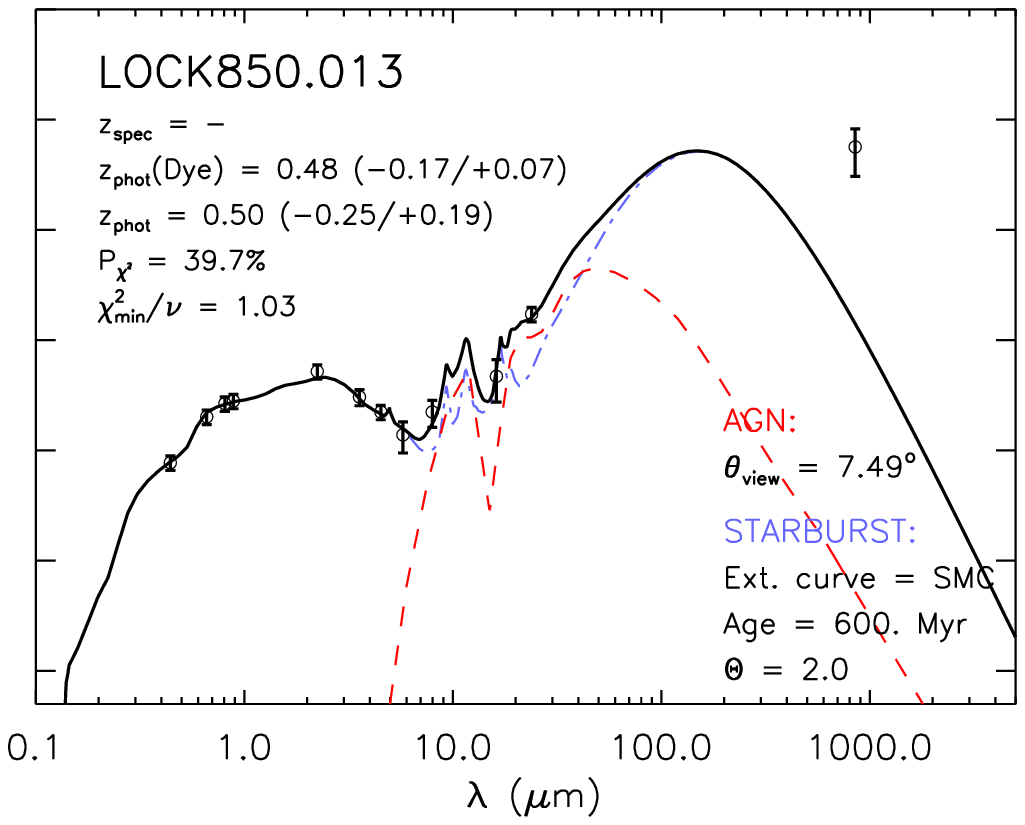}}\nolinebreak
\hspace*{-1.8cm}\resizebox{0.37\hsize}{!}{\includegraphics*{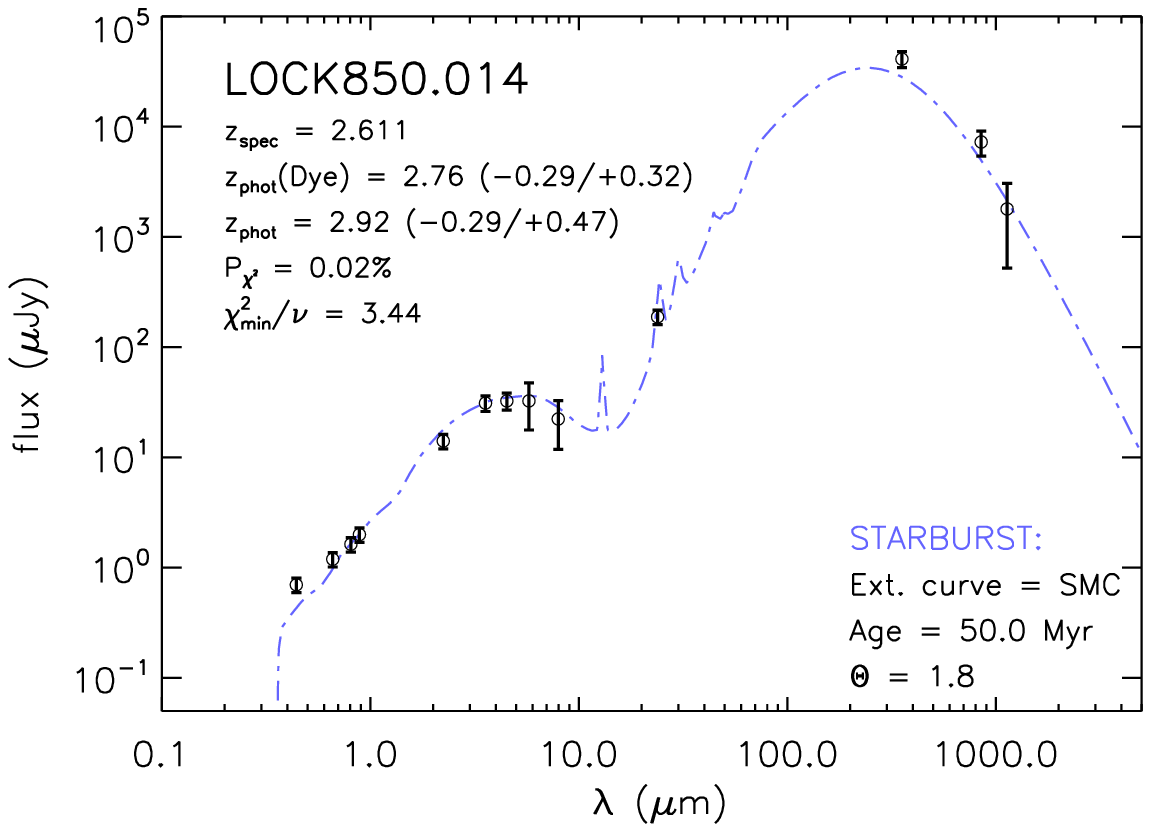}}\nolinebreak
\hspace*{-2.65cm}\resizebox{0.37\hsize}{!}{\includegraphics*{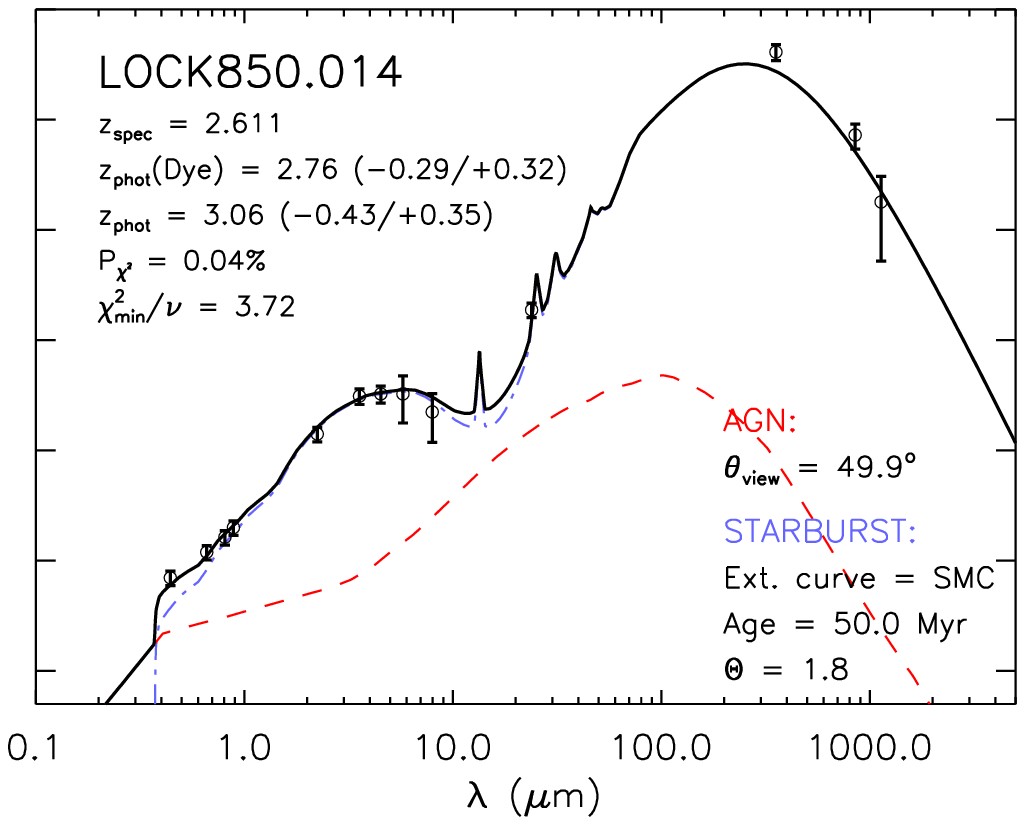}}\nolinebreak
\end{center}\vspace*{-1.3cm}
\begin{center}
\hspace*{-1.8cm}\resizebox{0.37\hsize}{!}{\includegraphics*{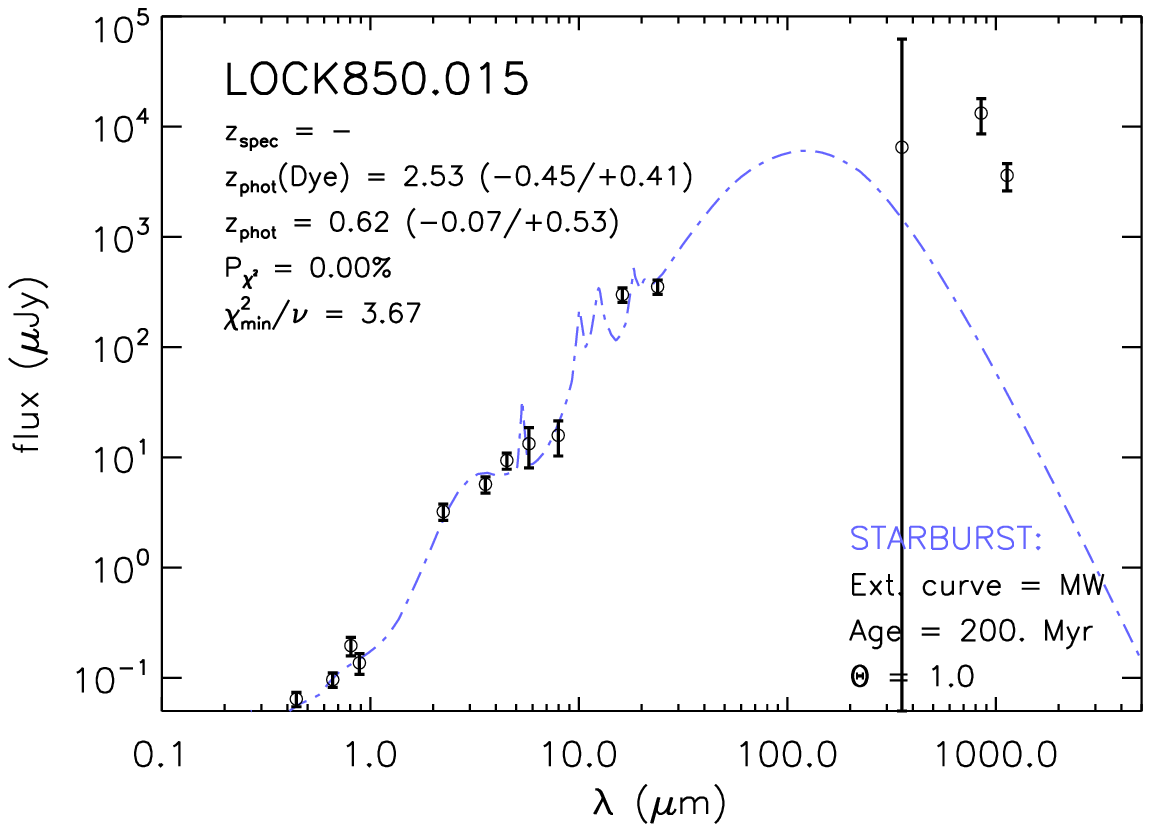}}\nolinebreak
\hspace*{-2.65cm}\resizebox{0.37\hsize}{!}{\includegraphics*{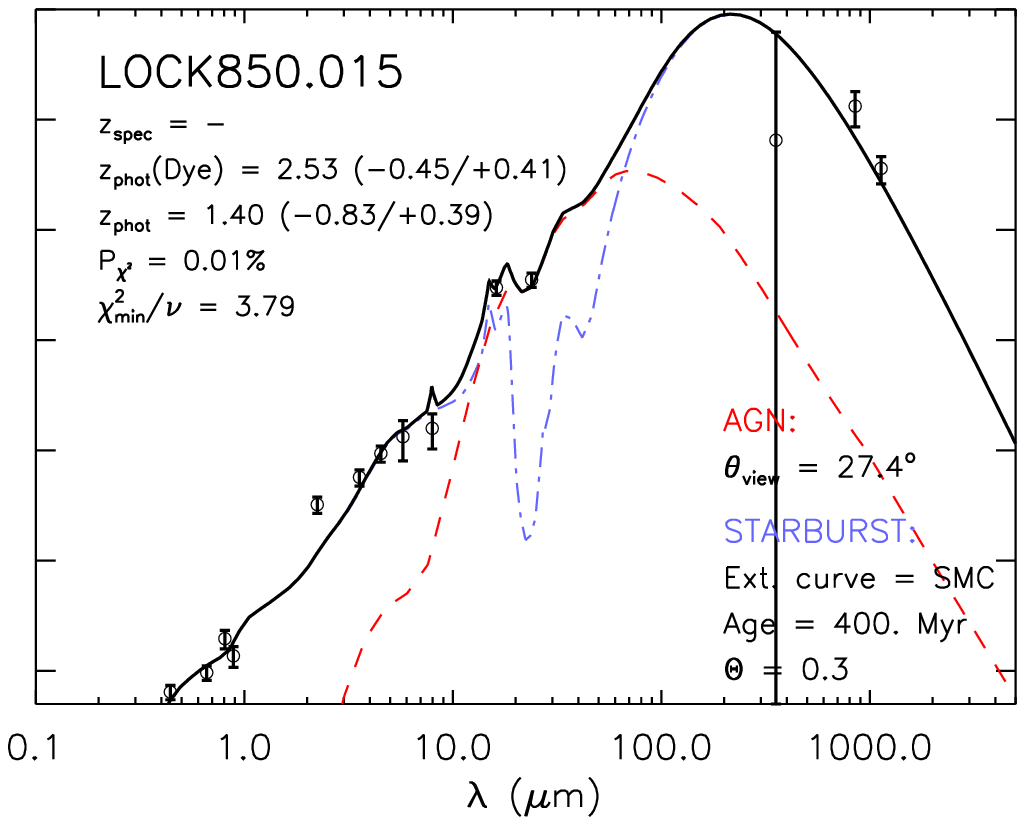}}\nolinebreak
\hspace*{-1.8cm}\resizebox{0.37\hsize}{!}{\includegraphics*{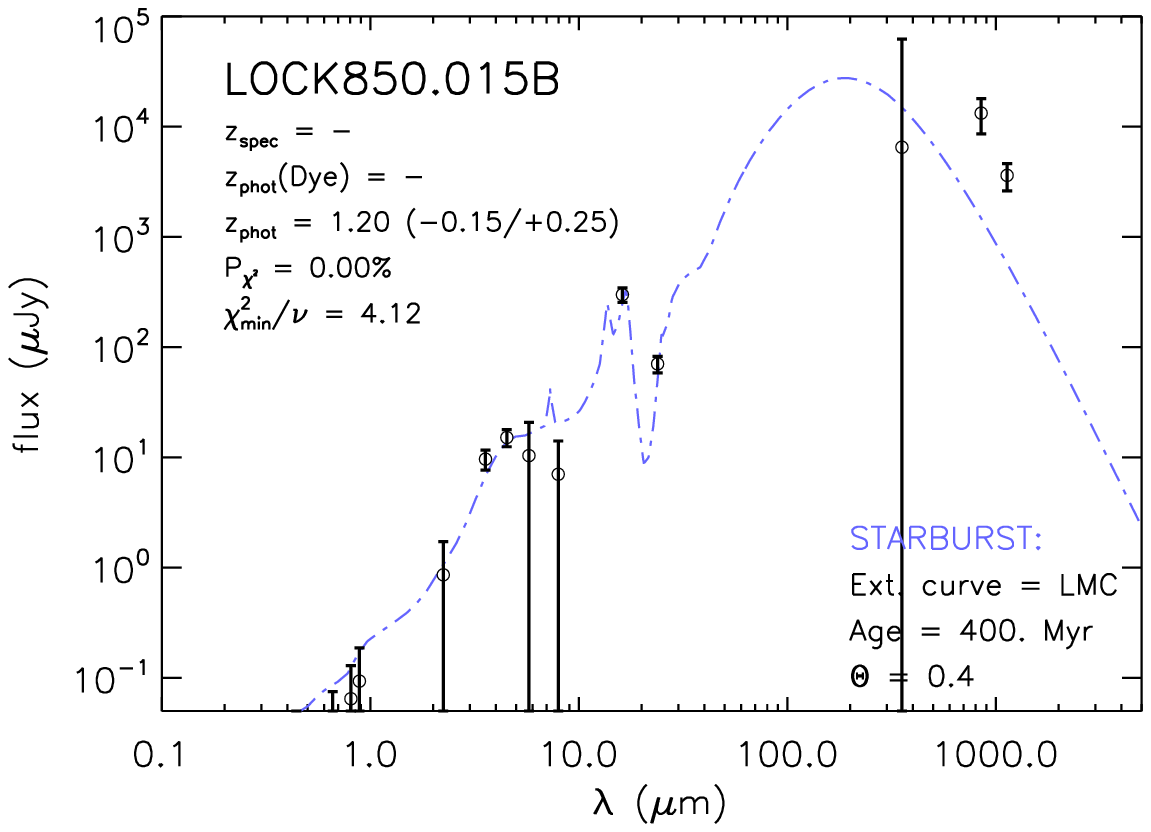}}\nolinebreak
\hspace*{-2.65cm}\resizebox{0.37\hsize}{!}{\includegraphics*{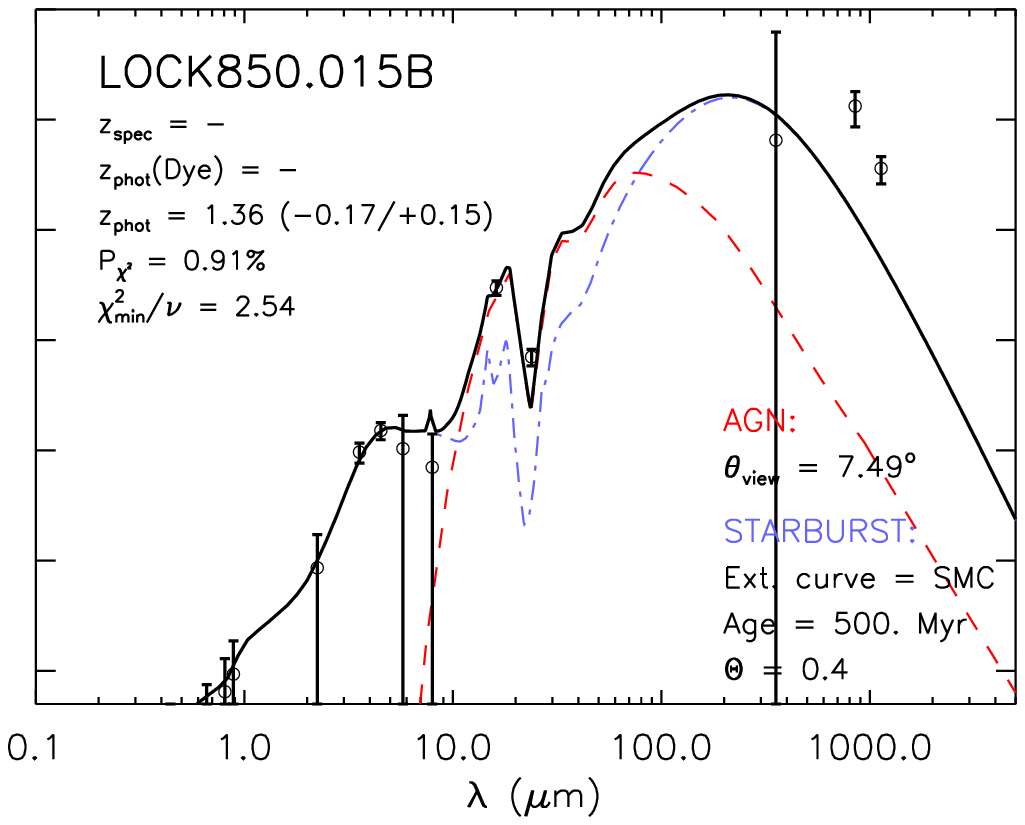}}\nolinebreak
\end{center}\vspace*{-1.3cm}
\begin{center}
\hspace*{-1.8cm}\resizebox{0.37\hsize}{!}{\includegraphics*{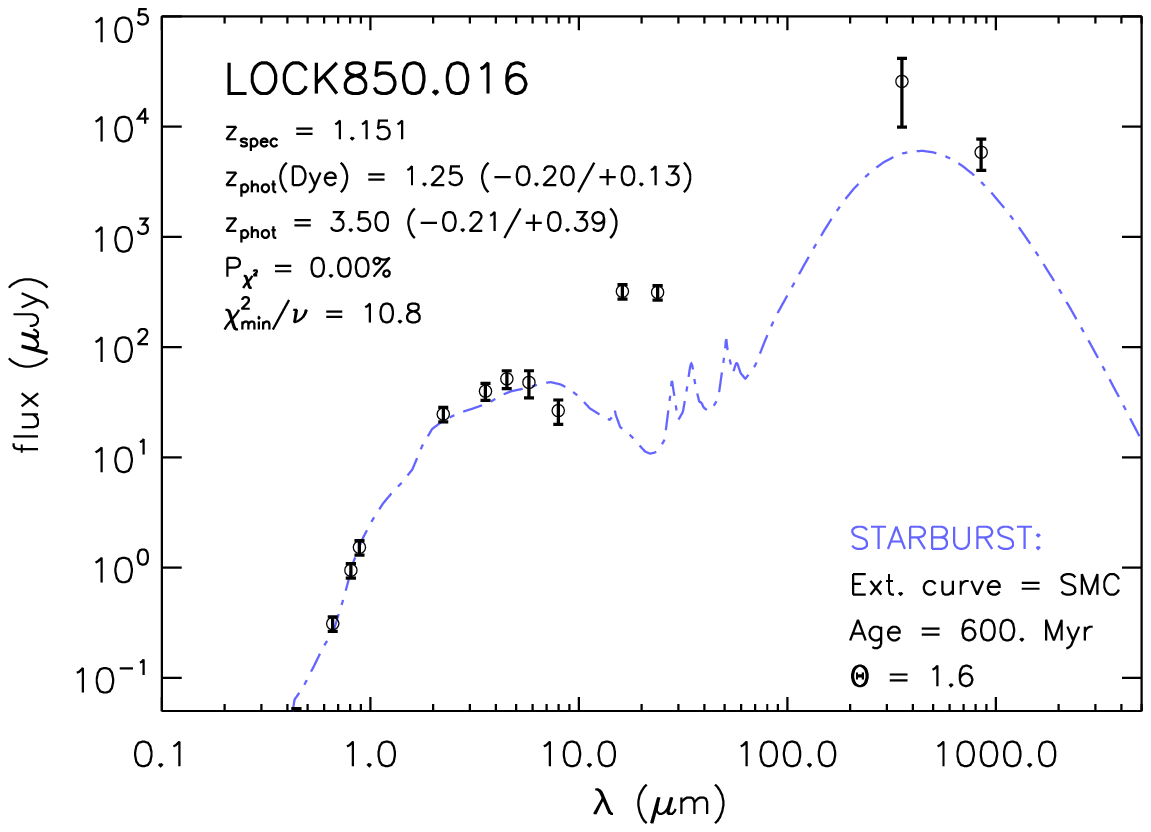}}\nolinebreak
\hspace*{-2.65cm}\resizebox{0.37\hsize}{!}{\includegraphics*{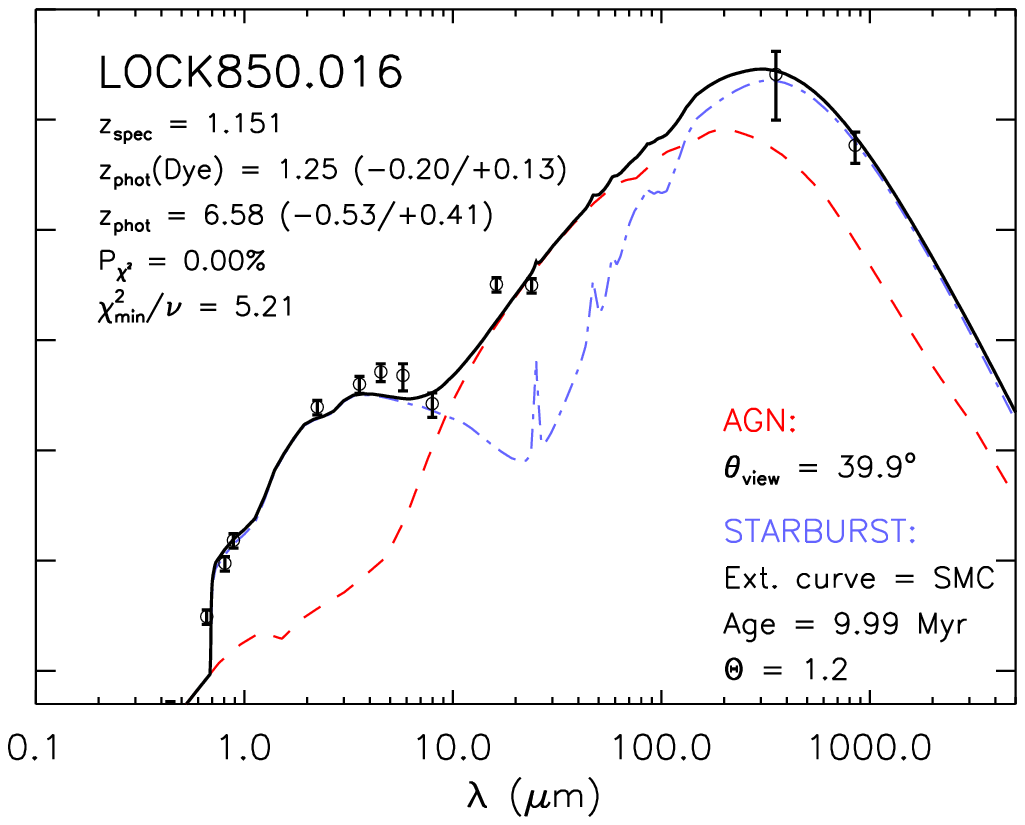}}\nolinebreak
\hspace*{-1.8cm}\resizebox{0.37\hsize}{!}{\includegraphics*{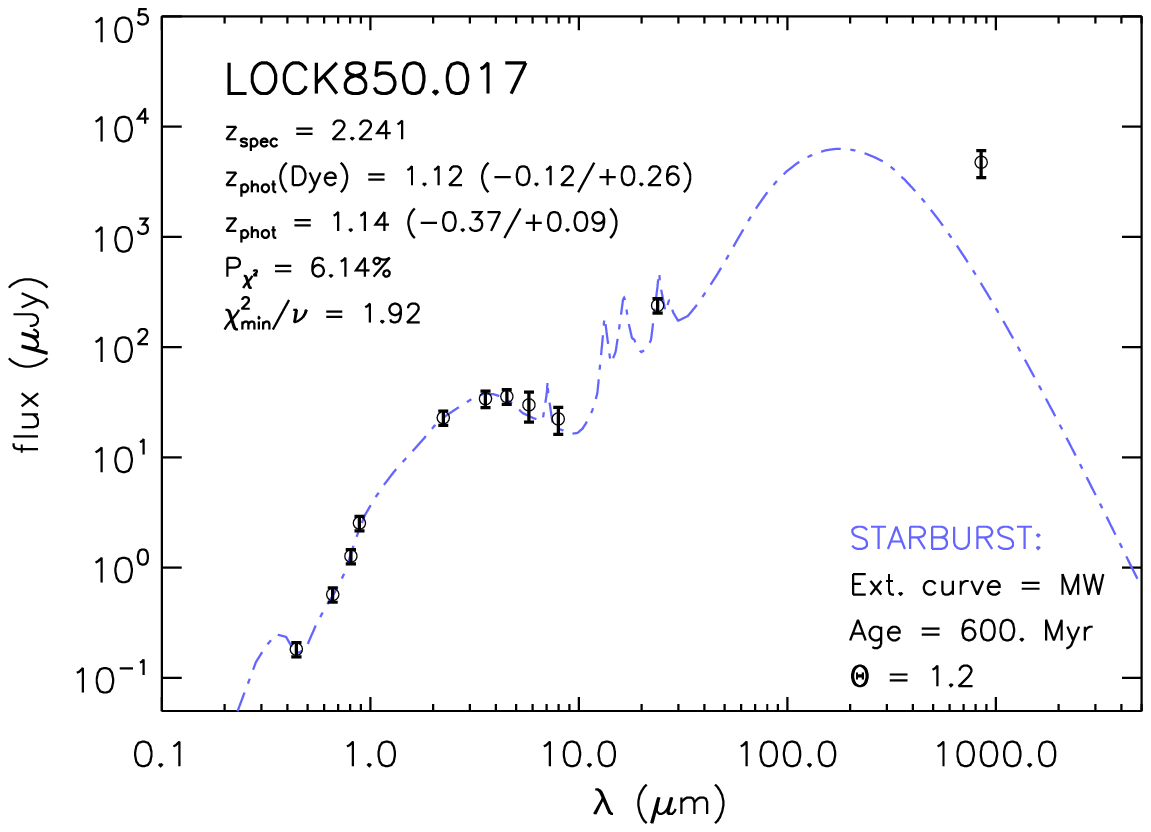}}\nolinebreak
\hspace*{-2.65cm}\resizebox{0.37\hsize}{!}{\includegraphics*{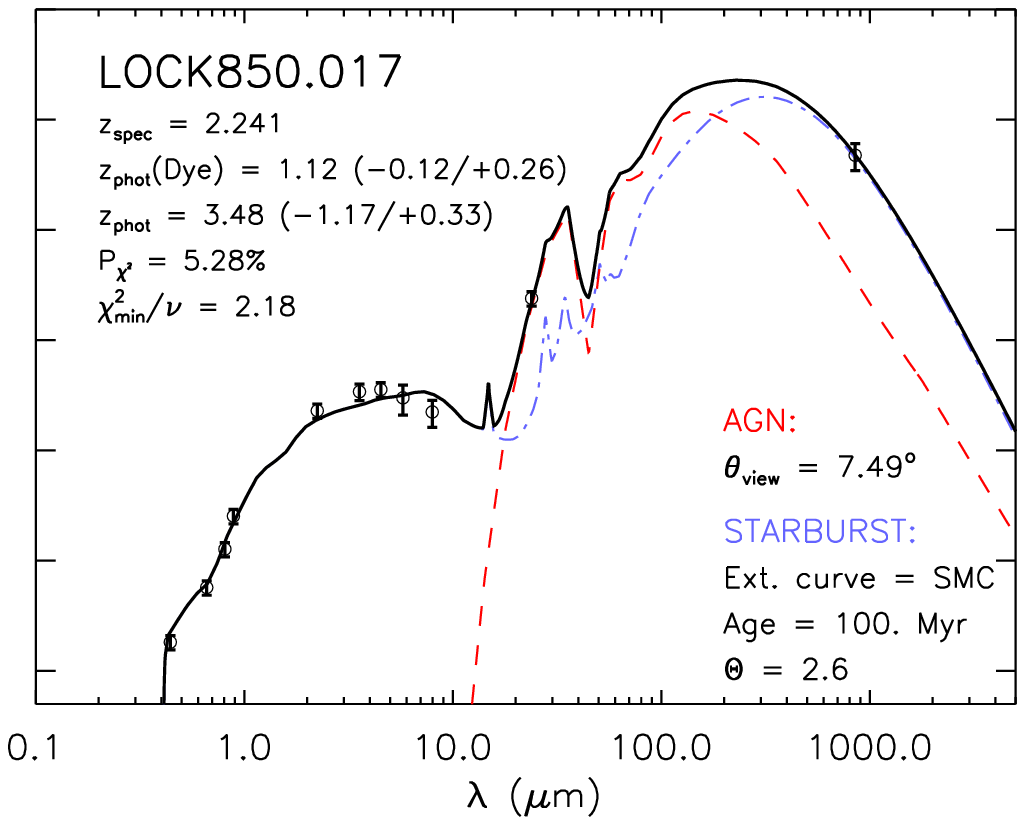}}\nolinebreak
\end{center}\vspace*{-1.3cm}
\begin{center}
\hspace*{-1.8cm}\resizebox{0.37\hsize}{!}{\includegraphics*{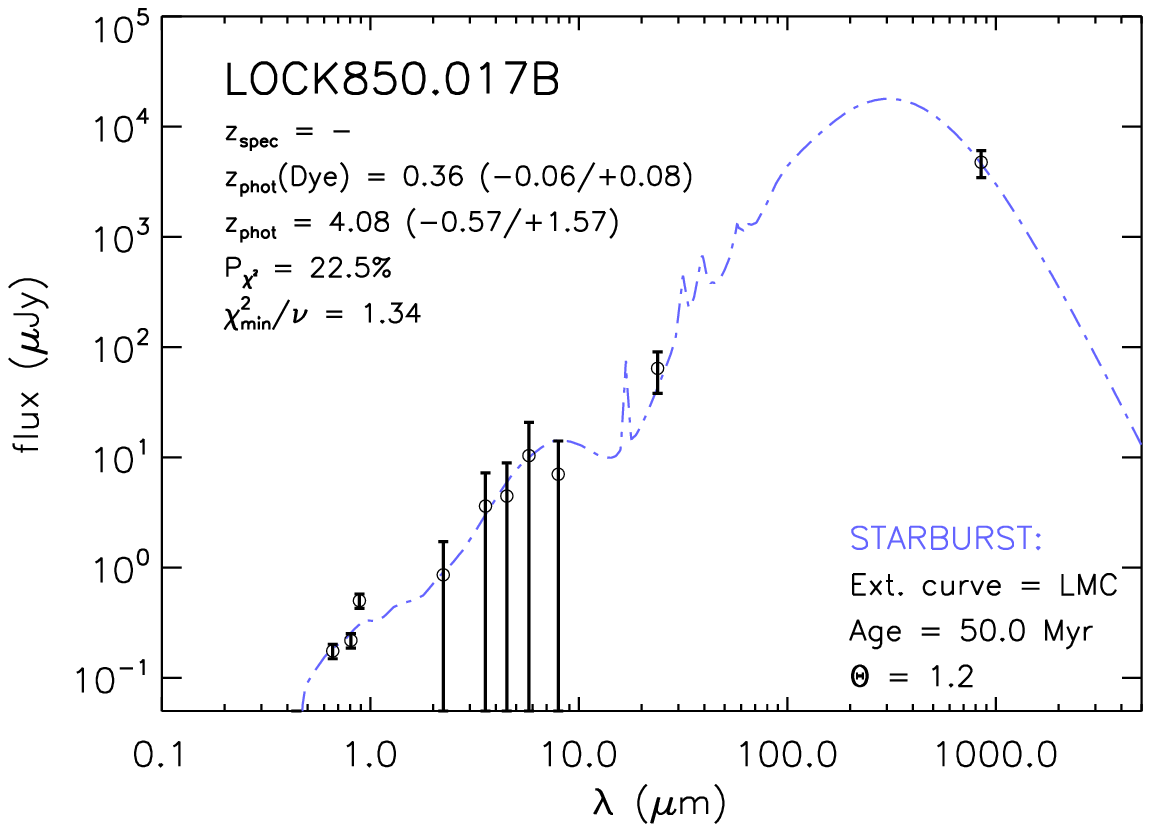}}\nolinebreak
\hspace*{-2.65cm}\resizebox{0.37\hsize}{!}{\includegraphics*{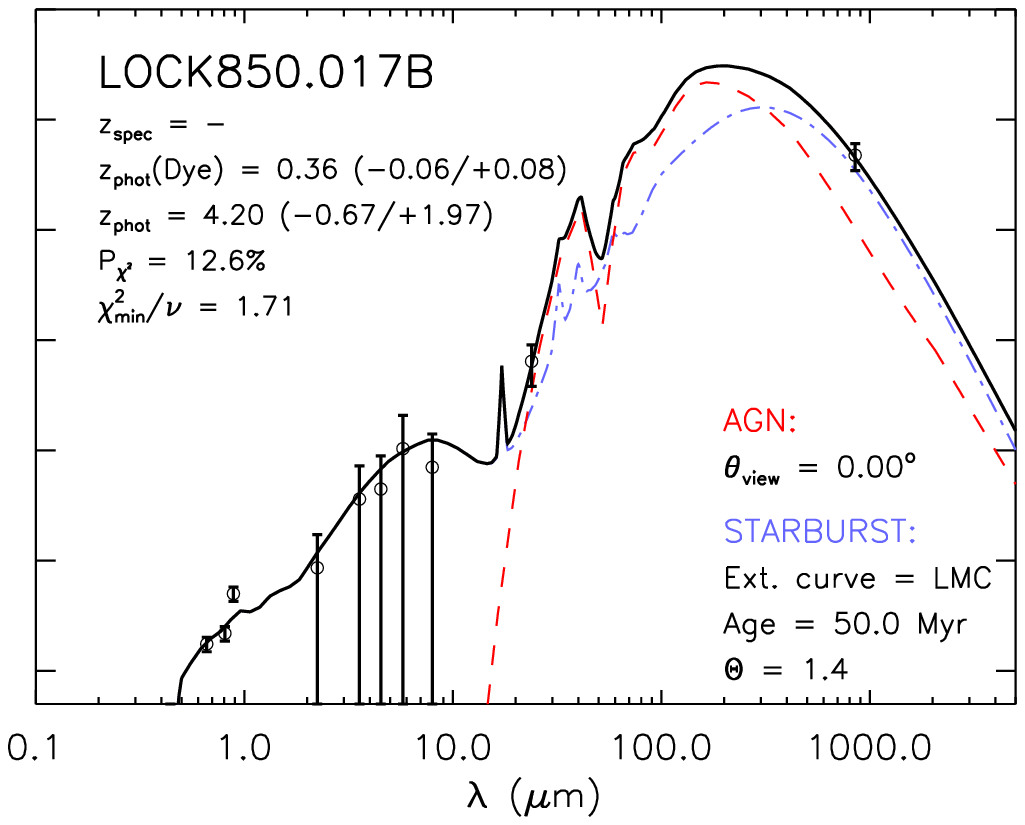}}\nolinebreak
\hspace*{-1.8cm}\resizebox{0.37\hsize}{!}{\includegraphics*{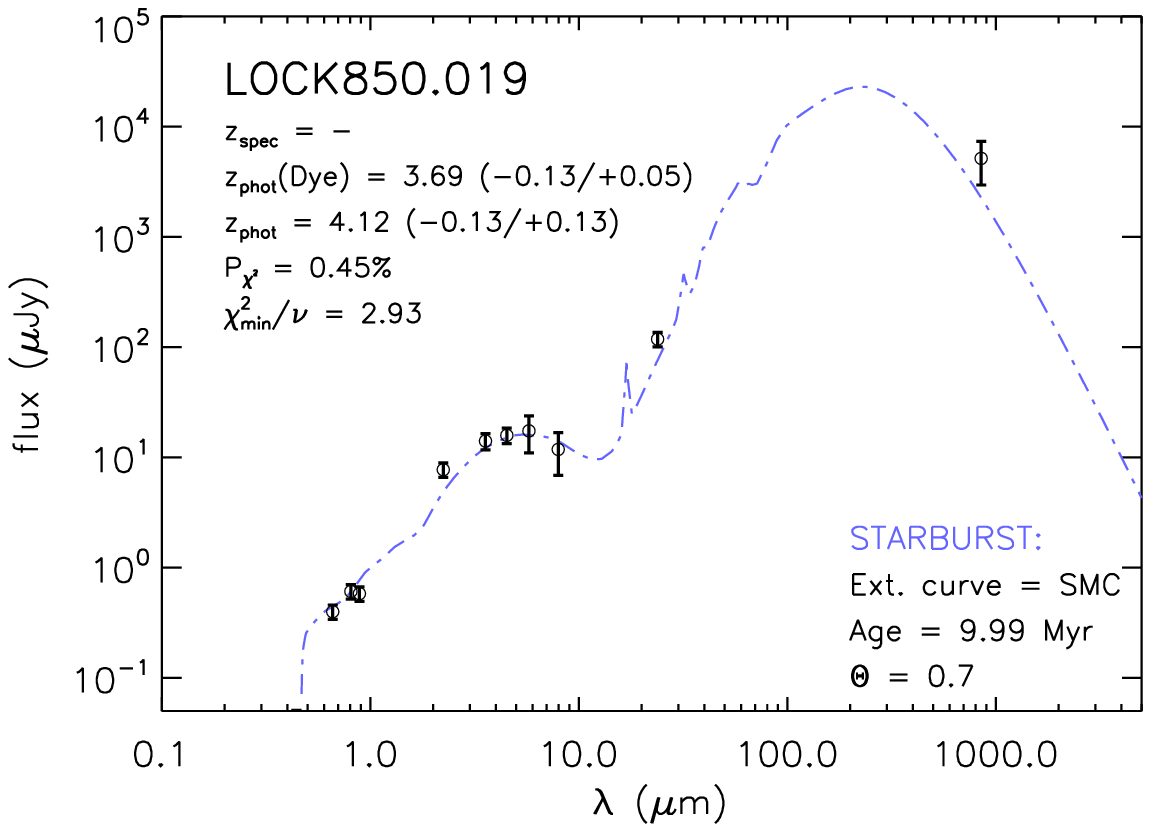}}\nolinebreak
\hspace*{-2.65cm}\resizebox{0.37\hsize}{!}{\includegraphics*{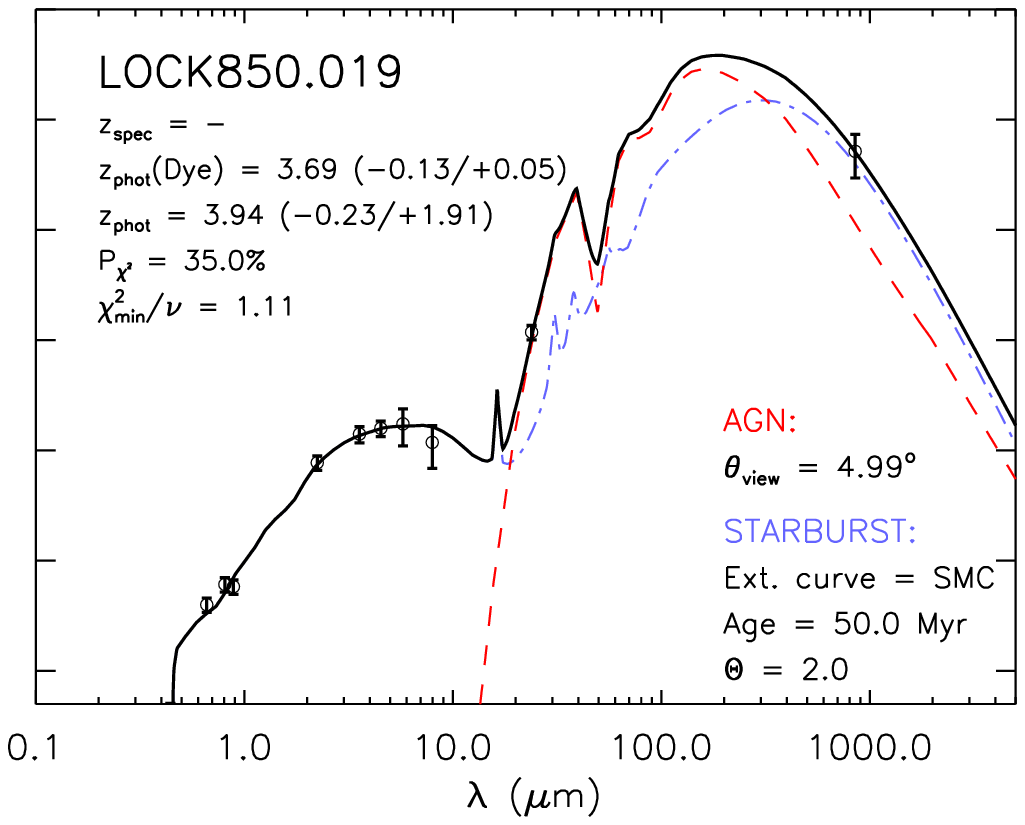}}\nolinebreak
\end{center}\vspace*{-1.3cm}
\begin{center}
\hspace*{-1.8cm}\resizebox{0.37\hsize}{!}{\includegraphics*{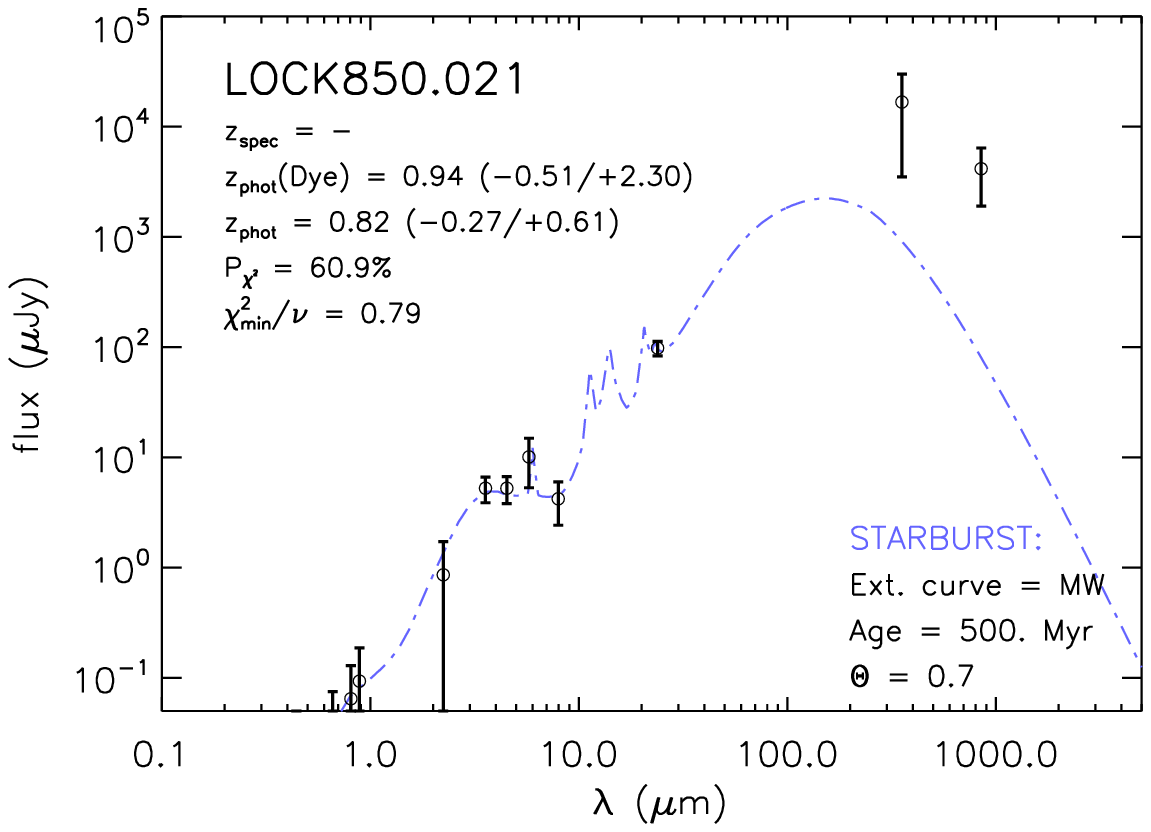}}\nolinebreak
\hspace*{-2.65cm}\resizebox{0.37\hsize}{!}{\includegraphics*{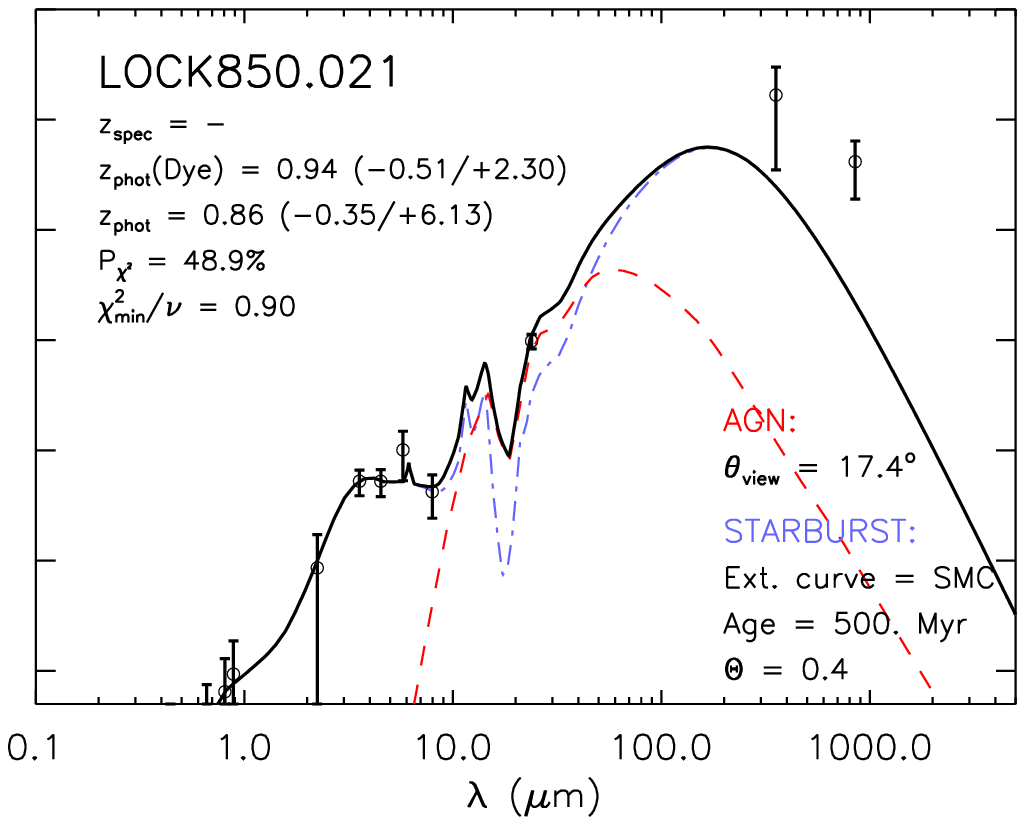}}\nolinebreak
\hspace*{-1.8cm}\resizebox{0.37\hsize}{!}{\includegraphics*{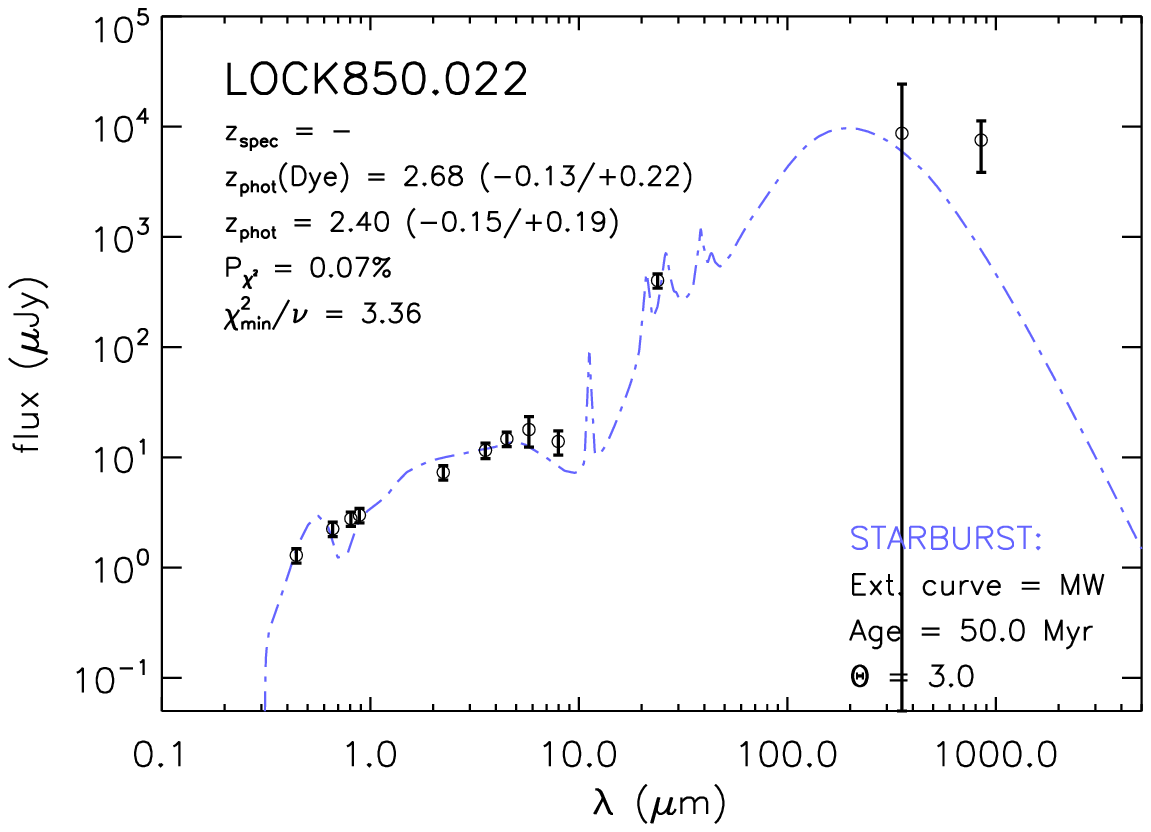}}\nolinebreak
\hspace*{-2.65cm}\resizebox{0.37\hsize}{!}{\includegraphics*{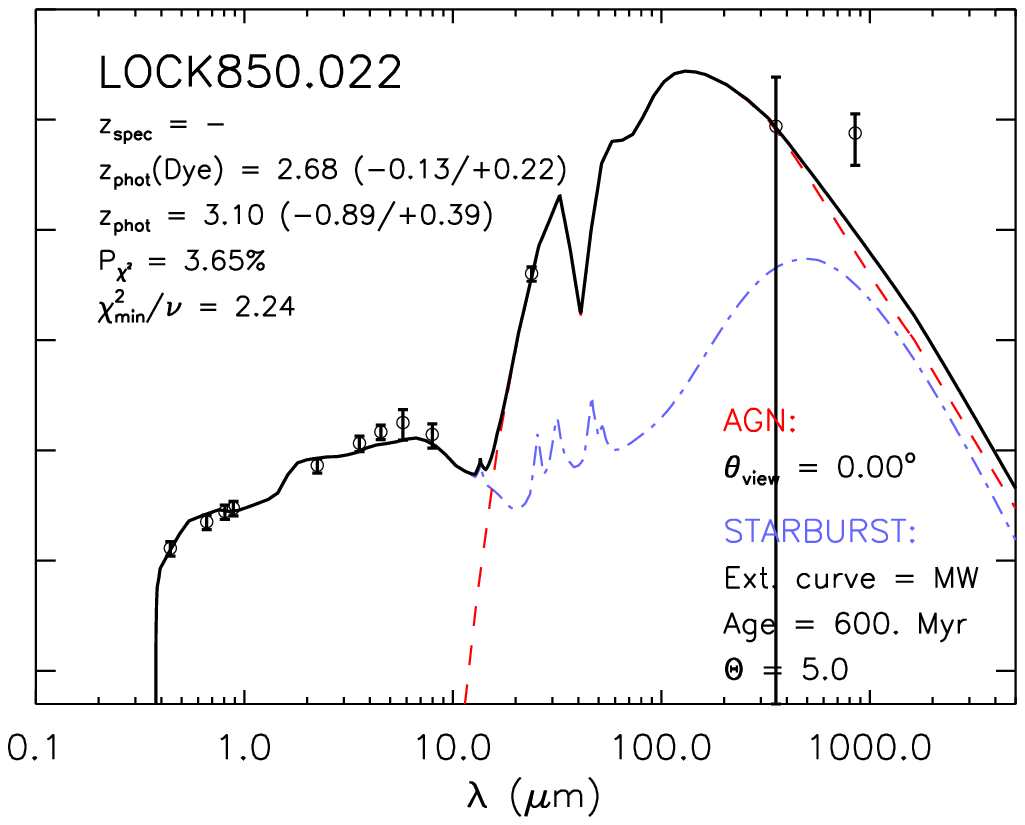}}\nolinebreak
\end{center}\vspace*{-1.3cm}
\begin{center}
\hspace*{-1.8cm}\resizebox{0.37\hsize}{!}{\includegraphics*{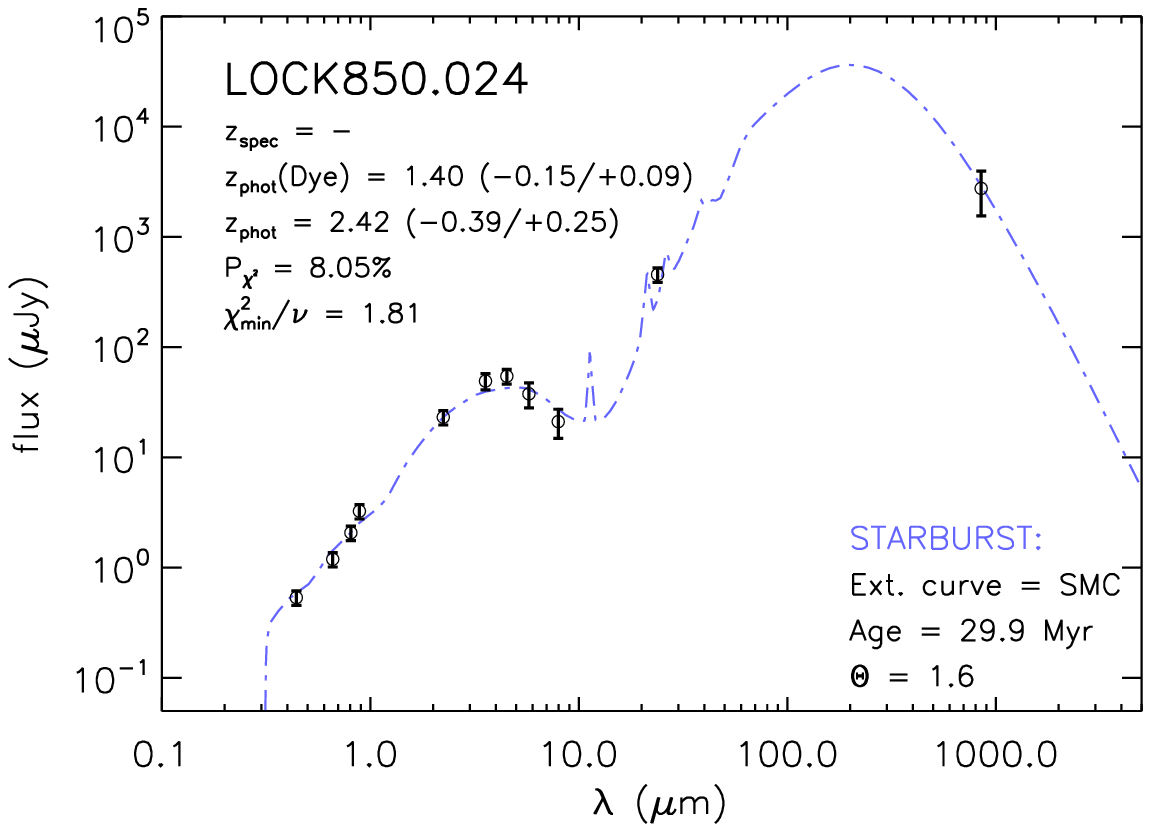}}\nolinebreak
\hspace*{-2.65cm}\resizebox{0.37\hsize}{!}{\includegraphics*{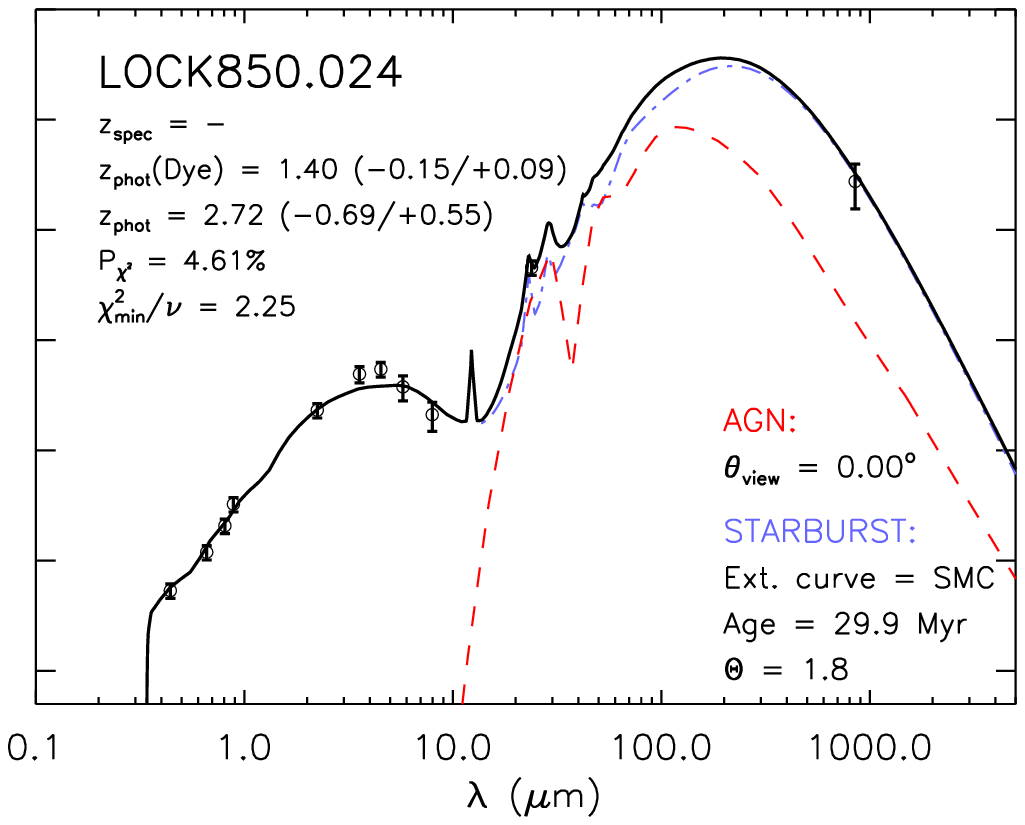}}\nolinebreak
\hspace*{-1.8cm}\resizebox{0.37\hsize}{!}{\includegraphics*{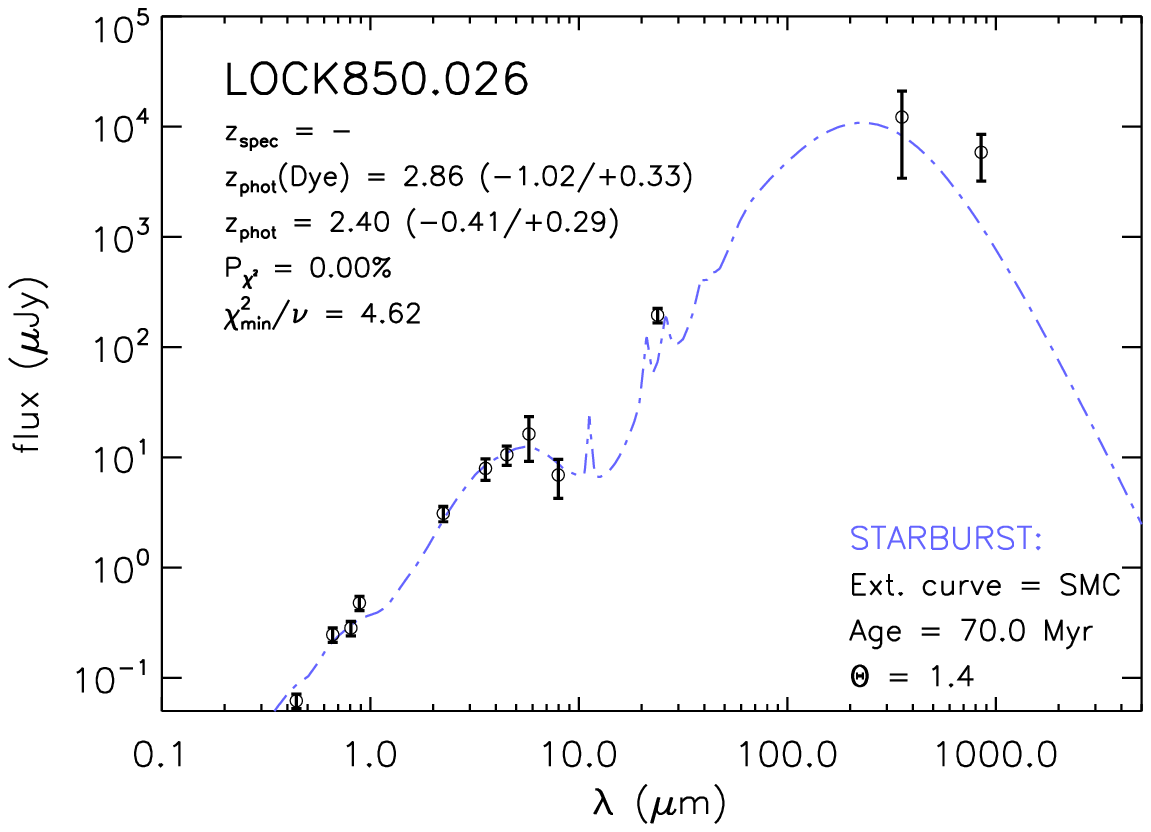}}\nolinebreak
\hspace*{-2.65cm}\resizebox{0.37\hsize}{!}{\includegraphics*{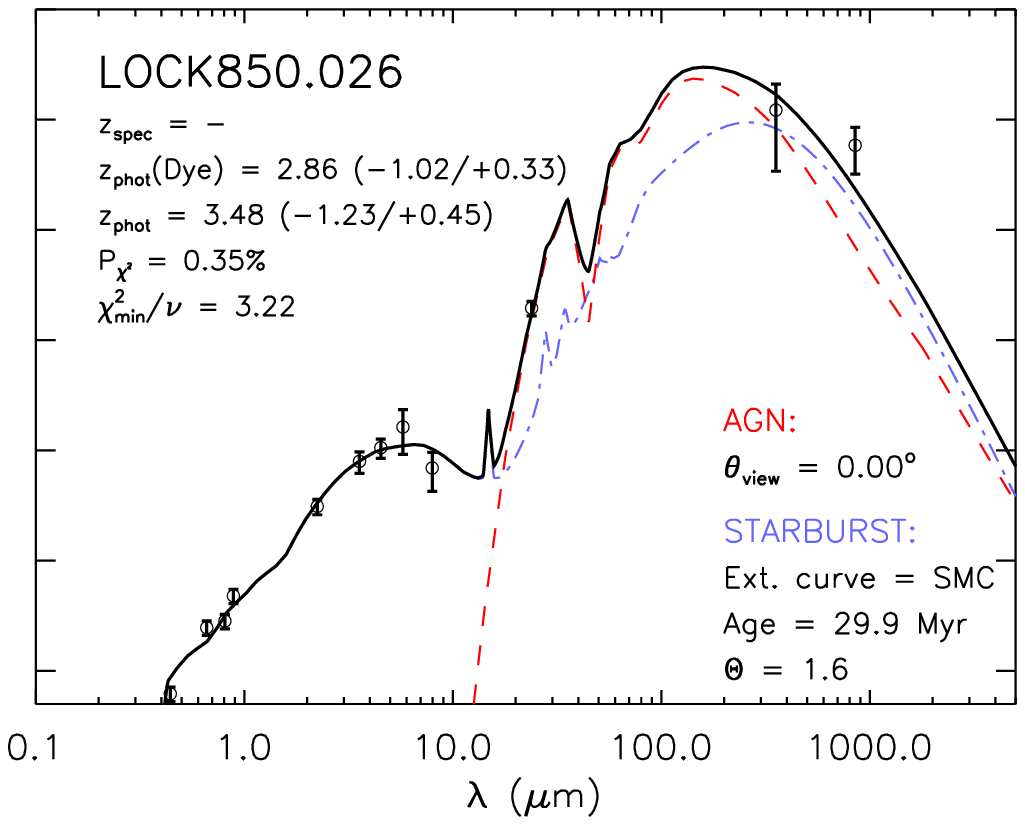}}\nolinebreak
\end{center}\vspace*{-1.3cm}
\begin{center}
\hspace*{-1.8cm}\resizebox{0.37\hsize}{!}{\includegraphics*{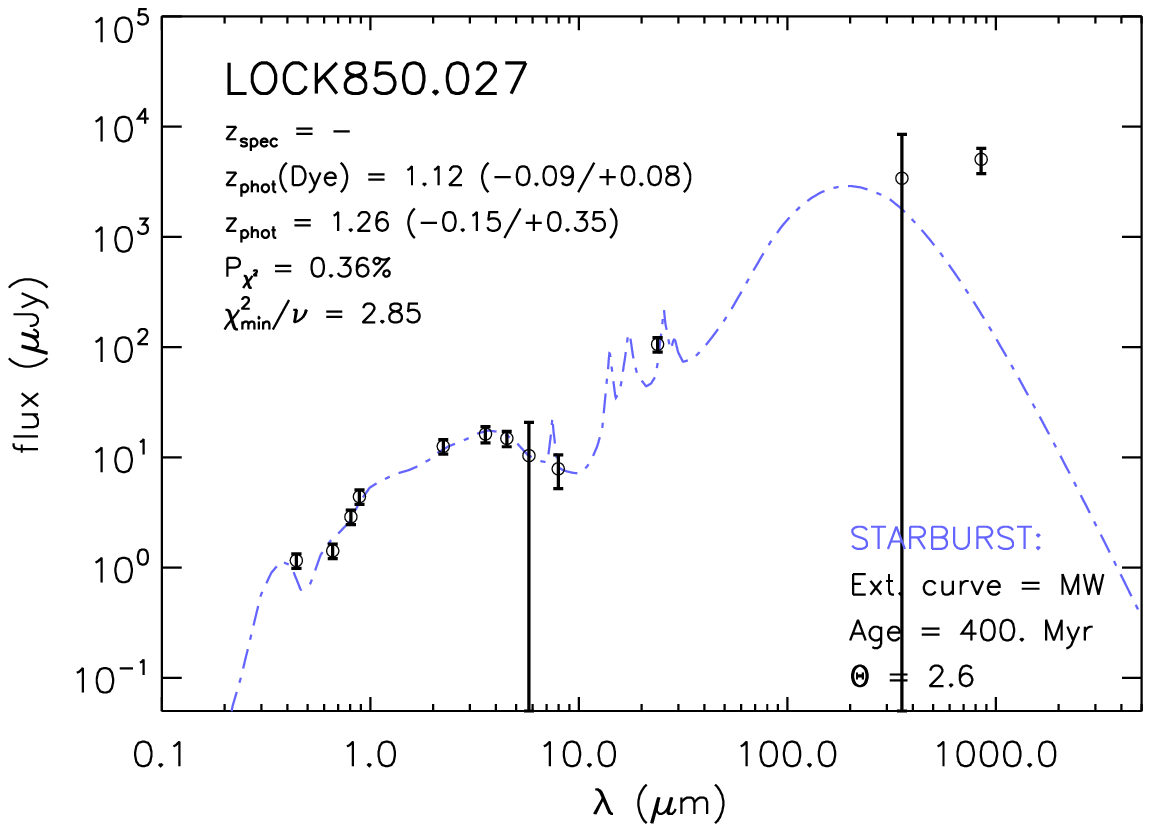}}\nolinebreak
\hspace*{-2.65cm}\resizebox{0.37\hsize}{!}{\includegraphics*{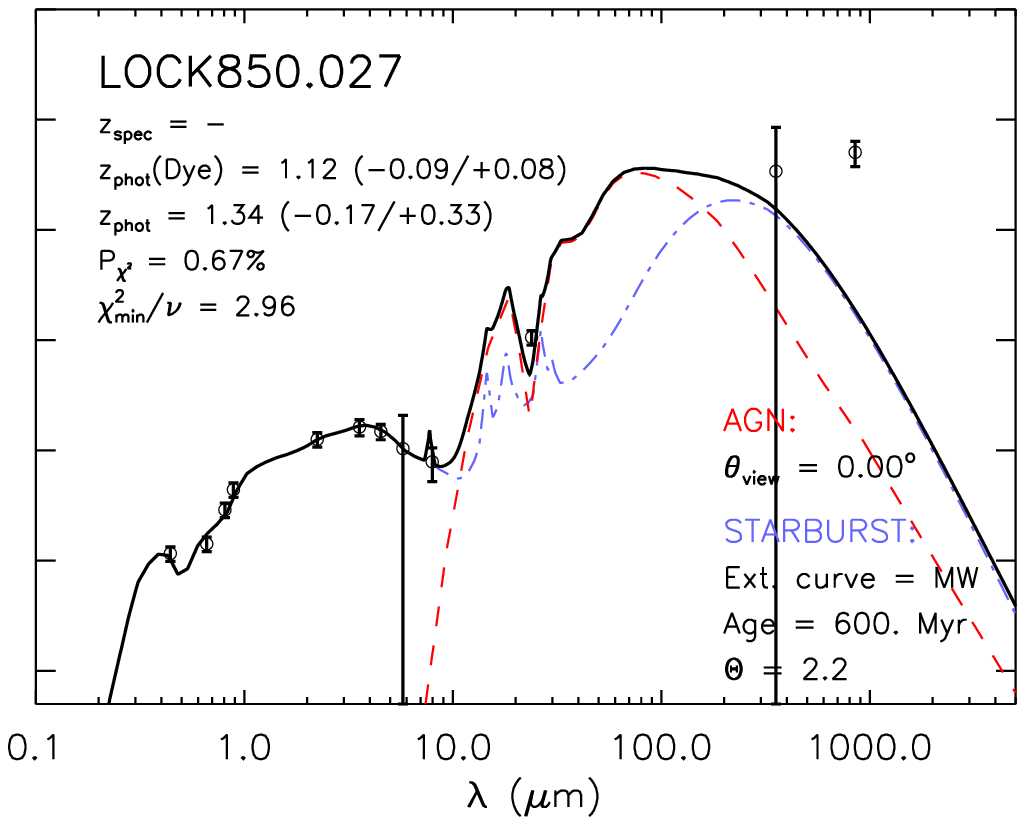}}\nolinebreak
\hspace*{-1.8cm}\resizebox{0.37\hsize}{!}{\includegraphics*{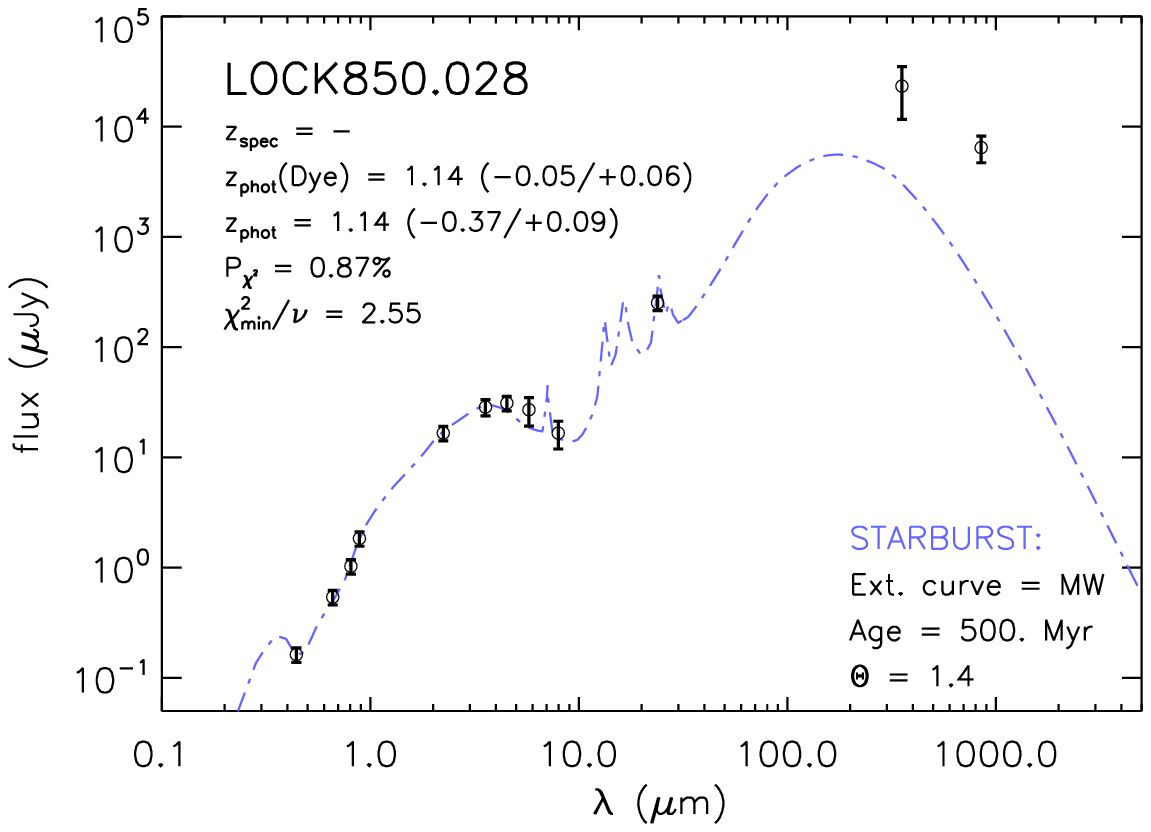}}\nolinebreak
\hspace*{-2.65cm}\resizebox{0.37\hsize}{!}{\includegraphics*{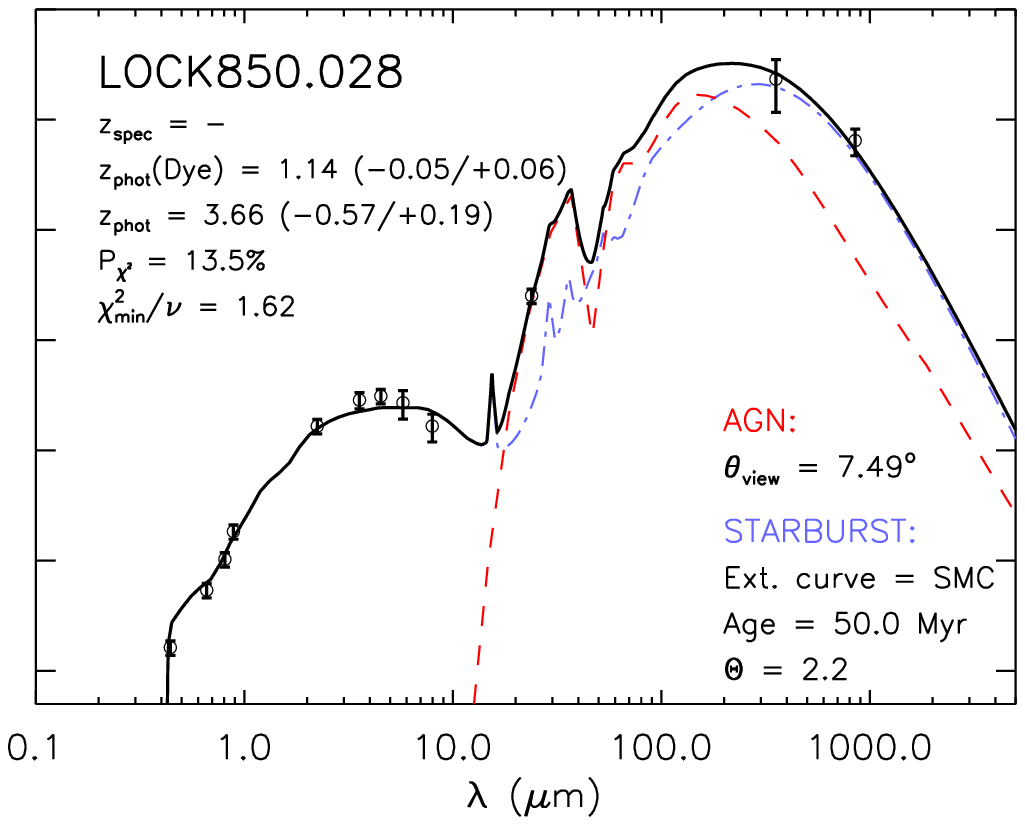}}\nolinebreak
\end{center}\vspace*{-1.3cm}
\vspace*{1.6cm}\caption{SED fits to submm-selected SHADES galaxies in the Lockman Hole East, using models from Takagi et al. (2003, 2004) and Efstathiou \& Rowan-Robinson (1995).}\label{fig:seds2}\end{figure*}

\begin{figure*}[!ht]
\begin{center}
\hspace*{-1.8cm}\resizebox{0.37\hsize}{!}{\includegraphics*{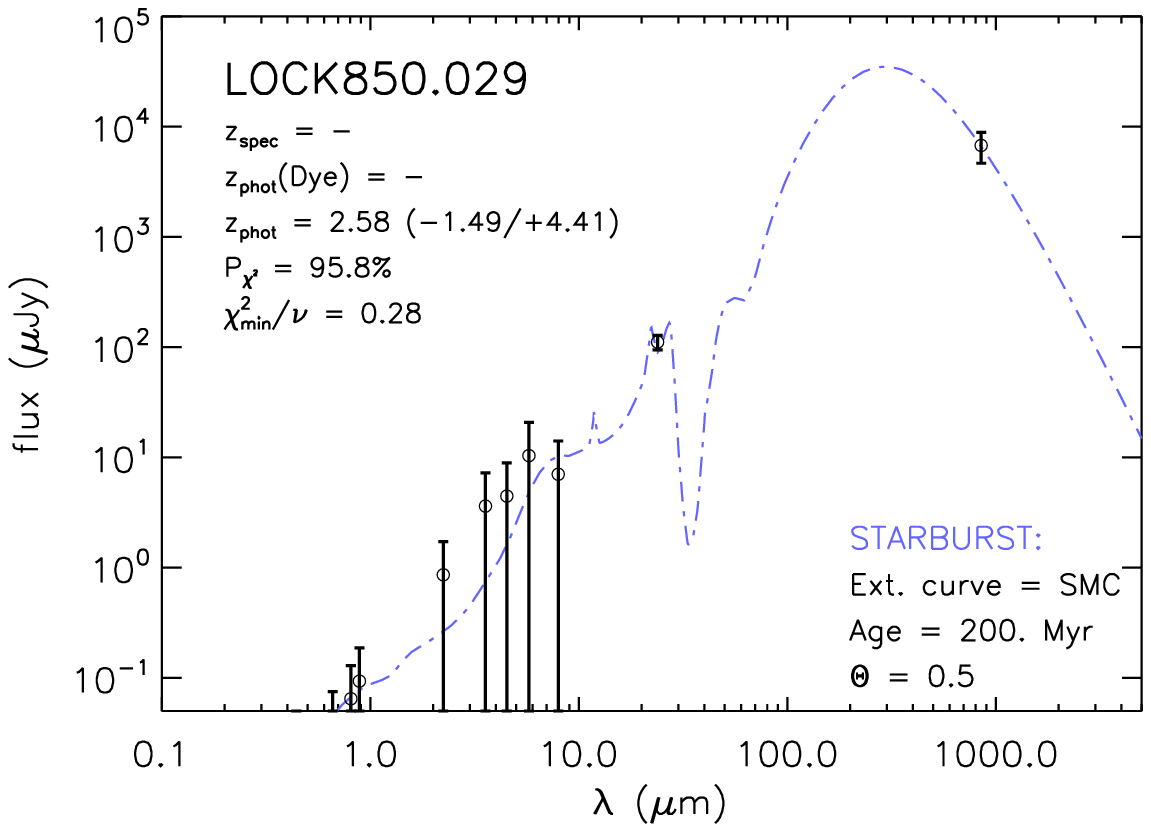}}\nolinebreak
\hspace*{-2.65cm}\resizebox{0.37\hsize}{!}{\includegraphics*{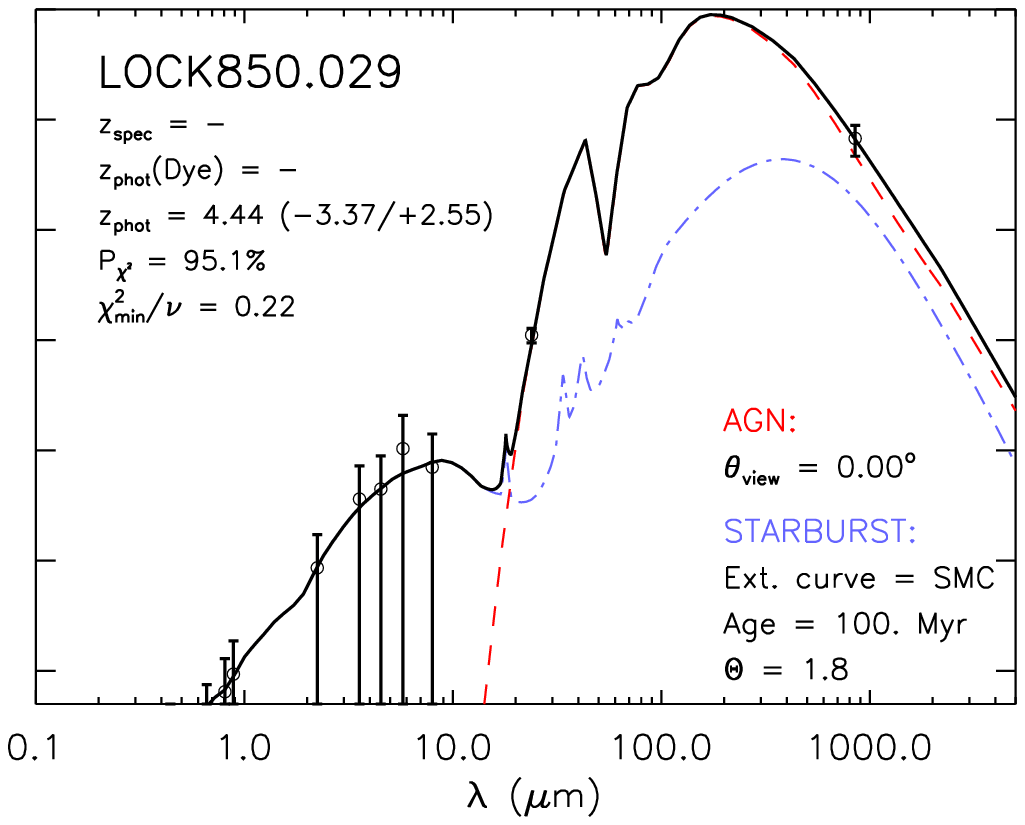}}\nolinebreak
\hspace*{-1.8cm}\resizebox{0.37\hsize}{!}{\includegraphics*{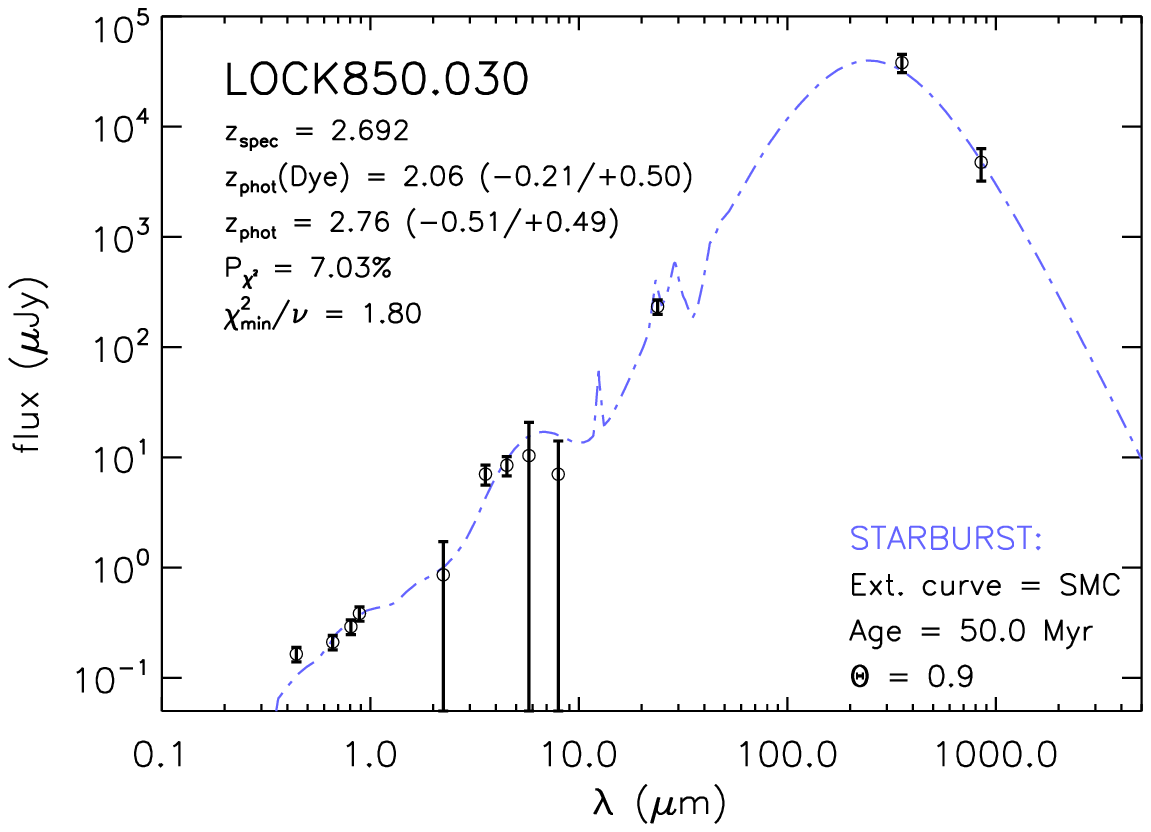}}\nolinebreak
\hspace*{-2.65cm}\resizebox{0.37\hsize}{!}{\includegraphics*{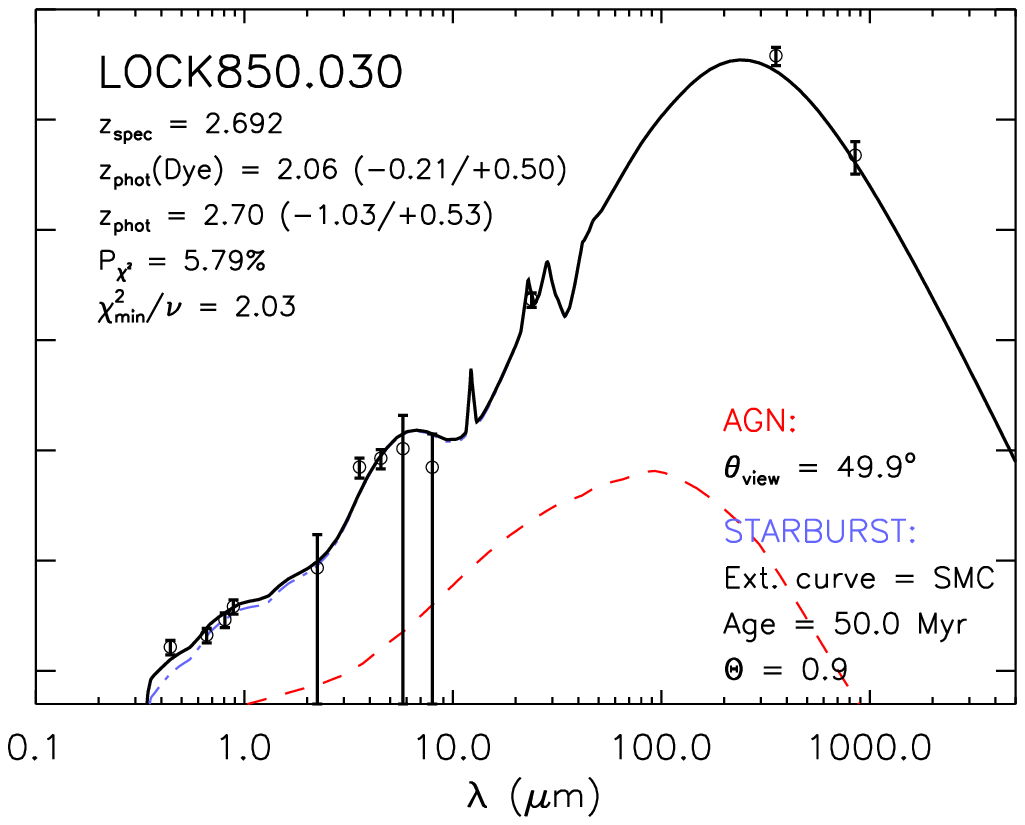}}\nolinebreak
\end{center}\vspace*{-1.3cm}
\begin{center}
\hspace*{-1.8cm}\resizebox{0.37\hsize}{!}{\includegraphics*{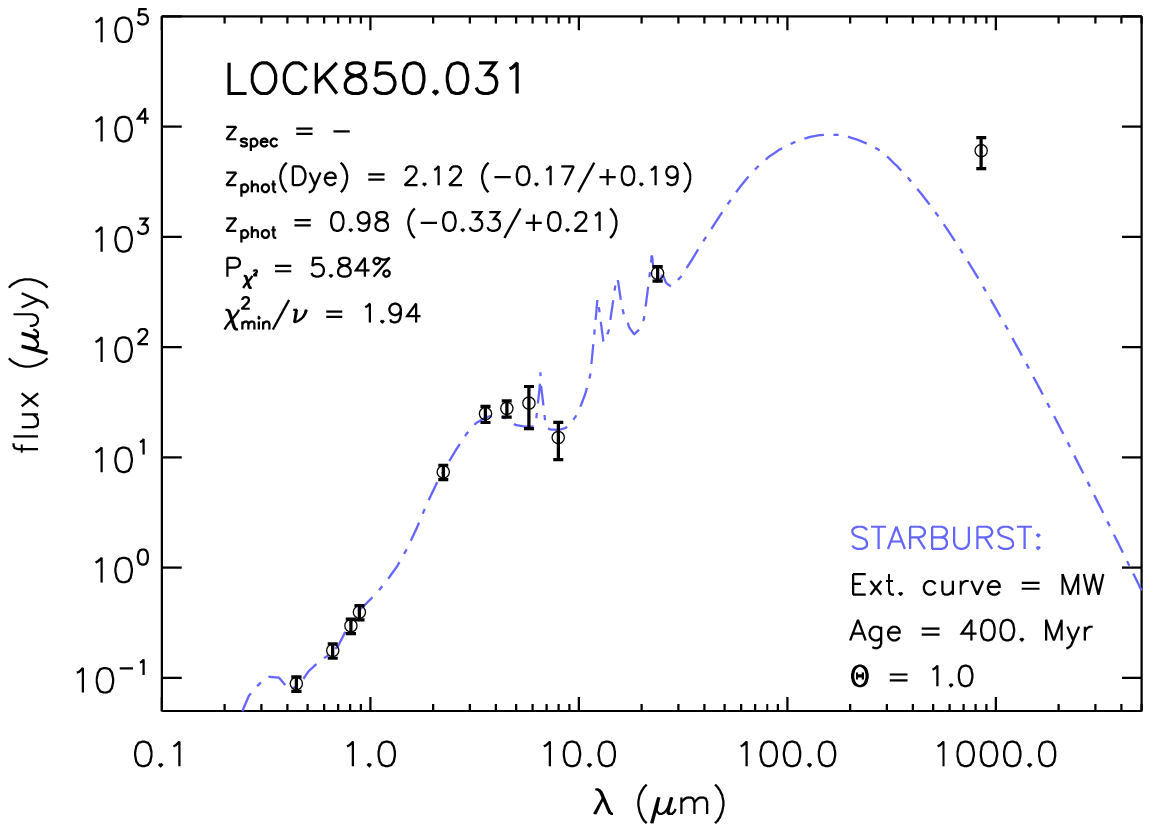}}\nolinebreak
\hspace*{-2.65cm}\resizebox{0.37\hsize}{!}{\includegraphics*{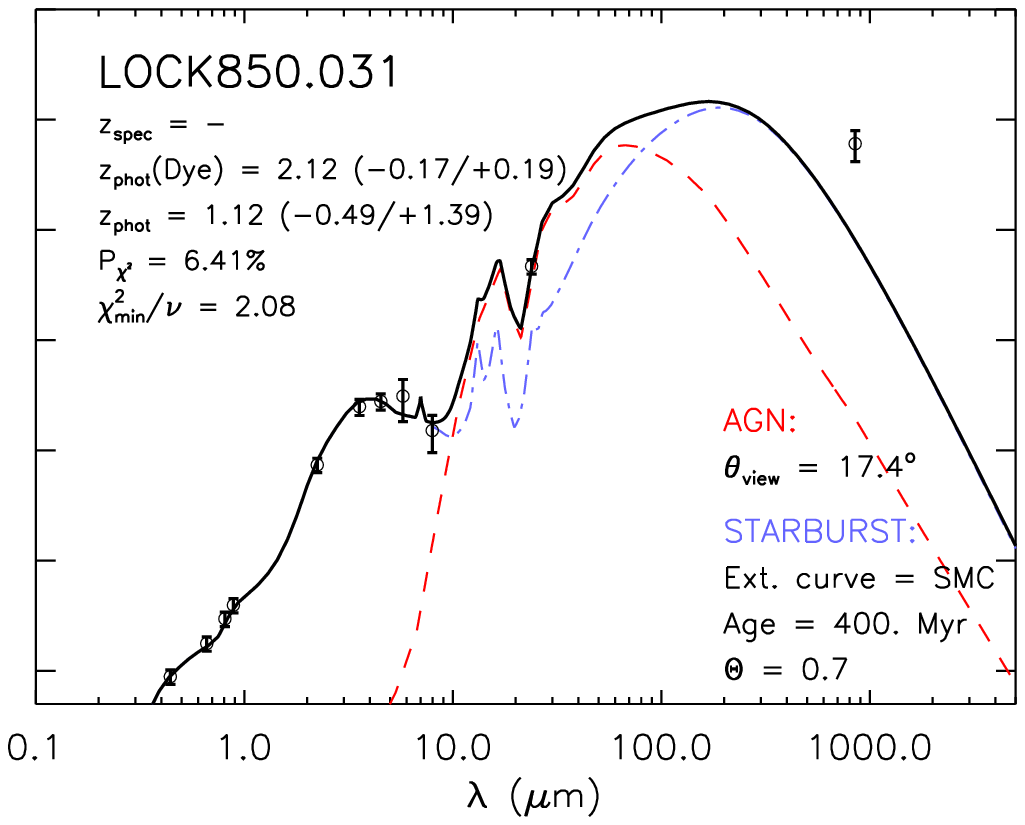}}\nolinebreak
\hspace*{-1.8cm}\resizebox{0.37\hsize}{!}{\includegraphics*{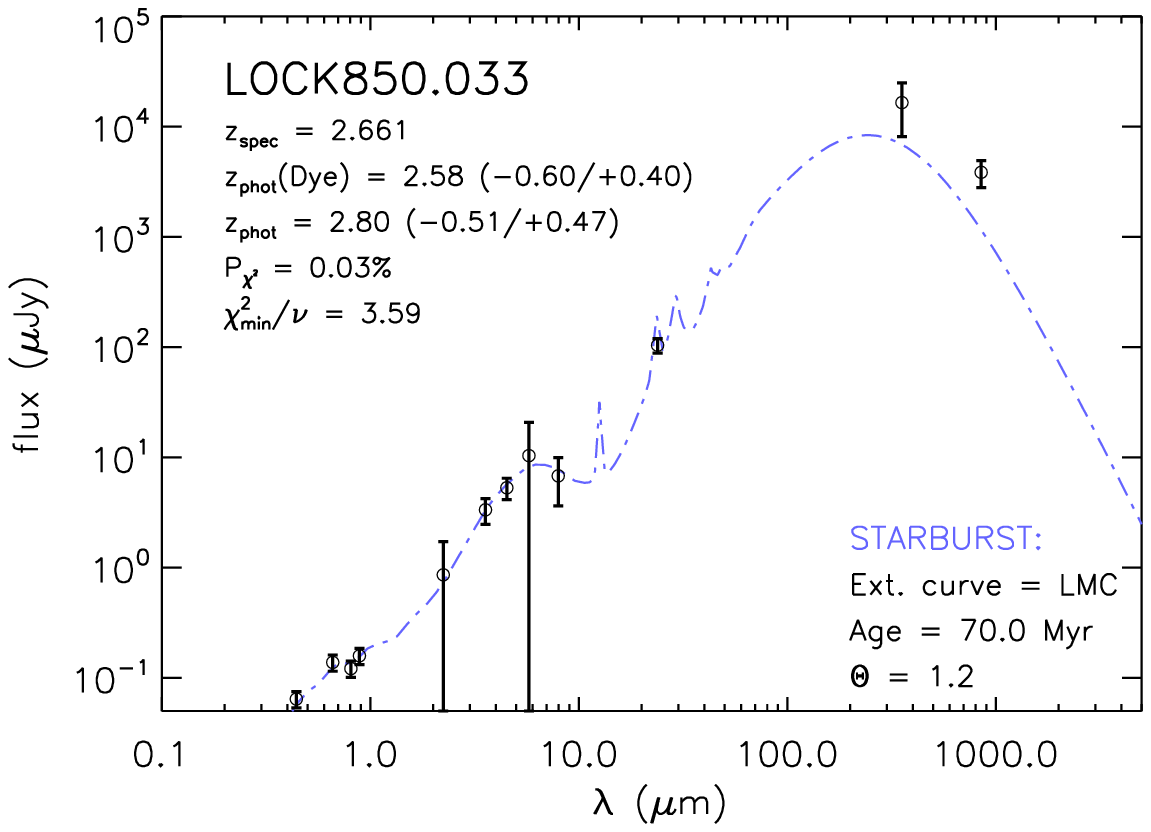}}\nolinebreak
\hspace*{-2.65cm}\resizebox{0.37\hsize}{!}{\includegraphics*{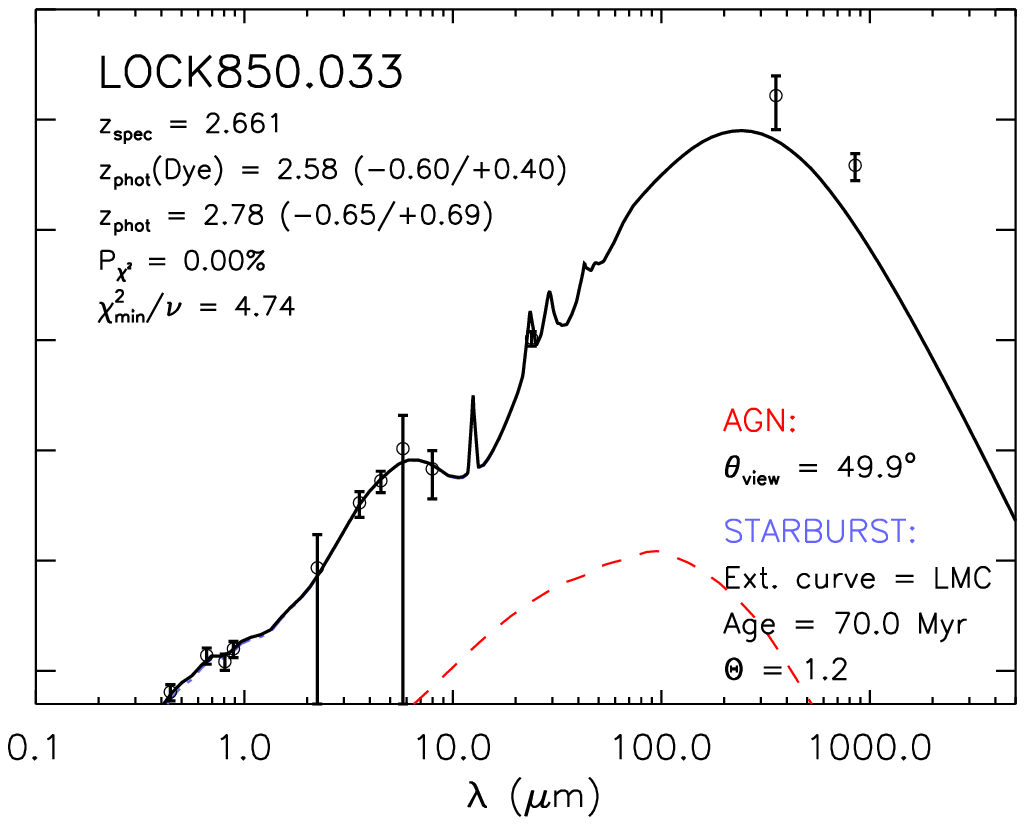}}\nolinebreak
\end{center}\vspace*{-1.3cm}
\begin{center}
\hspace*{-1.8cm}\resizebox{0.37\hsize}{!}{\includegraphics*{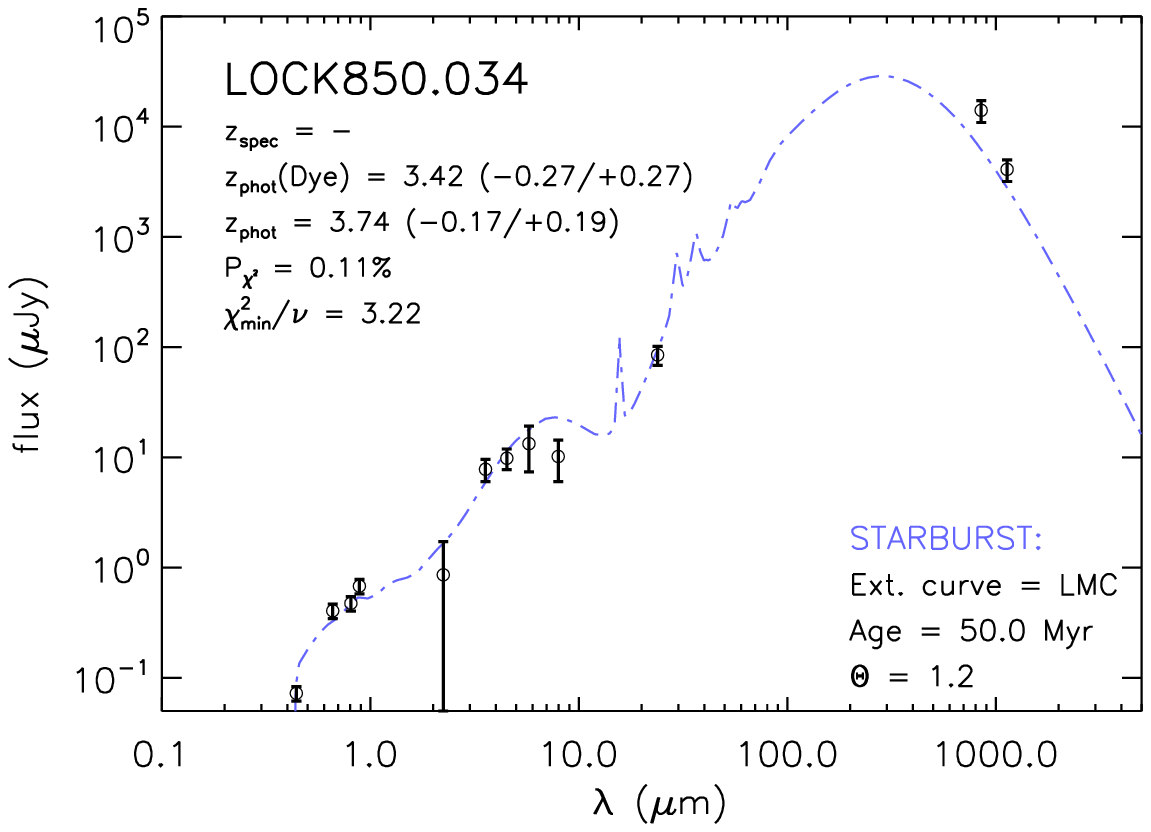}}\nolinebreak
\hspace*{-2.65cm}\resizebox{0.37\hsize}{!}{\includegraphics*{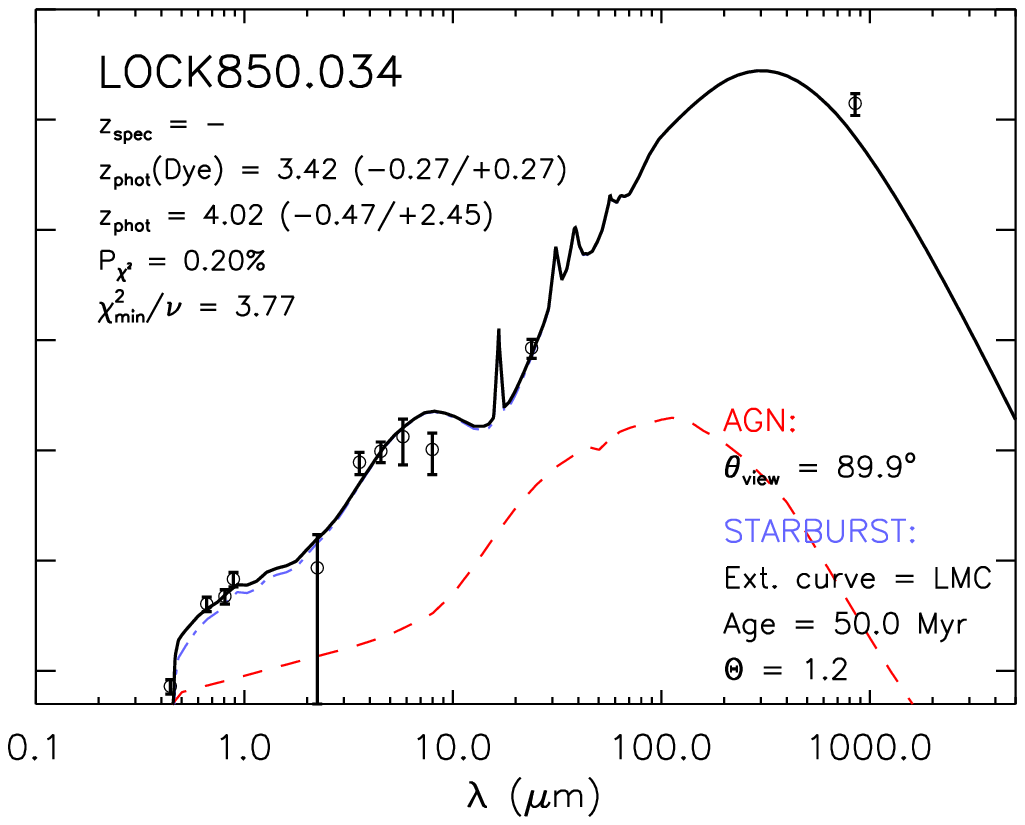}}\nolinebreak
\hspace*{-1.8cm}\resizebox{0.37\hsize}{!}{\includegraphics*{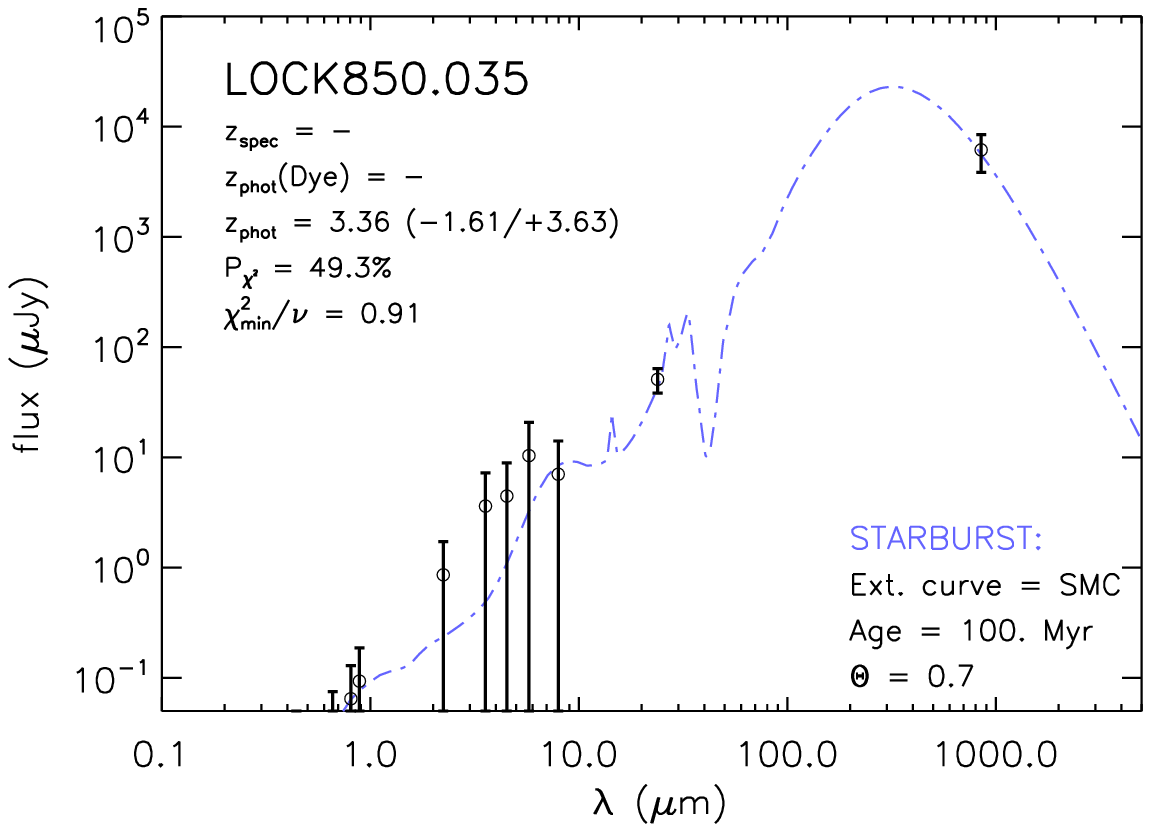}}\nolinebreak
\hspace*{-2.65cm}\resizebox{0.37\hsize}{!}{\includegraphics*{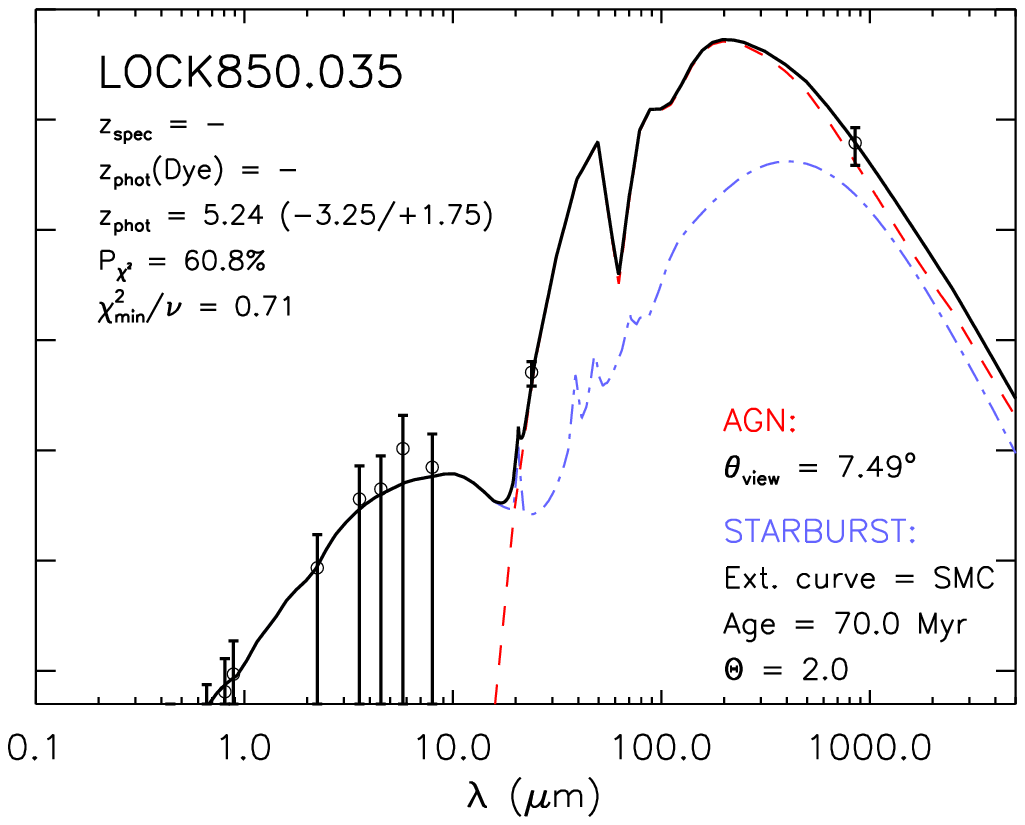}}\nolinebreak
\end{center}\vspace*{-1.3cm}
\begin{center}
\hspace*{-1.8cm}\resizebox{0.37\hsize}{!}{\includegraphics*{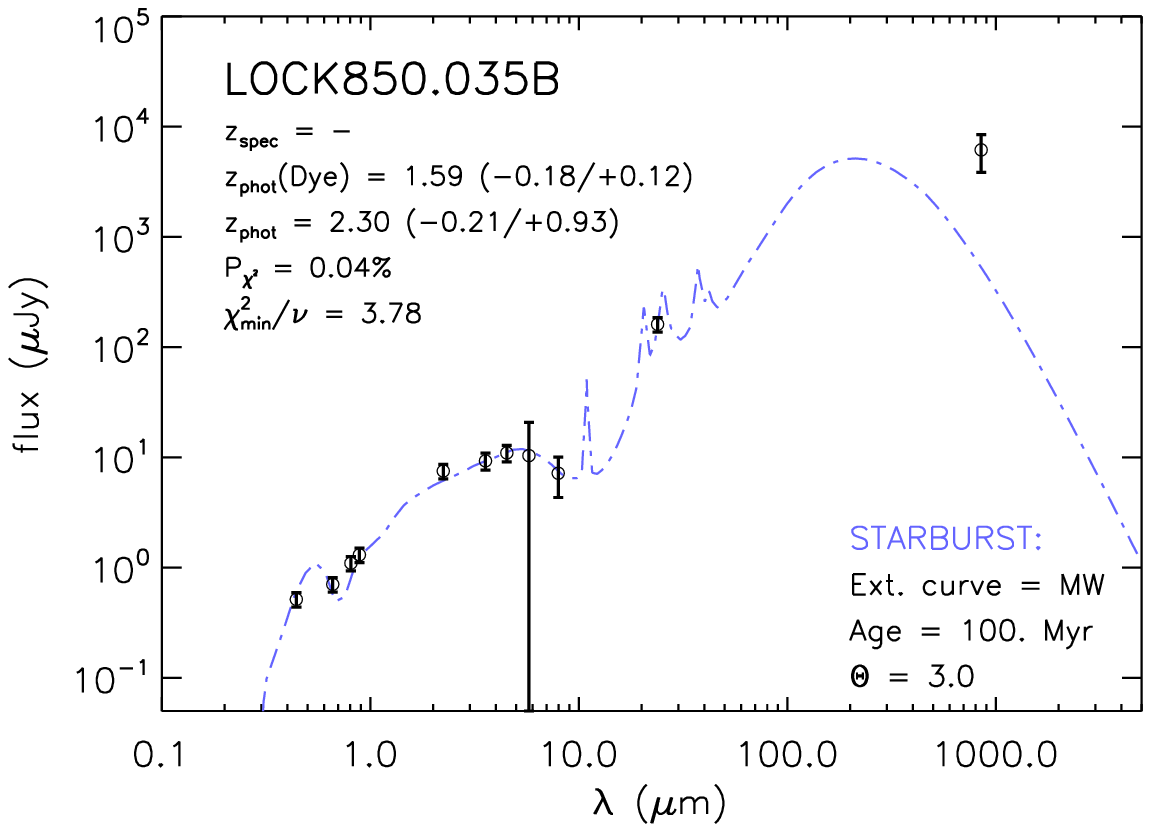}}\nolinebreak
\hspace*{-2.65cm}\resizebox{0.37\hsize}{!}{\includegraphics*{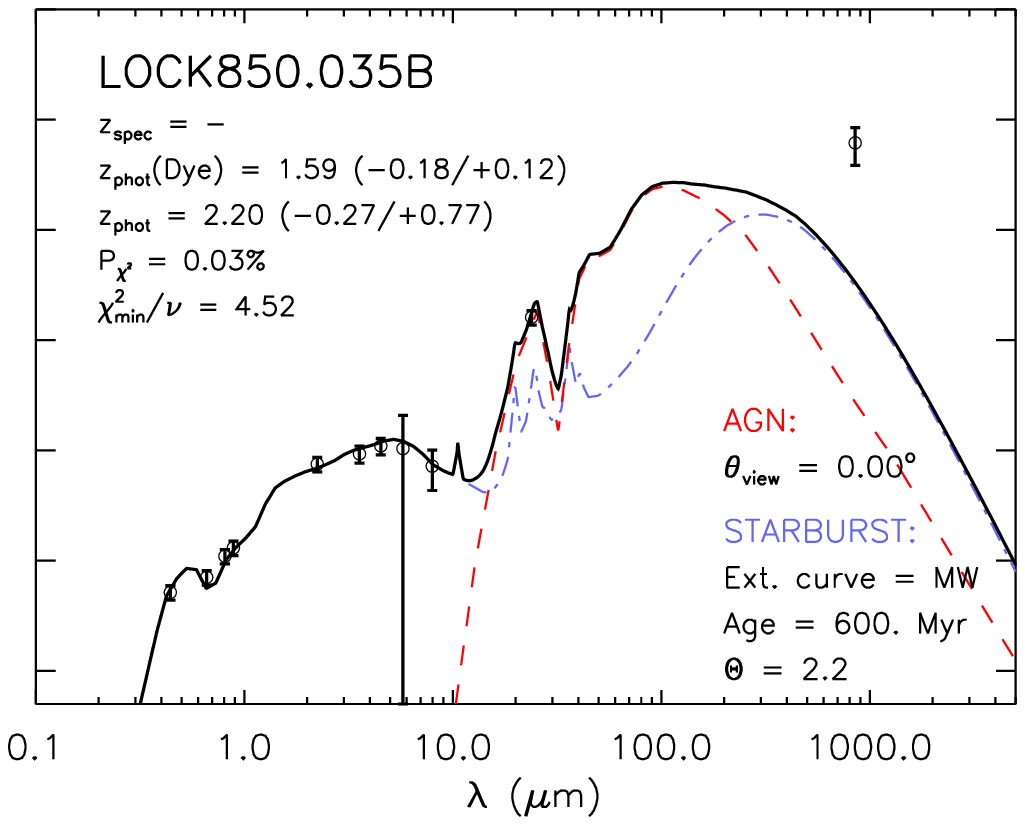}}\nolinebreak
\hspace*{-1.8cm}\resizebox{0.37\hsize}{!}{\includegraphics*{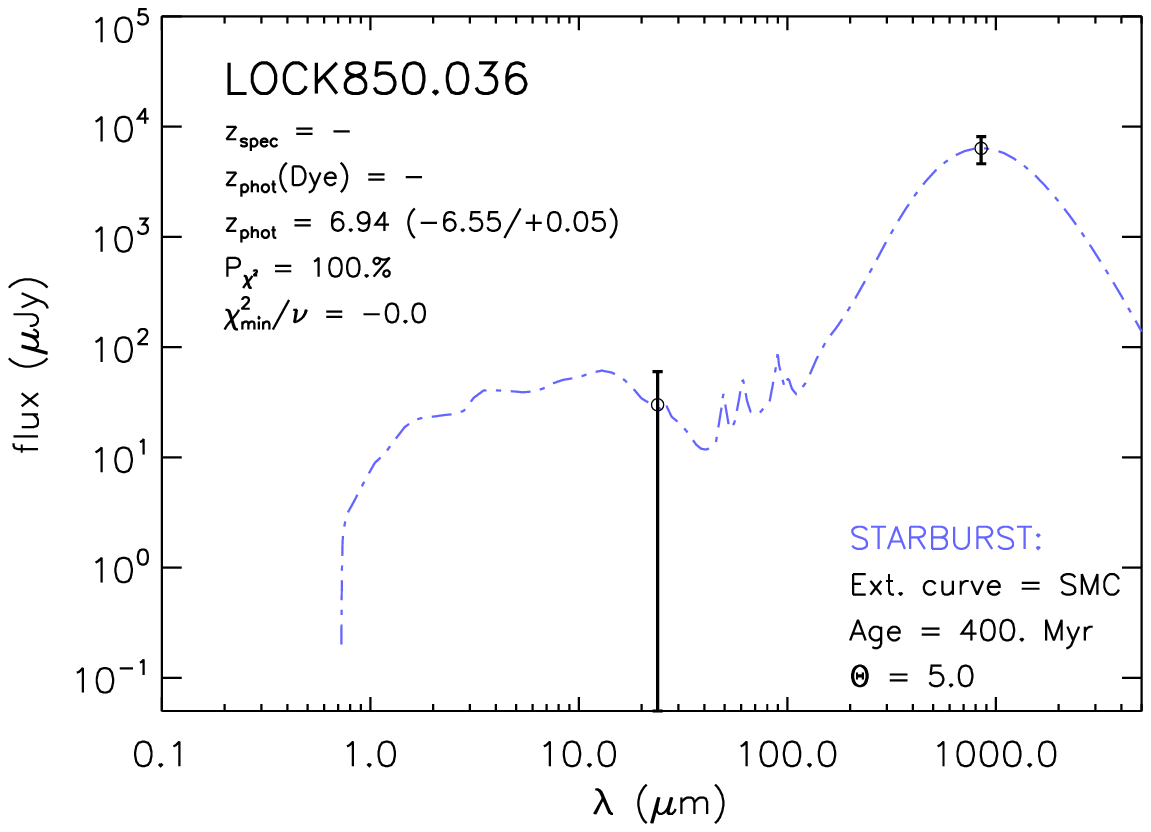}}\nolinebreak
\hspace*{-2.65cm}\resizebox{0.37\hsize}{!}{\includegraphics*{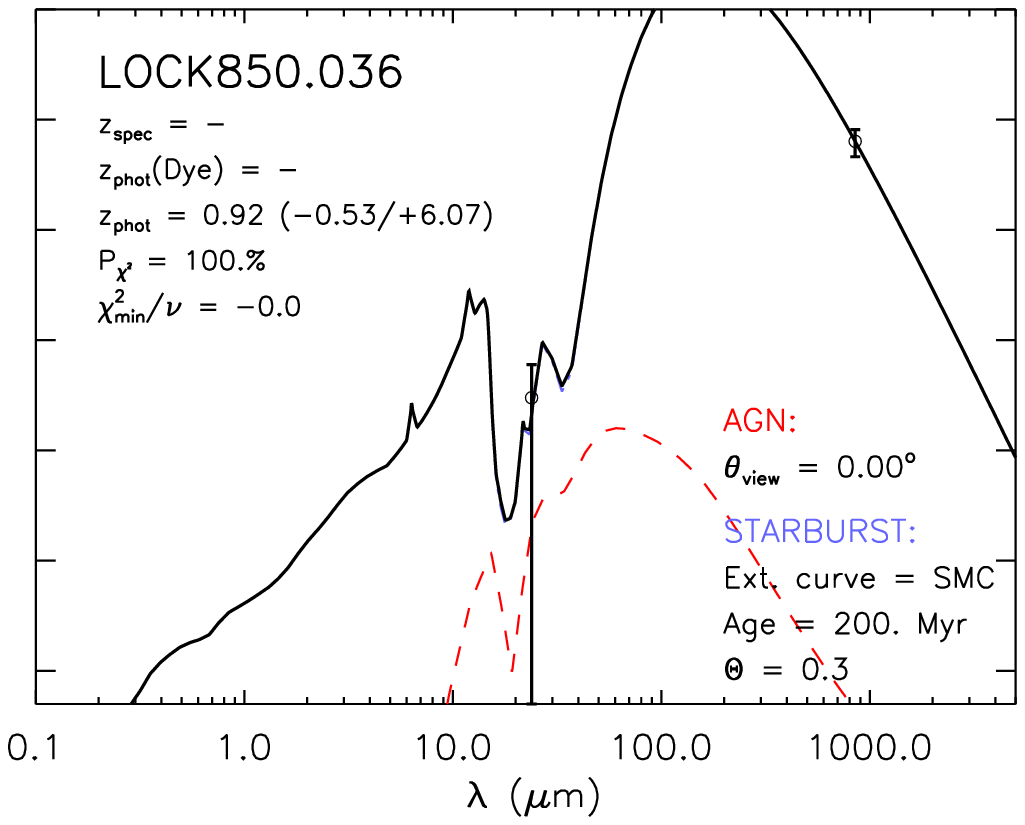}}\nolinebreak
\end{center}\vspace*{-1.3cm}
\begin{center}
\hspace*{-1.8cm}\resizebox{0.37\hsize}{!}{\includegraphics*{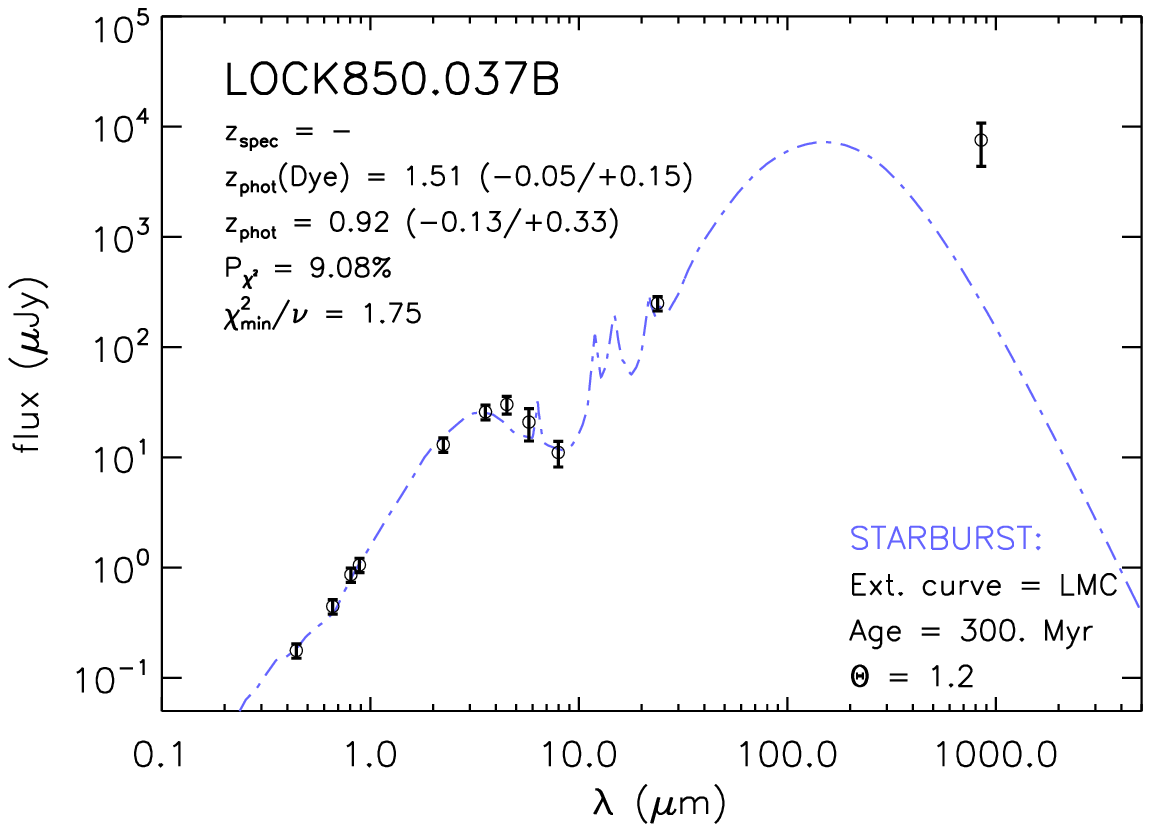}}\nolinebreak
\hspace*{-2.65cm}\resizebox{0.37\hsize}{!}{\includegraphics*{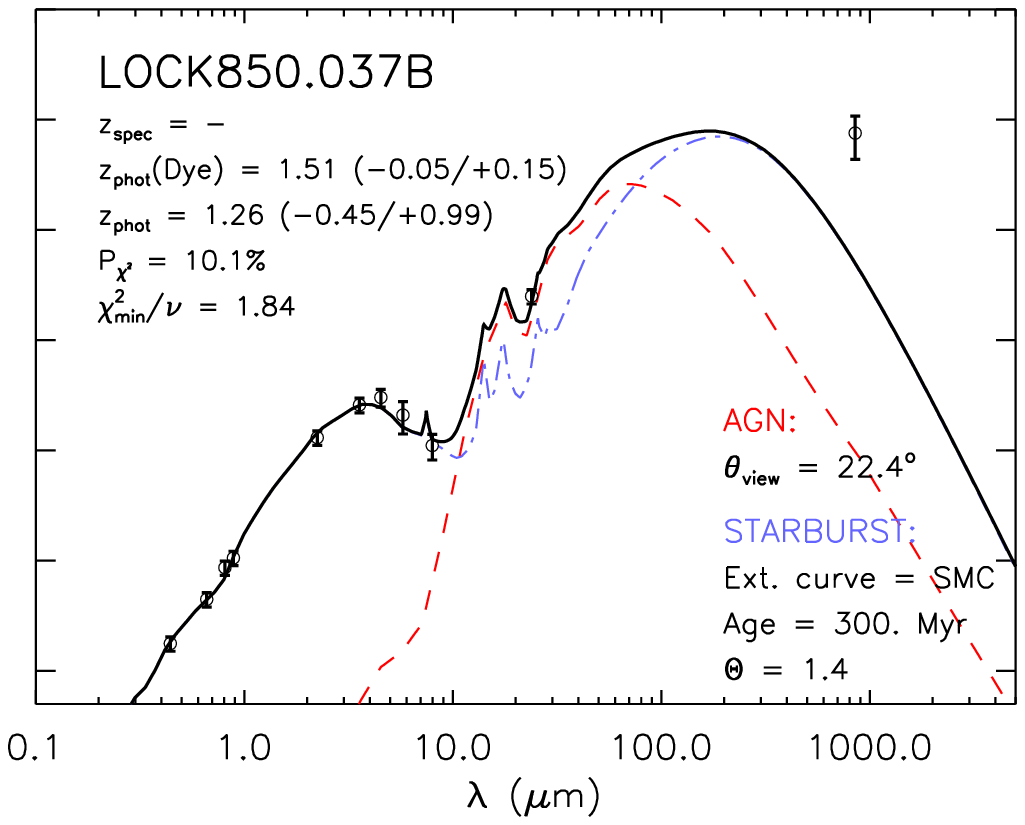}}\nolinebreak
\hspace*{-1.8cm}\resizebox{0.37\hsize}{!}{\includegraphics*{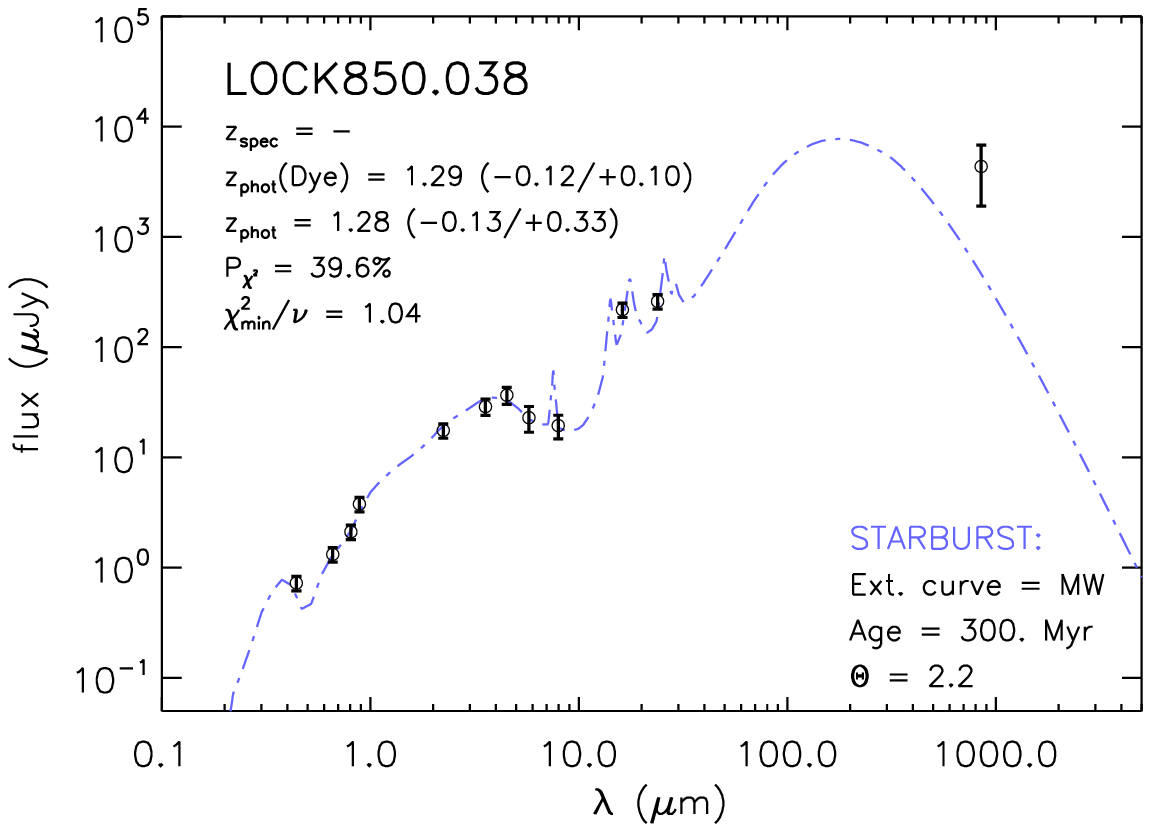}}\nolinebreak
\hspace*{-2.65cm}\resizebox{0.37\hsize}{!}{\includegraphics*{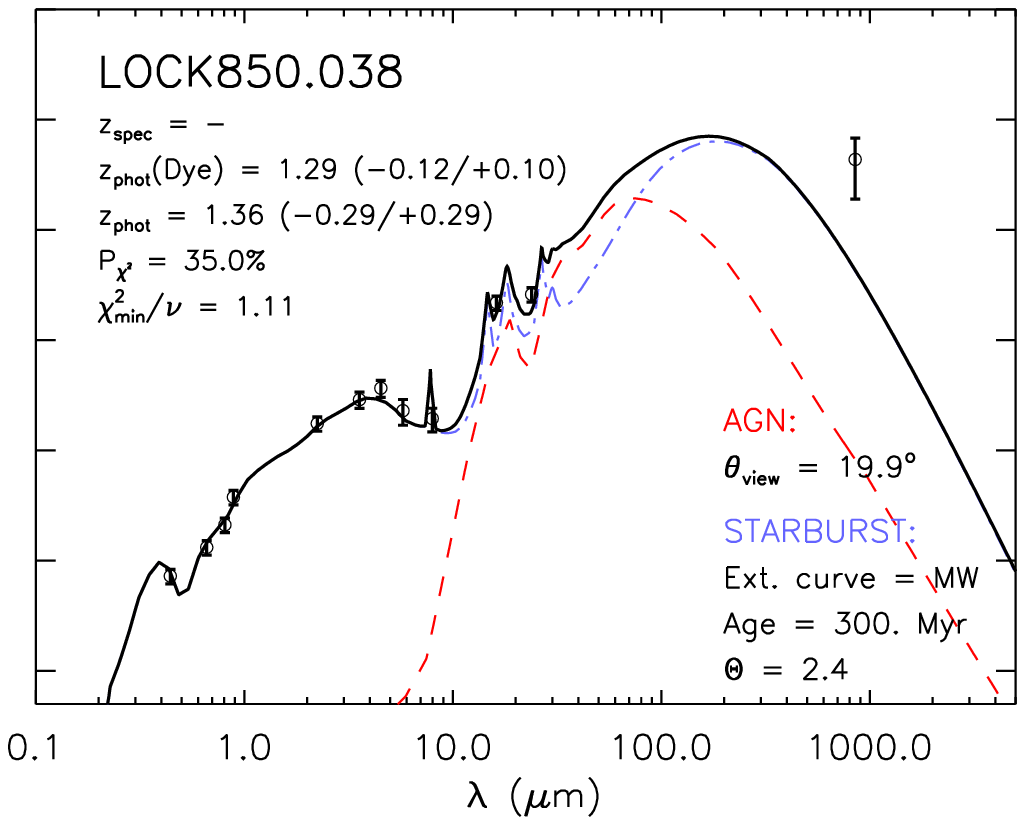}}\nolinebreak
\end{center}\vspace*{-1.3cm}
\begin{center}
\hspace*{-1.8cm}\resizebox{0.37\hsize}{!}{\includegraphics*{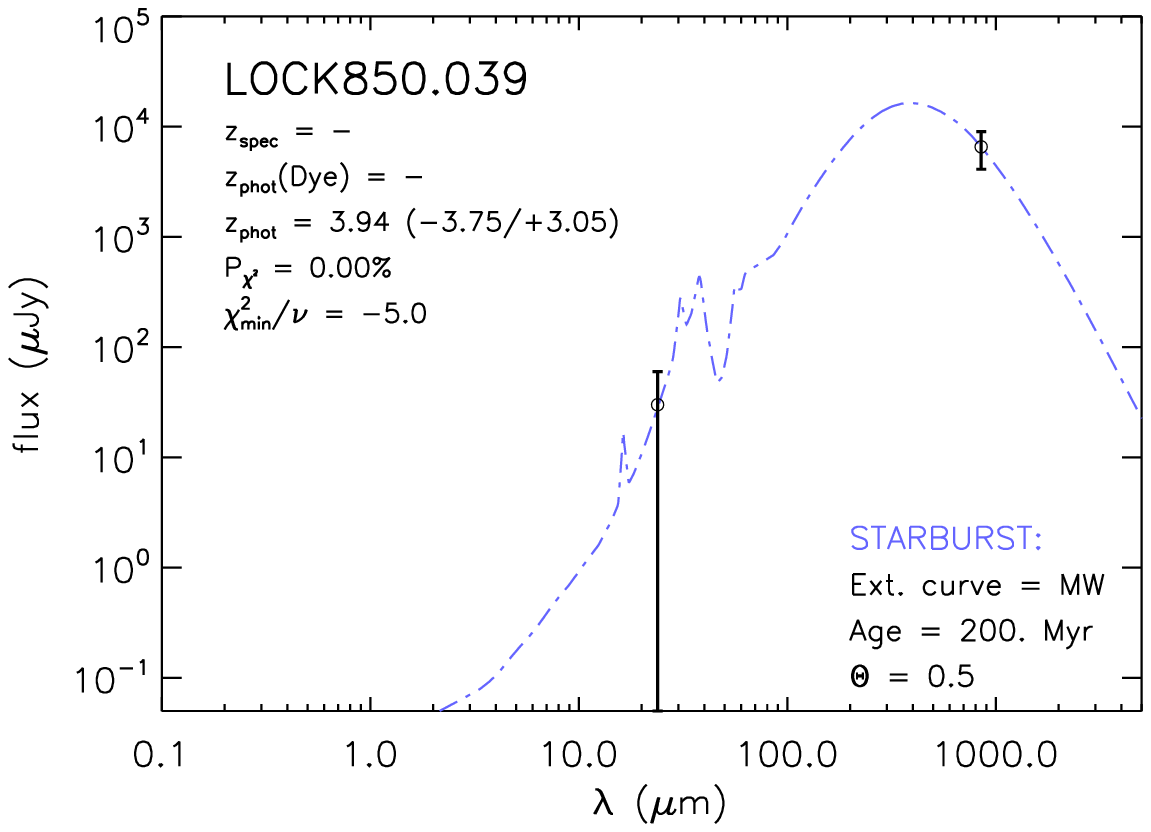}}\nolinebreak
\hspace*{-2.65cm}\resizebox{0.37\hsize}{!}{\includegraphics*{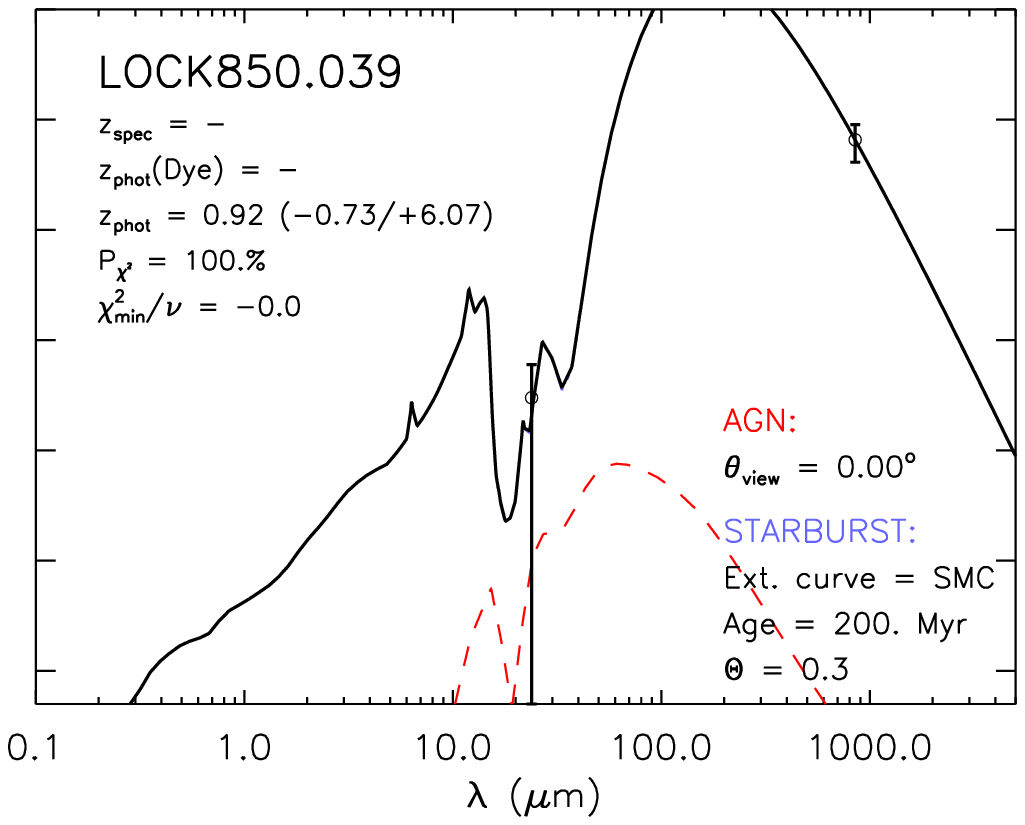}}\nolinebreak
\hspace*{-1.8cm}\resizebox{0.37\hsize}{!}{\includegraphics*{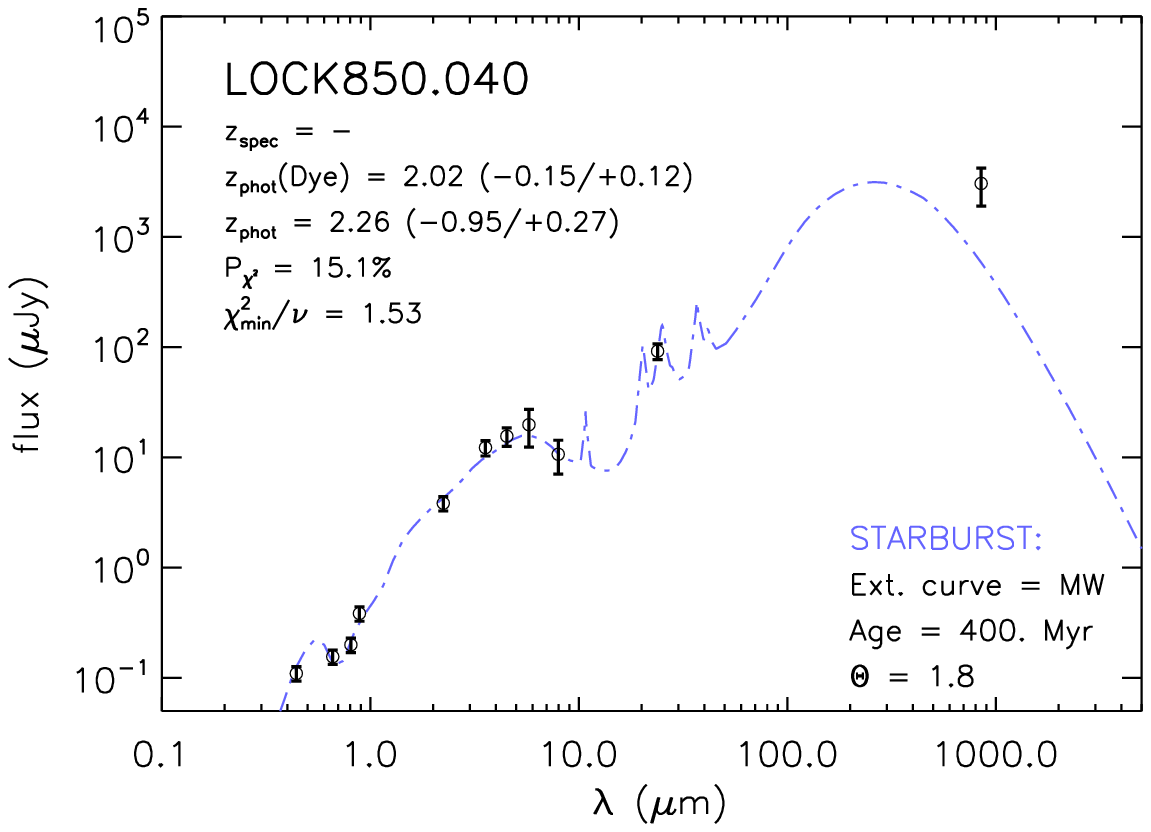}}\nolinebreak
\hspace*{-2.65cm}\resizebox{0.37\hsize}{!}{\includegraphics*{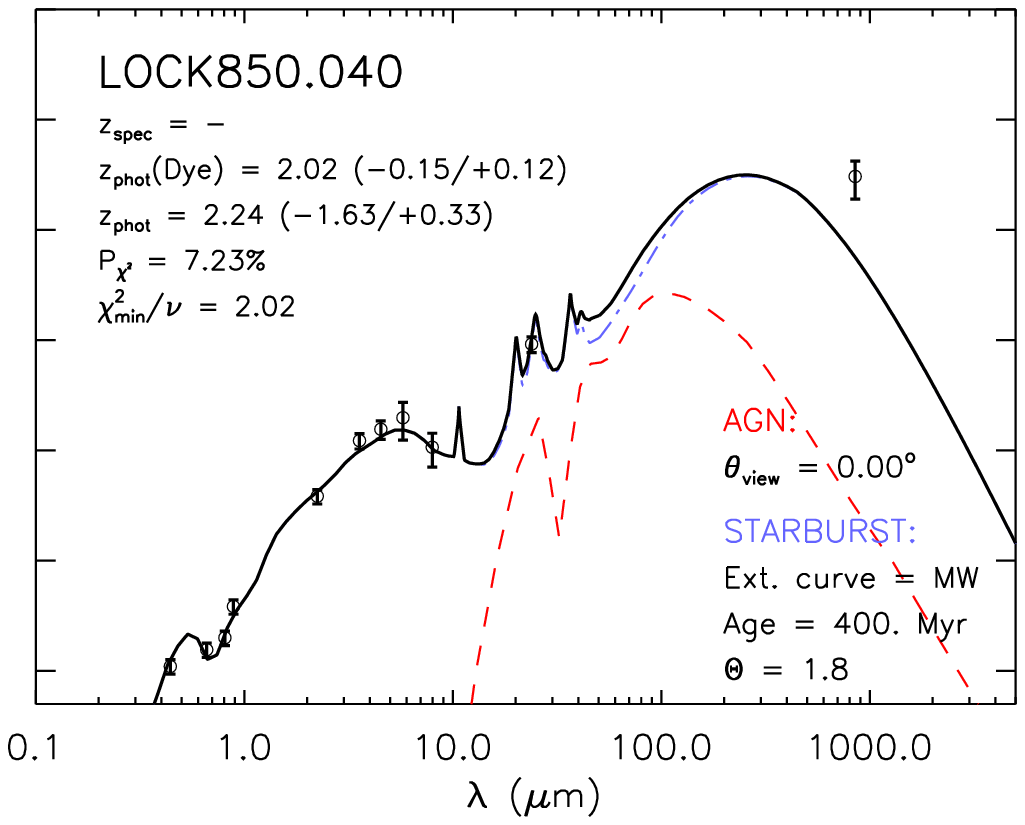}}\nolinebreak
\end{center}\vspace*{-1.3cm}
\begin{center}
\hspace*{-1.8cm}\resizebox{0.37\hsize}{!}{\includegraphics*{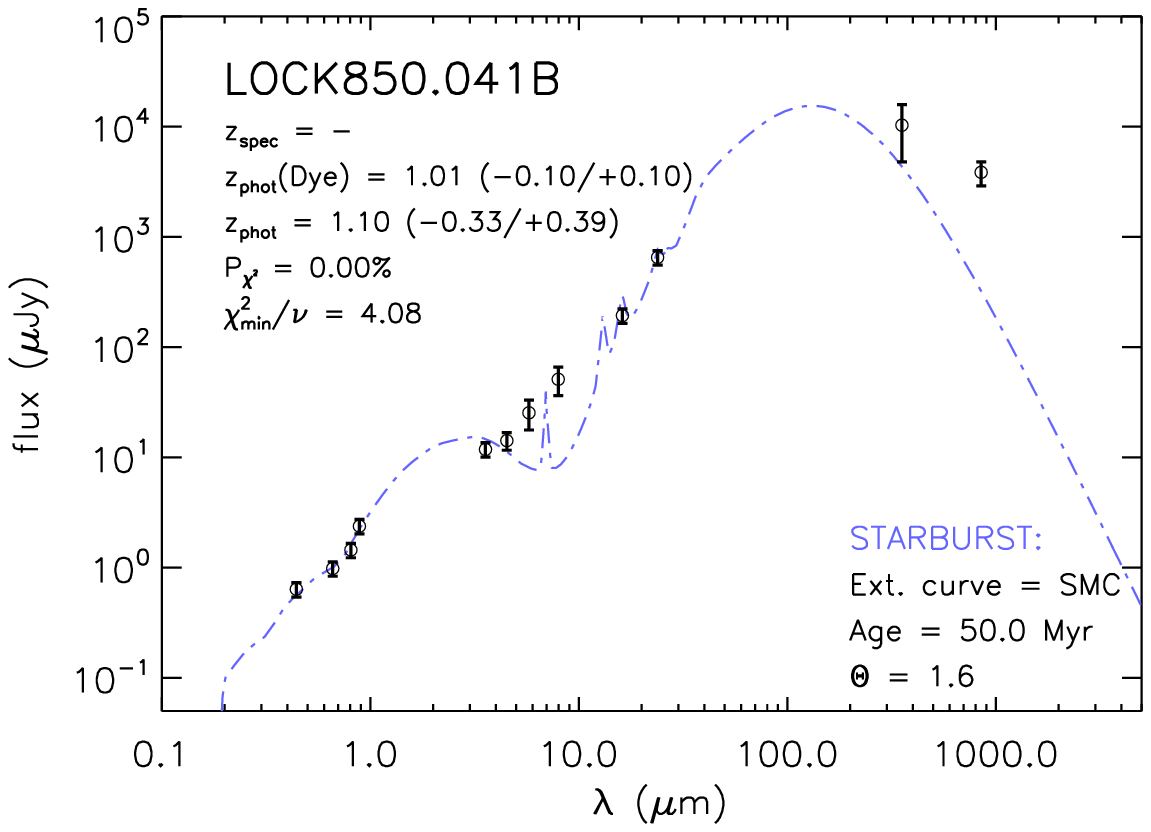}}\nolinebreak
\hspace*{-2.65cm}\resizebox{0.37\hsize}{!}{\includegraphics*{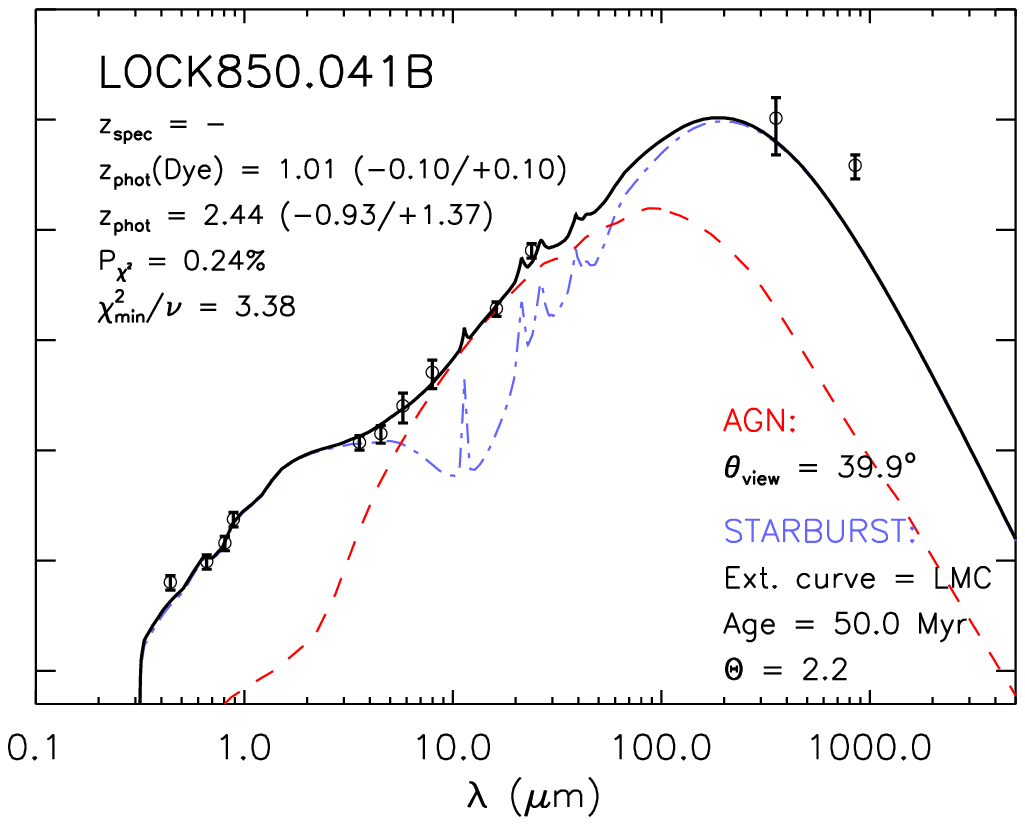}}\nolinebreak
\hspace*{-1.8cm}\resizebox{0.37\hsize}{!}{\includegraphics*{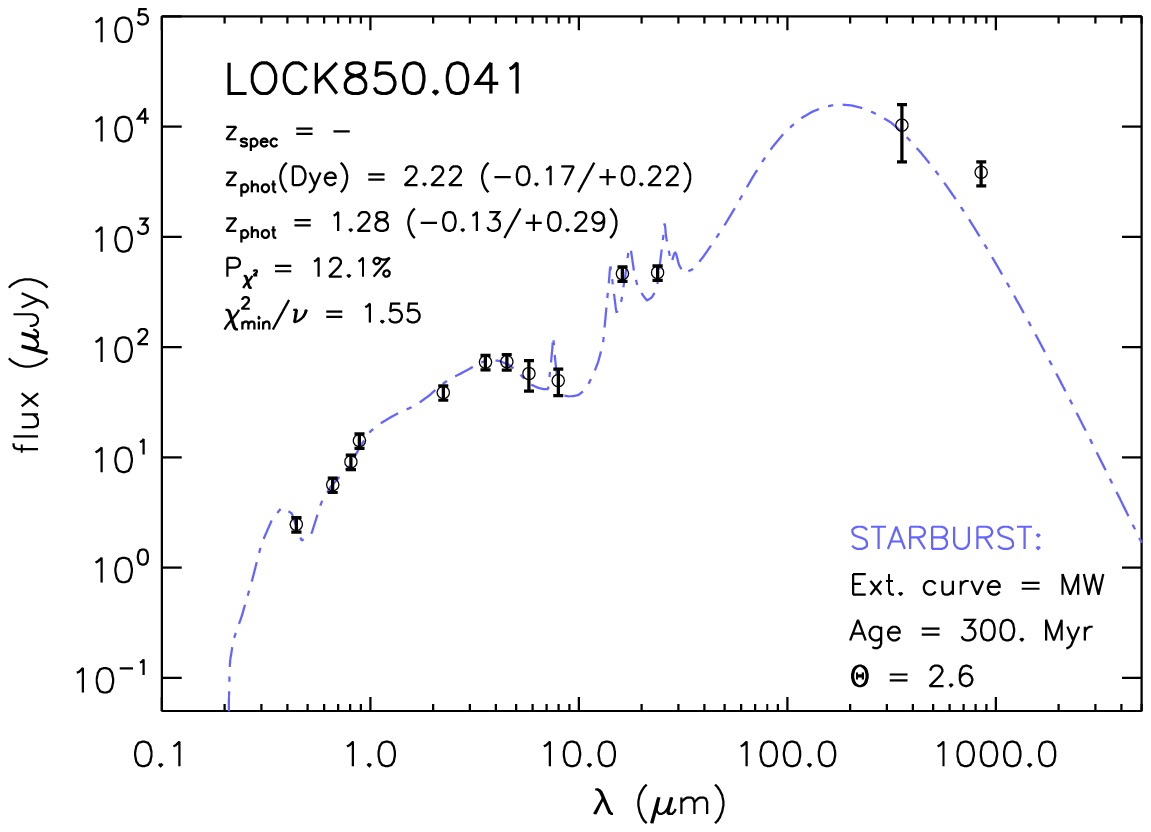}}\nolinebreak
\hspace*{-2.65cm}\resizebox{0.37\hsize}{!}{\includegraphics*{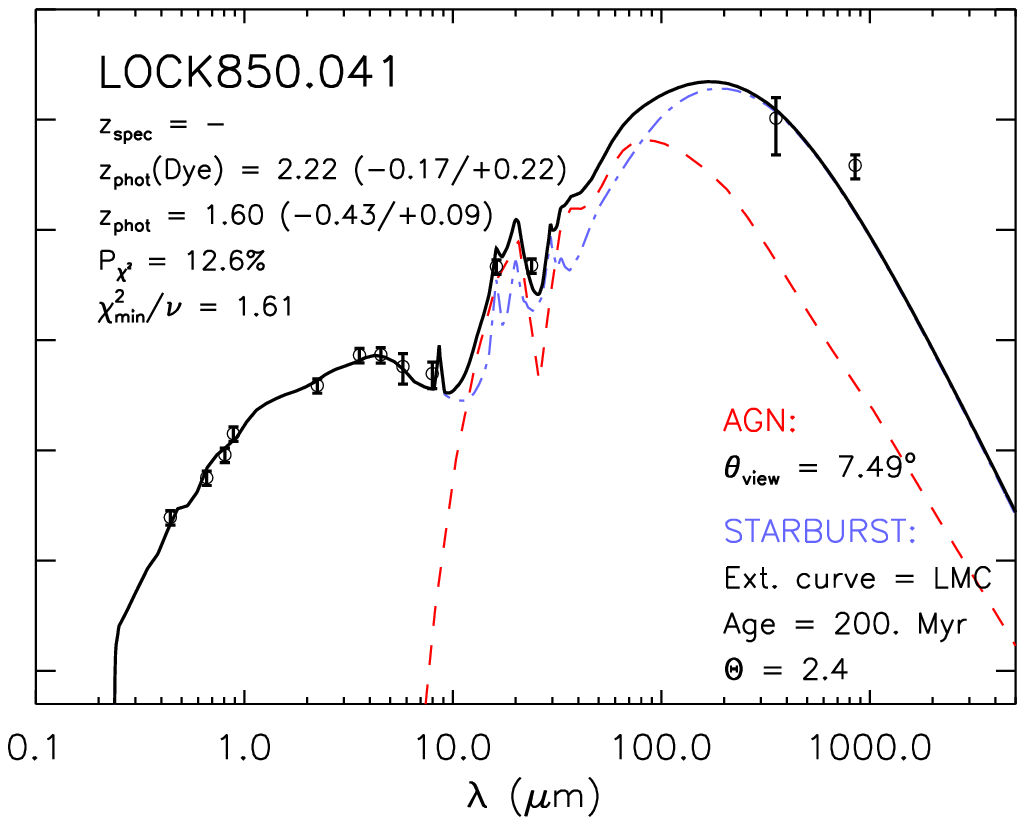}}\nolinebreak
\end{center}\vspace*{-1.3cm}
\vspace*{1.6cm}\caption{SED fits to submm-selected SHADES galaxies in the Lockman Hole East, using models from Takagi et al. (2003, 2004) and Efstathiou \& Rowan-Robinson (1995).}\label{fig:seds3}\end{figure*}

\begin{figure*}[!ht]
\begin{center}
\hspace*{-1.8cm}\resizebox{0.37\hsize}{!}{\includegraphics*{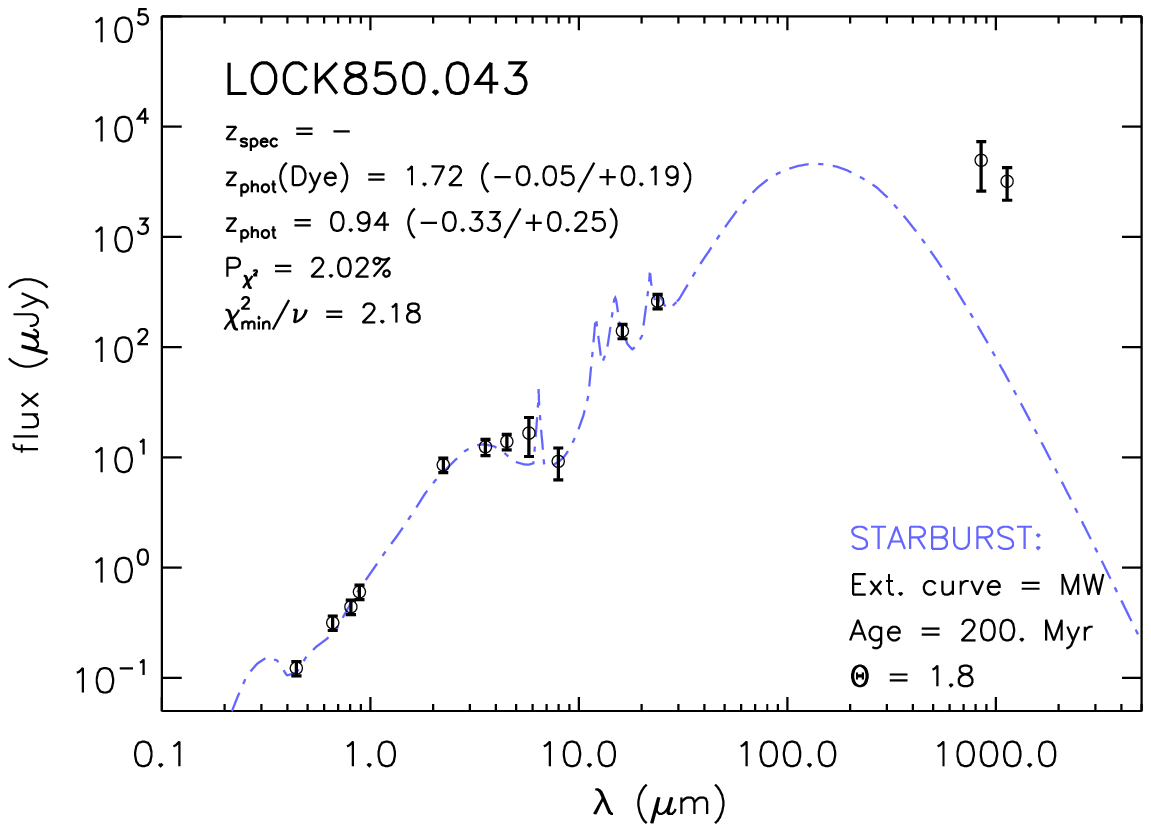}}\nolinebreak
\hspace*{-2.65cm}\resizebox{0.37\hsize}{!}{\includegraphics*{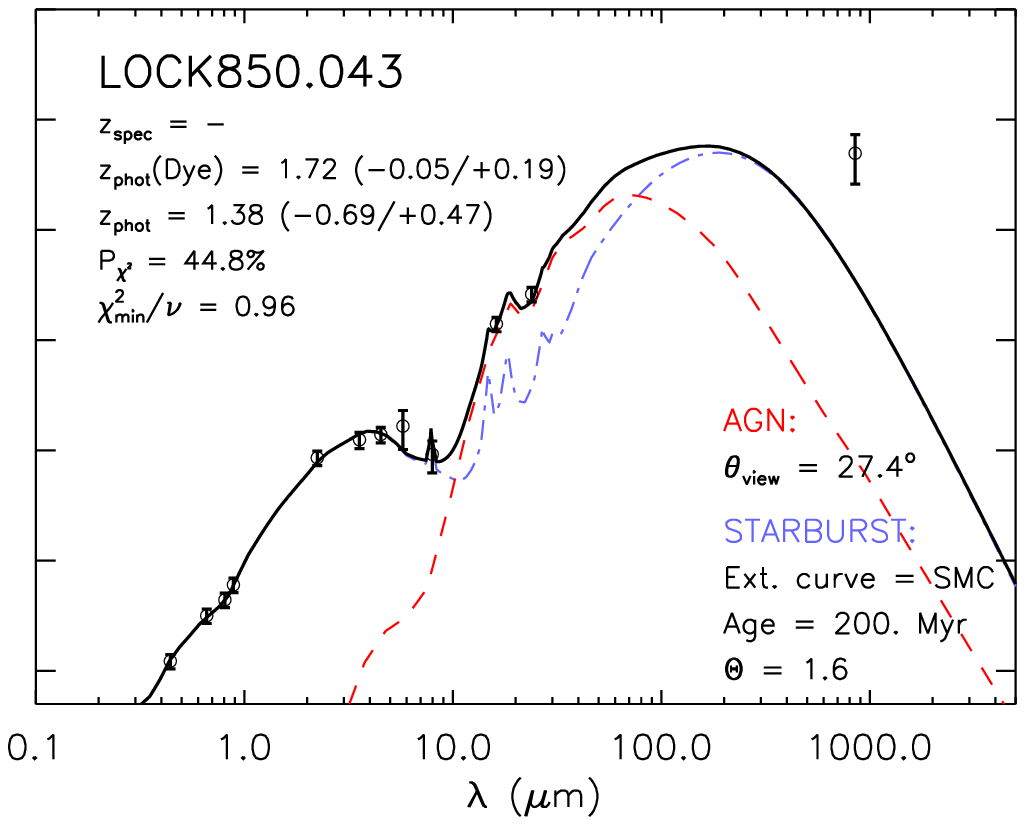}}\nolinebreak
\hspace*{-1.8cm}\resizebox{0.37\hsize}{!}{\includegraphics*{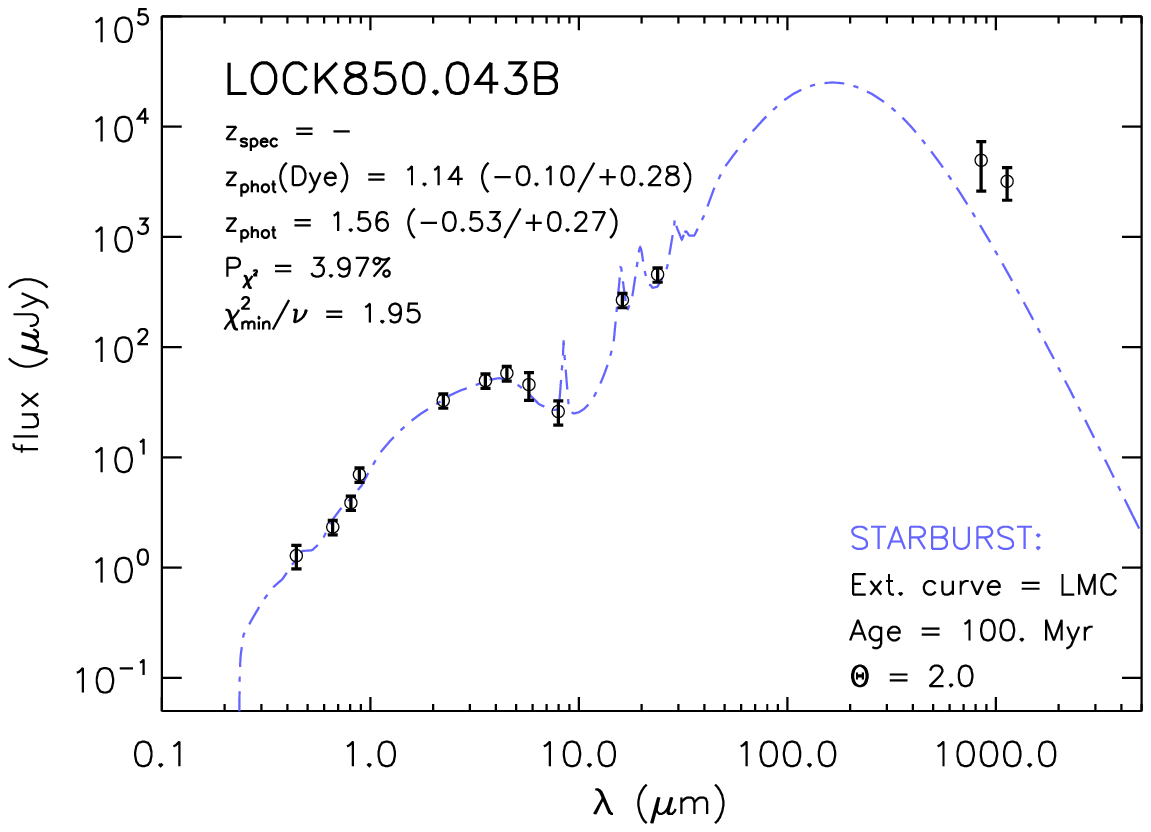}}\nolinebreak
\hspace*{-2.65cm}\resizebox{0.37\hsize}{!}{\includegraphics*{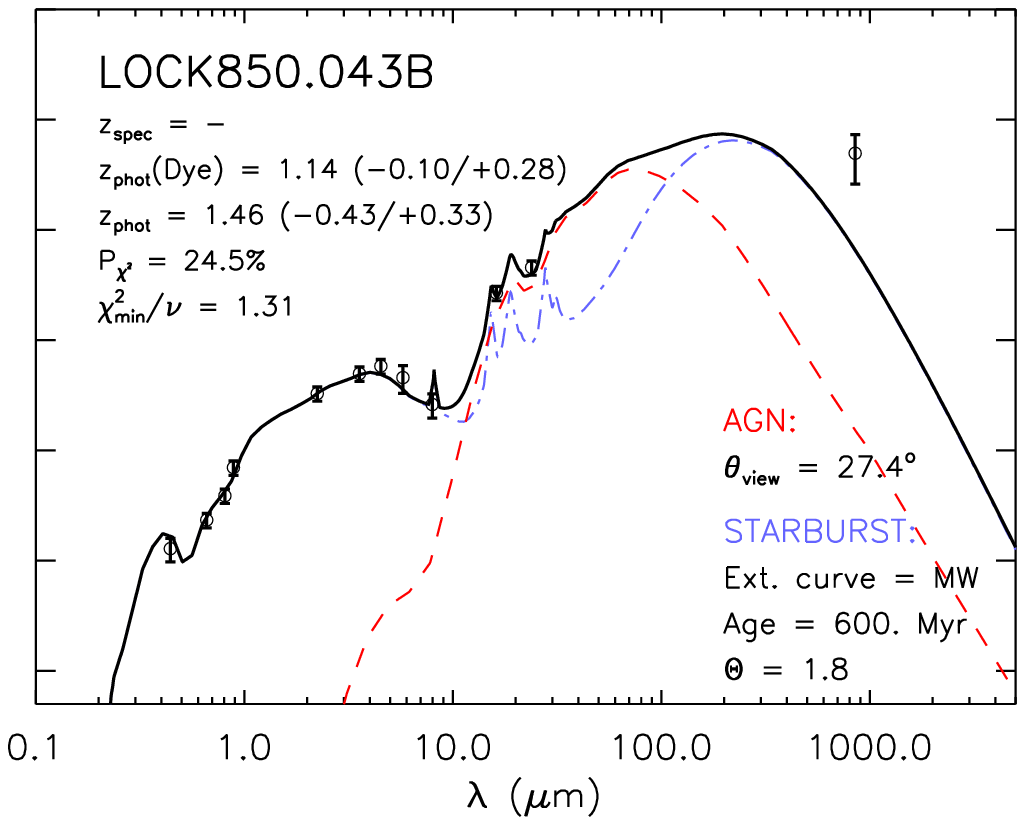}}\nolinebreak
\end{center}\vspace*{-1.3cm}
\begin{center}
\hspace*{-1.8cm}\resizebox{0.37\hsize}{!}{\includegraphics*{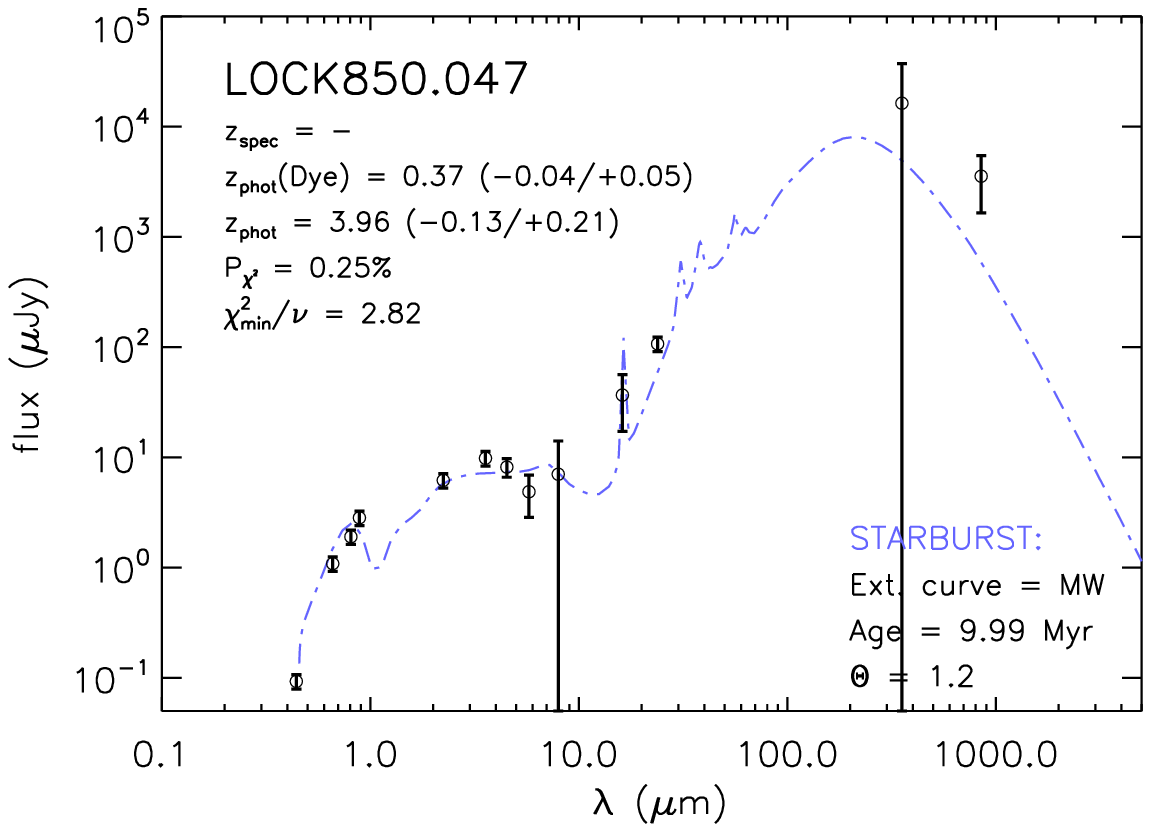}}\nolinebreak
\hspace*{-2.65cm}\resizebox{0.37\hsize}{!}{\includegraphics*{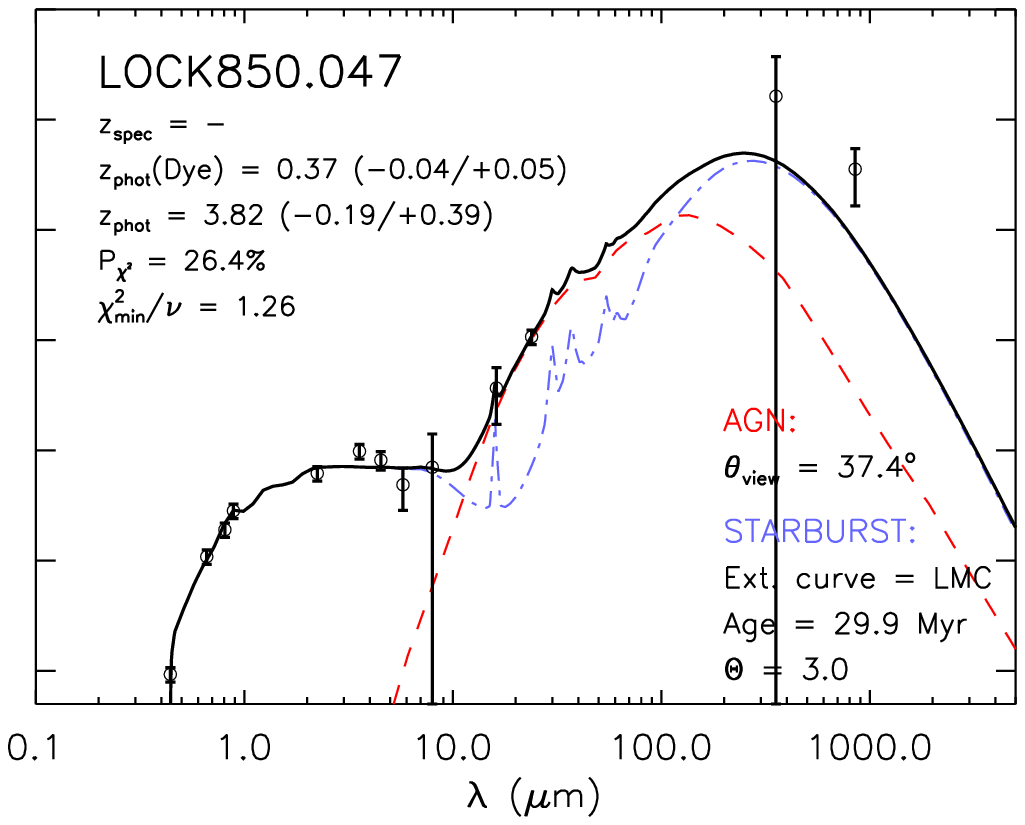}}\nolinebreak
\hspace*{-1.8cm}\resizebox{0.37\hsize}{!}{\includegraphics*{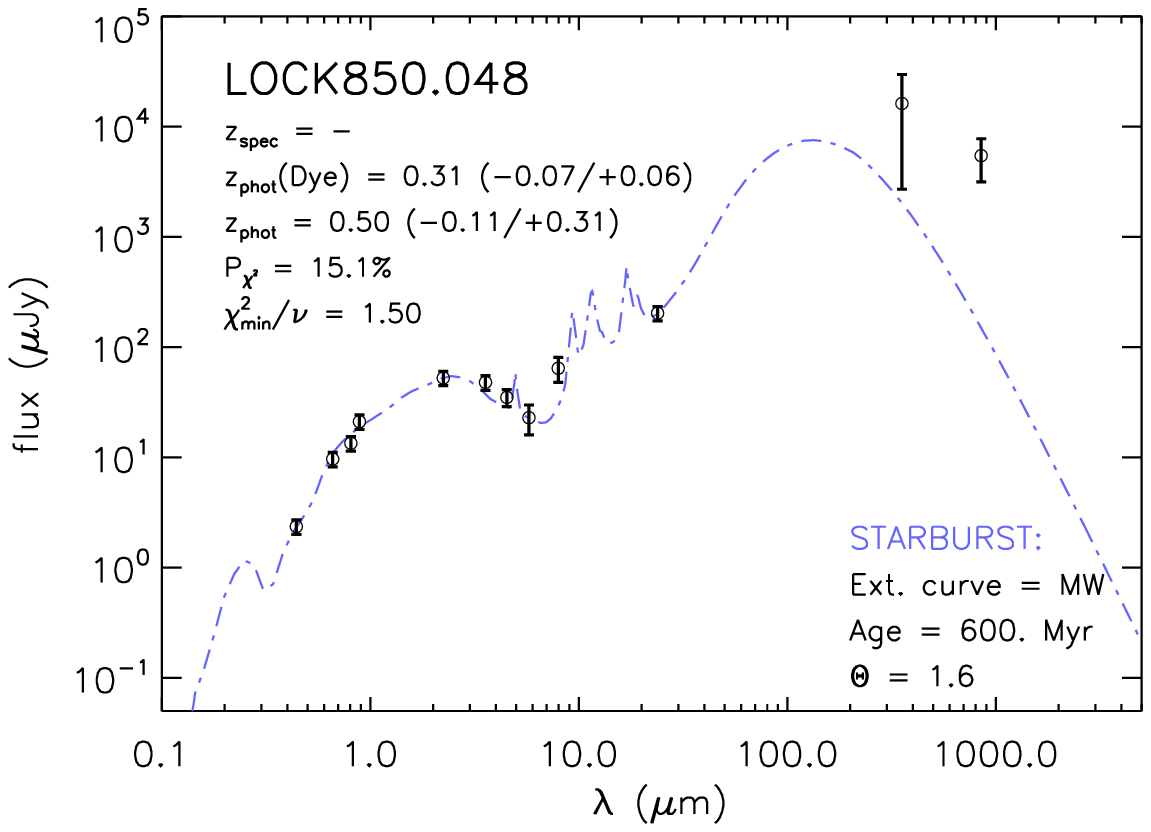}}\nolinebreak
\hspace*{-2.65cm}\resizebox{0.37\hsize}{!}{\includegraphics*{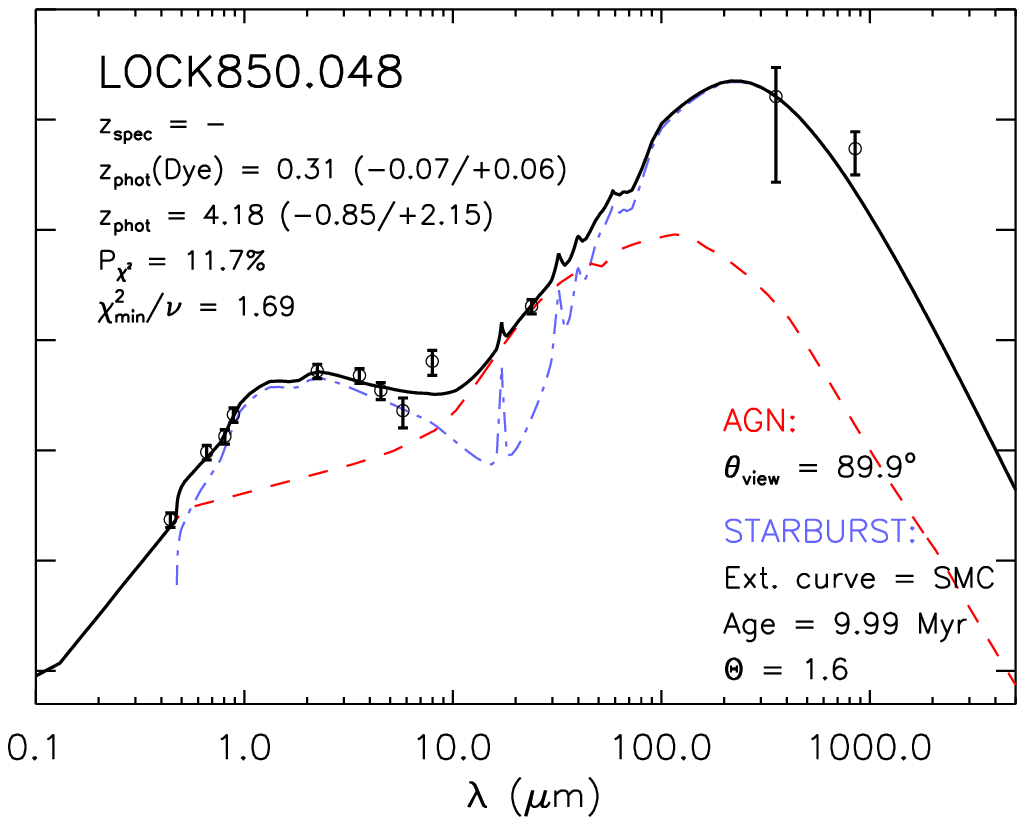}}\nolinebreak
\end{center}\vspace*{-1.3cm}
\begin{center}
\hspace*{-1.8cm}\resizebox{0.37\hsize}{!}{\includegraphics*{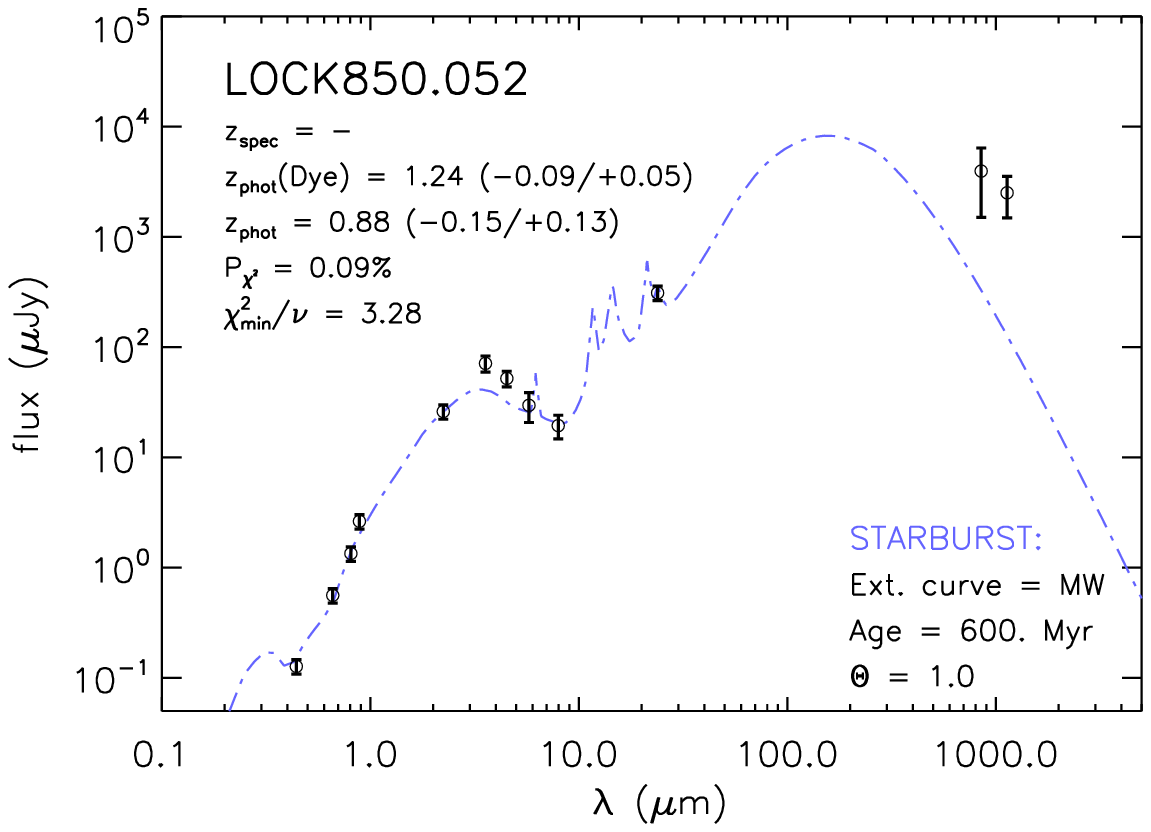}}\nolinebreak
\hspace*{-2.65cm}\resizebox{0.37\hsize}{!}{\includegraphics*{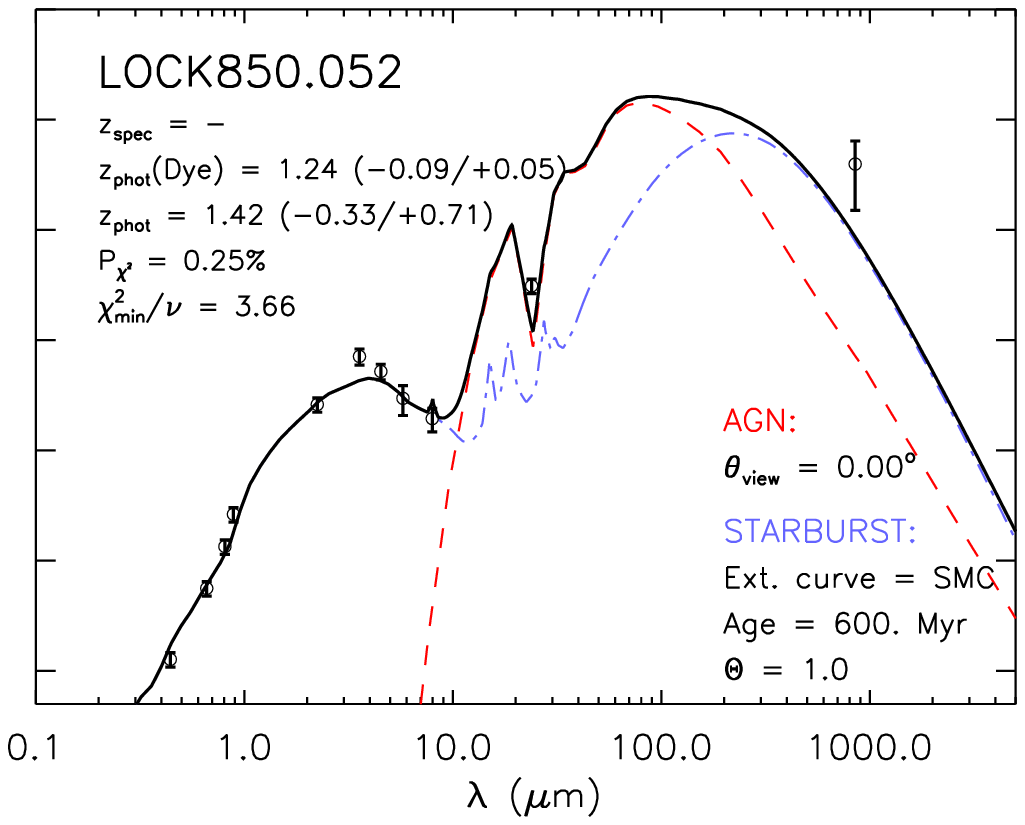}}\nolinebreak
\hspace*{-1.8cm}\resizebox{0.37\hsize}{!}{\includegraphics*{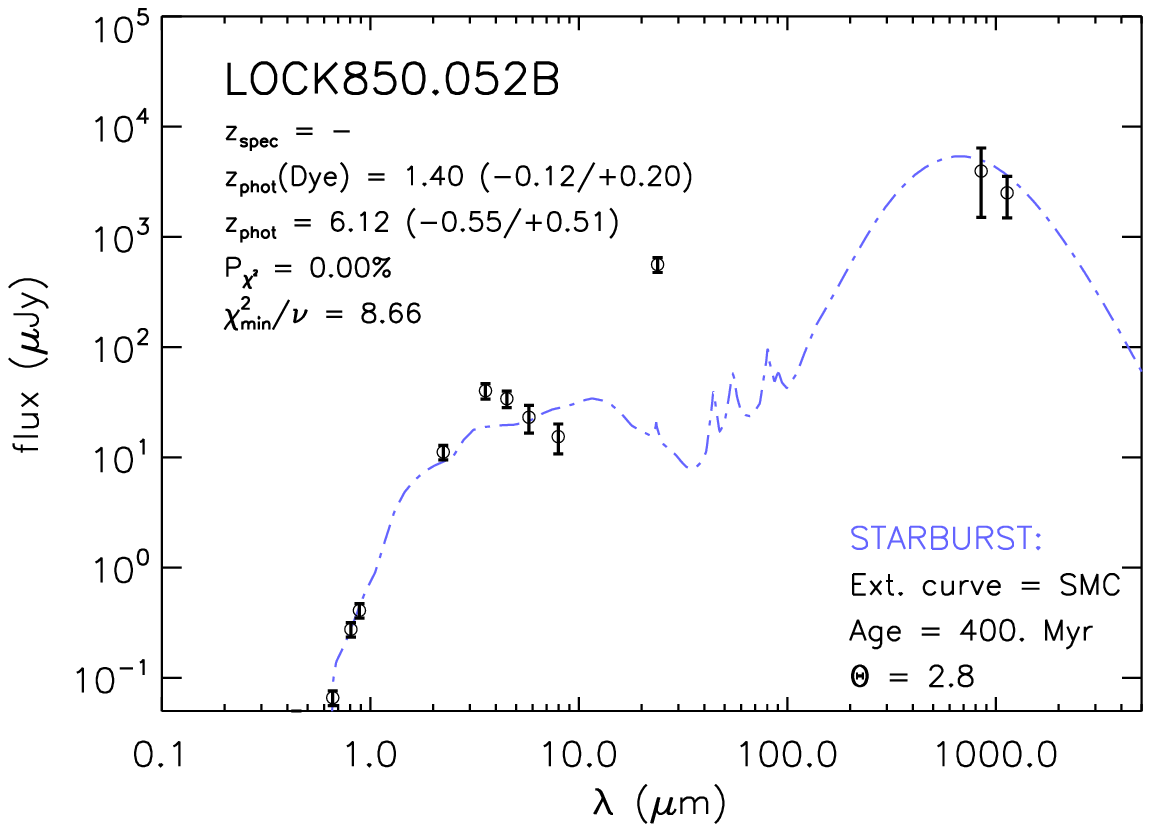}}\nolinebreak
\hspace*{-2.65cm}\resizebox{0.37\hsize}{!}{\includegraphics*{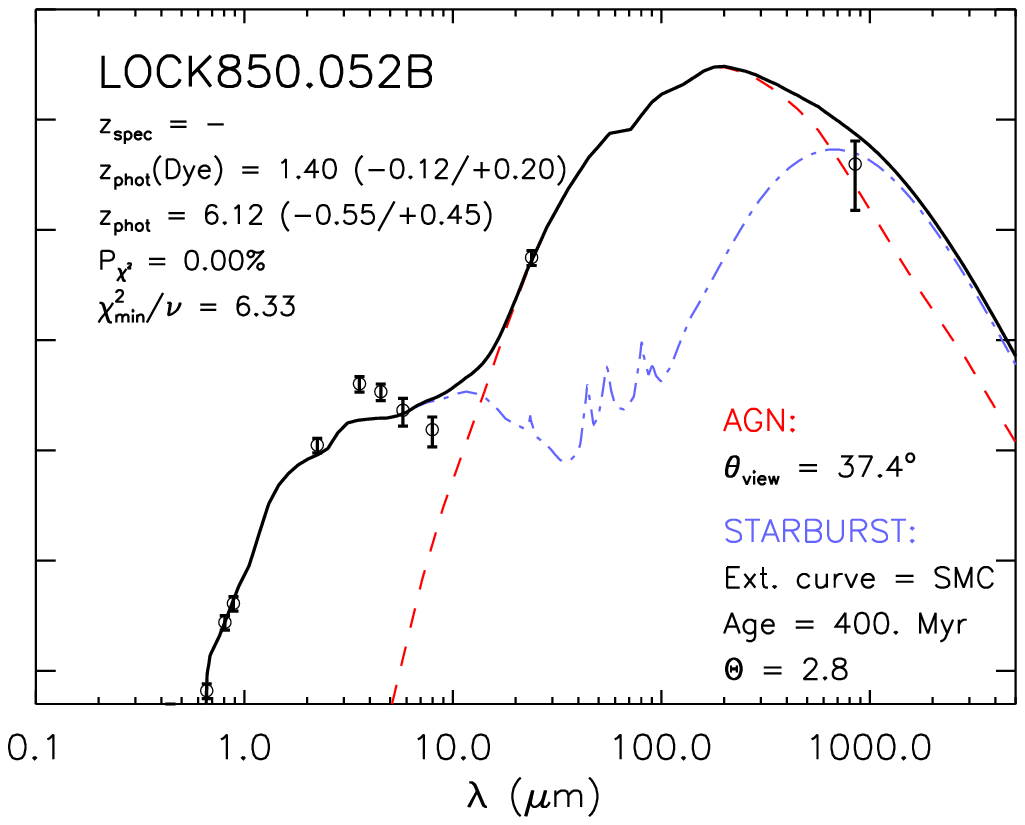}}\nolinebreak
\end{center}\vspace*{-1.3cm}
\begin{center}
\hspace*{-1.8cm}\resizebox{0.37\hsize}{!}{\includegraphics*{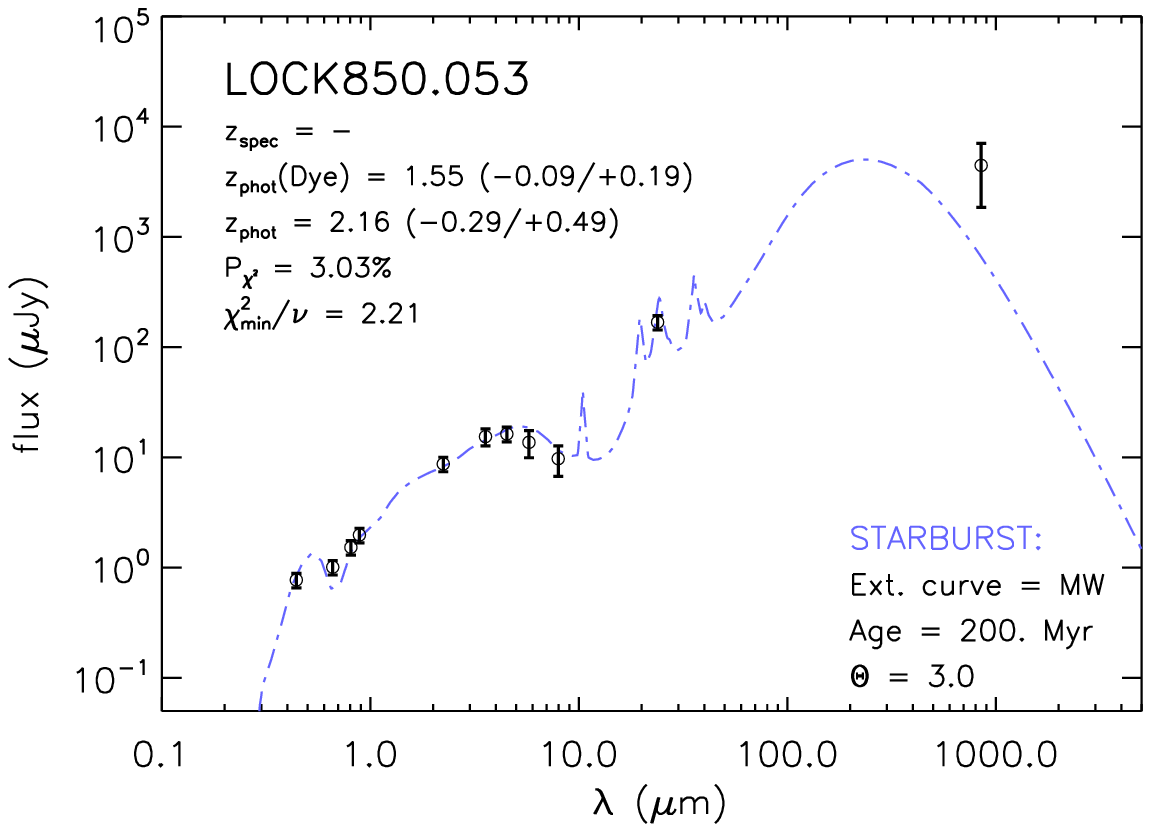}}\nolinebreak
\hspace*{-2.65cm}\resizebox{0.37\hsize}{!}{\includegraphics*{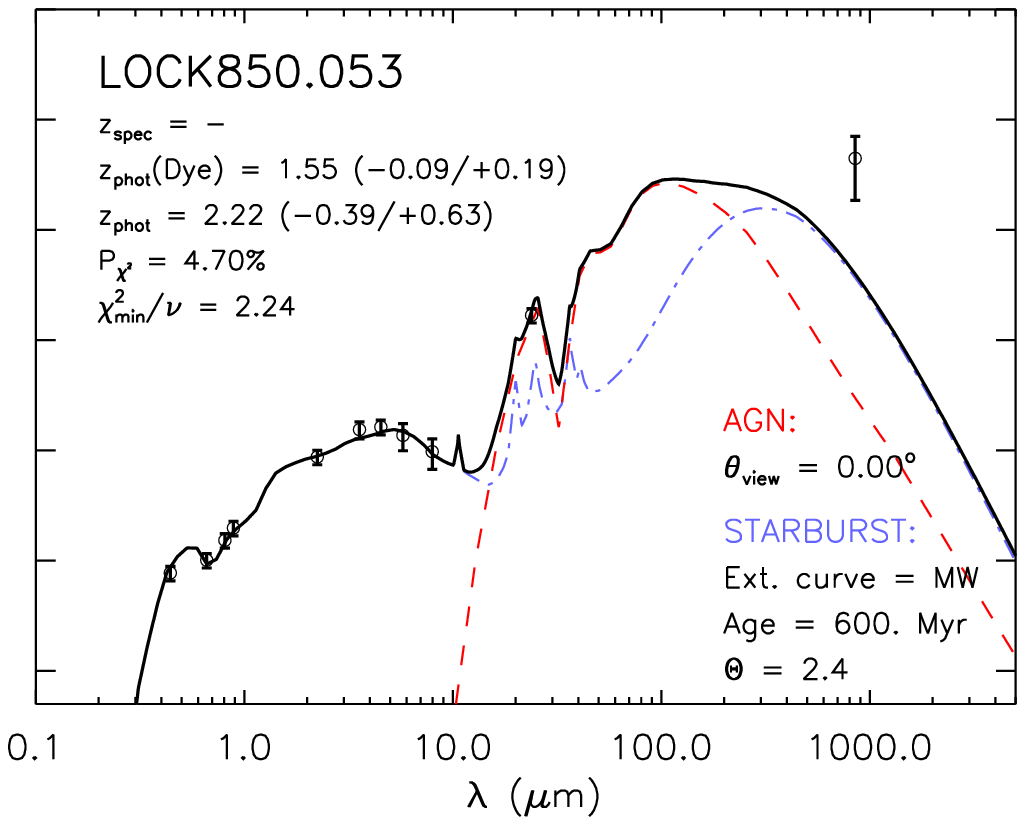}}\nolinebreak
\hspace*{-1.8cm}\resizebox{0.37\hsize}{!}{\includegraphics*{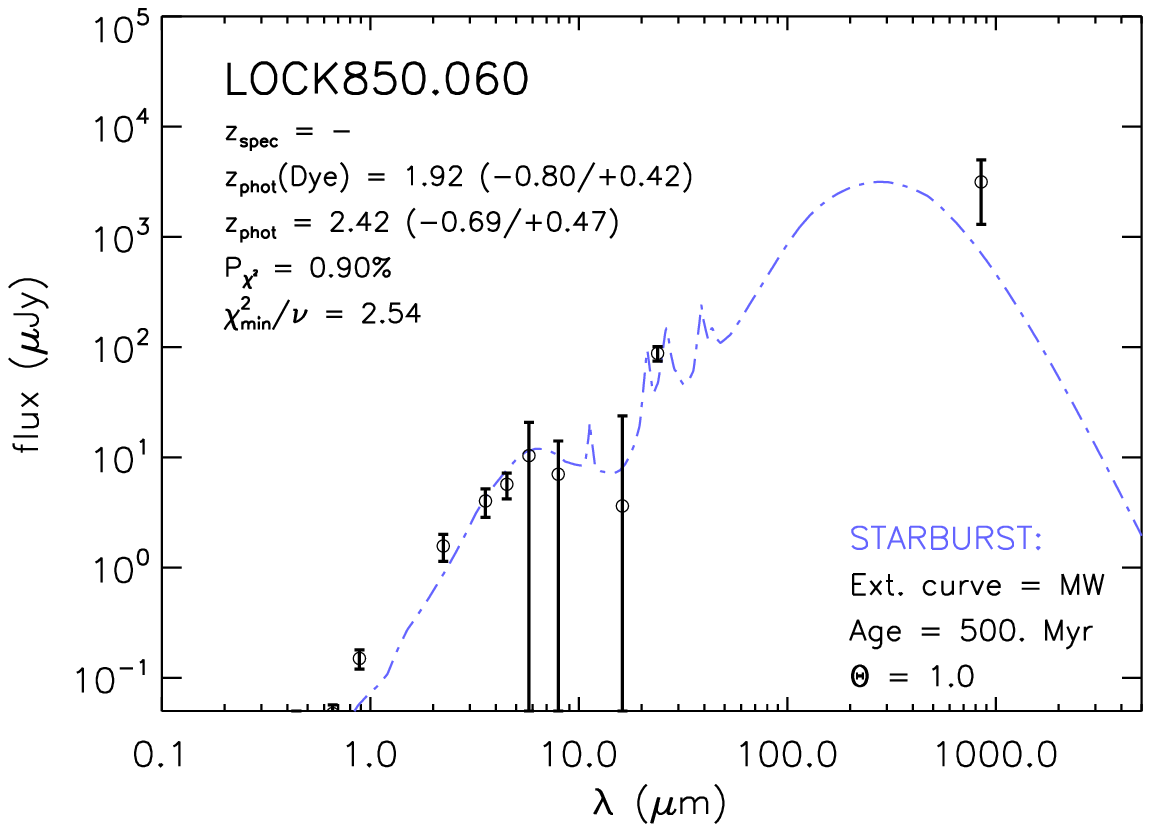}}\nolinebreak
\hspace*{-2.65cm}\resizebox{0.37\hsize}{!}{\includegraphics*{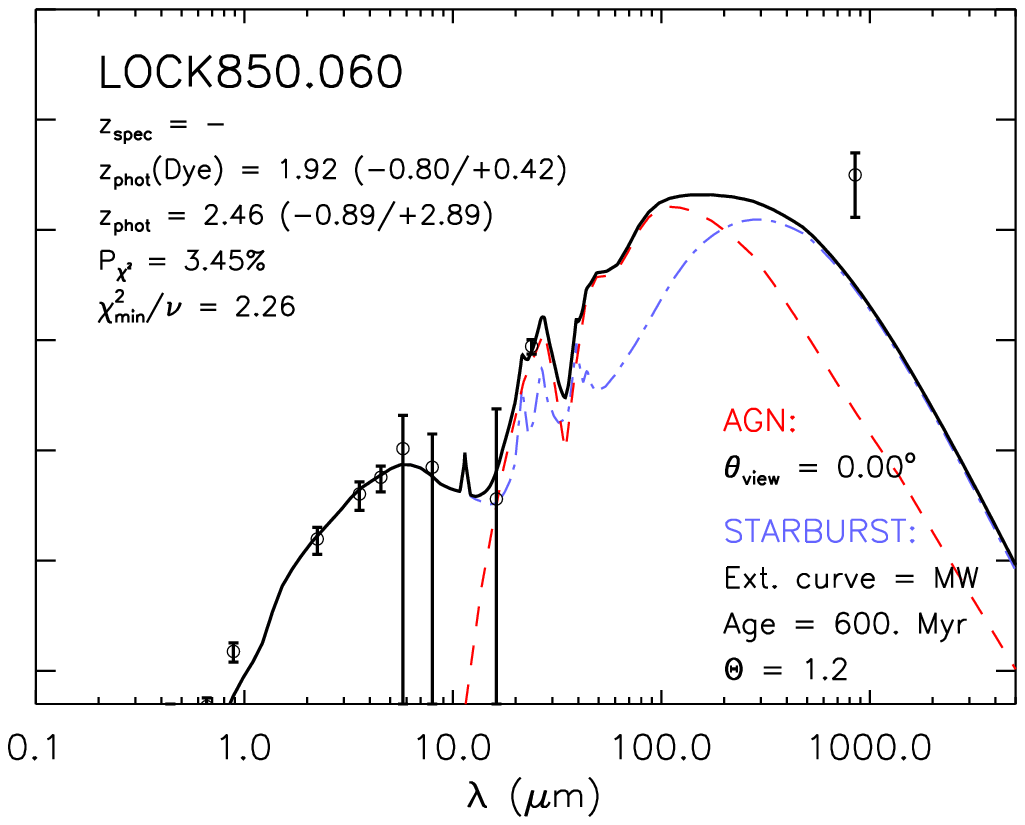}}\nolinebreak
\end{center}\vspace*{-1.3cm}
\begin{center}
\hspace*{-1.8cm}\resizebox{0.37\hsize}{!}{\includegraphics*{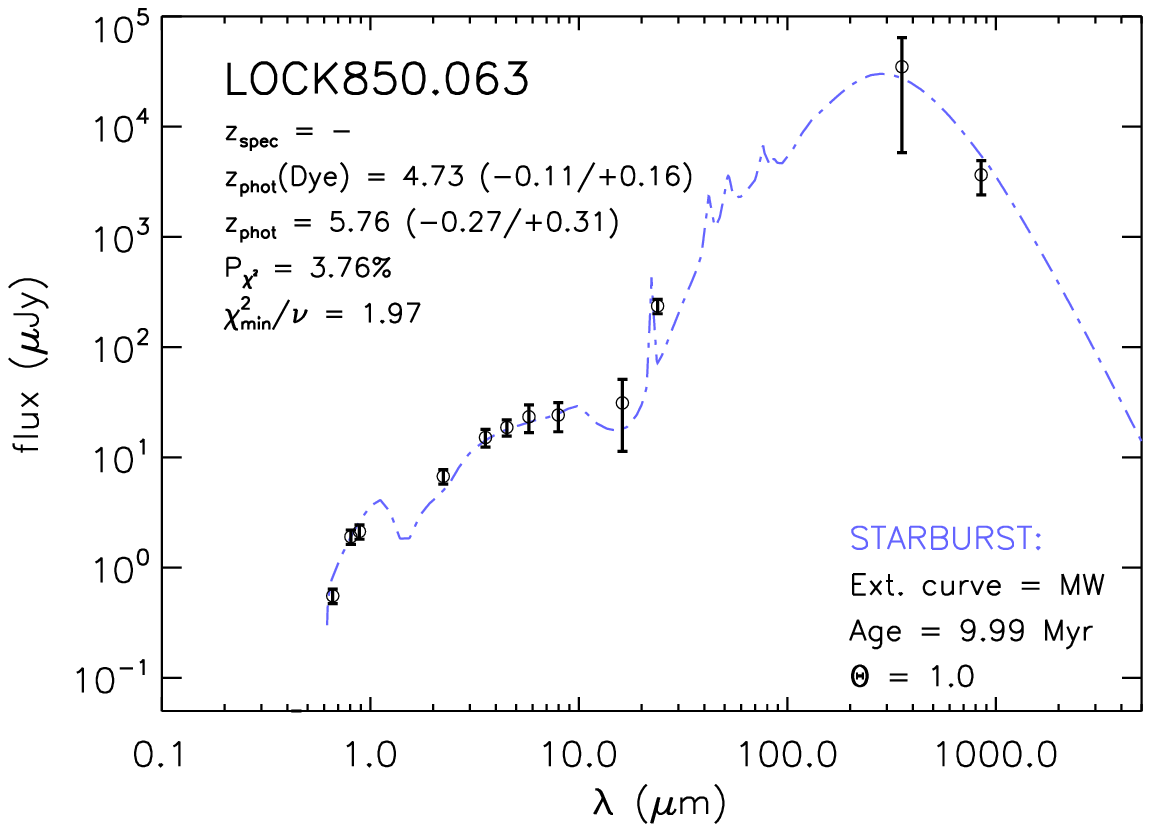}}\nolinebreak
\hspace*{-2.65cm}\resizebox{0.37\hsize}{!}{\includegraphics*{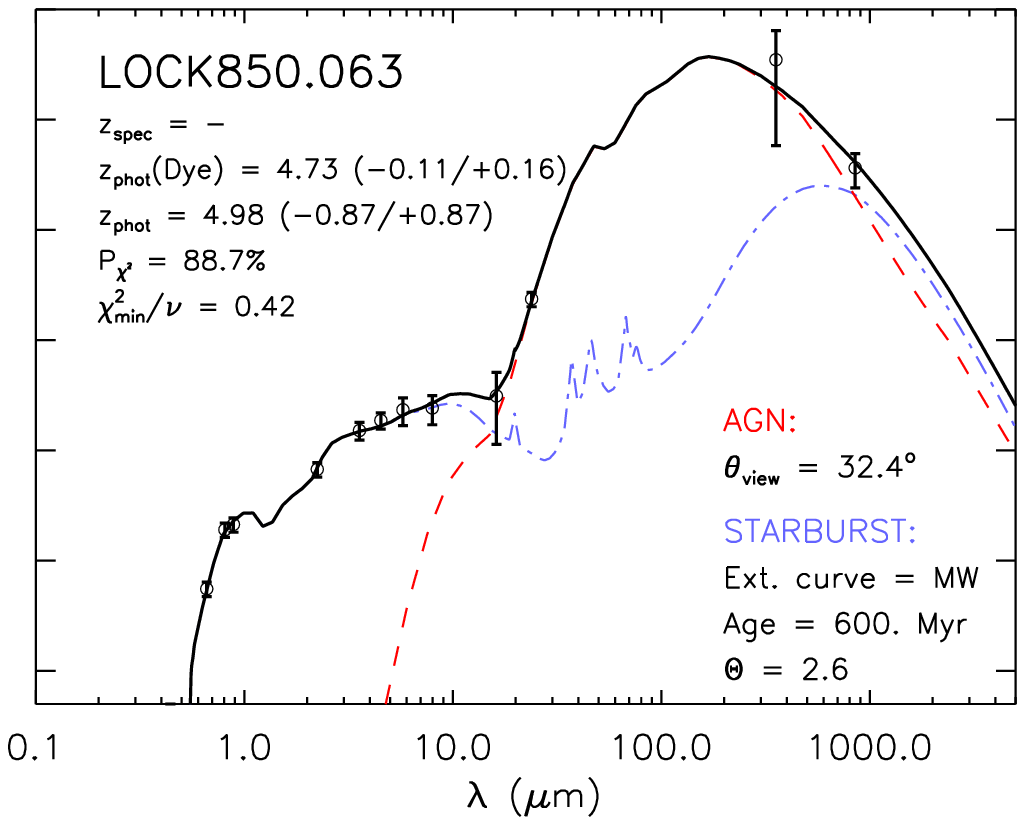}}\nolinebreak
\hspace*{-1.8cm}\resizebox{0.37\hsize}{!}{\includegraphics*{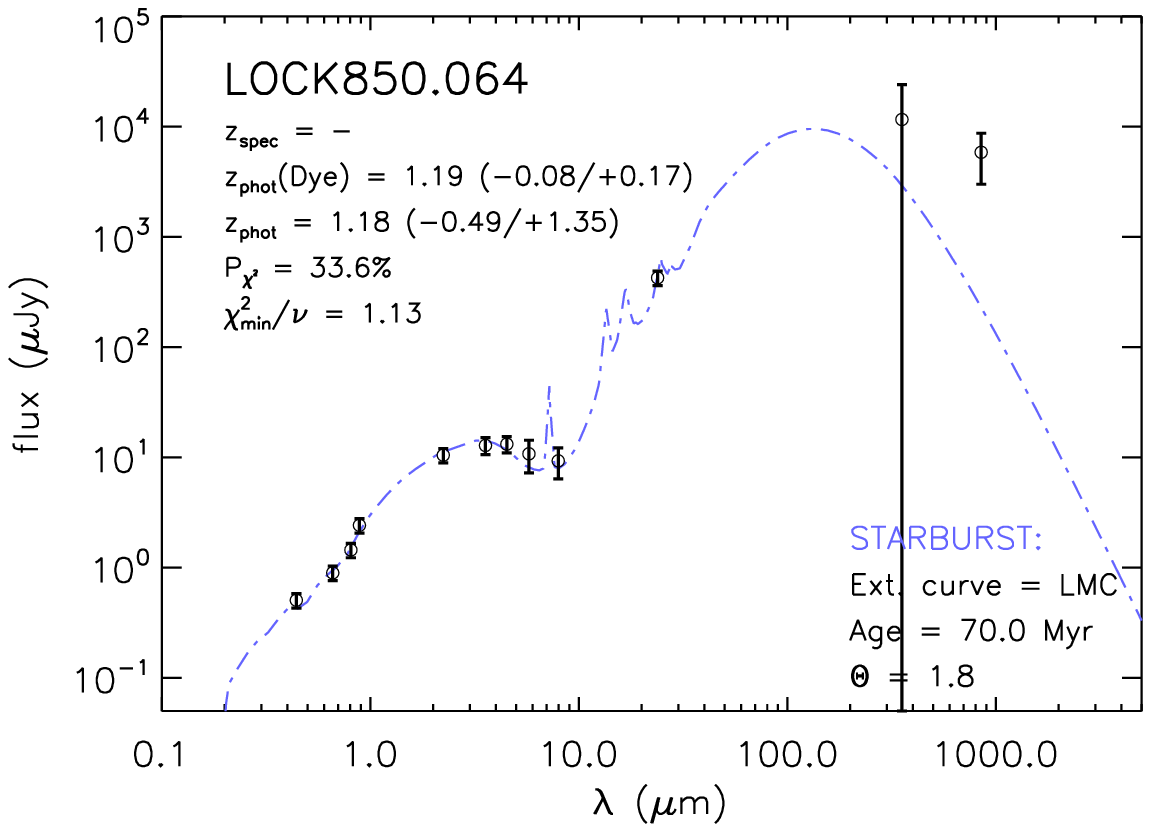}}\nolinebreak
\hspace*{-2.65cm}\resizebox{0.37\hsize}{!}{\includegraphics*{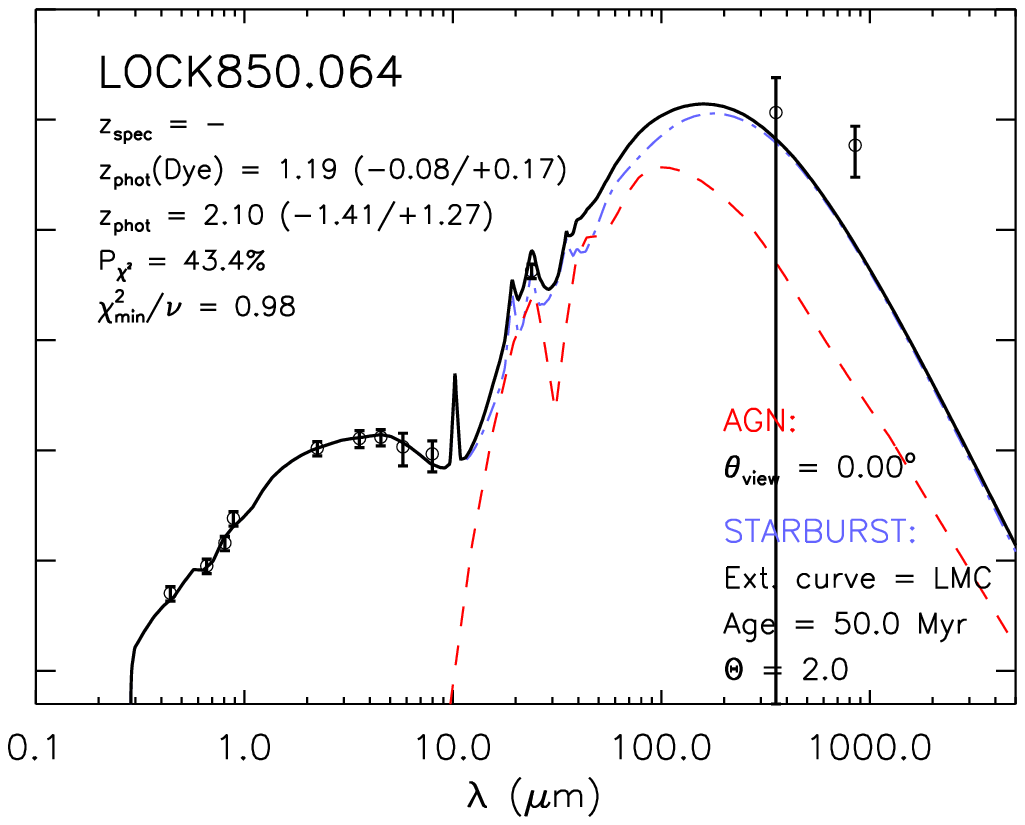}}\nolinebreak
\end{center}\vspace*{-1.3cm}
\begin{center}
\hspace*{-1.8cm}\resizebox{0.37\hsize}{!}{\includegraphics*{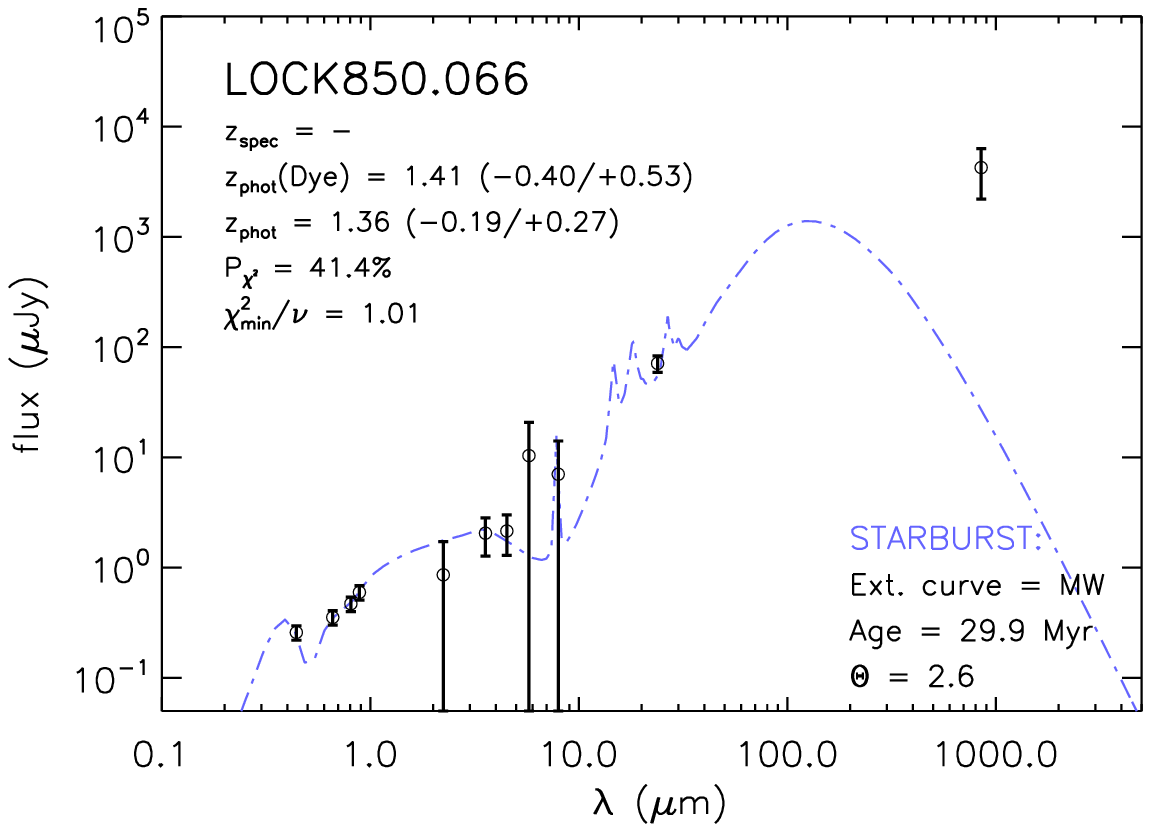}}\nolinebreak
\hspace*{-2.65cm}\resizebox{0.37\hsize}{!}{\includegraphics*{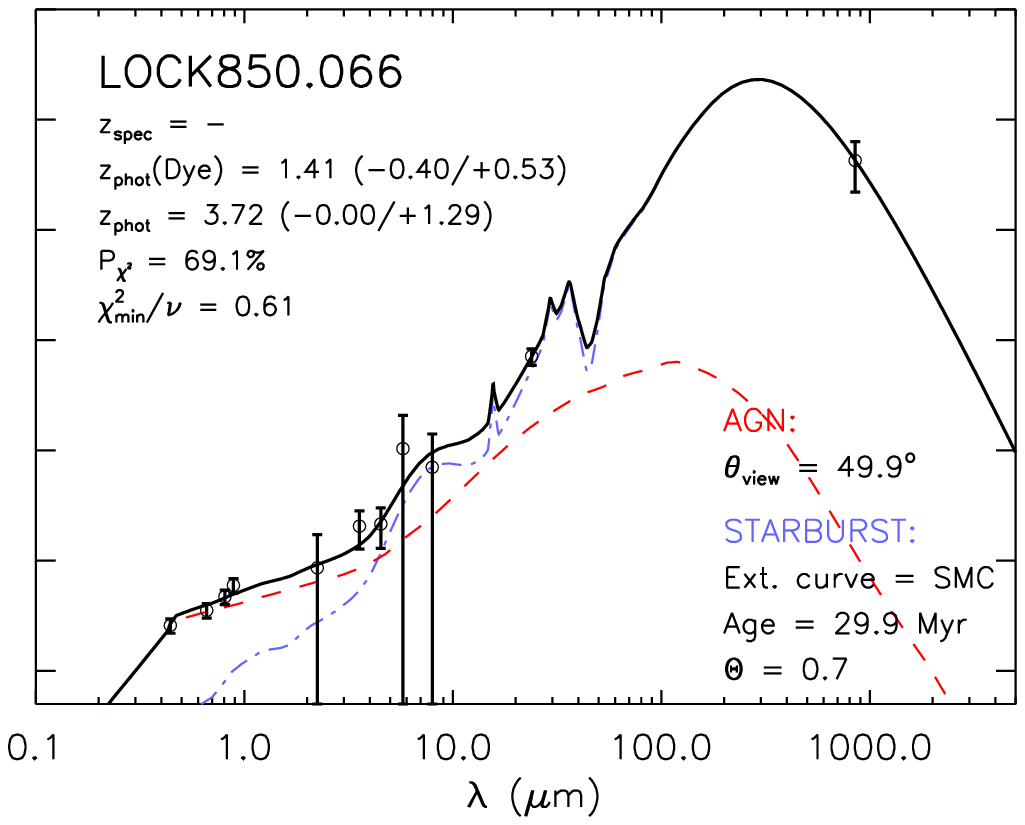}}\nolinebreak
\hspace*{-1.8cm}\resizebox{0.37\hsize}{!}{\includegraphics*{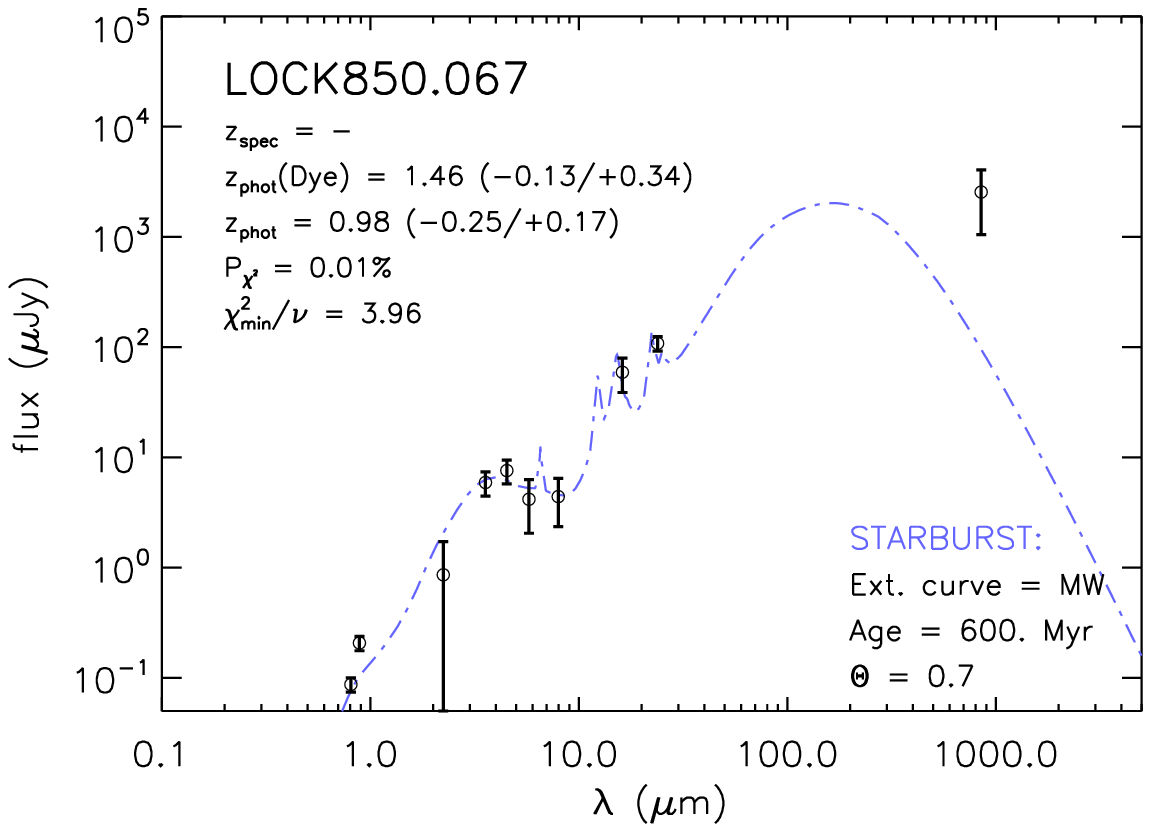}}\nolinebreak
\hspace*{-2.65cm}\resizebox{0.37\hsize}{!}{\includegraphics*{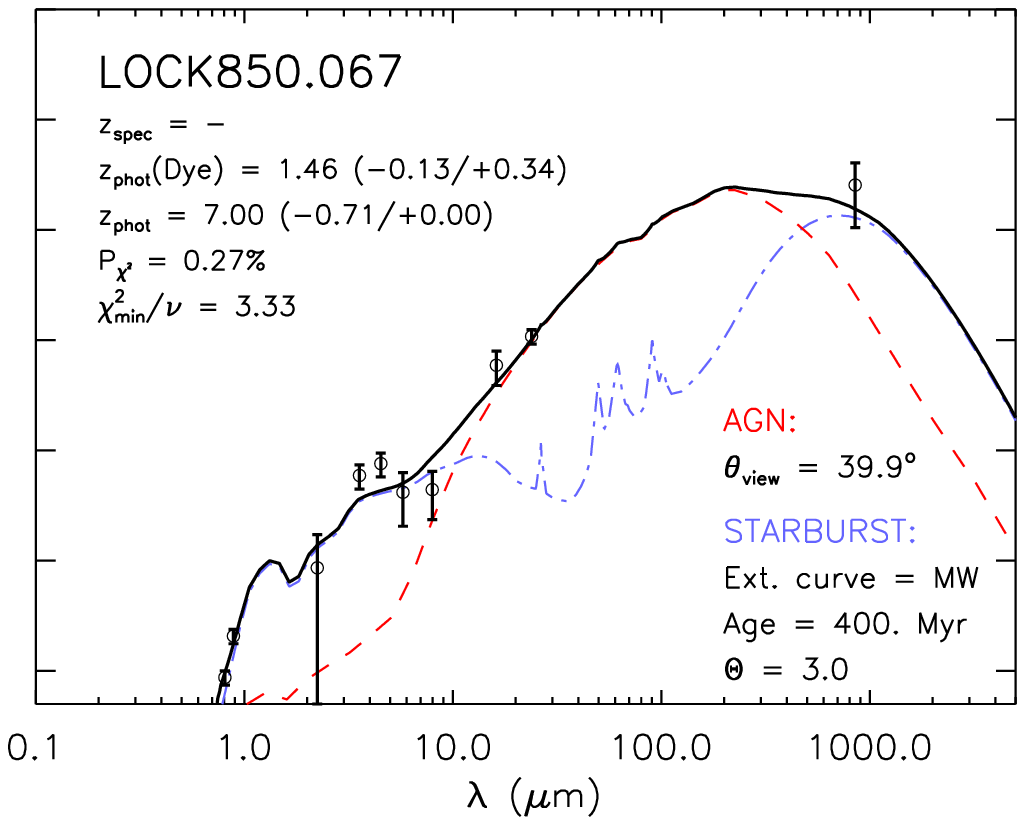}}\nolinebreak
\end{center}\vspace*{-1.3cm}
\begin{center}
\hspace*{-1.8cm}\resizebox{0.37\hsize}{!}{\includegraphics*{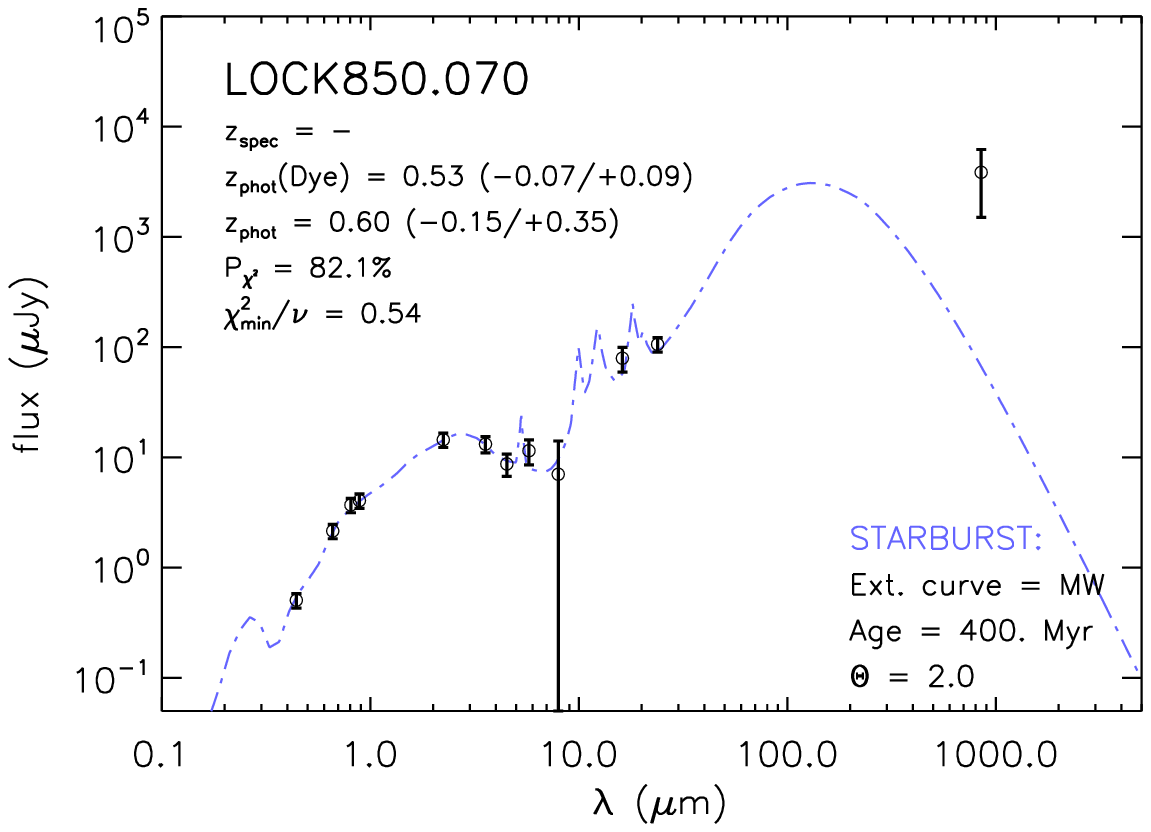}}\nolinebreak
\hspace*{-2.65cm}\resizebox{0.37\hsize}{!}{\includegraphics*{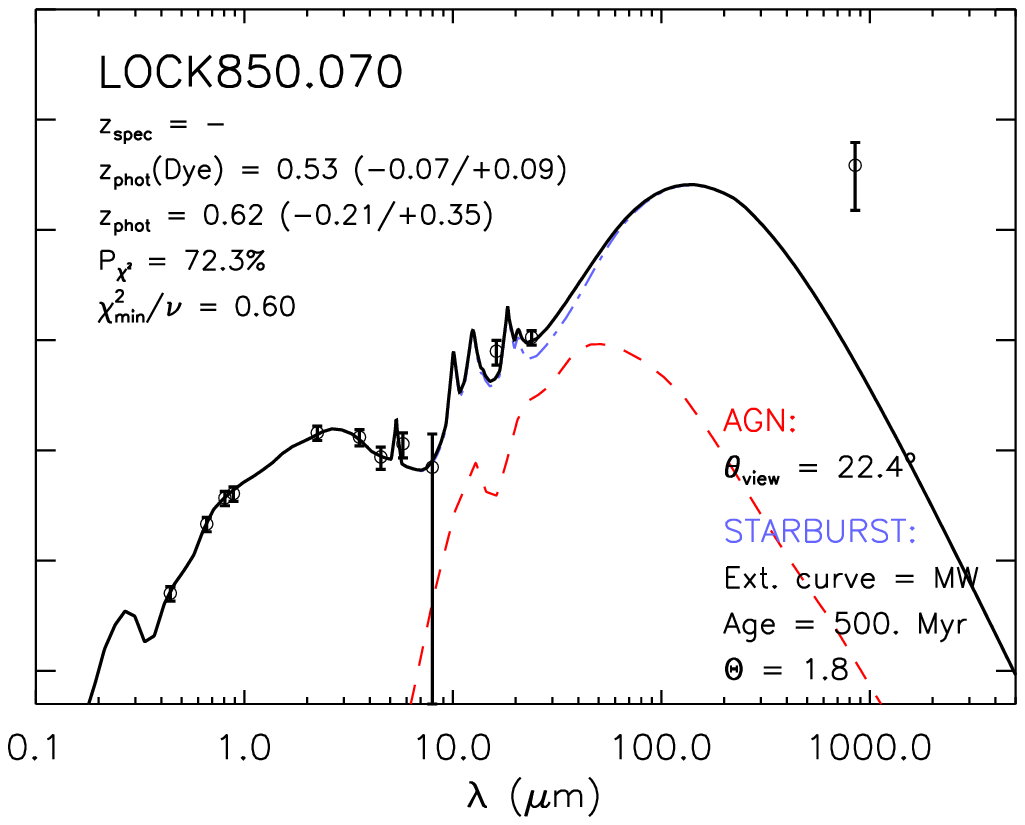}}\nolinebreak
\hspace*{-1.8cm}\resizebox{0.37\hsize}{!}{\includegraphics*{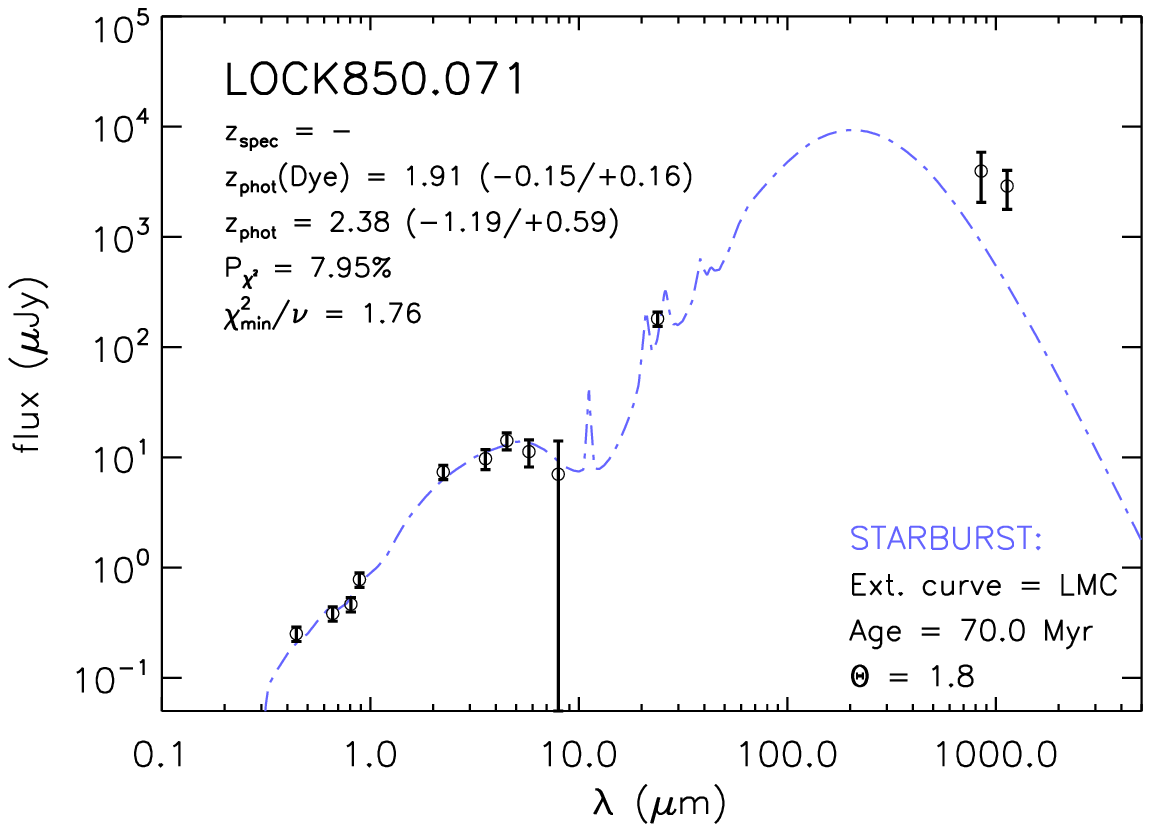}}\nolinebreak
\hspace*{-2.65cm}\resizebox{0.37\hsize}{!}{\includegraphics*{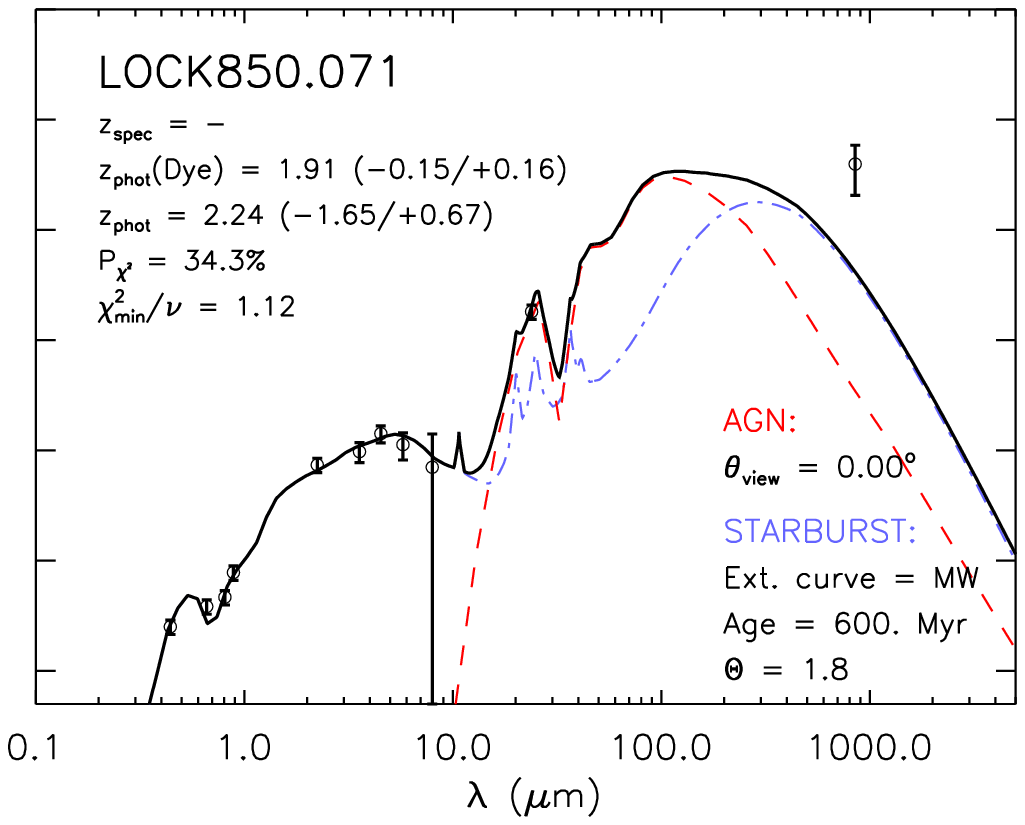}}\nolinebreak
\end{center}\vspace*{-1.3cm}
\vspace*{1.6cm}\caption{SED fits to submm-selected SHADES galaxies in the Lockman Hole East, using models from Takagi et al. (2003, 2004) and Efstathiou \& Rowan-Robinson (1995).}\label{fig:seds4}\end{figure*}

\begin{figure*}[!ht]
\begin{center}
\hspace*{-1.8cm}\resizebox{0.37\hsize}{!}{\includegraphics*{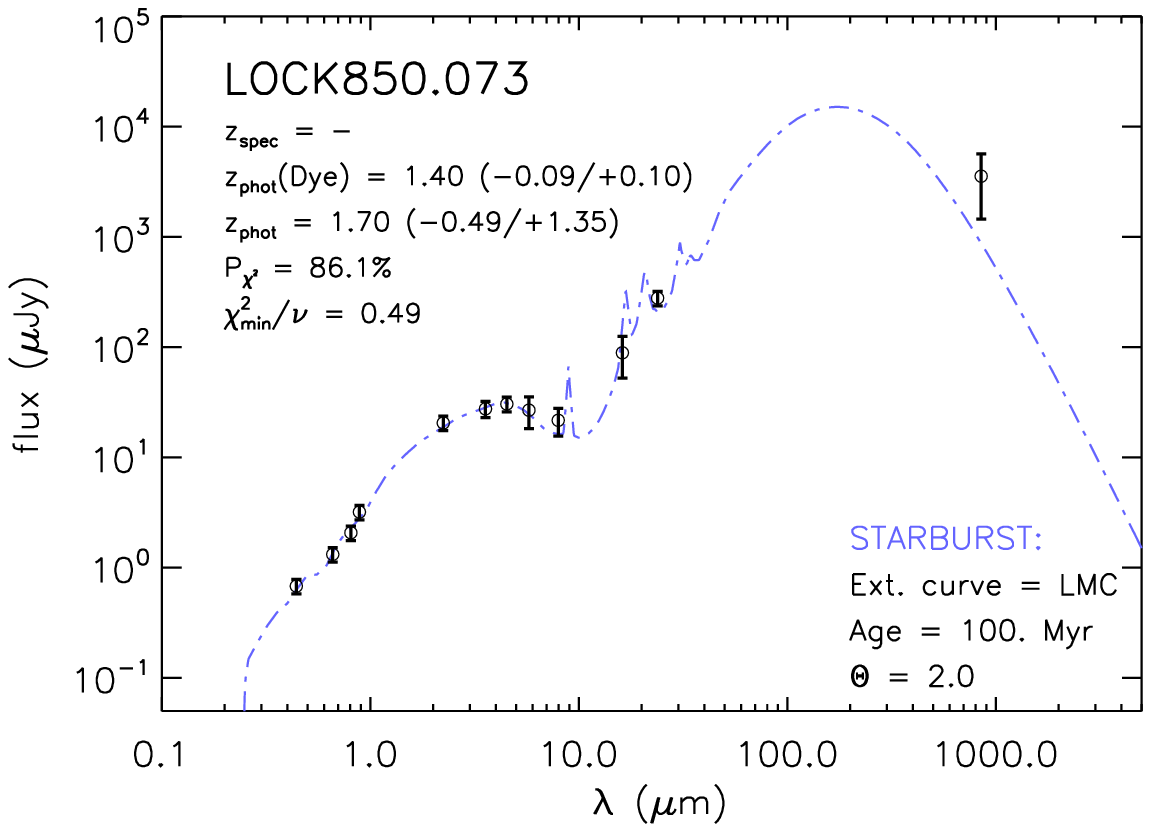}}\nolinebreak
\hspace*{-2.65cm}\resizebox{0.37\hsize}{!}{\includegraphics*{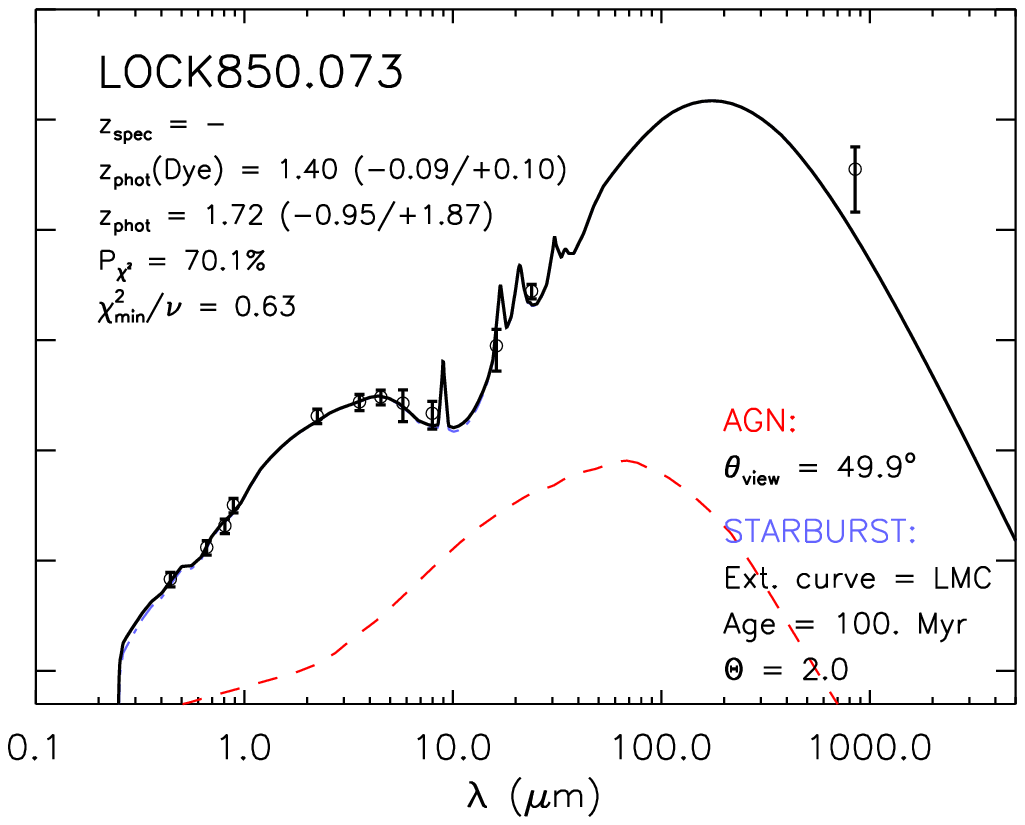}}\nolinebreak
\hspace*{-1.8cm}\resizebox{0.37\hsize}{!}{\includegraphics*{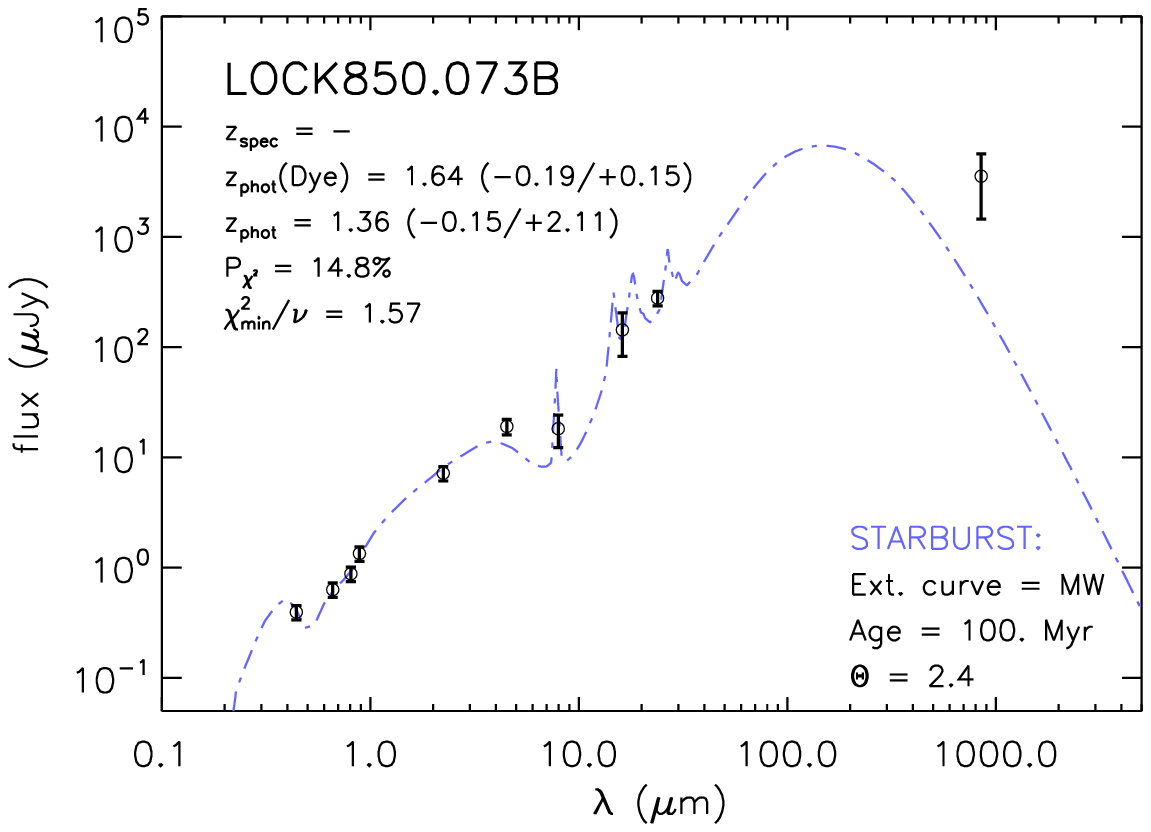}}\nolinebreak
\hspace*{-2.65cm}\resizebox{0.37\hsize}{!}{\includegraphics*{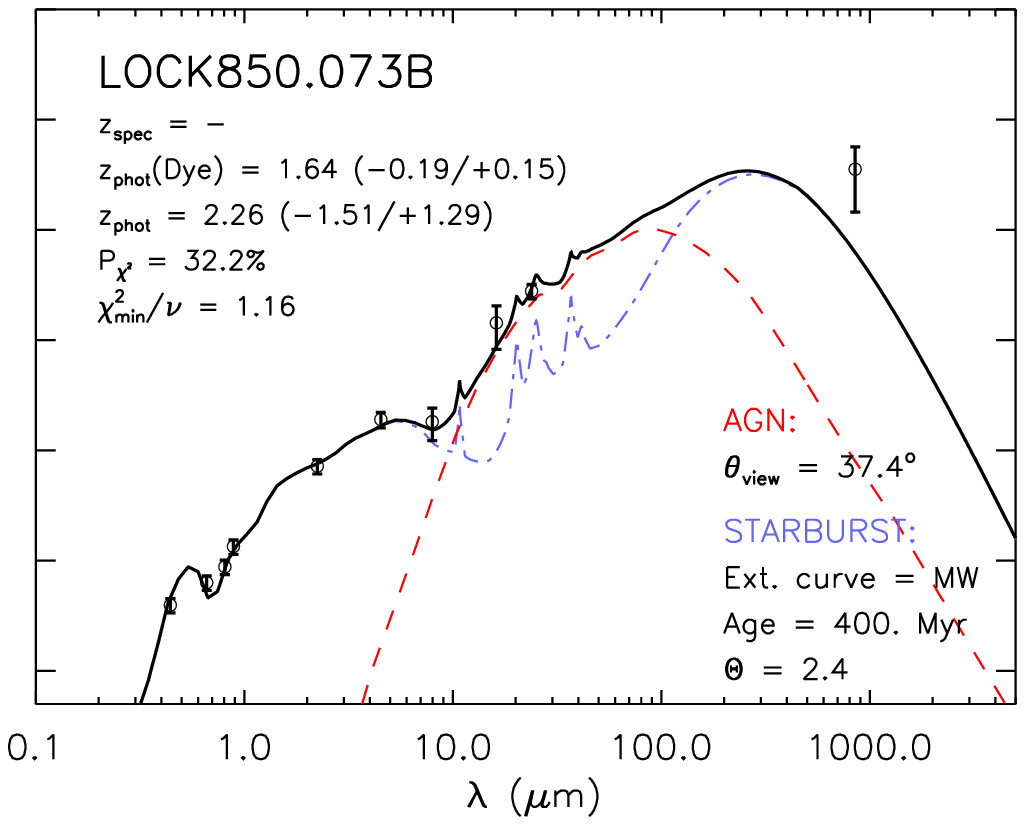}}\nolinebreak
\end{center}\vspace*{-1.3cm}
\begin{center}
\hspace*{-1.8cm}\resizebox{0.37\hsize}{!}{\includegraphics*{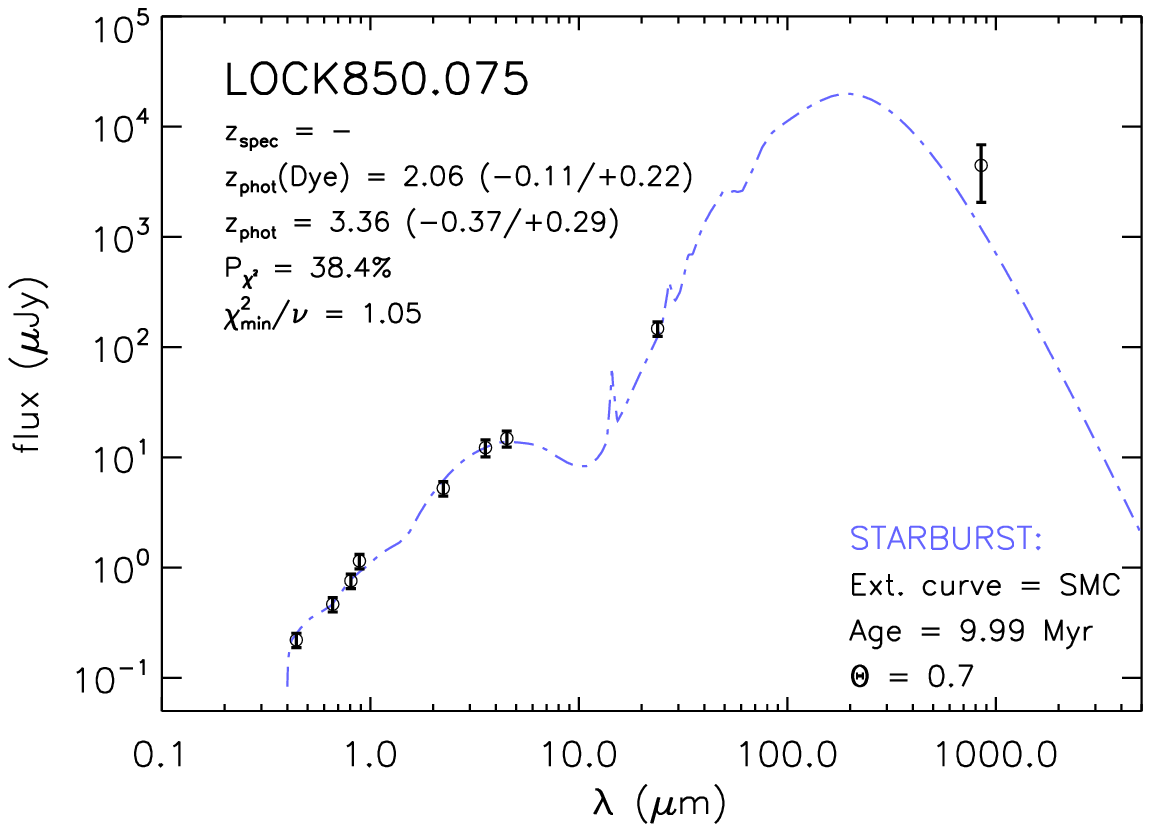}}\nolinebreak
\hspace*{-2.65cm}\resizebox{0.37\hsize}{!}{\includegraphics*{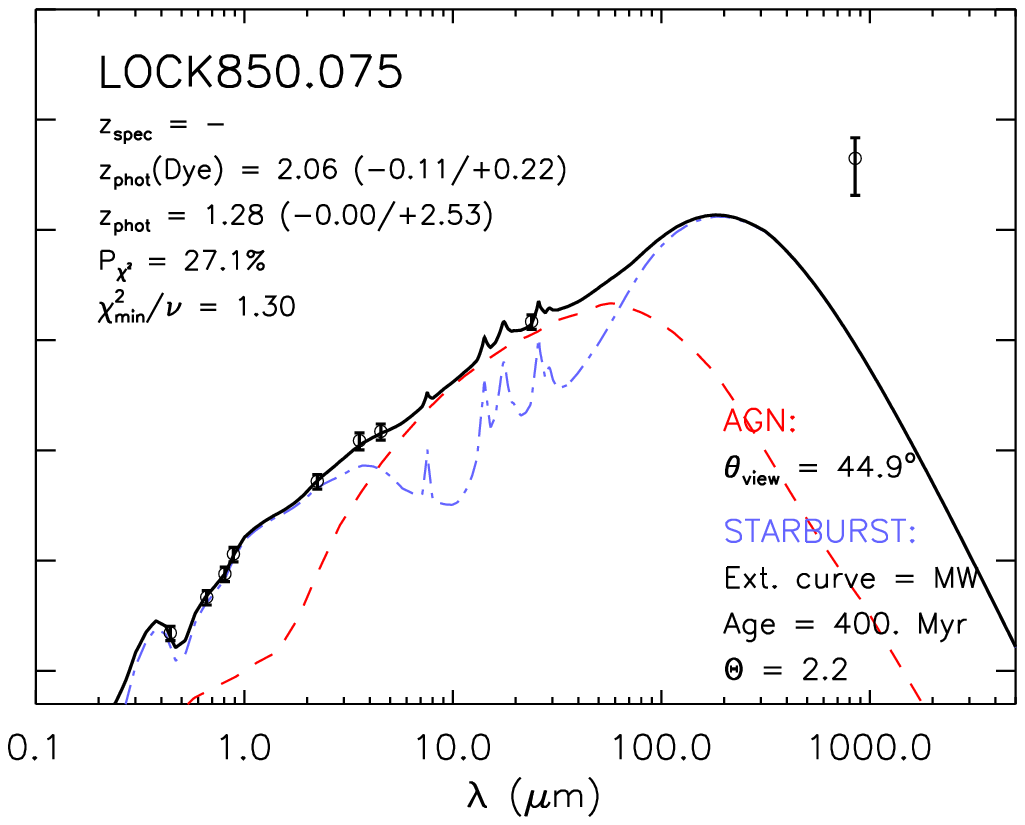}}\nolinebreak
\hspace*{-1.8cm}\resizebox{0.37\hsize}{!}{\includegraphics*{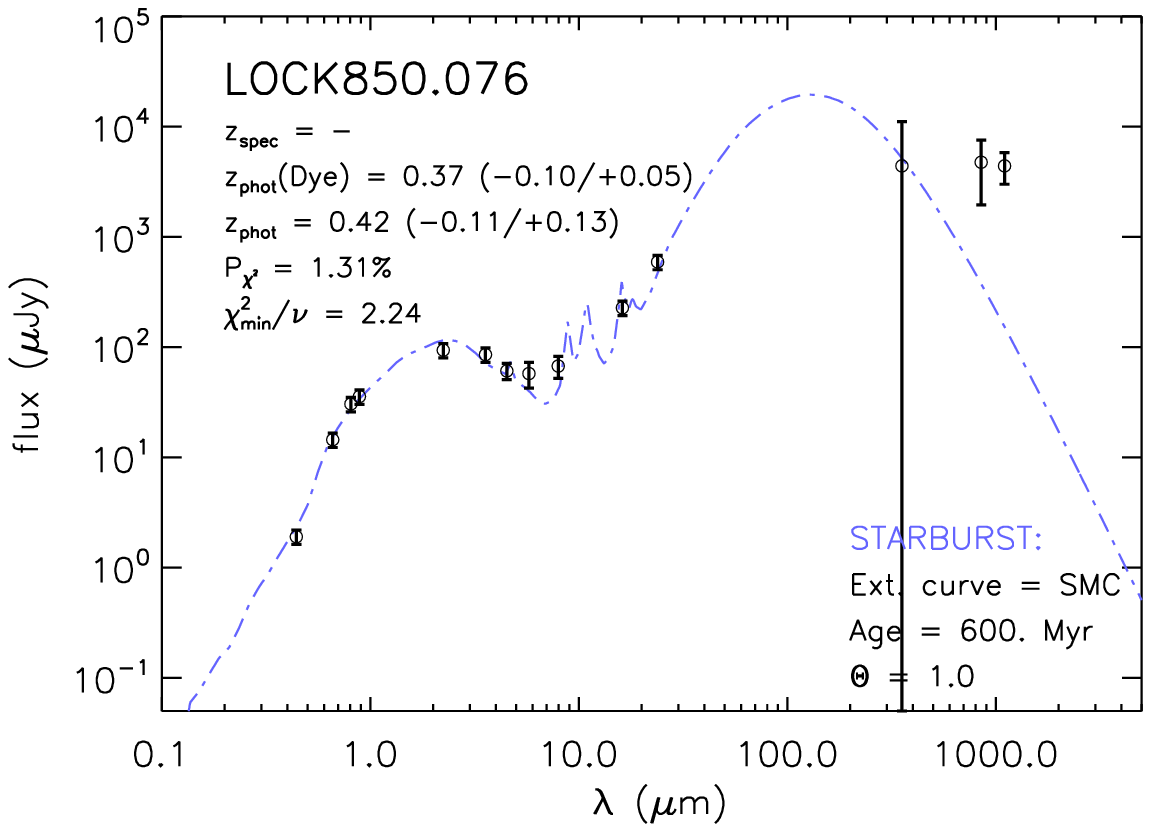}}\nolinebreak
\hspace*{-2.65cm}\resizebox{0.37\hsize}{!}{\includegraphics*{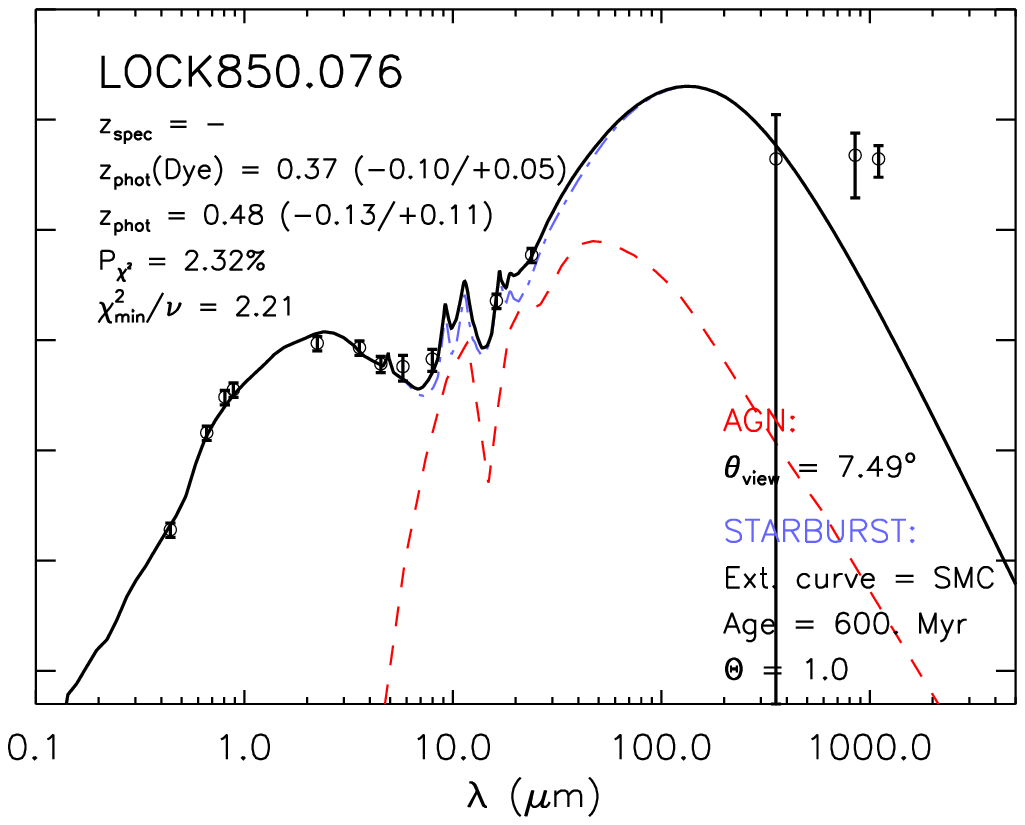}}\nolinebreak
\end{center}\vspace*{-1.3cm}
\begin{center}
\hspace*{-1.8cm}\resizebox{0.37\hsize}{!}{\includegraphics*{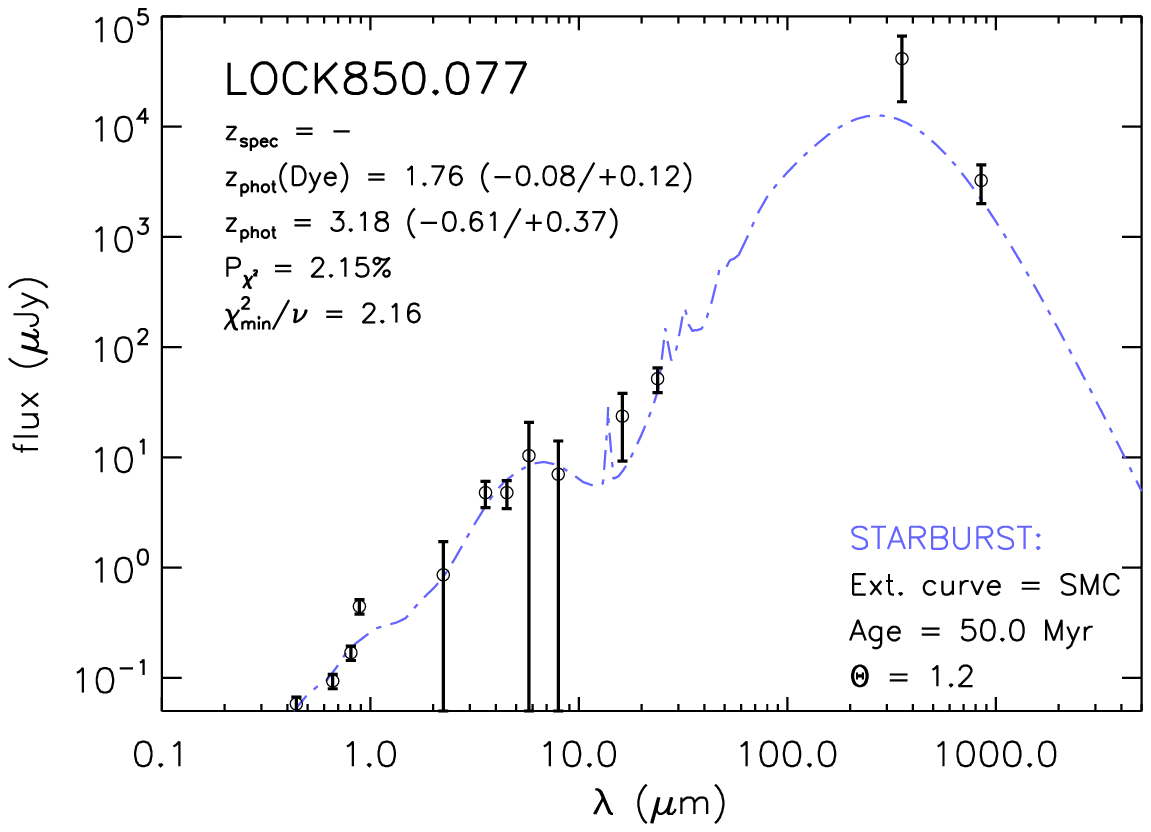}}\nolinebreak
\hspace*{-2.65cm}\resizebox{0.37\hsize}{!}{\includegraphics*{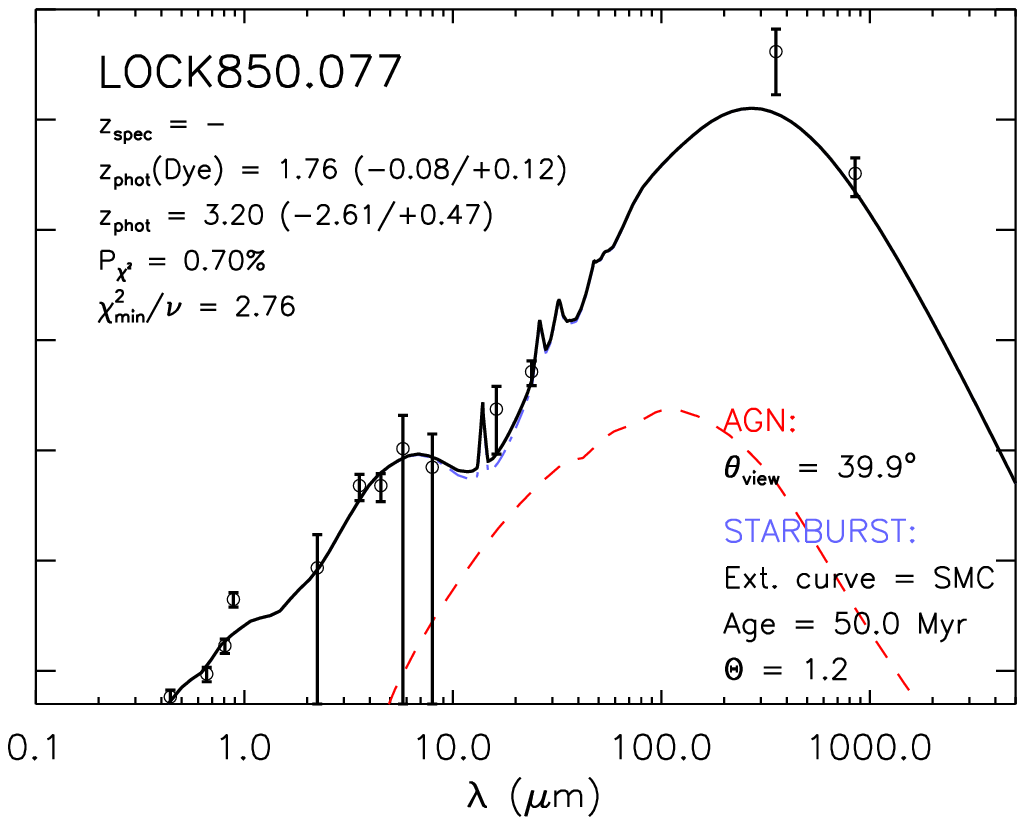}}\nolinebreak
\hspace*{-1.8cm}\resizebox{0.37\hsize}{!}{\includegraphics*{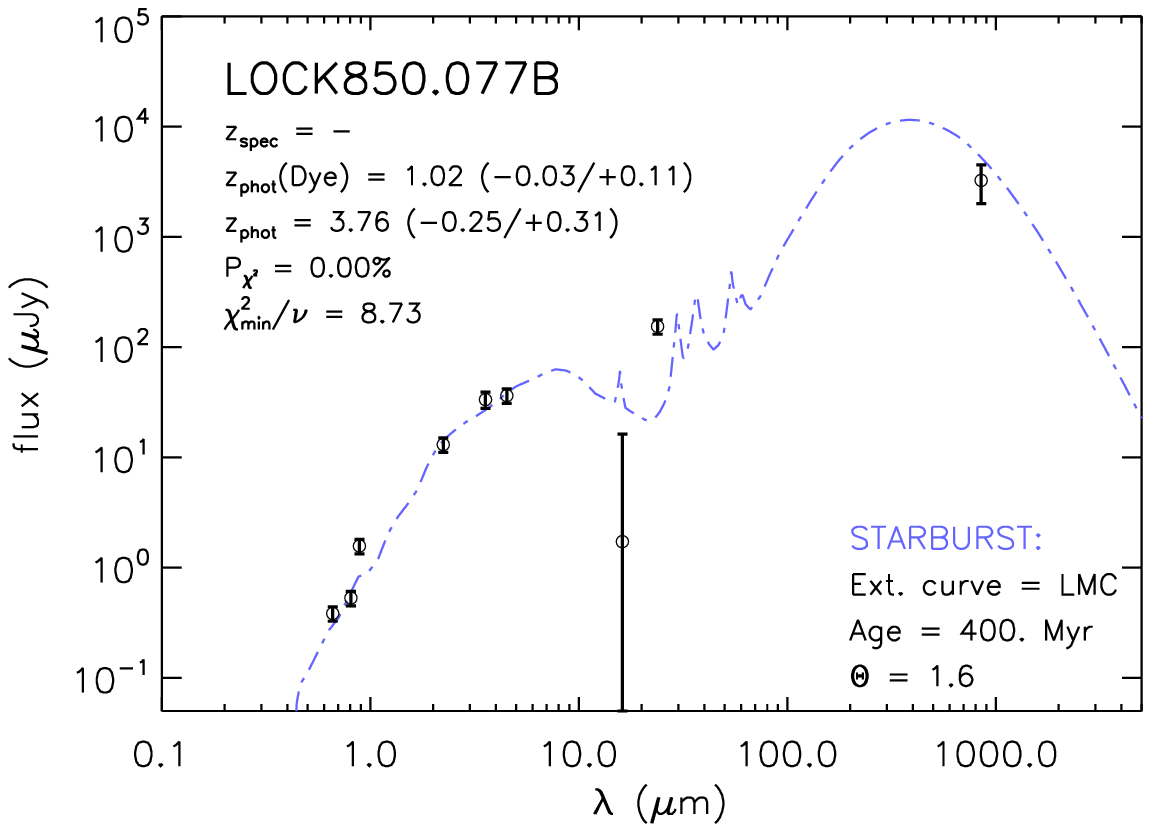}}\nolinebreak
\hspace*{-2.65cm}\resizebox{0.37\hsize}{!}{\includegraphics*{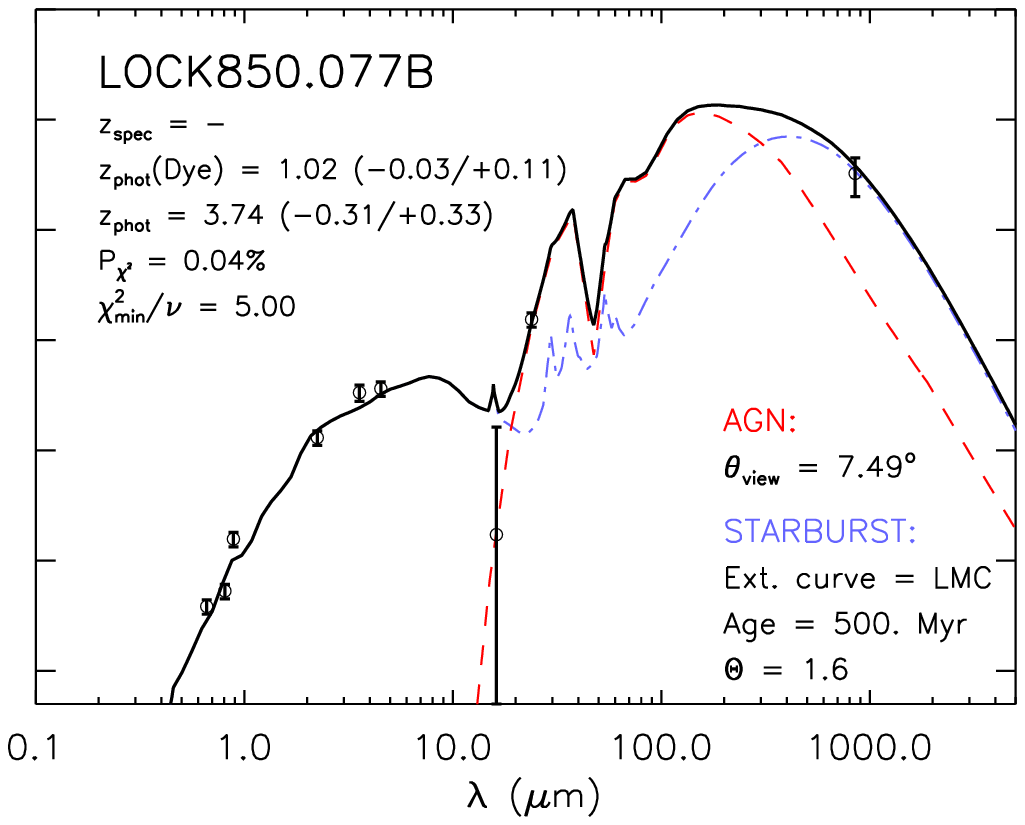}}\nolinebreak
\end{center}\vspace*{-1.3cm}
\begin{center}
\hspace*{-1.8cm}\resizebox{0.37\hsize}{!}{\includegraphics*{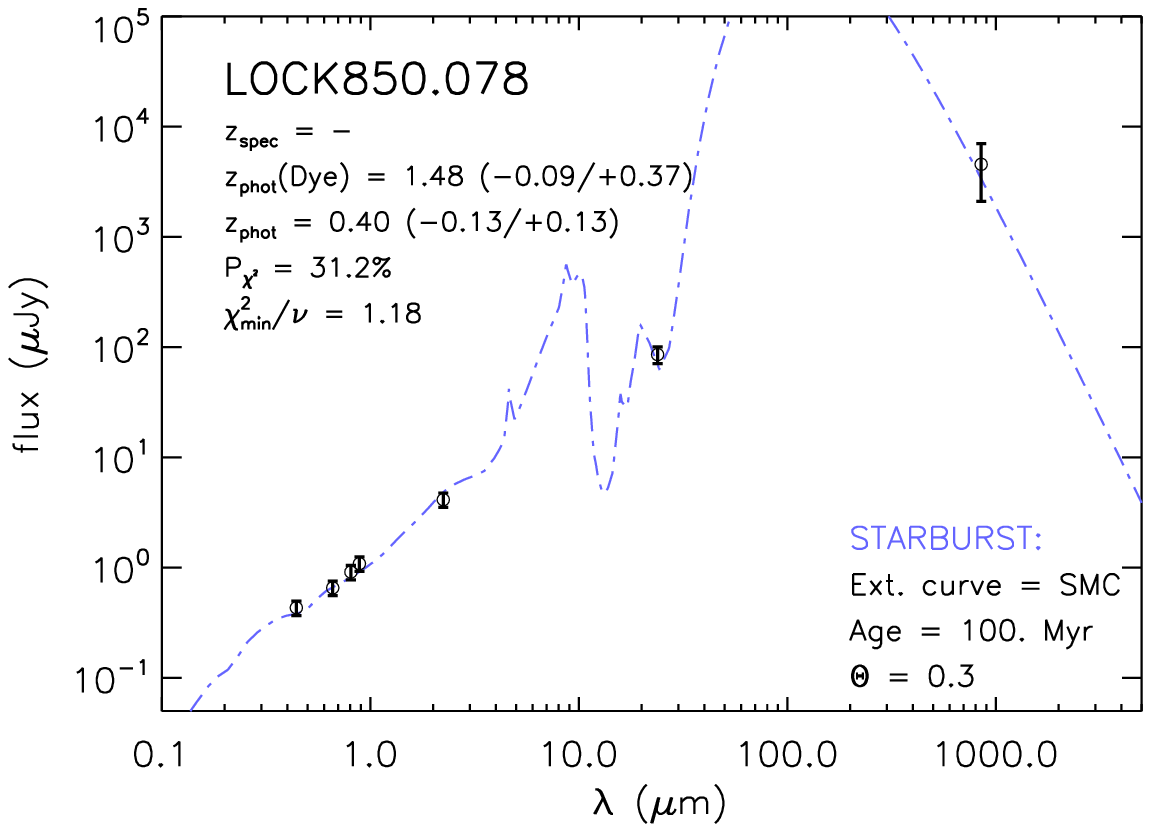}}\nolinebreak
\hspace*{-2.65cm}\resizebox{0.37\hsize}{!}{\includegraphics*{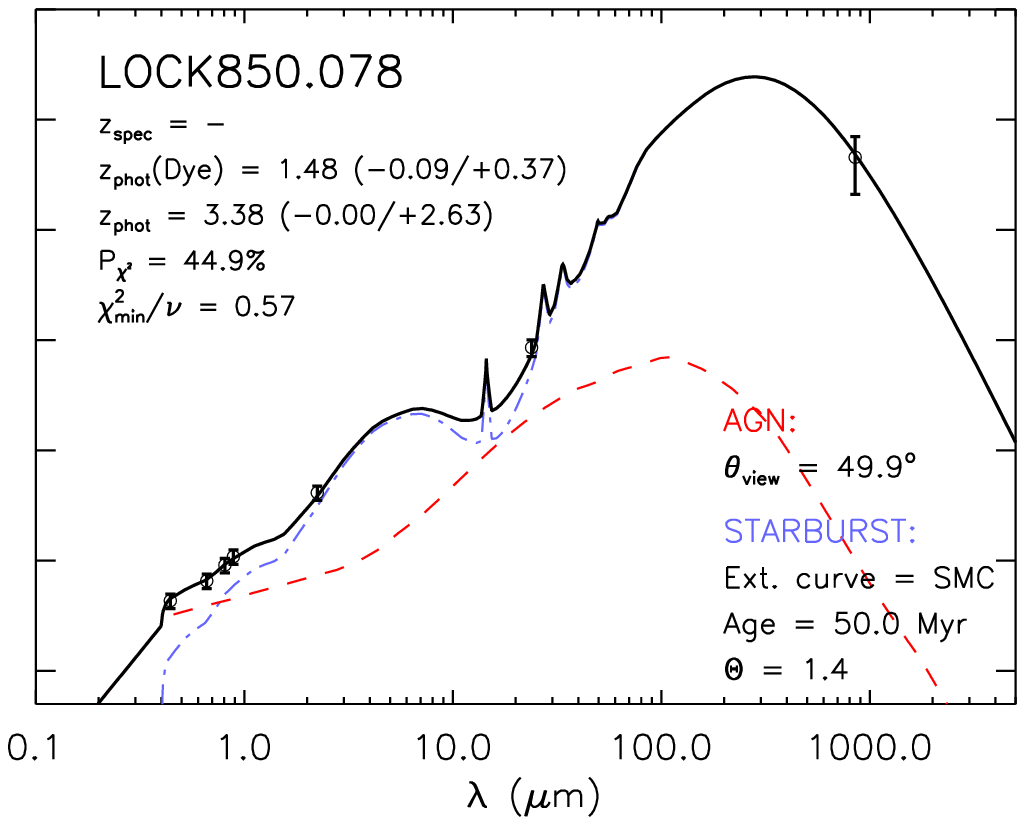}}\nolinebreak
\hspace*{-1.8cm}\resizebox{0.37\hsize}{!}{\includegraphics*{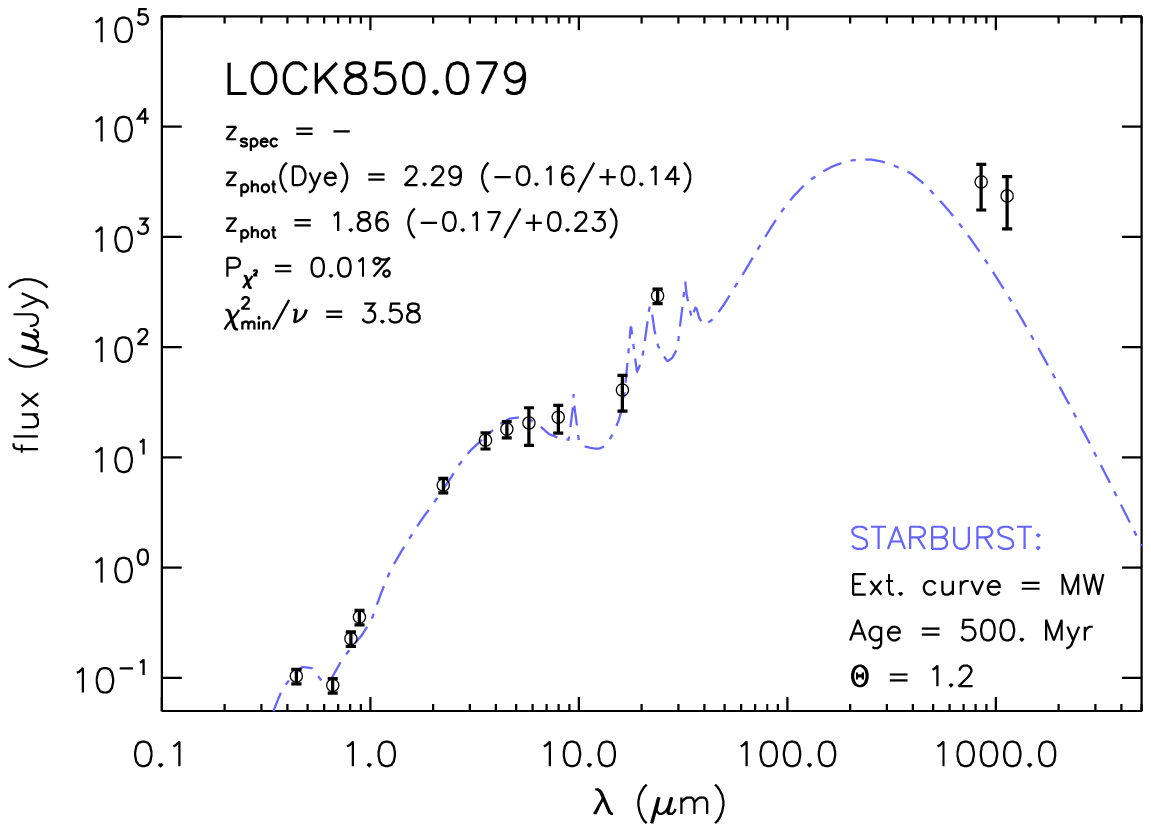}}\nolinebreak
\hspace*{-2.65cm}\resizebox{0.37\hsize}{!}{\includegraphics*{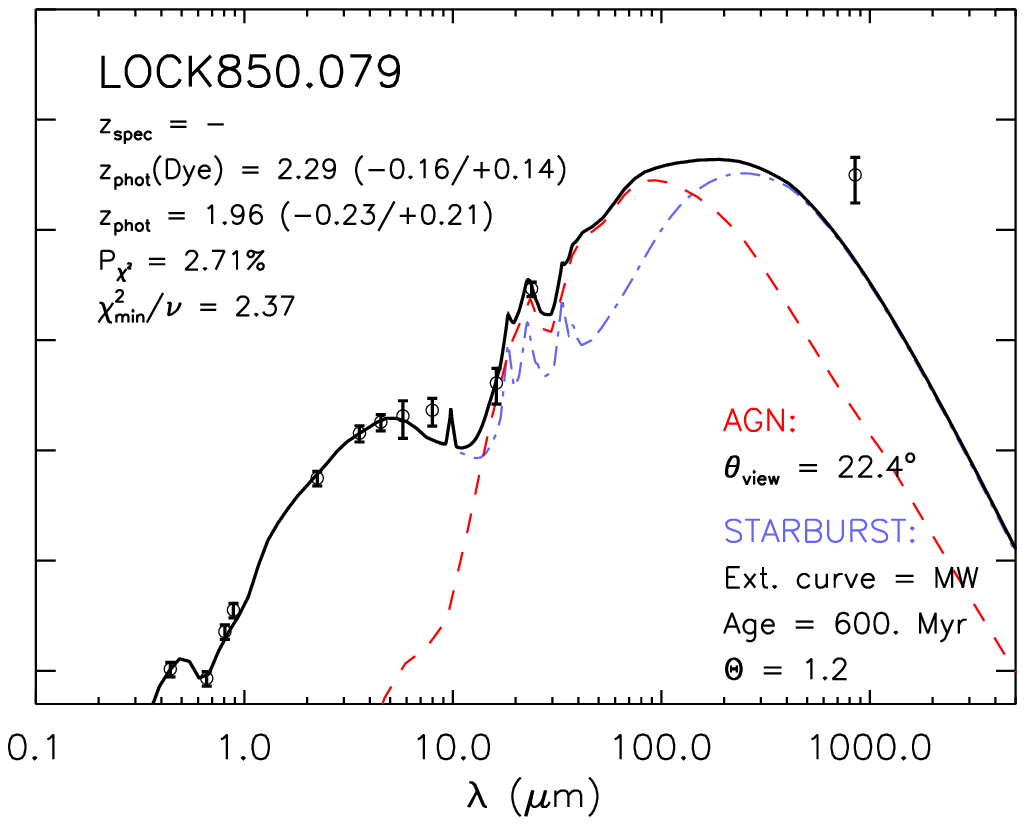}}\nolinebreak
\end{center}\vspace*{-1.3cm}
\begin{center}
\hspace*{-1.8cm}\resizebox{0.37\hsize}{!}{\includegraphics*{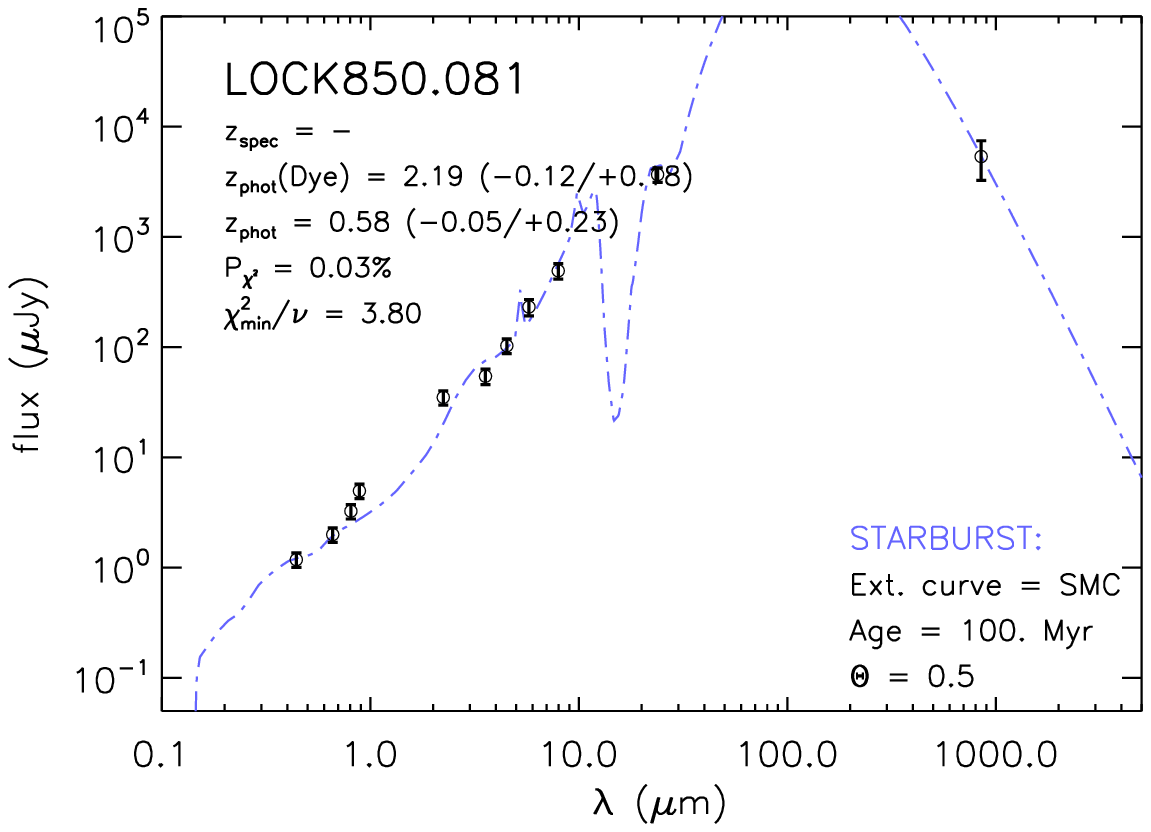}}\nolinebreak
\hspace*{-2.65cm}\resizebox{0.37\hsize}{!}{\includegraphics*{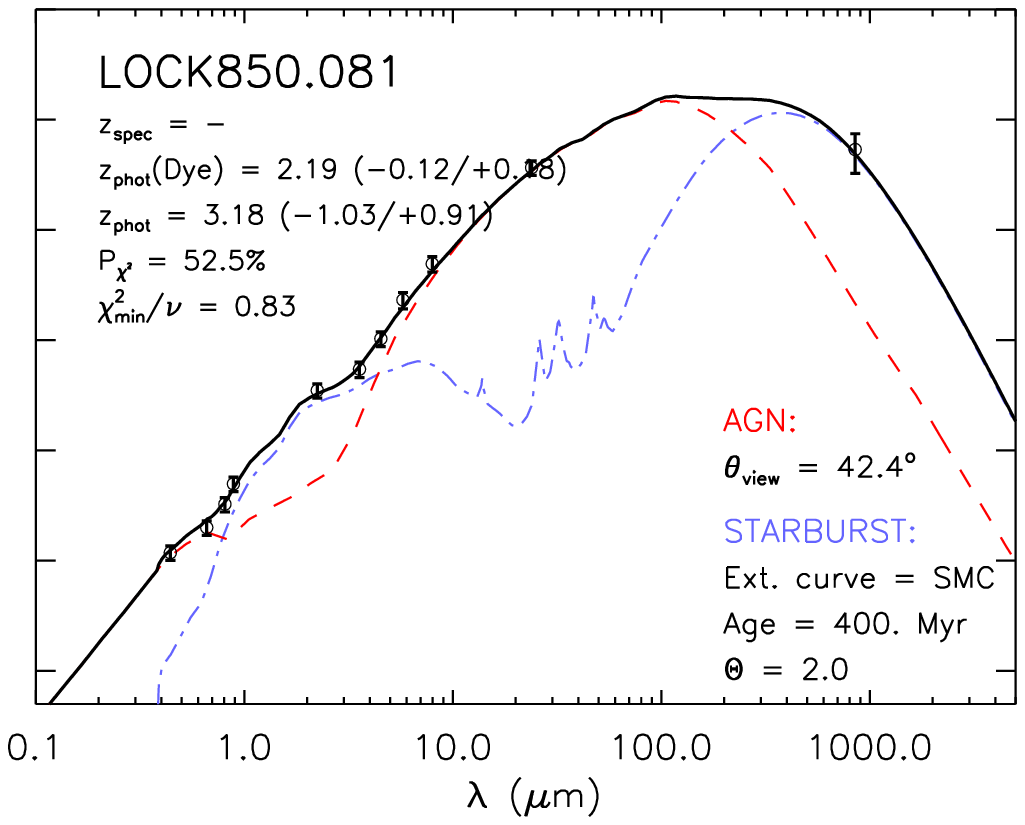}}\nolinebreak
\hspace*{-1.8cm}\resizebox{0.37\hsize}{!}{\includegraphics*{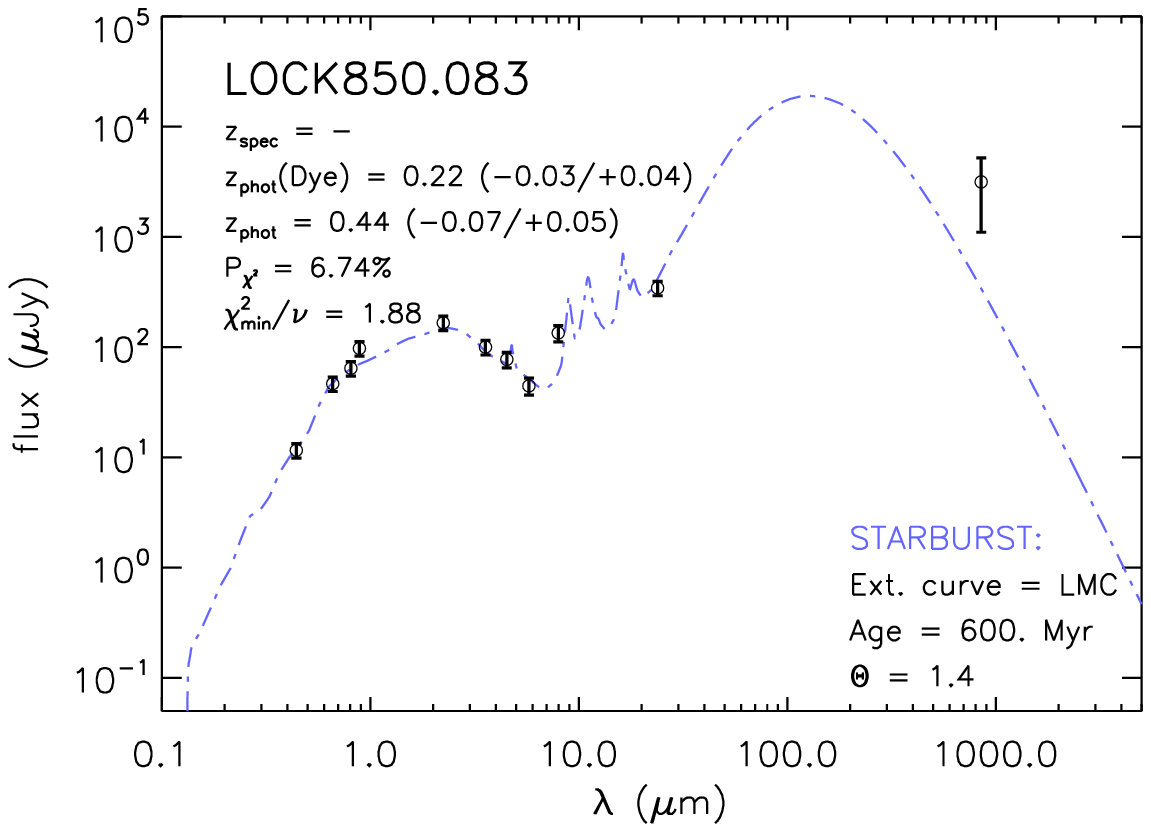}}\nolinebreak
\hspace*{-2.65cm}\resizebox{0.37\hsize}{!}{\includegraphics*{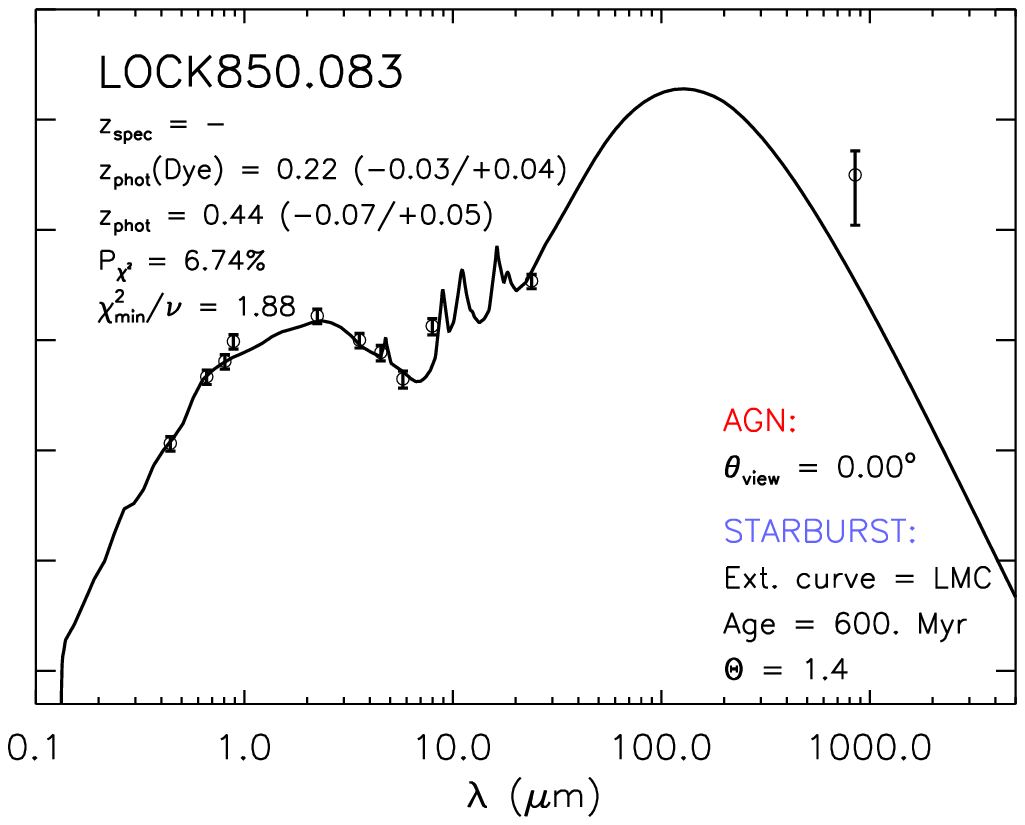}}\nolinebreak
\end{center}\vspace*{-1.3cm}
\begin{center}
\hspace*{-1.8cm}\resizebox{0.37\hsize}{!}{\includegraphics*{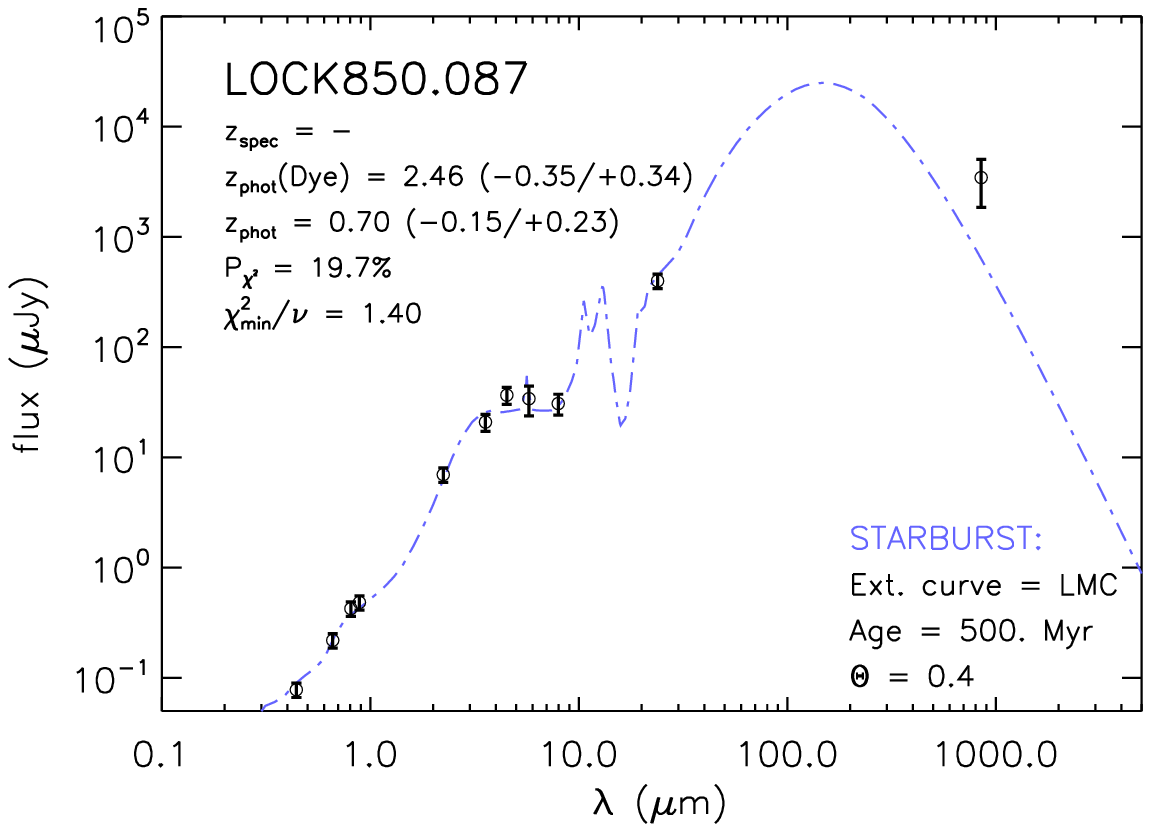}}\nolinebreak
\hspace*{-2.65cm}\resizebox{0.37\hsize}{!}{\includegraphics*{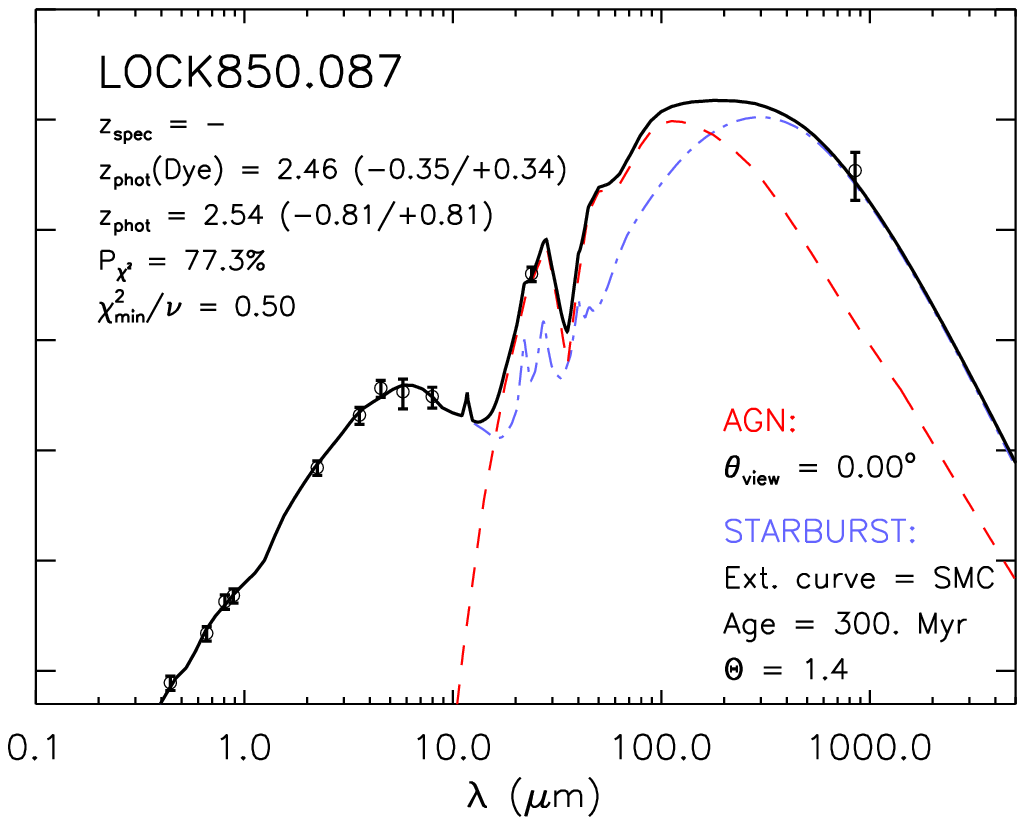}}\nolinebreak
\hspace*{-1.8cm}\resizebox{0.37\hsize}{!}{\includegraphics*{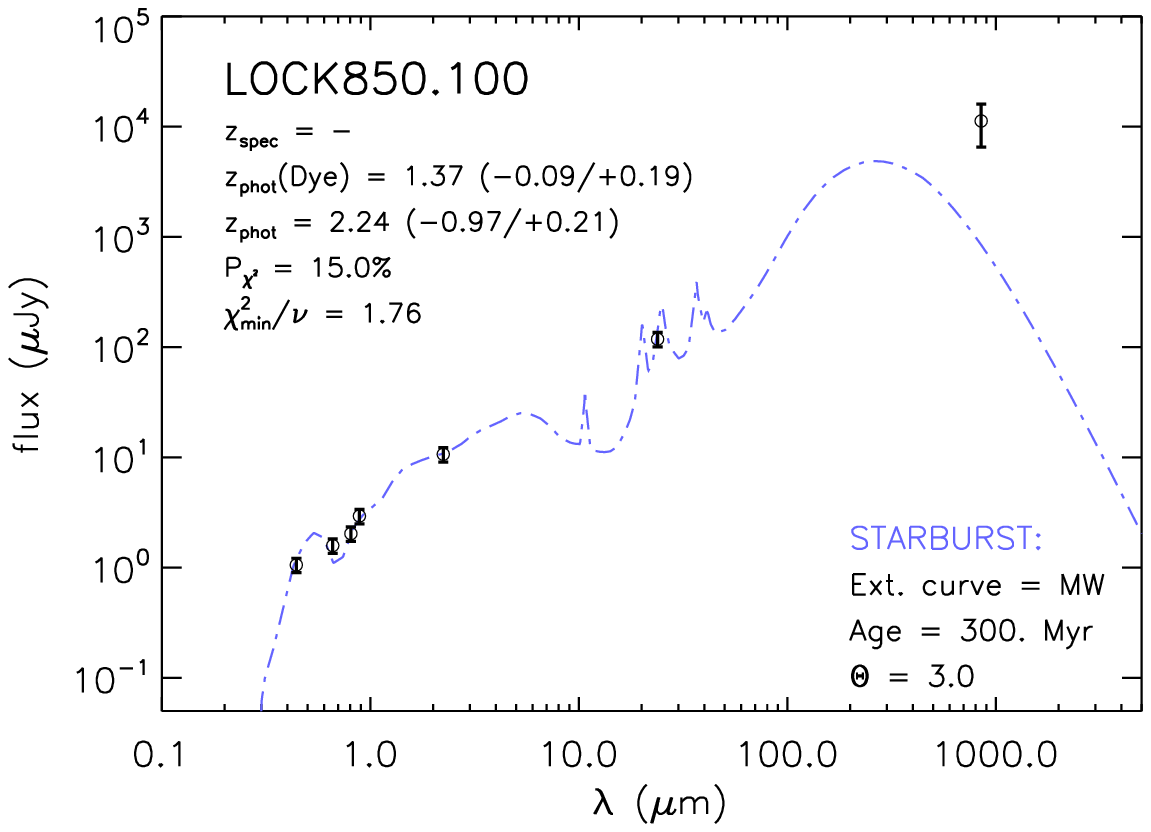}}\nolinebreak
\hspace*{-2.65cm}\resizebox{0.37\hsize}{!}{\includegraphics*{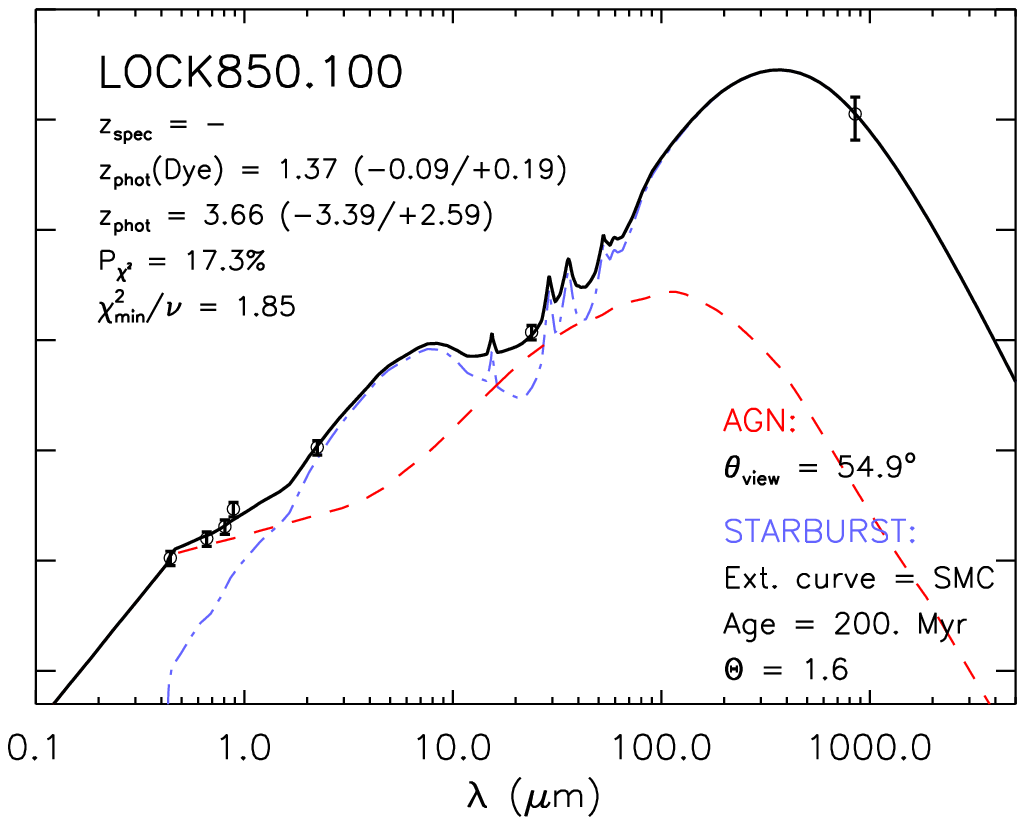}}\nolinebreak
\end{center}\vspace*{-1.3cm}
\vspace*{1.6cm}\caption{SED fits to submm-selected SHADES galaxies in the Lockman Hole East, using models from Takagi et al. (2003, 2004) and Efstathiou \& Rowan-Robinson (1995).}\label{fig:seds5}\end{figure*}

Following Negrello et al. (2009), we found the minimum $\chi^2$ solution for the available photometry 
assuming a mix of starburst and active nucleus, allowing the relative bolometric fractions of the components 
($f_{\rm SB}$ and $f_{\rm AGN}=1-f_{\rm SB}$ respectively) to vary. 
Table \ref{tab:sed_fitting_results} gives the best-fit
parameters of these two-component fits, excluding those for which no fit could be made, but including for completeness those for
which the fitting is under-constrained. 
Figs. \ref{fig:seds1} to \ref{fig:seds5} show the best-fit SEDs,
as well as the best-fit starburst-only model for each SHADES galaxy. In many cases, adding the active nucleus
is a requirement for reproducing the mid-infrared data. In several other cases however, no acceptable $\chi^2$
could be found, because the submm and mm-wave data exceeded the range given by the models. In these cases, 
a further cold cirrus component may contribute significantly to the longest wavelengths. 

Several galaxies are worth discussing in detail. The galaxies with an excess at submm/mm-wavelengths
are LOCK850.009, 011, 013, 031, 037B, 040, 043/043B, 053, 070, and 071. In addition, Lock850.076 
has peculiar submm/mm-wave colours, perhaps due to incorrect flux deboosting in either the submm or mm-wave data. 
The $15\,\mu$m flux in Lock850.003B may be contaminated by a nearby object. 
The SED fits in some
cases suggest particular identifications, e.g. LOCK850.010B is preferred over LOCK850.010 (not plotted since no acceptable
fit could be made to the data), as
is LOCK850.015 over LOCK850.015B, 
LOCK850.035 over LOCK850.035B, and LOCK850.073 over LOCK850.073B. The identifications for LOCK850.004, LOCK850.009, 
LOCK850.043 and LOCK850.77 remain ambiguous. 

\begin{table*}
\begin{tabular}{llllllllll llllllllll llllllllll llllllllll llll}
Name & B (AB) & R (AB) & I (AB) & z (AB) & K (AB) & $3.6\,\mu$m (AB) & $4.5\,\mu$m (AB) & $5.8\,\mu$m (AB) & $8\,\mu$m (AB) \\ 
Lock850.001&&26.29$\pm$0.1&25.52$\pm$0.07&25.7$\pm$0.18&&22.31$\pm$0.34&21.77$\pm$0.29 &              &21.78$\pm$1.12\\
Lock850.002&26.75$\pm$0.1&25.26$\pm$0.06&24.8$\pm$0.06&24.56$\pm$0.07&21.85$\pm$0.28&21.42$\pm$0.25 &21.38$\pm$0.52&21.29$\pm$0.45\\
Lock850.003&26.45$\pm$0.09&24.9$\pm$0.06&24.21$\pm$0.06&23.47$\pm$0.06&21.39$\pm$0.06&20.66$\pm$0.21&20.67$\pm$0.2 &20.47$\pm$0.39&21.09$\pm$0.44\\
Lock850.003b&&&25.56$\pm$0.07&25.05$\pm$0.1&&21.05$\pm$0.19&21.03$\pm$0.21 &20.9$\pm$0.44 &21.2$\pm$0.4  \\
Lock850.004&25.57$\pm$0.06&24.48$\pm$0.06&24.17$\pm$0.06&23.96$\pm$0.06&21.84$\pm$0.08&20.83$\pm$0.2&20.49$\pm$0.19 &20.64$\pm$0.41&20.48$\pm$0.4 \\
Lock850.004b&27.14$\pm$0.15&26.09$\pm$0.09&25.85$\pm$0.09&25.66$\pm$0.18&22.41$\pm$0.15&21.15$\pm$0.19&20.94$\pm$0.18 &20.79$\pm$0.45&20.45$\pm$0.43\\
Lock850.005&&25.66$\pm$0.06&24.67$\pm$0.06&23.77$\pm$0.06&21.16$\pm$0.06&20.62$\pm$0.2&20.61$\pm$0.21 &21.25$\pm$0.35&21.53$\pm$0.33\\
Lock850.006&28.08$\pm$0.34&26.2$\pm$0.1&25.81$\pm$0.08&25.22$\pm$0.12&&21.35$\pm$0.23&21.25$\pm$0.23 &20.93$\pm$0.58&21.42$\pm$0.46\\
Lock850.007& & & & & && &              &              \\
Lock850.008& & & & & && &              &              \\
Lock850.009&26.86$\pm$0.11&25.53$\pm$0.06&24.79$\pm$0.06&24.43$\pm$0.06&21.58$\pm$0.06&20.25$\pm$0.2&20.23$\pm$0.22 &20.43$\pm$0.54&20.91$\pm$0.45\\
Lock850.009b&24.04$\pm$0.06&23.09$\pm$0.06&22.3$\pm$0.06&21.8$\pm$0.06&20.42$\pm$0.06&19.83$\pm$0.19&20.26$\pm$0.16 &20.56$\pm$0.5 &21.21$\pm$0.46\\
Lock850.010&&&&&&22.71$\pm$0.4&22.27$\pm$0.35 &              &              \\
Lock850.010b&26.2$\pm$0.06&25.67$\pm$0.06&25.09$\pm$0.06&24.31$\pm$0.06&&21.95$\pm$0.3&21.67$\pm$0.29 &              &              \\
Lock850.011&26.91$\pm$0.12&26.09$\pm$0.09&25.37$\pm$0.06&24.63$\pm$0.07&22.17$\pm$0.11&21.53$\pm$0.19&21.56$\pm$0.18 &22.11$\pm$0.5 &21.57$\pm$0.41\\
Lock850.012&26.38$\pm$0.07&25.17$\pm$0.06&24.97$\pm$0.06&24.6$\pm$0.07&21.71$\pm$0.07&20.77$\pm$0.2&20.51$\pm$0.18 &20.52$\pm$0.46&20.96$\pm$0.39\\
Lock850.013&21.71$\pm$0.06&20.67$\pm$0.06&20.37$\pm$0.06&20.31$\pm$0.06&19.64$\pm$0.06&20.22$\pm$0.2&20.56$\pm$0.16 &21.08$\pm$0.41&20.56$\pm$0.35\\
Lock850.014&24.32$\pm$0.06&23.74$\pm$0.06&23.4$\pm$0.06&23.18$\pm$0.06&21.06$\pm$0.06&20.2$\pm$0.19&20.15$\pm$0.21 &20.15$\pm$0.66&20.56$\pm$0.69\\
Lock850.015&26.91$\pm$0.12&26.47$\pm$0.07&25.7$\pm$0.23&26.09$\pm$0.26&22.66$\pm$0.2&22.04$\pm$0.2&21.5$\pm$0.2 &21.12$\pm$0.55&20.93$\pm$0.47\\
Lock850.015b&&&&&&21.47$\pm$0.25&20.98$\pm$0.21 &              &              \\
Lock850.016&27.28$\pm$0.17&25.2$\pm$0.06&23.99$\pm$0.06&23.47$\pm$0.06&20.45$\pm$0.06&19.93$\pm$0.21&19.65$\pm$0.22 &19.73$\pm$0.35&20.37$\pm$0.31\\
Lock850.017&25.78$\pm$0.06&24.54$\pm$0.06&23.67$\pm$0.06&22.92$\pm$0.06&20.53$\pm$0.06&20.1$\pm$0.2&20.05$\pm$0.18 &20.24$\pm$0.39&20.56$\pm$0.35\\
Lock850.017b&&25.82$\pm$0.07&25.58$\pm$0.07&24.68$\pm$0.08&&& &              &              \\
Lock850.018&&&&&&& &              &              \\
Lock850.019&27.32$\pm$0.17&24.93$\pm$0.06&24.47$\pm$0.06&24.52$\pm$0.07&21.71$\pm$0.08&21.06$\pm$0.2&20.93$\pm$0.19 &20.83$\pm$0.5 &21.25$\pm$0.59\\
Lock850.021&&&&&&22.13$\pm$0.33&22.13$\pm$0.35 &21.42$\pm$0.7 &22.37$\pm$0.6 \\
Lock850.022&23.65$\pm$0.06&23.05$\pm$0.06&22.82$\pm$0.06&22.74$\pm$0.06&21.77$\pm$0.08&21.27$\pm$0.19&21.01$\pm$0.16 &20.8$\pm$0.4  &21.07$\pm$0.31\\
Lock850.023&25.47$\pm$0.06&22.38$\pm$0.06&21.25$\pm$0.06&20.7$\pm$0.06&19.06$\pm$0.06&18.89$\pm$0.2&19.51$\pm$0.21 &19.71$\pm$0.31&20.75$\pm$0.3 \\
Lock850.024&24.61$\pm$0.06&23.74$\pm$0.06&23.14$\pm$0.06&22.65$\pm$0.06&20.52$\pm$0.06&19.7$\pm$0.2&19.59$\pm$0.18 &19.99$\pm$0.32&20.62$\pm$0.38\\
Lock850.026&26.95$\pm$0.12&25.45$\pm$0.12&25.3$\pm$0.11&24.73$\pm$0.1&22.7$\pm$0.19&21.68$\pm$0.27&21.37$\pm$0.24 &20.9$\pm$0.62 &21.83$\pm$0.53\\
Lock850.027&23.77$\pm$0.06&23.55$\pm$0.06&22.78$\pm$0.06&22.32$\pm$0.06&21.18$\pm$0.06&20.9$\pm$0.2&21$\pm$0.19 &              &21.69$\pm$0.45\\
Lock850.028&25.9$\pm$0.06&24.6$\pm$0.06&23.9$\pm$0.06&23.27$\pm$0.06&20.88$\pm$0.06&20.29$\pm$0.2&20.2$\pm$0.17 &20.35$\pm$0.37&20.88$\pm$0.36\\
Lock850.029&&&&&&& &              &              \\
Lock850.030&25.89$\pm$0.06&25.62$\pm$0.1&25.27$\pm$0.1&24.97$\pm$0.08&&21.81$\pm$0.25&21.61$\pm$0.24 &              &              \\
Lock850.031&26.56$\pm$0.09&25.81$\pm$0.07&25.25$\pm$0.06&24.94$\pm$0.09&21.76$\pm$0.07&20.44$\pm$0.2&20.32$\pm$0.2 &20.2$\pm$0.58 &20.98$\pm$0.5 \\
Lock850.033&26.91$\pm$0.2&26.08$\pm$0.2&26.22$\pm$0.2&25.93$\pm$0.2&&22.62$\pm$0.33&22.12$\pm$0.27 &              &21.85$\pm$0.68\\
Lock850.034&26.78$\pm$0.11&24.91$\pm$0.06&24.74$\pm$0.06&24.35$\pm$0.06&&21.7$\pm$0.28&21.45$\pm$0.26 &21.12$\pm$0.64&21.41$\pm$0.57\\
Lock850.035&&&&&&& &              &              \\
Lock850.035b&24.65$\pm$0.06&24.31$\pm$0.06&23.83$\pm$0.06&23.64$\pm$0.06&21.74$\pm$0.08&21.51$\pm$0.21&21.33$\pm$0.2 &              &21.79$\pm$0.55\\
Lock850.036& & & & & && &              &              \\
Lock850.037&&&&&&& &              &              \\
Lock850.037b&25.81$\pm$0.06&24.81$\pm$0.06&24.09$\pm$0.06&23.87$\pm$0.06&21.14$\pm$0.06&20.4$\pm$0.18&20.23$\pm$0.22 &20.63$\pm$0.43&21.32$\pm$0.33\\
Lock850.038&24.28$\pm$0.06&23.63$\pm$0.06&23.12$\pm$0.06&22.49$\pm$0.06&20.82$\pm$0.06&20.28$\pm$0.2&20.02$\pm$0.21 &20.53$\pm$0.33&20.71$\pm$0.3 \\
Lock850.039& & & & & && &              &              \\
Lock850.040&26.33$\pm$0.07&25.95$\pm$0.08&25.68$\pm$0.08&24.97$\pm$0.1&22.47$\pm$0.16&21.21$\pm$0.19&20.95$\pm$0.23 &20.69$\pm$0.51&21.36$\pm$0.45\\
Lock850.041b&24.42$\pm$0.06&23.95$\pm$0.06&23.53$\pm$0.06&22.99$\pm$0.06& &21.25$\pm$0.1&21.05$\pm$0.22 &20.42$\pm$0.39&19.66$\pm$0.37\\
Lock850.041&22.95$\pm$0.06&22.05$\pm$0.06&21.53$\pm$0.06&21.05$\pm$0.06&19.96$\pm$0.06&19.27$\pm$0.16&19.26$\pm$0.19 &19.53$\pm$0.4 &19.69$\pm$0.34\\
Lock850.043&26.21$\pm$0.06&25.18$\pm$0.06&24.82$\pm$0.06&24.48$\pm$0.06&21.6$\pm$0.07&21.19$\pm$0.2&21.07$\pm$0.19 &20.88$\pm$0.53&21.52$\pm$0.42\\
Lock850.043b&23.66$\pm$0.3&23.01$\pm$0.15&22.46$\pm$0.11&21.82$\pm$0.06&20.14$\pm$0.06&19.69$\pm$0.16&19.52$\pm$0.18 &19.78$\pm$0.36&20.39$\pm$0.31\\
Lock850.047&26.51$\pm$0.08&23.84$\pm$0.06&23.23$\pm$0.06&22.8$\pm$0.06&21.95$\pm$0.1&21.45$\pm$0.18&21.65$\pm$0.23 &22.21$\pm$0.58&              \\
Lock850.048&23$\pm$0.06&21.47$\pm$0.06&21.11$\pm$0.06&20.62$\pm$0.06&19.63$\pm$0.06&19.73$\pm$0.18&20.07$\pm$0.21 &20.53$\pm$0.39&19.41$\pm$0.32\\
Lock850.052&26.17$\pm$0.06&24.56$\pm$0.06&23.61$\pm$0.06&22.88$\pm$0.06&20.39$\pm$0.06&19.3$\pm$0.2&19.64$\pm$0.19 &20.25$\pm$0.39&20.71$\pm$0.3 \\
Lock850.052b&&26.88$\pm$0.18&25.33$\pm$0.06&24.9$\pm$0.09&21.31$\pm$0.06&19.92$\pm$0.19&20.1$\pm$0.2 &20.52$\pm$0.36&20.96$\pm$0.39\\
Lock850.053&24.21$\pm$0.06&23.92$\pm$0.06&23.47$\pm$0.06&23.19$\pm$0.06&21.58$\pm$0.06&20.96$\pm$0.21&20.9$\pm$0.18 &21.09$\pm$0.35&21.46$\pm$0.4 \\
Lock850.060&&27.24$\pm$0.24&27.77$\pm$0.49&25.99$\pm$0.24&23.44$\pm$0.35&22.42$\pm$0.37&22.04$\pm$0.33 &              &              \\
Lock850.063&&24.57$\pm$0.15&23.23$\pm$0.06&23.11$\pm$0.06&21.86$\pm$0.08&20.98$\pm$0.22&20.75$\pm$0.2 &20.51$\pm$0.36&20.47$\pm$0.38\\
Lock850.064&24.67$\pm$0.06&24.05$\pm$0.06&23.53$\pm$0.06&22.97$\pm$0.06&21.38$\pm$0.06&21.16$\pm$0.21&21.13$\pm$0.2 &21.35$\pm$0.43&21.51$\pm$0.41\\
Lock850.066&25.4$\pm$0.06&25.06$\pm$0.06&24.75$\pm$0.06&24.49$\pm$0.06&&23.15$\pm$0.52&23.1$\pm$0.55 &              &              \\
Lock850.067&28.41$\pm$0.3&28.4$\pm$0.5&26.58$\pm$0.14&25.64$\pm$0.14&&22$\pm$0.31&21.73$\pm$0.3 &22.38$\pm$0.77&22.32$\pm$0.68\\
Lock850.070&24.67$\pm$0.05&23.1$\pm$0.06&22.51$\pm$0.06&22.41$\pm$0.06&21.03$\pm$0.06&21.13$\pm$0.2&21.58$\pm$0.28 &21.28$\pm$0.32&              \\
Lock850.071&25.43$\pm$0.06&24.97$\pm$0.06&24.76$\pm$0.06&24.2$\pm$0.06&21.76$\pm$0.06&21.46$\pm$0.25&21.05$\pm$0.21 &21.3$\pm$0.35 &              \\
Lock850.073&24.35$\pm$0.06&23.63$\pm$0.06&23.14$\pm$0.06&22.67$\pm$0.06&20.65$\pm$0.06&20.33$\pm$0.2&20.22$\pm$0.18 &20.36$\pm$0.42&20.59$\pm$0.36\\
Lock850.073b&24.94$\pm$0.06&24.43$\pm$0.06&24.07$\pm$0.06&23.61$\pm$0.06&21.79$\pm$0.08&&20.73$\pm$0.19 &              &20.78$\pm$0.43\\
Lock850.075&25.57$\pm$0.07&24.76$\pm$0.06&24.23$\pm$0.06&23.78$\pm$0.06&22.13$\pm$0.11&21.21$\pm$0.21&21$\pm$0.2 &              &              \\
Lock850.076&23.23$\pm$0.06&21.03$\pm$0.06&20.22$\pm$0.06&20.05$\pm$0.06&19$\pm$0.06&19.1$\pm$0.17&19.47$\pm$0.2 &19.53$\pm$0.33&19.36$\pm$0.28\\
Lock850.077&27.02$\pm$0.13&26.5$\pm$0.13&25.86$\pm$0.09&24.81$\pm$0.08&&22.23$\pm$0.34&22.23$\pm$0.36 &              &              \\
Lock850.077b&27.59$\pm$0.12&24.97$\pm$0.06&24.62$\pm$0.06&23.44$\pm$0.06&21.14$\pm$0.06&20.12$\pm$0.2&20.03$\pm$0.16 &              &              \\
Lock850.078&24.84$\pm$0.02&24.39$\pm$0.06&24.03$\pm$0.06&23.84$\pm$0.06&22.39$\pm$0.14&& &              &              \\
Lock850.079&26.39$\pm$0.07&26.6$\pm$0.14&25.54$\pm$0.07&25.05$\pm$0.1&22.06$\pm$0.12&21.04$\pm$0.2&20.79$\pm$0.2 &20.65$\pm$0.51&20.52$\pm$0.36\\
Lock850.081&23.75$\pm$0.05&23.18$\pm$0.06&22.65$\pm$0.06&22.19$\pm$0.06&20.07$\pm$0.06&19.59$\pm$0.19&18.9$\pm$0.18 &18.02$\pm$0.2 &17.2$\pm$0.19 \\
Lock850.083&21.27$\pm$0.06&19.76$\pm$0.06&19.41$\pm$0.06&18.96$\pm$0.06&18.38$\pm$0.06&18.93$\pm$0.18&19.21$\pm$0.19 &19.81$\pm$0.21&18.61$\pm$0.2 \\
Lock850.087&26.7$\pm$0.1&25.58$\pm$0.07&24.86$\pm$0.06&24.72$\pm$0.08&21.82$\pm$0.08&20.63$\pm$0.21&20.02$\pm$0.21 &20.1$\pm$0.39 &20.21$\pm$0.26\\
Lock850.100&23.87$\pm$0.06&23.43$\pm$0.06&23.16$\pm$0.06&22.76$\pm$0.06&21.36$\pm$0.06&& &              &              \\
\end{tabular}
\caption{\label{tab:photometry_opt_ir} Optical and near-infrared AB magnitude photometry of SHADES galaxies in the Lockman Hole East.
Blank spaced indicated non-detections or lack of observations. The data are taken from Dye et al. 2008.}
\end{table*}

\begin{table*}
\begin{tabular}{llllllllll llllllllll llllllllll llllllllll}
Name        & $15\,\mu$m ($\mu$Jy)  & $24\,\mu$m  ($\mu$Jy) & $350\,\mu$m (mJy) & $850\,\mu$m (mJy) & $1.1$\,mm (mJy) & $1.2$\,mm (mJy) & $1.4\,$GHz  ($\mu$Jy) \\
Lock850.001 & 25$\pm$20   &217$\pm$16&24.1$\pm$5.5& $8.85^{+1.0}_{-1.0}$ & 4.65$_{-0.98}^{+0.97}$ &3.6$\pm$0.5 &78.9$\pm$4.7\\
Lock850.002 & 13$\pm$20   &545$\pm$31&25.3$\pm$10.3& $13.45^{+2.1}_{-2.1}$ & 6.21$_{-0.92}^{+1.05}$ &5.7$\pm$1.0 &52.4$\pm$5.6\\
Lock850.003 & 208$\pm$20  &175$\pm$23&40.5$\pm$6.5& $10.95^{+1.8}_{-1.9}$ & 3.01$_{-1.06}^{+1.04}$ &4.6$\pm$0.4 &25.8$\pm$4.9\\
Lock850.003b& 438$\pm$20  &183$\pm$33&40.5$\pm$6.5& $10.95^{+1.8}_{-1.9}$ & 3.01$_{-1.06}^{+1.04}$ &4.6$\pm$0.4 &35.0$\pm$5.2\\
Lock850.004 & 118$\pm$19  &179$\pm$68&24.9$\pm$9.1& $10.65^{+1.7}_{-1.8}$ & 4.75$_{-1.06}^{+0.89}$ &3.7$\pm$0.4 &32.0$\pm$5.1\\
Lock850.004b& 59$\pm$19   &261$\pm$73&& $10.65^{+1.7}_{-1.8}$ & 4.75$_{-1.06}^{+0.89}$ & &73.0$\pm$5.0\\
Lock850.005 & 770$\pm$20  &58.6$\pm$15.1&& $8.15^{+2.0}_{-2.1}$ && &$<$22\\
Lock850.006 & 94$\pm$20   &75.1$\pm$12.7&38.0$\pm$37.6& $6.85^{+1.3}_{-1.3}$ & 2.13$_{-1.26}^{+1.12}$ & &15$\pm$4.8\\
Lock850.007 & 35$\pm$19   &341$\pm$21&& $8.55^{+1.8}_{-1.9}$ && &42.6$\pm$5.8\\
Lock850.008 &             &481$\pm$25&& $5.45^{+1.1}_{-1.2}$ && &$<$22\\
Lock850.009 & 87$\pm$20   &466$\pm$74&& $5.95^{+1.6}_{-1.6}$ && &52.6$\pm$4.7\\
Lock850.009b& 87$\pm$20   &159$\pm$73&& $5.95^{+1.6}_{-1.6}$ & &
\\
Lock850.010 &             &
&24.1$\pm$12.0& $9.15^{+2.7}_{-2.9}$ & & &25.5$\pm$6.3\\
Lock850.010b&             &79.6$\pm$10.8& & $9.15^{+2.7}_{-2.9}$ & & &
\\
Lock850.011 & 105$\pm$20  &112$\pm$57&& $6.25^{+1.7}_{-1.8}$ & & &
\\
Lock850.012 & 9$\pm$19    &263$\pm$19&29.3$\pm$16.0& $6.15^{+1.7}_{-1.7}$ &4.1$\pm$1.3&2.6$\pm$0.4 &44.3$\pm$5.1\\
Lock850.013 & 47$\pm$20   &172$\pm$14&& $5.65^{+2.3}_{-2.9}$ & 3.07$_{-0.97}^{+0.91}$ & &$<$28\\
Lock850.014 & -45$\pm$20  &188$\pm$16&41.0$\pm$6.8& $7.25^{+1.8}_{-1.9}$ & 1.79$_{-1.57}^{+0.97}$ &3.4$\pm$0.6 &37.4$\pm$4.2\\
Lock850.015 & 299$\pm$19  &353$\pm$20&6.5$\pm$55.7& $13.25^{+4.3}_{-5.0}$ & 3.61$_{-0.98}^{+1.02}$ &4.1$\pm$0.7 &61.5$\pm$7.6\\
Lock850.015b& 299$\pm$19  &70.4$\pm$12.1&6.5$\pm$55.7& $13.25^{+4.3}_{-5.0}$ & 3.61$_{-0.98}^{+1.02}$ &4.1$\pm$0.7 &43.9$\pm$7.8\\
Lock850.016 & 321$\pm$19  &314$\pm$24&25.7$\pm$15.8& $5.85^{+1.8}_{-1.9}$ &&1.8$\pm$0.5 &106$\pm$6\\
Lock850.017 &             &239$\pm$18&& $4.75^{+1.3}_{-1.3}$ && &92.3$\pm$4.5\\
Lock850.017b&             &64.2$\pm$26.1&& $4.75^{+1.3}_{-1.3}$ && &Confused\\
Lock850.018 & 40$\pm$19   &
&7.5$\pm$6.7& $6.05^{+1.9}_{-2.1}$ &5.1$\pm$1.3&3.4$\pm$0.6 &29.4$\pm$4.4\\
Lock850.019 &             &118$\pm$15&& $5.15^{+2.0}_{-2.4}$ && &$<$27\\
Lock850.021 &             &97.9$\pm$14.1&16.7$\pm$13.2& $4.15^{+2.0}_{-2.5}$ &&1.6$\pm$0.4 &$<$30\\
Lock850.022 & -15$\pm$20  &402$\pm$21&8.7$\pm$15.6& $7.55^{+3.2}_{-4.2}$ && &$<$30\\
Lock850.023 &             &
&& $4.35^{+1.9}_{-2.4}$ && &$<$25\\
Lock850.024 & -1$\pm$30   &455$\pm$21&& $2.75^{+1.2}_{-1.2}$ && &28.5$\pm$4.8\\
Lock850.026 & -80$\pm$28  &195$\pm$16&12.2$\pm$8.8& $5.85^{+2.4}_{-2.9}$ && &31.4$\pm$5.2\\
Lock850.027 &             &106$\pm$15&3.4$\pm$5.1& $5.05^{+1.3}_{-1.3}$ &5.2$\pm$1.4&3.2$\pm$0.7 &
\\
Lock850.028 &             &252$\pm$14&23.3$\pm$11.7& $6.45^{+1.7}_{-1.8}$ && &63.0$\pm$8.2\\
Lock850.029 &             &111$\pm$14&& $6.75^{+2.0}_{-2.2}$ && &23.7$\pm$4.9\\
Lock850.030 &             &233$\pm$19&38.0$\pm$7.2& $4.75^{+1.5}_{-1.6}$ &&0.4$\pm$0.8 &245$\pm$13\\
Lock850.031 &             &467$\pm$19&& $6.05^{+1.8}_{-2.0}$ && &43.0$\pm$4.7\\
Lock850.033 & -71$\pm$20  &104$\pm$14&16.5$\pm$8.4& $3.85^{+1.0}_{-1.1}$ &&2.8$\pm$0.6 &51.0$\pm$4.3\\
Lock850.034 &             &84.9$\pm$16.7&& $14.05^{+3.1}_{-3.2}$ & 4.09$_{-0.92}^{+0.90}$ & &58.4$\pm$8.5\\
Lock850.035 & -25$\pm$20  &51.0$\pm$12.7&& $6.15^{+2.2}_{-2.4}$ && &17.4$\pm$5.0\\
Lock850.035b& -53$\pm$20  &161$\pm$14&& $6.15^{+2.2}_{-2.4}$ && &
\\
Lock850.036 &             &$<$60&& $6.35^{+1.7}_{-1.8}$ && &$<$20\\
Lock850.037 &             &
&& $7.55^{+2.9}_{-3.5}$ && &41.8$\pm$8.7\\
Lock850.037b&             &250$\pm$17&& $7.55^{+2.9}_{-3.5}$ && &14.8$\pm$5.4\\
Lock850.038 & 218$\pm$19  &260$\pm$16&& $4.35^{+2.2}_{-2.7}$ && &24.4$\pm$6.7\\
Lock850.039 &             &$<$60&& $6.55^{+2.2}_{-2.7}$ && &$<$20\\
Lock850.040 &             &91.9$\pm$15.0&& $3.05^{+1.1}_{-1.2}$ && &16.2$\pm$4.3\\
Lock850.041b& 193$\pm$20  &651$\pm$46&10.3$\pm$5.5& $3.85^{+0.9}_{-1.0}$ &4.0$\pm$1.3&2.4$\pm$0.5 &22.1$\pm$4.8\\
Lock850.041 & 464$\pm$20  &475$\pm$37&10.3$\pm$5.5& $3.85^{+0.9}_{-1.0}$ &4.0$\pm$1.3&2.4$\pm$0.5 &43.6$\pm$4.7\\
Lock850.043 & 140$\pm$19  &261$\pm$24&& $4.95^{+2.1}_{-2.6}$ & 3.19$_{-0.99}^{+1.10}$ & &25.4$\pm$5.4\\
Lock850.043b& 267$\pm$19  &456$\pm$35&& $4.95^{+2.1}_{-2.6}$ & 3.19$_{-0.99}^{+1.10}$ & &40.8$\pm$5.9\\
Lock850.047 & 37$\pm$20   &107$\pm$16&16.3$\pm$20.9& $3.55^{+1.7}_{-2.1}$ && &$<$22\\
Lock850.048 &             &203$\pm$17&16.2$\pm$13.5& $5.45^{+2.1}_{-2.5}$ &&1.6$\pm$0.4 &43.7$\pm$10.0\\
Lock850.052 &             &310$\pm$35&& $3.95^{+2.2}_{-2.7}$ & 2.51$_{-1.02}^{+1.03}$ & &38.7$\pm$8.0\\
Lock850.052b&             &561$\pm$86&& $3.95^{+2.2}_{-2.7}$ & 2.51$_{-1.02}^{+1.03}$ & &Confused\\
Lock850.053 & -26$\pm$20  &168$\pm$15&& $4.45^{+2.3}_{-2.9}$ && &$<$21\\
Lock850.060 & 4$\pm$20    &87.8$\pm$12.0&& $3.15^{+1.7}_{-2.0}$ && &
\\
Lock850.063 & 31$\pm$20   &236$\pm$17&34.9$\pm$29.1& $3.65^{+1.2}_{-1.3}$ && &22.6$\pm$4.8\\
Lock850.064 &             &425$\pm$25&11.6$\pm$12.4& $5.85^{+2.5}_{-3.2}$ &&1.7$\pm$0.4 &45.5$\pm$7.4\\
Lock850.066 &             &71.2$\pm$12.1&& $4.25^{+1.9}_{-2.2}$ && &$<$21\\
Lock850.067 & 59$\pm$20   &108$\pm$14&& $2.55^{+1.5}_{-1.5}$ && &$<$21\\
Lock850.070 & 79$\pm$20   &106$\pm$12&& $3.85^{+2.2}_{-2.5}$ && &21.9$\pm$7.2\\
Lock850.071 &             &181$\pm$20&& $3.95^{+1.8}_{-2.0}$ & 2.89$_{-1.13}^{+1.10}$ & &95.8$\pm$4.6\\
Lock850.073 & 89$\pm$36   &278$\pm$19&& $3.55^{+1.9}_{-2.3}$ && &27.3$\pm$4.8\\
Lock850.073b& 143$\pm$61  &278$\pm$19&& $3.55^{+1.9}_{-2.3}$ && &26.7$\pm$4.6\\
Lock850.075 &             &147$\pm$17&& $4.45^{+2.2}_{-2.6}$ && &
\\
Lock850.076 & 227$\pm$20  &592$\pm$26&4.4$\pm$6.7& $4.75^{+2.5}_{-3.1}$ &4.4$\pm$1.4& &48.0$\pm$6.0\\
Lock850.077 & 24$\pm$14   &51.7$\pm$13.1&41.4$\pm$24.6& $3.25^{+1.2}_{-1.3}$ && &15.5$\pm$4.4\\
Lock850.077b& 2$\pm$15    &154$\pm$15&& $3.25^{+1.2}_{-1.3}$ && &39.5$\pm$7.8\\
Lock850.078 &             &85.6$\pm$14.7&& $4.55^{+2.2}_{-2.7}$ && &$<$23\\
Lock850.079 & 41$\pm$15   &292$\pm$18&& $3.15^{+1.3}_{-1.5}$ & 2.35$_{-1.22}^{+1.12}$ & &22.41$\pm$4.5\\
Lock850.081 &             &3667$\pm$51&& $5.35^{+1.9}_{-2.3}$ && &55.2$\pm$5.3\\
Lock850.083 &             &344$\pm$25&& $3.15^{+2.0}_{-2.1}$ && &$<$28\\
Lock850.087 &             &399$\pm$22&& $3.45^{+1.5}_{-1.7}$ && &84.5$\pm$5.3\\
Lock850.100 &             &118$\pm$13&& $11.25^{+4.2}_{-5.3}$ && &19.8$\pm$6.3\\
\end{tabular}
\caption{\label{tab:photometry_fir} Mid-infrared to radio photometry of SHADES galaxies in the Lockman Hole East. 
The $15\,\mu$m data are presented for the first time here, and the remaining
data are taken from Coppin et al. 2006, Austermann et al. 2009, Ivison et al. 2007, Coppin et al. 2008 and Negrello et al. 2009.}
\end{table*}

\begin{table*}
\begin{tabular}{lllllllllllll}
Name & $z_{\rm phot}$ & $\delta z_{\rm phot,min}$ & $\delta z_{\rm phot,max}$ & $\chi^2_\nu$ & $N_{\rm DOF}$ & Dust   & Age & $\Theta$ & $\theta_{\rm view}$ & $f_{\rm AGN}$ & Notes\\
    LOCK850.001 &  5.30 &  1.10 &  0.46 &   0.98 &   8 & SMC &   50 &    1.6 &    35  & 0.73 & \\
    LOCK850.002 &  3.68 &  0.34 &  0.24 &   4.15 &   8 & LMC &   50 &    1.2 &     5  & 0.77 & I\\ 
    LOCK850.003 &  3.80 &  0.22 &  0.20 &   6.67 &   8 & SMC &   50 &    2.0 &     0  & 0.70 & \\ 
   LOCK850.003B &  6.56 &  1.14 &  0.44 &   7.15 &   8 &  MW &  600 &    2.2 &    35  & 0.98 & D,C \\ 
    LOCK850.004 &  3.76 &  0.56 &  2.90 &   2.77 &   9 & SMC &   50 &    1.6 &    45  & 0.08 & U \\ 
   LOCK850.004B &  6.54 &  1.88 &  0.46 &   3.54 &   7 & SMC &  400 &    1.8 &    45  & 0.42 & U \\ 
    LOCK850.005 &  6.14 &  0.60 &  0.44 &   10.02&   6 & SMC &   10 &    1.2 &    30  & 0.86 & C \\
    LOCK850.006 &  5.58 &  1.94 &  1.41 &   3.31 &   8 & SMC &   70 &    2.0 &    42.5& 0.20 & I \\
    LOCK850.007 &  2.56 &  0.60 &  4.44 &   0    &  -3 & SMC &   50 &    0.3 &    10  & 0.18 & I \\
    LOCK850.008 &  0.22 &  0.06 &  6.78 &   0    &  -4 & SMC &  200 &    0.3 &     0  & 0.00 & I \\
    LOCK850.009 &  2.30 &  0.54 &  0.70 &   2.36 &   6 & SMC &  400 &    1.4 &    32.5& 0.56 & U,X \\
   LOCK850.009B &  1.18 &  0.46 &  0.42 &   2.10 &   8 &  MW &  600 &    1.8 &     0  & 0.00 & U\\
   LOCK850.010B &  3.28 &  0.68 &  0.52 &   3.05 &   5 & SMC &   50 &    1.2 &    90 & 0.00\\
    LOCK850.011 &  1.24 &  0.32 &  0.40 &   2.38 &   6 &  MW &  500 &    1.4 &    15 & 0.55 & X \\
    LOCK850.012 &  2.26 &  0.36 &  1.45 &   2.03 &   7 & SMC &  200 &    1.6 &     0 & 0.37 & I \\
    LOCK850.013 &  0.50 &  0.26 &  0.20 &   2.36 &   7 & SMC &  600 &    2.0 &    7.5& 0.09 & X \\
    LOCK850.014 &  3.06 &  0.44 &  0.36 &   3.72 &   7 & SMC &   50 &    1.8 &    50 & 0.01\\
    LOCK850.015 &  1.40 &  0.84 &  0.40 &   3.79 &   8 & SMC &  400 &    0.3 &    27.5& 0.11\\
   LOCK850.015B &  1.36 &  0.18 &  0.16 &   2.54 &   8 & SMC &  500 &    0.4 &    7.5& 0.32 & D\\ 
    LOCK850.016 &  6.58 &  0.54 &  0.42 &   5.22 &   7 & SMC &   10 &    1.2 &    40 & 0.44\\
    LOCK850.017 &  3.48 &  1.18 &  0.34 &   2.19 &   5 & SMC &  100 &    2.6 &    7.5& 0.50 & U\\
   LOCK850.017B &  4.20 &  0.68 &  1.98 &   1.72 &   5 & LMC &   50 &    1.4 &     0 & 0.65 & U\\
    LOCK850.019 &  3.94 &  0.24 &  1.92 &   1.11 &   5 & SMC &   50 &    2.0 &     5 & 0.69 & I \\
    LOCK850.021 &  0.86 &  0.36 &  6.14 &   0.91 &   6 & SMC &  500 &    0.4 &    17.5 & 0.15 & I \\ 
    LOCK850.022 &  3.10 &  0.90 &  0.40 &   2.24 &   6 &  MW &  600 &    5.0 &     0 & 0.96 \\
    LOCK850.024 &  2.72 &  0.70 &  0.56 &   2.26 &   5 & SMC &   30 &    1.8 &     0 & 0.24\\
    LOCK850.026 &  3.48 &  1.24 &  0.46 &   3.23 &   6 & SMC &   30 &    1.6 &     0 & 0.74 & I\\
    LOCK850.027 &  1.34 &  0.18 &  0.34 &   2.97 &   6 &  MW &  600 &    2.2 &     0 & 0.66\\
    LOCK850.028 &  3.66 &  0.58 &  0.20 &   1.63 &   6 & SMC &   50 &    2.2 &    7.5& 0.50\\
    LOCK850.029 &  4.44 &  3.38 &  2.56 &   0.23 &   5 & SMC &  100 &    1.8 &     0 & 0.96 & I \\
    LOCK850.030 &  2.70 &  1.04 &  0.53 &   2.03 &   6 & SMC &   50 &    0.9 &    50 & 0.00\\
    LOCK850.031 &  1.12 &  0.50 &  1.40 &   2.08 &   5 & SMC &  400 &    0.7 &   17.5& 0.46 & X \\
    LOCK850.033 &  2.78 &  0.66 &  0.70 &   4.75 &   6 & LMC &   70 &    1.2 &    50 & 0.00\\
    LOCK850.034 &  4.14 &  0.45 &  2.02 &   3.19 &   6 & LMC &   50 &    1.2 &    90 & 0.00\\
    LOCK850.035 &  5.24 &  3.26 &  1.76 &   0.72 &   5 & SMC &   70 &    2.0 &    7.5& 0.94 & I\\
   LOCK850.035B &  2.20 &  0.28 &  0.78 &   4.53 &   5 &  MW &  600 &    2.2 &     0 & 0.66 & D\\
    LOCK850.036 &  0.92 &  0.54 &  6.08 &   0    &  -4 & SMC &  200 &    0.3 &     0 & 0.00 & I\\
   LOCK850.037B &  1.26 &  0.46 &  1.00 &   1.84 &   5 & SMC &  300 &    1.4 &   22.5& 0.38 & X \\
    LOCK850.038 &  1.36 &  0.30 &  0.30 &   1.11 &   6 &  MW &  300 &    2.4 &    20 & 0.30\\
    LOCK850.039 &  0.92 &  0.74 &  6.08 &   0    &  -4 & SMC &  200 &    0.3 &     0 & 0.00 & I\\
    LOCK850.040 &  2.24 &  1.64 &  0.34 &   2.02 &   5 &  MW &  400 &    1.8 &     0 & 0.10 & X \\
   LOCK850.041B &  2.44 &  0.94 &  1.38 &   3.38 &   6 & LMC &   50 &    2.2 &    40 & 0.31 \\ 
    LOCK850.041 &  1.60 &  0.44 &  0.10 &   1.61 &   7 & LMC &  200 &    2.4 &    7.5& 0.31\\
    LOCK850.043 &  1.38 &  0.70 &  0.66 &   2.03 &   7 & SMC &  200 &    1.6 &   27.5& 0.42 & U,X \\
   LOCK850.043B &  1.60 &  0.58 &  0.40 &   2.21 &   7 & SMC &  100 &    2.0 &     0 & 0.22 & U,X \\  
    LOCK850.047 &  3.82 &  0.20 &  0.40 &   1.26 &   7 & LMC &   30 &    3.0 &   37.5& 0.40\\
    LOCK850.048 &  4.18 &  0.86 &  2.16 &   1.70 &   6 & SMC &   10 &    1.6 &    90 & 0.15 \\
    LOCK850.052 &  1.42 &  0.32 &  0.76 &   3.77 &   6 & SMC &  600 &    1.0 &     0 & 0.73 & U\\
   LOCK850.052B &  3.64 &  0.68 &  0.58 &   5.73 &   5 & SMC &  600 &    1.4 &     5 & 0.97 & U\\
    LOCK850.053 &  2.22 &  0.40 &  0.64 &   2.25 &   5 &  MW &  600 &    2.4 &     0 & 0.64 & X\\
    LOCK850.060 &  2.46 &  0.90 &  2.90 &   2.27 &   6 &  MW &  600 &    1.2 &     0 & 0.64 \\
    LOCK850.063 &  4.98 &  0.88 &  0.88 &   0.42 &   7 &  MW &  600 &    2.6 &   32.5& 0.96 & I\\
    LOCK850.064 &  2.10 &  1.41 &  1.28 &   0.98 &   6 & LMC &   50 &    2.2 &     0 & 0.27 & I \\ 
    LOCK850.066 &  3.72 &  0.00 &  1.30 &   0.61 &   5 & SMC &   30 &    0.7 &    50 & 0.03\\
    LOCK850.067 &  7.00 &  0.72 &  0.00 &   3.33 &   6 &  MW &  400 &    3.0 &    40 & 0.83\\
    LOCK850.070 &  0.62 &  0.22 &  0.36 &   0.61 &   6 &  MW &  500 &    1.8 &   22.5& 0.05 & X \\
    LOCK850.071 &  2.24 &  1.48 &  0.88 &   1.91 &   6 &  MW &  600 &    1.8 &     0 & 0.67 & X \\
    LOCK850.073 &  1.72 &  0.96 &  1.88 &   0.64 &   6 & LMC &  100 &    2.0 &    50 & 0.00\\
   LOCK850.073B &  2.26 &  1.52 &  1.30 &   1.17 &   4 &  MW &  400 &    2.4 &   37.5& 0.45 & D \\
    LOCK850.075 &  1.28 &  0.00 &  2.54 &   1.30 &   3 &  MW &  400 &    2.2 &    45 & 0.42 \\
    LOCK850.076 &  0.48 &  0.14 &  0.12 &   2.22 &   8 & SMC &  600 &    1.0 &    7.5& 0.06 & X?\\
    LOCK850.077 &  3.20 &  2.62 &  0.48 &   2.77 &   7 & SMC &   50 &    1.2 &    40 & 0.00 & U\\
   LOCK850.077B &  3.74 &  0.32 &  0.34 &   5.01 &   4 & LMC &  500 &    1.6 &    7.5& 0.69 & U\\
    LOCK850.078 &  3.38 &  0.00 &  2.64 &   0.57 &   1 & SMC &   50 &    1.4 &    50 & 0.02\\
    LOCK850.079 &  1.96 &  0.24 &  0.22 &   2.50 &   7 &  MW &  600 &    1.2 &   22.5& 0.57\\
    LOCK850.081 &  3.18 &  1.04 &  0.92 &   0.83 &   5 & SMC &  400 &    2.0 &   42.5& 0.84 & I\\
    LOCK850.083 &  0.44 &  0.08 &  0.06 &   1.89 &   7 & LMC &  600 &    1.4 &     0 & 0.00\\
    LOCK850.087 &  2.54 &  0.82 &  0.82 &   0.50 &   5 & SMC &  300 &    1.4 &     0 & 0.57\\
    LOCK850.100 &  3.66 &  3.40 &  2.60 &   1.85 &   1 & SMC &  200 &    1.6 &    55 & 0.08\\
\end{tabular}
\caption{\label{tab:sed_fitting_results} Results of the SED fitting, with the parameters defined in the text.
All parameters are dimensionless except the age which is in Myr. In the final column, 
``C'' notes possibly contaminated flux at $15\,\mu$m, 
``D'' notes a depreciated identification, 
``I'' insufficient data to determine the AGN bolometric fraction, 
``U'' an uncertain identification, 
and ``X'' a far-infrared excess.}
\end{table*}


\section{Discussion and conclusions}

Following the methodology of Serjeant et al. (2008), we can estimate the contribution that clustering of the mid-infrared population
makes to the submm and mm-wave stacks, by integrating the correlation function $w(\theta)$ with respect to angle $\theta$:
\begin{equation}\label{eqn:clustering}
S = I\int_0^\infty w(\theta) \, B(\theta) \, 2\pi \theta \, {\rm d}\theta
\end{equation}
where $B$ is the submm or mm-wave beam and $I$ is the total background intensity. We argued in Serjeant et al. (2008) that
the clustered contributions at $450\,\mu$m and $850\,\mu$m can be neglected. At $1100\,\mu$m equation \ref{eqn:clustering} gives
$S=0.12(\theta_0/{\rm arcsec})^{0.8}$ for an assumed $w(\theta)= (\theta/\theta_0)^{-0.8}$. 
The value of $\theta_0$ is not well-determined for the mid-infrared population at these
flux densities (typically $\stackrel{>}{_\sim}200\,\mu$Jy), but 
Oliver et al. (2004) measured $\theta_0=1.24''$ for the faintest $3.6\,\mu$m-selected populations
(note that nearly all $15\,\mu$m-selected galaxies are detectable
at $3.6\,\mu$m with Spitzer in even the deepest $15\,\mu$m exposures, e.g. Hopwood et al. 2009). 
To avoid introducing model-dependences in our
quoted results, we opt not to correct for clustering in the stacked flux numbers quoted here. 
Nevertheless, the total background contributed by the $15\,\mu$m population in this paper is considerably less than the integrated
background, so the clustered contribution estimated by this methodology is a very conservative upper limit. 

The observed $1100\,\mu$m stacked signal of $\langle S_{{1100\,\mu}\rm m}\rangle = (0.148\pm 0.052)\,$mJy at a $15\,\mu$m source
combined with our observed surface density of $4950\pm170$ $15\,\mu$m galaxies per square degree (Poisson errors)
corresponds to a $1100\,\mu$m background contribution of $(0.73\pm 0.26)$\,Jy\,deg$^{-2}$. The estimated total background at $1100\,\mu$m 
is around $18$\,Jy\,deg$^{-2}$ (e.g. Gispert et al. 2000). 
Given the contribution from clustering, we can set an upper limit of $4\%$ to the contributions from
these $15\,\mu$m-selected galaxies to the $1100\,\mu$m extragalactic background light. 
Our $15\,\mu$m-selected sample is not as deep as the $24\,\mu$m-selected
population used by Serjeant et al. (2008), but following the same methodology as we have applied at $1100\,\mu$m, we estimate
background contributions of $(1.02\pm0.42)$\,Jy\,deg$^{-2}$ at $850\,\mu$m and $18.3\pm5.4$\,Jy\,deg$^{-2}$ and $450\,\mu$m, 
corresponding to $(3\pm 1)\%$ of the $850\,\mu$m background and $(11\pm3)\%$ of the $450\,\mu$m background. This is consistent
with our results in Serjeant et al. (2008), implying the populations which dominate the ultra-deep confusion-limited maps
expected with the SCUBA-2 instrument will overlap substantially with populations already detected with AKARI and Spitzer. 

The small background contribution at $1100\,\mu$m is in stark contrast to our earlier
results for the submm backgrounds (Serjeant et al. 2008) 
in which we found the majority of the $450\,\mu$m background is attributable to the 
$24\,\mu$m-selected population, and to claims from the Balloon-Borne Large Area Submm Telescope (BLAST) data 
that the mid-infrared population is responsible for all the $250-500\,\mu$m
background (Devlin et al. 2009, Marsden et al. 2009). 
Taken in combination with the results in Serjeant et al. (2008), our 
results suggest strongly that other populations undetected by AKARI or Spitzer must dominate the $850\,\mu$m and mm-wave background.

Not all our sample has sufficient mid-infrared data to distinguish starburst and AGN mid-infrared contributions, but table 
\ref{tab:sed_fitting_results} gives the results where available. In total $41$ SHADES galaxies have sufficient data for this
constraint, not counting multiple identifications. 
Of these, we find that our models cannot reproduce the far-infrared luminosity in 
10 unambiguous identifications (LOCK850.013, 031, 037B, 038, 040, 053, 060, 070, 071, 075) 
and in two further ambiguous identifications (LOCK850.009, 043). 
One possibility is that the spectral energy distributions have
an additional cool cirrus component not accounted for in the models, which would further reduce an AGN bolometric contribution. 
Another possibility is that the identifications are wrong, and the submm emission comes from objects with larger far-infrared to optical/near-infrared
luminosity ratios. Of the remaining (apparently) unambiguously-identified objects, 12 have AGN fractions $>0.3$ 
(LOCK850.001, 003, 016, 022, 027, 028, 041, 047, 052, 067, 079, 087), 
while 14 have starburst fractions $>0.7$ (LOCK850.010B, 014, 015, 024, 030, 033, 034, 048, 066, 073, 076, 078, 083, 100). 
For LOCK850.004 and LOCK850.077 the
two candidate identifications result in one starburst  ($f_{\rm AGN}<0.3$) and one AGN ($f_{\rm AGN}>0.3$) interpretation for each galaxy. 
Finally two further cases merit individual attention. In LOCK850.060, the addition of the AGN component created a worse fit in the submm;
we opt to attribute greater weight to this long-wavelength data point and include this galaxy among the starburst. In LOCK850.076 there is 
additional mm-wave flux, which we attribute to errors in the deboosting the submm or mm-wave data 
in this individual object; this galaxy is also counted among the starbursts. 

In summary therefore, considering only those galaxies in which sufficient data is available, 
we have 12 galaxies with an unknown far-infrared excess relative to our models, 12 with AGN fractions $f_{\rm AGN}>0.3$, 
16 starburst-dominated ($f_{\rm AGN}<0.3$). Note that we previously found
that AGN bolometric fractions above $0.3$ cannot be reliably measured in this broad-band fitting (Negrello et al. 2009). If we exclude
the poorly-fit SEDs, we find that $(43\pm9)\%$ of our submm-selected galaxies have $f_{\rm AGN}>0.3$; alternatively, if we
conservatively include the galaxies with far-infrared excesses as starburst-dominated, we find
$(30\pm7)\%$ have AGN bolometric contributions above this threshold. We therefore treat the $1\sigma$ range to 
be $(23-52)\%$. Pope et al. (2008) used mid-infrared spectroscopy of submm-selected
galaxies and found that a fraction $(15\pm10)\%$ (i.e. $2/13$ galaxies
in their sample) had an active nucleus which contributes significantly to the mid-infrared. This is slightly lower than
found in our study, though only marginally inconsistent given the small number statistics in both studies. 

Eight of our SHADES galaxies have been identified as having anomalous $610-1400$\,MHz spectral indices (Ibar et al. in preparation):
LOCK850.001, 015, 018, 022, 024, 033, 040, 087. Of these, LOCK850.001, 024 and 033 have flat radio spectra ($-{\rm d}\ln S_\nu /{\rm d}\ln\nu<0.5$), while
the remainder have unusually steep spectra, consistent with synchrotron ageing of electrons in the radio lobes from active nuclei. 
We have identified three of these eight securely as having AGN bolometric contributions
$f_{\rm AGN}>0.3$, and three with $f_{\rm AGN}<0.3$. This suggests that evidence of AGN in the radio is not necessarily an indicator
of bolometrically-dominant active nuclei.

The high star formation rates (several hundred $M_\odot$/year, e.g. Hughes et al. 1998) and low number density of submm-selected
galaxies (e.g. Scott et al. 2006, Coppin et al. 2006) suggest that submm galaxies are short-lived. Indeed the steep number 
counts of submm-selected galaxies necessarily imples that we sample submm galaxies at
around their peak phase of far-infrared luminosity. Even a slowly-varying monochromatic luminosity will induce this effect: in
the models of Takagi et al. (2003, 2004), the submm luminosity varies by around a factor of $2-3$ over a timescale of $\sim 200$\,Myr. 
Denoting this for simplicity as $S_\nu(t)\propto e^{-0.5(\tau/\sigma)^2}$, where $\tau$ measures the secular evolution and 
$\sigma\simeq (200/2.35)$\,Myr is the 
timescale of the variation, 
we find that a power-law differential source counts of $dN/dS\propto S^{-\alpha}$ gives an observed age distribution in a 
flux-limited sample of 
\begin{equation}
{\rm Pr}(\tau)\propto e^{-\left (\alpha+\frac{1}{2}\right )\left(\frac{\tau}{\sigma}\right)^2} 
    = e^{-\frac{1}{2}\left (\frac{\tau}{\sigma'}\right )^2}
\end{equation}
where $(\sigma')^2=\sigma^2/(2\alpha+1)$. This is a far shorter timescale than the intrinsic variation, given the
observed counts slope of $\alpha\simeq3.5-4$ (Coppin et al. 2006). 

There have been suggestions from numerical simulations (e.g. di Matteo et al. 2005) that feedback from active nuclei is capable of
truncating star formation activity. However quasars also have star-forming hosts (eg Serjeant \& Hatziminaoglou 2009) which
may pose challenges for models in which star formation is truncated too abruptly by active nuclei (e.g. Narayanan et al. 2009). 
Therefore we suggest that the energy input from active nuclei
through quasar-mode or radio-mode feedback does not immediately truncate star formation,
but rather suppresses it on the same timescales as the quasar lifetime itself. We may also find that star-forming far-infrared-luminous
populations selected at shorter wavelengths than SHADES have warmer colour temperatures (e.g. Blain et al. 2003) and higher AGN bolometric
contributions. In this interpretation, these would be later phases in the co-evolution of active nucleus and starburst. This would also
be consistent with the observation that the $K$-band Hubble diagram of hyperluminous starbursts is tight when the hyperluminous 
galaxies are selected at $60\,\mu$m (Serjeant et al. 2003b), but has a much higher dispersion when the hyperluminous galaxies are 
selected at submm-wavelengths (e.g. Smail et al. 2004). 

Finally, our models predict submm-selected galaxies have 
fluxes of $\simeq 1-10$\,mJy at an observed frame wavelength of $70\,\mu$m. This was beyond the capabilities
of both AKARI and Spitzer for direct detections, but is consistent with the Spitzer stacking analysis detection 
by Dye et al. (2007) of $(3.63\pm 0.77)$\,mJy. Direct detections at this depth may be accessible to Herschel at shorter wavelengths, and
would probe the hot dust components in submm-selected galaxies, currently not strongly constrained. 

\begin{acknowledgements} This research is based on observations with AKARI, a JAXA project with the participation of ESA. 
This work was funded in part by STFC (grant PP/D002400/1), the Royal Society (2006/R4-IJP) and the Sasakawa Foundation (3108).
JSD thanks the Royal Society for a Wolfson Research Merit Award. 
MI was supported by the Korea Science and Engineering Foundation(KOSEF) grant
No. 2009-0063616, funded by the Korea government(MEST).
SK were supported by the Basic Science Research Programme through the National Research Foundation of Korea (NRF)
funded by the Ministry of Education, Science and Technology 2009-0066892.
We thank the anonymous referee for helpful comments. 
 \end{acknowledgements}


\begin{thebibliography}{}


\bibitem[]{}Alexander, D., et al., 2005, ApJ, 632, 736 
\bibitem[]{}Austermann, J., et al., 2009, MNRAS, in press
\bibitem[]{}Aretxaga, I., et al., 1998, MNRAS, 296, 643
\bibitem[]{}Aretxaga, I., et al., 2007, MNRAS, 379, 1571
\bibitem[]{}Barger, A.J., et al., 1998, Nature, 394, 248
\bibitem[]{}Baugh, C., et al., 2005, MNRAS, 356, 1191
\bibitem[]{}Bertoldi, F., \& Cox, P., 2002, A\&A, 384, L11
\bibitem[]{}Bertoldi, F., et al., 2003, A\&A, 406, L55
\bibitem[]{}Blain, A.W., Barnard, V.E., Chapman, S.C., 2003, MNRAS, 338, 733
\bibitem[]{}Brotherton, M.S., et al., 1999, ApJL, 520, 87
\bibitem[]{}Carilli, C.L., et al., 2001, ApJ, 555, 625
\bibitem[]{}Clements, D., et al., 2008, MNRAS, 387, 247 
\bibitem[]{}Coppin, K., et al., 2004, MNRAS, 354, 193 
\bibitem[]{}Coppin, K., et al., 2006, MNRAS, 372, 1621
\bibitem[]{}Coppin, K., et al., 2008, MNRAS, 384, 1597
\bibitem[]{}Devlin, M.J., et al., 2009, Nature, 458, 737
\bibitem[]{}Dey, A., et al., 2008, ApJ, 677, 943 
\bibitem[]{}di Matteo, T., Springel, V., Hernquist, L., 2005, Nature, 433, 604
\bibitem[]{}Dye, S., et al., 2007, MNRAS, 375, 725
\bibitem[]{}Dye, S., et al., 2008, MNRAS, 386, 1107
\bibitem[]{}Eddington, A.S., 1913, MNRAS, 73, 359
\bibitem[]{}Efstathiou, A., \& Rowan-Robinson M., 1995, MNRAS, 273, 649
\bibitem[]{}Elbaz, D., et al., 1999, A\&A, 351, L37
\bibitem[]{}Fazio, G.G., et al., 2004, ApJS, 154, 10 
\bibitem[]{}Ferrarese, L., Merritt, S., 2000, ApJ, 539, L9 
\bibitem[]{}Genzel, R., \& Cezarsky, C., 2005, ARA\&A, 38, 761 
\bibitem[]{}Gispert, R., Lagache, G., Puget, J.L., 2000, A\&A,360, 1 
\bibitem[]{}Granato, L., et al., 2006, MNRAS, 368, L72 
\bibitem[]{}Greve, T.R., et al., 2009, preprint (arXiv:0904.0028)
\bibitem[]{}Hopwood, R.H., et al., 2009, ApJL, submitted 
\bibitem[]{}Hughes, D.H., Dunlop, J.S., Rawlings, S., 1997, MNRAS, 289, 766
\bibitem[]{}Hughes, D.H., et al., 1998, Nature, 394, 241 
\bibitem[]{}Isaak, K.G., et al., 2002, MNRAS, 329, 149
\bibitem[]{}Ivison, R.J., et al., 2007, MNRAS, 380, 199 
\bibitem[]{}Lacey, C.G., et al., 2008, MNRAS, 385, 1155
\bibitem[]{}Lehnert, M.D., et al., 1992, ApJ, 393, 68
\bibitem[]{}Lutz, D., et al., 2008, ApJ, 684, 853
\bibitem[]{}Magorrian, J., et al., 1998, AJ, 115, 2285
\bibitem[]{}Marsden, G., et al., 2009, preprint (arXiv:0904.1205)
\bibitem[]{}Mart\'{i}nez-Sansigre, A., et al., 2005, Nature, 436, 666
\bibitem[]{}Men\'{e}ndez-Delmestre, K., et al., 2009, ApJ, 699, 667
\bibitem[]{}Mortier, A.M.J., et al., 2005, MNRAS, 363, 563
\bibitem[]{}Mullaney, J.R., et al., 2009, preprint (arXiv:0909.3842)
\bibitem[]{}Narayanan, D., et al., 2009, preprint (arXiv:0904.0004)
\bibitem[]{}Negrello, M., et al., 2009, MNRAS, 394, 375
\bibitem[]{}Oliver, S., et al., 2002, MNRAS, 332, 536
\bibitem[]{}Omont, A., Cox, P., Bertoldi, F., McMahon, R.G., Carilli, C., Isaak, K.G., 2001, A\&A, 374, 371
\bibitem[]{}Omont, A., et al., 2003, A\&A, 398, 857
\bibitem[]{}Peacock, J.A., 1999, Cosmological Physics, Cambridge University Press 
\bibitem[]{}Peacock, J.A., et al., 2000, MNRAS, 318, 535
\bibitem[]{}Pearson, C.P.., 2005, MNRAS, 358, 1417
\bibitem[]{}Pearson, C.P., et al., 2009, A\& A, submitted 
\bibitem[]{}Pope, A., et al., 2008, ApJ, 675, 1171
\bibitem[]{}Priddey, R.S., et al., 2003a, MNRAS, 339, 1183
\bibitem[]{}Priddey, R.S., et al., 2003b, MNRAS, 344, L74
\bibitem[]{}Rieke, G., et al., 2004, ApJS, 154, 25
\bibitem[]{}Rodighiero,  G., Lari, C., Fadda, D., Franceschini, A.,
  Elbaz, D., Cesarsky, C., 2004, A\&A, 427, 773
\bibitem[]{}Scott, S.E., et al., 2006, MNRAS, 370, 1057
\bibitem[]{}Scott, K.S., et al., 2008, MNRAS, 385, 2225
\bibitem[]{}Serjeant, S., et al., 2003, MNRAS, 344, 887
\bibitem[]{}Serjeant, S., et al., 2003b, MNRAS, 346, L51
\bibitem[]{}Serjeant, S., et al., 2004, ApJS, 154, 118 
\bibitem[]{}Serjeant, S., et al., 2008, MNRAS, 386, 1907
\bibitem[]{}Serjeant, S., \& Hatziminaoglou, E., 2009, MNRAS, 397, 265
\bibitem[]{}Shi, Y., et al., 2009, preprint (arXiv:0908.0952)
\bibitem[]{}Smail, I., Ivison, R.J., Blain, A.W., 1997, ApJ, 490, L5
\bibitem[]{}Smail, I., et al., 2004, ApJ, 616, 71
\bibitem[]{}Takagi T., Arimoto N., Hanami H., 2003, MNRAS, 340, 813 
\bibitem[]{}Takagi T., Hanami H., Arimoto N., 2004, MNRAS, 355, 424 
\bibitem[]{}Takagi, T., et al., 2007, MNRAS, 381, 1154
\bibitem[]{}Veilleux, S., et al., 2009, ApJS, 182, 628
\bibitem[]{}Wang, W.-H., Barger, A.J., Cowie, L.L., 2007, ApJ, 690, 319 
\bibitem[]{}Willott, C., Rawlings, S., Grimes, J.A., 2003, ApJ, 598, 909
\bibitem[]{}Yun, M., et al., 2008, MNRAS, 389, 333

%
%
%
%
%
%
%
%
%
%


\end{thebibliography}
\end{document}